\documentclass[a4paper, 12pt, twoside]{book}
\pdfoutput=1
\usepackage{varioref}
\usepackage{graphicx}
\usepackage[pagewise,switch]{lineno}
\usepackage{color}
\usepackage{chngcntr}

\usepackage{subfigure}
\usepackage{fancyhdr}
\usepackage{verbatim}
\usepackage{pdflscape}
\makeatletter
\def\tagform@#1{\maketag@@@{[\ignorespaces#1\unskip\@@italiccorr]}}
\makeatother

\usepackage[printonlyused]{acronym}
\usepackage{epstopdf}

\usepackage{ctable}
\usepackage[super]{nth}
\usepackage[latin1]{inputenc}
\usepackage[numbers,square]{natbib}
\usepackage[T1]{fontenc}
\usepackage{yfonts}
\usepackage{hepparticles}
\usepackage{hepnames}
\usepackage{hepunits}
\let\HepTo\undefined
\usepackage{array}
\usepackage[english]{babel}
\usepackage{epsfig}
\usepackage{latexsym}
\usepackage{mathptmx}

\newcommand*\fermi{\textit{Fermi}}

\newcolumntype{s}{>{\footnotesize}{c}}

\graphicspath{{./figures/},{./immagini/}}

\newcommand{\HepTo}{\to}
\newcommand{\micro}{\mycro}
\renewcommand{\sectionmark}[1]{%
\markboth{\thesection.\ #1}{}}
\title{\LARGE A review of Cosmic-ray electrons and \fermi-LAT}
\author{\Large Marco Tinivella \\ email m.tinivella@gmail.com}
\date{\today}

\makeatletter
\@addtoreset{footnote}{section}
\makeatother

\usepackage[colorlinks=false, urlcolor=black,linkcolor=black]{hyperref}
\usepackage[english]{cleveref}
\usepackage{amssymb}
\usepackage{amsmath}

\let\originaleqref=\eqref
\renewcommand{\eqref}{\originaleqref}
\begin{document}
\maketitle
{\large \bfseries Abstract}

The study of Galactic \acfp{cre} saw important developments in recent years, with the assumption of \APelectron production only in interaction of hadronic Cosmic-rays with interstellar matter challenged by new measurements of \ac{cre} spectrum and related quantities. Indeed, all recent experiments seem to  confirm an hardening in the \APelectron, a feature that is totally in contrast with the all-secondaries hypothesis, even if significant disagreements are present about the \ac{cre} spectral behavior  and the possible presence of spectral features. Together with insufficient precision of current measurements, these disagreements prevent the identification of the primary \APelectron source, with models involving Dark matter or astrophysical sources like \acfp{snr} and \acfp{pwn} all able to explain current data. 

The \fermi-LAT contribution to the \ac{cre} study was fundamental, with the 2009 measurement of the \APelectron $+$\Pelectron spectrum extended to the 7~MeV -- 1~TeV range with a statistics already exceeding previous results by many order of magnitude; since then, the last statistic has largely increased, while the LAT event reconstruction was significantly improved.

In this article the reader will find an  extensive and historical review of the \ac{cre} science, a summary of the history of gamma astronomy before  \fermi, an accurate description of the LAT and of its data analysis, a review of the present knowledge about \ac{cre} spectrum and on the theories that try to explain it, and finally a description of the changes and improvements introduced in the LAT event reconstruction process at the beginning of 2015.

\tableofcontents 
\chapter*{Introduction\markboth{Introduction}{}}
\addcontentsline{toc}{chapter}{Introduction}
The study of  \acfp{cre} is  fundamental for the knowledge of the \acf{cr} population of the Galaxy, of their sources and of the way they travel to reach Earth.

The study of Galactic \acp{cre} experienced  important progress in recent years,  with new experiments (\fermi-LAT, ATIC, HESS, AMS-2)  providing measurements of the \Pelectron and \APelectron spectra, with statistics and precision never reached before. Also,  spectra of particles that are strictly related with   \Pelectron and \APelectron spectra, like light nuclei (B,Be,Li)  and \APproton saw new and significantly more precise measurements. 

These measurements  led to a very important change in the Galactic \ac{cre} paradigm, that assumed the  production of \APelectron only in  interaction of hadronic \acp{cr} with the \acf{ism}. In this scenario, the constraints posed on the hadronic \acp{cr} with the \acf{ism} by the measurements of secondary species like light nuclei or \APproton led to a predicted fraction of \APelectron that decreases with energy, while all recent measurements point out that this fraction, and also the same \APelectron spectrum, increase with energy. For this reason, the  existence of sources of primary \APelectron is  now commonly accepted, while  their nature is uncertain. 
Models involving \acf{dm} decay or annihilation, acceleration of secondary particles in \acfp{snr} and \APelectron\Pelectron acceleration in \acfp{pwn} are all able to reproduce actual data. Furthermore,  there is disagreement about the spectral behavior and on the presence of features in the \APelectron $+$ \Pelectron spectrum, and further measurements are required to solve this problem.

 In 2009, \fermi-LAT measured the $\Pelectron + \Ppositron$ spectrum between 7~GeV and 1~TeV, exploring an almost unknown energy range and exploiting the large \ac{lat} effective area to obtain the first $\Pelectron + \Ppositron$ measurement not dominated by statistical uncertainties. The \acf{lat}, on board of the \fermi\ observatory, is a pair-conversion telescope designed to reconstruct electro-magnetic events (\Pphoton but also \Pepm) from \-100~MeV to $>300~\GeV$, with a sensitivity and an effective area that exceed those of  previous instruments like EGRET by more than one order of magnitude.
With 8 years of data, and exploiting the updated event reconstruction process released in 2015, \fermi-LAT can still play a fundamental role in solving some of the \ac{cre} questions. 

The purpose of this work is to give a review  of \acp{cre} science, including updated theories about their production,  recent measurements of \ac{cre} spectrum and  related quantities. A specific focus is made on the \fermi-LAT measurement, with a large  review of the instrument technical characteristic, of the data-reconstruction process and of the new event-reconstruction process released in 2015. 

This review is organized as follow:

\S~\ref{cr-chapter} is an extensive review of the Galactic \ac{cr} and \ac{cre}, describing their diffusion in the Galaxy with focus on the peculiarities of \acp{cre} diffusion like their radiative energy losses, the \ac{snr} paradigm that identifies \acp{snr} as the most probable source of \acp{cr}, the production of secondary \APelectron via \acp{cr} interactions with \ac{ism} and the most probable sources of primary \APelectron;

 \S~\ref{storia-gamma} presents a brief review of the evolution of the space \Pphoton-ray astronomy, including an overview of the most important experiments before \fermi\ and their major results;

\S~\ref{capitolo-fermi} contains the description of the \fermi\ experiment, the \acf{lat}, its technical characteristics and performances, the event reconstruction process and the major updates introduced since operation started in August 2008;

 \S~\ref{CRE-fermi} contains a detailed review of  the 2009 $\Pelectron + \Ppositron$ measurement made by \fermi-LAT, focusing both on the experimental technique and on the result;

\S~\ref{interpretazioni-spettro} compares the measurement made by \fermi-LAT with similar measurements made by recent experiments and describes the current status of \ac{cre} studies with possible future improvements;
 
  \S~\ref{pass8} describes the new Pass~8 event reconstruction, focusing on the most important improvements introduced and their impact on the \ac{lat} performances.

\chapter{Cosmic-Ray Electrons}\label{cr-chapter}

\section{Cosmic Rays}
\acfp{cr} are energetic particles consisting mainly of protons ($\sim87 \%$) and alpha particles (helium nuclei, $\sim9 \%$), the rest being  electrons ($\sim2 \%$), heavy ions ($\sim1 \%$), positrons and anti-protons. They were discovered in 1912 by Victor Hess,  during observations conducted on flying balloon up to heights of 5 \kilo\meter. Hess noticed~(\citet{hess-cr}) that the flux of ionizing particles, observed on ground, after an initial decrease  increased with altitude above 100~\meter, therefore revealing an extra-terrestrial origin. In the same year, Domenico Pacini reached the same conclusion~(\citet{pacini}) noticing that the flux of charged particles in water of the lake of Bracciano and in front of the harbor of Livorno was not related with the distance from the  bottom of the water,  suggesting that ground radio-activity was not the source of the observed radiation. In 1930 Bruno Rossi hypothesized that (\citet{effetto-e-w1}), if \acp{cr} were charged particles, a dependence of the \ac{cr} flux on the Earth magnetic field would be observed, with particles of a specific sign prevented from coming from specific directions because of the presence of the Earth in their trajectory. In 1933-1935, the observation of this ``East-West'' effect  (\citet{effetto-e-w2,east-west1,east-west2,east-west4,east-west5}), a difference in the flux coming from these directions, proved  that \acp{cr} are mainly composed by positively charged particles.

\begin{figure}[htb!]
\begin{center}
\includegraphics[width=.9\textwidth]{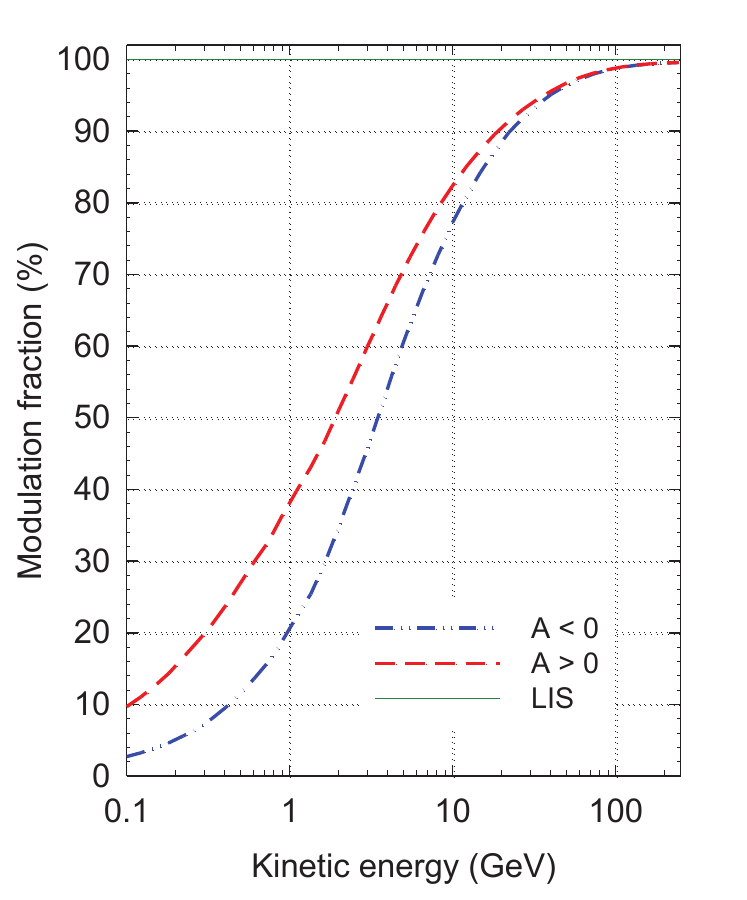}
\caption{Fraction of protons entering the heliosphere that reaches the Earth, from the simulation performed in~\cite{modulazione}. Solid green line (100\%) is the assumed Local Interstellar Spectrum (LIS), namely the spectrum of \ac{cr} beyond the heliopause. Calculation performed in condition of minimum solar activity, A is the polarity of the solar magnetic field, that reverts on a 21~\yr\ cycle.  }
	\label{solar-modulation}

\end{center}
\end{figure}

\ac{cr} spectrum has been observed from $\sim$ 10 --100~\MeV/nucleon, where the spectrum is strongly distorted  by the solar wind, up to energies of $10^{20}$~\eV. At low energies,  the spectrum is modified by the energy loss caused by the interaction of particles with the solar wind. Furthermore, the \ac{cr} flux at low energy is influenced by the 11~\yr\ solar cycle, during which the Sun significantly changes its activity, therefore modifying the minimum energy needed by a particle to reach the Earth.
An example of the effect of  solar modulation on \acp{cr} is presented in figure~\ref{solar-modulation}, showing  the fraction of protons that, after passing the heliopause, reaches  Earth. The figure is taken from~\citet{modulazione}, who present a possible simulation of the interaction of \acp{cr} with solar wind and solar magnetic field. Another numerical code, named HELIOPROP, to simulate the propagation of charged particles in the heliosphere, is presented in~\citet{helioprop}.

 Above some \GeV/nucleon, when the effect of solar wind  becomes small, the spectrum follows a power law $dN=N_0\cdot E^{-k}\cdot dE$, with an index $k\simeq 2.7$ -- 2.8 until $3\times 10^{15}$~\eV. 

\label{discrepanza-gev-tev}Recent measurements from CREAM~\cite{protoni-cream}, PAMELA~\cite{pamela-p-and-he} and ATIC~\cite{atic-nuclei} have shown an hardening of the \ac{cr} spectrum above 250~GeV/nucleon -- 1~TeV/nucleon, hardening that decreases for increasing particle charge. Approximately at $10^{15}$~\eV\ (the ``knee") the spectrum slope changes from 2.7 to $\sim 3$, while around  $\sim 5\times 10^{18}$~\eV\ Auger, TA,Yakutsk report that  the spectrum flattens again at $k\sim2.7$ (the ``ankle"~\citet{cr-tutti,cr-news}). Recently, a steepening of the spectrum from $\sim3$ to $\sim 3.3$ (``second knee") was discovered at $\sim  4\times10^{17}$~\eV~(\citet{Lipari}). Finally, above $\sim  6\times10^{19}$~\eV\ a strong steepening of the spectrum was reported in~\citet{gzk,gzk2,cr-tutti}, confirming (even if other explanations are still possible~\cite{cr-news,gzk-forse}) the prediction  of a flux suppression caused by the interaction of protons with the \ac{cmb} via \Ppi production \HepProcess{\Pp+\Pphoton \HepTo \Pp+\Ppizero} or \HepProcess{\Pp+\Pphoton \HepTo\Pn+\Ppiplus}~(\citet{longair,Lipari,gzk3,gzk4}) and by the photo-disintegration of nuclei impacting on \ac{cmb} photons (\citet{fotodisintegrazione,fotodisintegrazione2}).
 
\begin{figure}[htb!]
\begin{center}
\includegraphics[width=.9\textwidth]{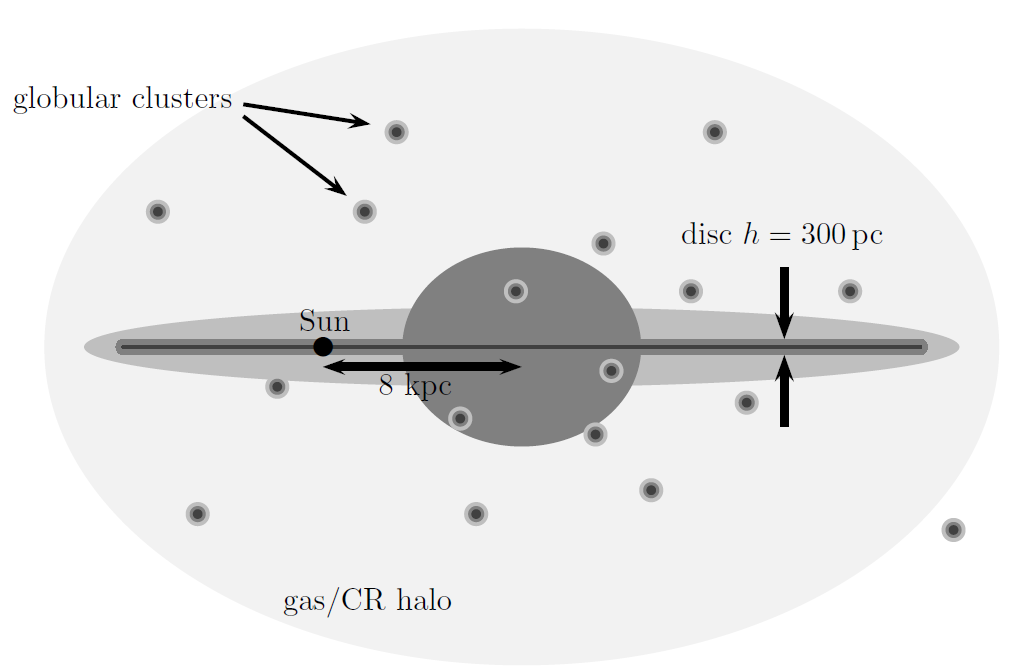}
\caption{Schematic representation of the galaxy from \cite{Kachelriess}: a thin gas and dust disc with a central bulge is surrounded by a halo of gas and \ac{cr},  that also contains globular clusters of stars.}
	\label{galassiaCR}

\end{center}
\end{figure}

 When they travel in the Galaxy, \acp{cr} interact with the matter and with the magnetic field of the Milky Way. Because of these interactions their motion is almost a random walk (until they have very high energy): because this walk is different for particles of different charge and energy,   the final  energetic distribution and also  composition can be very different with respect to the injection ones. Figure \ref{galassiaCR} shows a schematic representation of the Milky Way, that is composed by a disc of $300~\pc$ of heights  and 15~\kilo\pc\ of radius, surrounded by a halo extending to  10~--15~\kilo\pc. Stars are concentrated in a central bulge and in the disc, that is filled with gas with average density $n\sim1$~particle~\cm\rpcubed, while the matter density in halo is $n\sim0.01~\cm\rpcubed$. The galactic disc contains a magnetic field, oriented following the arms of the galaxy's spiral, of $\sim 3~\mycro\gauss$; all the halo is filled by a turbulent magnetic field of $\sim$ 3 -- $4~\mycro\gauss$ that is incoherent on distance $>55$ --$150~\pc~$(\citet[page.~81]{vietri},\citet{B-galattico}). If we simply assume that \acp{cr} rotate in the Galaxy under the effect of the Lorentz force (so ignoring the diffusion processes that have small impact only on high energy \ac{cr}) the Larmor radius $R_L$ of a particle of momentum p and charge Ze  is 

 \begin{equation} 
 R_L=\frac{cp}{ZeB} \sim 100\,\pc\frac{3\mycro\gauss}{B}\, \frac{E}{Z\times 10^{18}\eV}
 \end{equation} 
When energy is so high that $R_L$ becomes comparable with the width of the disk, that happens  between the ankle and the second knee, the motion of particles starts to be rectilinear and not diffusive anymore; in the energy range between the ankle and the second ankle it is also expected to be the maximum energy of \ac{cr} of Galactic origin, above which    \acp{cr} of extra-galactic origin become dominant.

The sources of \acp{cr}  are still unclear, because \acp{cr} are deflected by inhomogeneities in the galactic magnetic field, therefore they assume  a strongly isotropic distribution that contains almost no information about their original direction. However, theoretical models  strongly suggest that \acp{cr} below $\sim 10^{15}~\eV$ are accelerated in \acfp{snr}, by interaction of the expanding shock generated by a  supernova with the surrounding medium. This hypothesis is supported by the observation of many \acp{snr}  in the X-ray and Gamma-ray band, with the detection  of the synchrotron X emission of high energy  electrons~(\citet{snr-review}),  the detection of high energetic gamma-rays~(\citet{elettroniHess}) emitted via \ac{ic} of electrons or via \Ppizero production in hadron scattering, and recently by the detection by \fermi\ of the characteristic signature of the \Ppizero production~(\citet{fermipi0}).

 The main source of \acp{cr} in the galaxy are probably \acp{snr}, nevertheless the \ac{cr} population could have other contributions, like the acceleration of particles in the strong electromagnetic fields of pulsars, or the decay or annihilation of dark matter particles.

The study of the  \aclp{cre}\footnote{In this work, the term electrons will refer to the sum of electrons and positrons components.} is very important because electrons can be produced in sources different from that of other \acp{cr} and their diffusion in the \ac{ism} is strongly different from that of hadrons. For these reasons the study of \acp{cre} can provide very valuable information about the \ac{ism}, the diffusion of other \ac{cr} species through it and about the production of Gamma-rays through the interactions between \acp{cr} and \ac{ism}.

This chapter will describe the diffusion of \acp{cr} in the galaxy (section~\ref{CRE-diffusione}), focusing on the energy loss processes that affect \acp{cre}, making their spectrum very different from other \acp{cr}. Section~\ref{CRE-sorgenti} will depict the potential sources of \acp{cre}.

\section{Diffusion of Cosmic Rays}
\label{CRE-diffusione}

\subsection{Propagation of Cosmic Rays}\label{CR-propag}
The interaction of \acp{cr} with the \ac{ism} produces secondary particles that change the composition of the \ac{cr} population. Of particular interest for the study of \acp{cr} is the production of anti-particles (\APelectron and \APproton),  of   light elements  (Li, Be, B are very rare, because stars quickly consume these elements via sub-branches of the proton-proton cycle, see~\citet{castellani}), of rare isotopes ($^3$He is quickly burned to $^4$He in stars) and of radioactive isotopes. The last three groups of elements are typically produced via spallation processes, during which the primary nucleus is transformed in a lighter secondary  one without significantly losses in energy/nucleon. If we describe the decrease in particles number caused by interaction with \ac{ism} with an interaction length $\lambda$ (expressed in unit of traversed matter, $\gram~ \centi\meter\rpsquared$), we can write the production of secondary particles via spallation process

\begin{eqnarray}
\frac{dn_p}{dX}&=& -\frac{n_p}{\lambda_p}\label{primary-spallation}\\
\frac{dn_s}{dX}&=& -\frac{n_s}{\lambda_s} + \frac{p_{sp}n_p}{\lambda_p} \label{secondary-spallation}\\
\end{eqnarray}

In the previous equations, in which X is the amount of traversed matter ($\gram~\centi\meter\rpsquared$), a flux of primary particles $n_p$ decreases for the interaction with \ac{ism} (equation~\eqref{primary-spallation}); a fraction of the interacting primaries ($p_{sp}=\sigma_{sp}/\sigma_{tot}$, $\sigma$ are the cross sections of spallation production and of all involved processes respectively)  is transformed in the analyzed secondary species (equation~\eqref{secondary-spallation}). In this equation we have used the Feynman scaling hypothesis (see e.g. \citet{scaling1,scaling2}), that states that when the rest-mass frame energy of colliding particles is significantly higher than their masses, the inelastic cross section of nuclear reactions is almost energy-independent and the energy distribution of produced particle is also energy independent. Analyzing elements like Li,Be,B, that are almost absent in the production sites (so $n_s(0)\simeq 0$) and therefore mainly produced via spallation of C-N-O elements, we can solve the equation, obtaining 

\begin{equation} \frac{n_s}{n_p}=\frac{p_{sp}\lambda_s}{\lambda_s-\lambda_p}\left[\mathrm{exp}\left(\frac{X}{\lambda_p}-\frac{X}{\lambda_s}\right)-1\right]
 \end{equation}  
Inserting the observed secondary/primary ratio $\sim 0.25$ and the measures made at accelerators of $\lambda_{CNO}\sim 6.7~\gram~\centi\meter\rpsquared$, $\lambda_{LiBeB}\sim 10~\gram~\centi\meter\rpsquared$ and $p_{sp}\sim0.35$~(\citet{Kachelriess}), the traversed matter result $X\sim 4.3~\gram~\centi\meter\rpsquared$ or $d\sim1.3~\mega\pc$ in the disk of the Milky Way (1~\centi\meter\rpcubed density) or $\tau_{CR}\sim 5\times 10^6 \yr$ of residence time in the galaxy. Because the galactic disk is only $\approx 200 $-- $300~\pc$ thick, the only possible explanation to this result is that the propagation of \acp{cr} in the Galaxy is  a diffusive random-walk.

Furthermore, the confinement time allows to calculate the luminosity $L_{CR}$ of the sour\-ces that inject \ac{cr} in the Galaxy~(\citet{Kachelriess}):

\begin{equation}
\label{LumCr}
L_{CR}=\frac{\rho_{CR}Vol}{\tau_{CR}} \sim 5\times 10^{40}\erg~\reciprocal\second
\end{equation}
for an observed energy density of \acp{cr} $\rho_{CR}\sim 1~\eV~\centi\meter\rpcubed$, a disk of height$\sim 200~\pc$ and 15~\kilo\pc\ of radius.

These last results use the over-simplified assumption that \acp{cr} are confined in the Galactic plane. From measurement of the abundances of radioactive secondaries like $^{10}$Be (lifetime 1.51$\times10^6~\yr$, see table I and II of~\citet{referenze-berillio} for a list of measures of the $^{10}$Be/$^{9}$Be ratio), that are produced through spallation processes with well known cross-sections, it is possible to directly  determine $\tau$, obtaining values in the range 2 -- 3 $\times 10^7~\yr$~(\citet{giuseppe}), or 1.5 -- 6$\times 10^7~\yr$ when considering more complex diffusion models~(\citet{lifetime-cr}). These ranges are not compatible with a propagation occurring only in the galactic plane: in typical models,  \acp{cr} are confined in a roughly cylindrical region above and below the galactic plane, with height varying between 3 and 7~\kilo\pc, because of the uncertainties on the radioactive secondaries abundances.\label{berillio}

Finally, we have to consider  that the spectrum of secondary particles is significantly steeper than that of primaries, $\propto E^{-3.3}$ instead of  $\propto E^{-2.7}$. This observation is in complete agreement with the hypothesis that \acp{cr} are confined in the Galaxy by the galactic magnetic field, therefore is expected that with increasing energy \acp{cr} confinement time should decrease and less secondary particles should be produced. 

\subsection{Diffusion of Cosmic Rays}\label{cr-diffusion-section}
The diffusion of \acp{cr} is described by the \textit{transport equation}~(e.g. \citet{CR-propag})

\begin{equation}
\label{diffeqCR}
 \frac{\partial \psi(p,t,\mathbf{r})}{\partial t} = D_{xx}\nabla^2\psi - \frac{\partial}{\partial p}(\dot p \psi) + \mathcal{Q}(E,t,\mathbf{r}) + \frac{\partial}{\partial p}p^2 D_{pp}\frac{\partial}{\partial p}\frac{1}{p^2}\psi -\frac{1}{\tau_f}\psi  -\frac{1}{\tau_d}\psi 
\end{equation}
where the left term  is the variation of the number of particles of momentum $p$ in a volume around the position $\mathbf{r}$ at the time t, $D_{xx}\nabla^2\psi$ represents the spatial diffusion of particles assuming that the diffusion coefficient $D_{xx}$ is constant in space , $\frac{\partial}{\partial p}(\dot p \psi)$ represents the effect of energy losses (that will be treated specifically for electrons in section~\ref{eloss-section}), $\mathcal{Q}(E,t,\mathbf{r})$ describes the injection of particles, either by primary sources or from process of spallation/decay etc, $D_{pp}$ describes the effect of diffusive re-acceleration, $\tau_f$ is the nuclear fragmentation timescale and $\tau_d$ is the radioactive decay timescale. For simplicity we have neglected terms related to convection of particles by flows of material, because they are not relevant for this analysis.

The spatial diffusion of particles is caused by the interaction with random \acp{mhw}, that arise in magnetized plasma in response to perturbations. A fundamental quantity to describe these interactions is the magnetic giro-radius of a particle with momentum $p$ and charge $Ze$ in a magnetic field of strength B: $R_g=pc/ZeB$ from which we derive the \textit{particle magnetic rigidity} $\mathcal{R}=pc/Ze$. Particles with the same $\mathcal{R}$ will have the same behaviour in a magnetic field. From the definition of $\mathcal{R}$ it follows that at high energies, when $E\gg m$, the rigidity of \Pepm and \Pproton or \APproton is the same, and therefore their behaviour in a magnetic field will be the same (except for the  radiative energy losses).

Particles principally interact with \acp{mhw} with wavelength $k\sim 1/R_g \mu$, where $\mu$ is the pitch angle of the particle with respect to the wave. On microscopic level, particles diffuse principally along the field lines, therefore creating strong local  anisotropies. However, the galactic magnetic field has a random component that can be larger than the regular magnetic field and that fluctuates on $L\sim 100~\pc$ scale: these large fluctuations lead to the complete isotropization of the particles.

The diffusion coefficient $D_{xx}$ can be calculated~(\citet{CR-propag}), resulting in

\begin{equation} 
D_{xx}\simeq  \left(\frac{\delta B_{res}}{B}\right)^{-2}\frac{vR_g}{3}
\end{equation}
where $B_{res}$ is the resonant component ($k=1/R_g$) of the magnetic field fluctuations and $v$ the particle velocity. Typically, $\delta B\simeq 5~\mycro\gauss$ and  the spectrum of \acp{mhw} can be described as a power-law 

\begin{equation}\label{mhw-spettro} W_k = \frac{wB^2L}{4\pi(1-a)}(kL)^{-2+a},\; kL\geq 1,\; a=constant 
\end{equation} 
where w describes the level of turbulence and L is the principal scale of the turbulence~(\citet{CR-propag,riaccelerazione}). The resulting diffusion coefficient is $D_{xx}\simeq2\times 10^{27}v/c \mathcal{R}_{GV}^a~\squaren\cm~\second$, where the magnetic rigidity $\mathcal{R}$ is expressed in \giga\volt\ and $a$, from equation \eqref{mhw-spettro}, is typically comprised between 0.3 and 0.6, depending on different spectral model for magnetic turbulence. \citet{trotta-D0} suggested a slightly higher value, with $D(4~\giga\volt)=8\times 10^{28}~\squaren\cm~\second$ instead of $3\times 10^{27}$.
This result is in good agreement with the result of  \S~\ref{CR-propag} about the dependence of the confining time of particles on energy. In a pure diffusive model, $\tau \propto 1/D_{xx}$: the observed ratio between the primary and secondary particle index suggested $\tau\propto E^{-0.6}$, that is consistent with the calculated energy dependence for $D_{xx}$. \citet{dibernardo}, using the ratio between different \ac{cr} species (B/C,N/O,C/O) and the \APproton/\Pproton ratio, claim that values in the higher part of the range are more probable.  Recently, \citet{min-max-diffusion-coeff} calculated from the B/C ratio that the diffusion coefficient should be comprised in the 0.46 -- 0.85 range, with the most probable value at 0.70, slightly higher than previous calculations.

The described model is valid until the particle giro-radius $R_g$ is smaller than the large  scale fluctuation distance of the galactic magnetic field (100~\pc), verified for particles with magnetic rigidity $\mathcal{R}\geqslant 10^{17}~\volt$. Above this threshold, resonant wavelengths of  the spectrum of \acp{mhw} are suppressed and the deflection caused by the random component of the magnetic field becomes small, $\delta \sim (L/R_g)$. 
 This results, for random scattering ($<\delta> = 0$), in $<\delta^2>\sim  N(L/R_g)^2$, with N number of suffered deflection from the magnetic field, that is one every L traveled distance. The free mean path $(l_0)$ of a random walk is the (average) distance traveled by a particle before its direction is completely changed by the random interaction suffered: in this picture, if \textit{r} is the traveled distance after N interactions, $<r^2>\sim Nl_0^2$. Because in a random walk the mean traveled distance is $\propto\sqrt{Dt}$ where D is the diffusion coefficient and t the elapsed time, connecting the two pictures we obtain~(\citet{Kachelriess}) $D\sim Nl_0^2/t=vl_0$, where v is the particle velocity. A more precise analysis of a 3-dimensional random walk leads to $D=l_0v/3$. By definition, $l_0$ is the distance after which $<\delta^2>\sim1$: inserting this result in the diffusion coefficient, the result leads to $D(E)\propto R_g^2 \propto E^2$. Therefore, above energies of $\sim10^{17}\eV/nucleon$ there is an abrupt decrease of the confinement time of \acp{cr} and also of their isotropy.
 
 Recently, \citet{self-diffusion} proposed that the scattering of \acp{cr} on \acp{mhw} could lead to significant excitation of streaming instability. This self-generated turbulence increases the diffusion coefficient, that for energy below 1~\TeV/n is  almost dominated by this process.
 
This model seems able to explain some  features in the spectrum of \acp{cr} and in the primary-to-secondary ratios, without the necessity of introducing \textit{ad hoc} breaks in the energy dependence of the diffusion coefficient and/or in the injection spectrum of \acp{cr}. 

\subsubsection{Re-acceleration}\label{riaccelerazione}
\paragraph{Diffusive re-acceleration}
Diffusive re-acceleration is caused by stochastic acceleration of particles interacting with \acp{mhw} and is described in the transport equation~\eqref{diffeqCR} by diffusion in the momentum space with a coefficient $D_{pp}$~(\cite{CR-propag}). This coefficient can be related to the diffusion coefficient, obtaining  $D_{pp}=p^2V_a^2/9D_{xx}$ where $V_a$ is the Alfv\`en velocity, the characteristic velocity of propagation of weak disturbances of the magnetic field~(\cite{CR-propag,riaccelerazione}) and $p$ the momentum of the particle. The effect of the re-acceleration is to increase the escape time of high energy particles, that remain for more time confined in the galaxy; in other word, the index of the energy dependence of the diffusion coefficient $D_{xx}\propto E^a$ is decreased. 

The increase of the particle escape time at high energies leads to an increase of the relative abundance of secondary particles up to energies of 30~GeV/nucleon. \citet{riaccelerazione} and~\citet{riaccelerazione2} show that diffusive models with weak re-accelera\-tion are completely compatible with the observed abundances of secondaries, while high re-acceleration models produce an excess of secondaries, both hadronic or leptonic~ (~\citet{modello-con-radio,Pamela-3D-modello}) at low energy. Furthermore, the inclusion of re-acce\-leration in diffusive models allows to describe the diffusion coefficient as a simple rigidity-depen\-dent power law $D\propto \mathcal{R}^a$ without the necessity of introducing a spectral break to describe    secondary to primary ratios below 1~\giga\volt\  rigidity. In agreement with these results, \citet{riaccelerazione-E} calculated the energetic impact of re-acceleration on \ac{cr} energetic, obtaining a re-acceleration time-scale.

\begin{equation}
\tau_{Re}=\frac{E_{CR}}{P_{Re}}=\frac{9 D_{xx}}{4 V_a^2}
\end{equation} 
where $P_{Re}$ is the power transferred from the interstellar turbulence to \acp{cr}. When inserting reasonable values of $D_{xx}$ and $V_a$, the impact or re-acceleration at 4~\giga\volt\ results of the order of 1 -- 10\%, rapidly increasing at lower energy.

\paragraph{Re-acceleration at weak shocks}
Recently, in an effort to explain the spectral discrepancy between GeV and TeV energies mentioned at pag~\pageref{discrepanza-gev-tev}, \citet{shock-reacceleration}  suggested that in addition to diffusive re-acceleration, \acp{cr} can also suffer re-acceleration when interacting with \ac{snr} shocks, through the same process that initially produced them. Because the probability of encountering a shock during the diffusion in the Galaxy is dependent on the extension of the shock, this process would be dominated by old \acp{snr}, whose shocks are significantly more extended and weaker than those of young \acp{snr},  where \acp{cr} are probably produced. 

The calculation performed in~\cite{shock-reacceleration} shows that this process could be dominating up to some TeV, and its effect would be a  steepening in the spectrum of \Pproton, He and light nuclei, while it should be negligible for heavier nuclei and for \Pepm.

\section{Electron energy losses}\label{eloss-section}
Among the \ac{cr} species, the lepton component is subjected to unique processes  during the diffusion in the Galaxy: the  particular features that result from these processes make the \acp{cre}  study  an independent way of testing models of \acp{cr} diffusion. Because of their low mass, \acp{cre} are subject to important electromagnetic energy losses  that cause highly energetic particles to rapidly loose their energy. The result of this higher energy loss rate is that \acp{cre} have a steeper spectrum, $\sim E^{-3.0}\,-\,E^{-3.1}$  (this is true for energies above some \GeV: below, energy dependence of the diffusion coefficient becomes very low, and the spectrum  turns to  $\sim E^{-1.5}$ -- $E^{-1.2}$, \cite{eliopausa,modello-con-radio} ) compared to $E^{-2.73}$ for protons, and that the distance that they can travel through \ac{ism} is strongly reduced with respect to other \ac{cr} species. 

The energy losses for  an electron of energy E  are:
\begin{equation}\label{Eloss-all}  -\frac{dE}{dt}=A_{Ion}\cdot\ln{E}+A_{Brem}\cdot E+A_{IC+Sync}\cdot E^2
\end{equation}
The first term describes ionization losses in \ac{ism} and  is dominant for energies up to a few tens of \MeV. The second term is due to  bremsstrahlung, adiabatic losses and pair production in \Pelectron-\Pphoton interaction, while the last term represents losses by synchrotron emission and \acf{ic} under the Thomson approximation, that holds very well for electrons up to a few \TeV\ of energy. The last term is the dominating factor for energy greater than a few \GeV, resulting therefore the only interesting term in the present \ac{cre} analysis.
Equation \eqref{Eloss-all} can be therefore approximated as

\begin{equation}\label{Eloss-simple}\frac{dE}{dt}=-kE^2\end{equation}
The \textit{k} term includes the contribution from \ac{ic} of electrons on photons of the \ac{cmb},  on photons of the \ac{isrf} and infrared photons and the emission of synchrotron radiation from electrons in the galactic magnetic field, that can be seen as the interaction of electrons with the virtual photons of the magnetic field.
All the processes contribute to k with a term 

 \begin{equation}
k_i=\frac{4}{3}\sigma_{Th}c\omega_{ph,i}\left(\frac{1}{m_e c^2}\right)^2
\end{equation} 
where $\sigma_{Th}$ is the Thomson cross section, $\omega_{ph,i}$ is the energy density of the interacting radiation field  (the energy density of the magnetic field is $\frac{B^2}{8\pi}$). Estimated value of  \textit{k} is $\sim(1.4\pm0.2)10^{-16}\reciprocal\GeV\reciprocal\second $. Inserting the numerical values of the constants, the energy loss of \acp{cre} is: 

\begin{equation}\label{Eloss} \frac{dE}{dt}=-8\cdot 10^{-17}\left(\omega_{ph}+\frac{B^2}{8\pi}\right)E^2~\GeV~\reciprocal\second
\end{equation} 
with the energy density of the photon fields expressed in \eV~\cm\rpcubed. Both $\omega_{ph}$ and $\frac{B^2}{8\pi}$ are $\sim 1~\eV~\cm\rpcubed$.
Solving equation \eqref{Eloss-simple}, the energy E of an electron starting with energy $E_0$ at time $t=0$ is 

\begin{equation}
E(t)=\frac{E_0}{(1+kE_0t)}
\end{equation}
We can define the \textit{radiative lifetime} 

\begin{equation}
 \tau(E_0)=\frac{1}{kE_0}
\end{equation}
so that after a time $t\gg\tau$ an electron has lost almost all of its initial energy ($E_0\gg E$).

The ubiquitous presence of the \ac{cmb} in the whole Universe establishes a maximum limit to the electron lifetime: in equation \eqref{Eloss-simple} k cannot vanish, but it has a minimum, $k\geq \frac{4}{3}\sigma_{Th}c\omega_{CMB}\left(\frac{1}{m_e c^2}\right)^2$. That results, for a 100~\GeV\ electron, in a  maximum lifetime in the universe $\tau_{extragal} \lesssim 10^7~\yr$. If we insert in equation \eqref{Eloss} the typical values of the galactic magnetic and photon fields, that are both $\sim1~\eV~\cm\rpcubed$ 

\begin{equation}
\tau_{gal}(E)\sim 3\times 10^5\left(\frac{1\,\TeV}{E}\right)~\yr
\end{equation}
For energy above $\sim 10~\GeV$, this value is significantly shorter than the  escape time caused by diffusion, $\tau_{esc}\propto 1/D$,  calculated in \S~\ref{cr-diffusion-section}.\cite{Blasi}

 The average distance \textit{d} traveled  in a time $\tau$ by a particle moving on a random walk is 
 
\begin{equation}
 d=\sqrt{2D_0\tau}
\end{equation}

If we insert as the diffusion coefficient $D_0$ a simple constant with  a value near the typical value at 1~\TeV, $D_0=10^{29}~\squaren\cm~\reciprocal\second$, we obtain 

\begin{equation}
d=3\times 10^2\left(\frac{1\,\TeV}{E}\right)^{\frac{1}{2}}~\pc
\end{equation}

The correct calculation, performed using an energy dependent $D_0$ (that is treated in \S~\ref{cr-diffusion-section}), does not change significantly the result: high energy electrons observed on Earth are accelerated in sources with distances $d\lesssim 600~\pc$ and emitted less than $10^5~\yr$ ago~(\citet{longair}). 

\section{Spectrum of Cosmic-ray electrons}
Equation \eqref{diffeqCR} that describes the diffusion of \acp{cr} in the galaxy, can be approximated for electrons as
\label{diffusion-CRE}

\begin{equation}
\label{diffeq}
 \frac{\partial N_e(E,t,\mathbf{r})}{\partial t} = D(E)\nabla^2N_e + \frac{\partial}{\partial E}(b(E)N_e) + \mathcal{Q}(E,t,\mathbf{r})
\end{equation}
where terms describing decaying or fragmentation of particles are removed and the term describing re-acceleration is not included for simplicity (in section~\ref{cr-diffusion-section} was shown that the main effect of re-acceleration in the analyzed energy range is a flattening of the energy dependence of the diffusion coefficient). 
The source term $\mathcal{Q}(E,t,\mathbf{r})$, that describes the injection of electrons in the \ac{ism},  takes account of all the sources in the galaxy, including the production of secondary electrons and positrons by \ac{cr} interaction with \ac{ism}. The principal sources of HE electrons will be described in section~\ref{CRE-sorgenti}. 

Eq~\ref{diffeq} does not include the effect of  solar modulation on the electron flux, which can not be neglected below 30~\GeV\ and which is dominant below 5~\GeV.  The solar wind strongly reduces the low-energy electrons flux: \citet{eliopausa} gives an estimate of this effect, combining the recent PAMELA measurement with data from Voyager~1, which recently measured the 6 -- 100~\MeV\ electron flux   at a distance of 119~AU from the Sun, very close to the heliopause. \citet{modello-con-radio} made an estimation of the \ac{cre} spectrum in the galaxy using the diffuse radio synchrotron emission, obtaining similar results. 

A typical source of primary electrons (\acf{snr}, pulsar) can be represented by

\begin{equation}
 \mathcal{Q}_i=\mathcal{Q}_0\cdot E^{-\gamma}e^{-E/E_{cut}}\delta(t-t_0)\delta(\mathbf{r}-\mathbf{r_0})
\end{equation}
This equation describes a source emitting electrons with a power-law spectrum of index $\gamma$ up to a maximum cut-off energy $E_{cut}$. The source is burst-like, that is the emission time is much shorter than the typical lifetime of the emitted electrons, $t_s\ll\tau_{e}$ and the emission can be considered as instantaneous. Finally, the source is point-like. In the case of \acp{snr}, the sources are typically assumed to be  uniformly  distributed  in the galactic plane and regularly recurrent in time ($\mathcal{Q}(E,t,\mathbf{r})=\mathcal{Q}(E)\delta(z)$, where the galactic plane is supposed to be infinitely thin with respect to the galactic diffusion halo). When inserting in the diffusion equation \eqref{diffeq} a source term that assume sources continuously distributed in the galactic plane, and requiring that electrons are confined for a long time in the magnetic halo of the galaxy before being dispersed in the outer space, with $z_{halo}\gg z_{disk}$, it is possible to solve analytically the diffusion equation~(\citet{giuseppe}). The resulting \ac{cre} spectrum in the solar system position is dependent by the relation between the radiative average free path $\lambda(E)$ and the scale quantities $ z_{halo}, z_{disk}$. For electron in the energy range 5\GeV--5\TeV, where $z_d\ll \lambda(E)\ll z_h$, the resulting electron spectrum is

\begin{equation}
 N_e(E)\varpropto E^{-\left(\gamma_{inj} +\frac{1}{2}+\frac{\delta}{2}\right)}
\end{equation}
where $\gamma_{inj}$ is the index of the source spectrum and $\delta$ is the index of the energy dependence of the diffusion coefficient $D(E)\propto E^\delta$.

However, the short propagation distance of \acp{cre} leads to inhomogeneities in the electrons distribution that have to be accounted for.  A standard approach till recent times  (used for example in the GALPROP code~\cite{galprop}) was to add to the source distribution a radial dependence, to be estimated from observations, therefore obtaining $\mathcal{Q}(E,t,\mathbf{r})=\mathcal{Q}(E)\delta(z)\rho(r)$). More recently,  the code DRAGON~(\citet{dragon2}) was developed to simulate the \ac{cre} flux originated by sources distributed in the galactic spiral arms.   In the case of a pulsar contribution to the source term, the  source term can be specified as the sum of singularly computed  pulsar contributions. The electron flux on Earth could have a dominant contribution from the emission of a few near ($d<1~\kilo\pc$)  pulsars of medium age ($3\times10^6~\yr>T>5\times10^4~\yr$, see section~\ref{section-pulsar} for the reason of this choice). Because there are only two known objects that satisfy this prescription, Geminga (PSR J0633+1746) and Monogem (PSR B0656+14),  the resultant spectrum could show a potentially observable anisotropy,however  searches for it (for example~\citet{cre-anisotropia}) have until now shown no results.

\begin{figure}[htb!]
\begin{center}
\includegraphics[width=.9\textwidth]{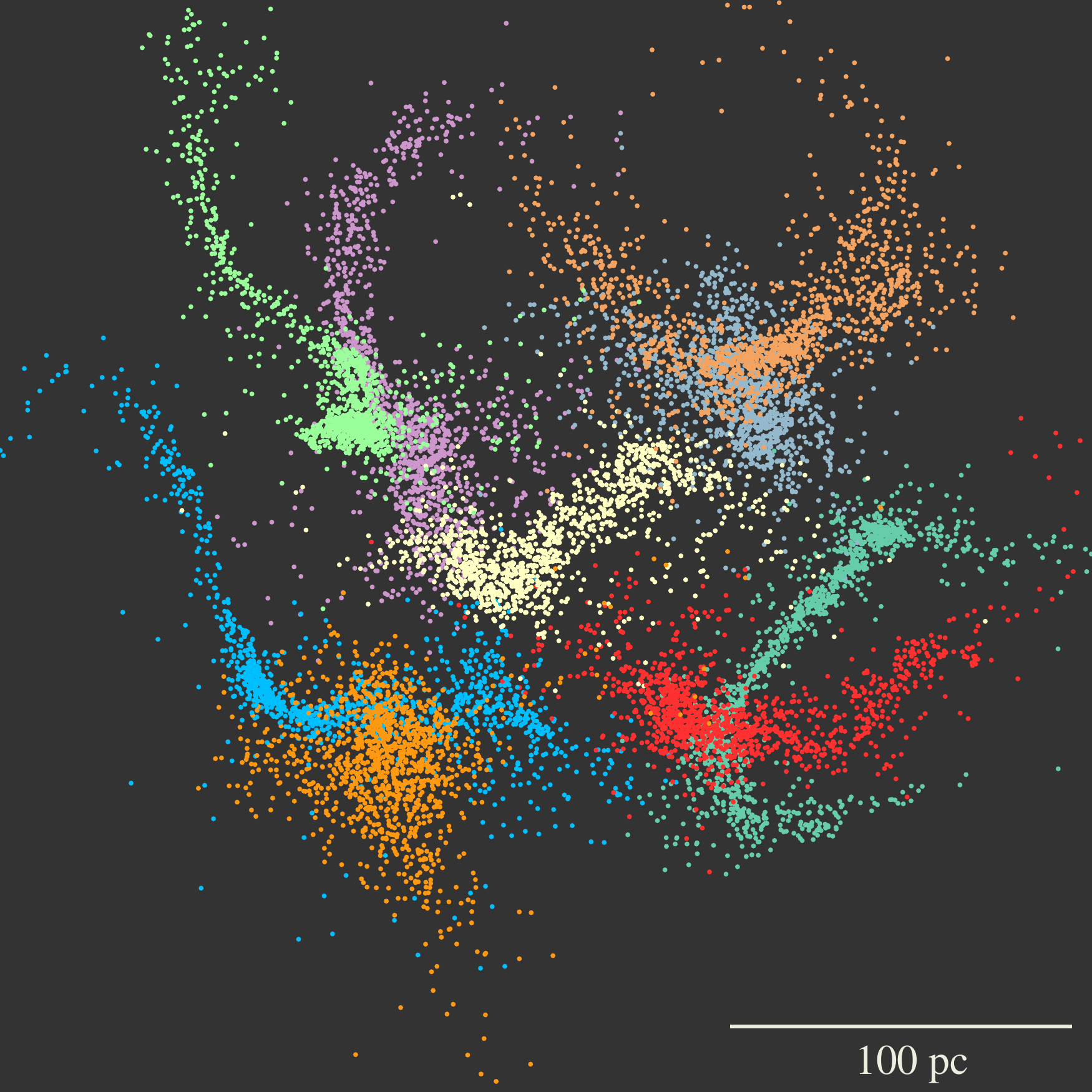}
\caption{Distribution of electrons with initial energy 1~PeV after 5000~yr of diffusion in a 3~\mycro\gauss\ random magnetic field. Electrons are originated in 9 different locations, identified with different colors.\cite{anisotropia-diffusione-cre}}

	\label{stream-cre}

\end{center}
\end{figure}

All these calculations suppose that \acp{cr} diffuse in the Galaxy and are isotropized by the interaction with the turbulent component of the Galactic magnetic field,  that has a coherence distance of 50 -- 150~\pc. While this is almost sure for heavy \acp{cr}, \citet{anisotropia-diffusione-cre} suggested that high energy leptons, because of their heavy radiation losses,  can lose all their energy before reaching an isotropic diffusion, spending all their life confined in particles streams.
 The impact of this model on the \ac{cre} physics is unclear, because the authors have calculated the transition energy $E_t$ between the stream and the isotropic diffusion with a large uncertainty, constraining its value in the 10~\GeV\ -- 1~TeV range, however  it would possible that the flux from potential \ac{cre} sources, identified via \Pphoton or radio observation, would not reach the Earth or, on the other hand, would be concentrated on the Earth by a magnetic stream, as shown in \cref{stream-cre}. Another potential effect of this hypothetical scenario is that, even if the \acp{cre} flux reaching the Earth is dominated by a single source, it could not show an excess in direction of  the source, as expected for standard diffusion scenario.

\section{Production of Cosmic-ray Electrons}
\label{CRE-sorgenti}
\subsection{Secondary electrons and positrons and the positron fraction}
The \acp{cre} can be accelerated in astrophysical sources and then injected in the interstellar space (primary electrons) or produced via interactions of other \acp{cr} with \ac{ism}.

\begin{figure}[htb!]
\begin{center}
\includegraphics[width=.9\textwidth]{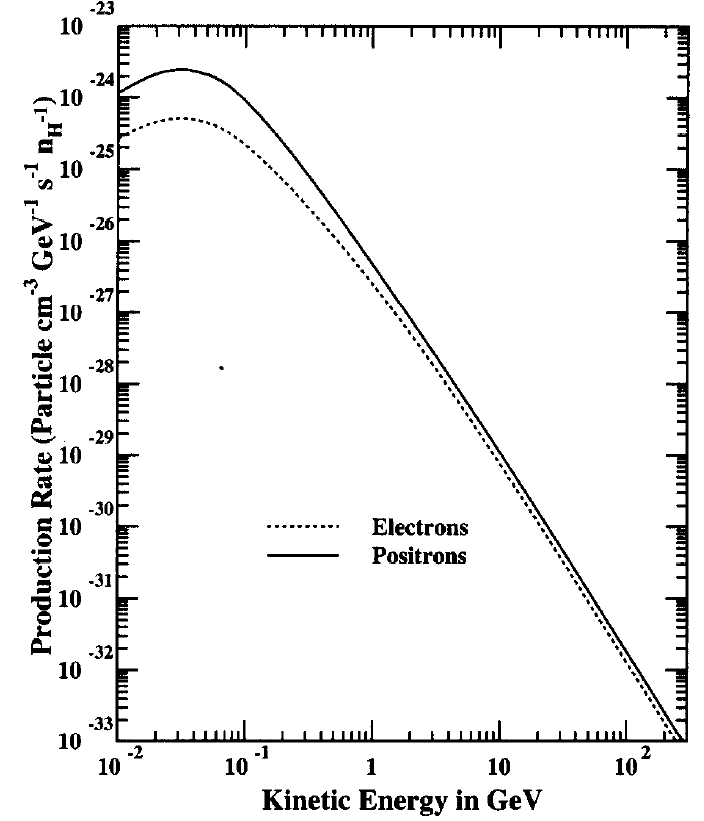}
\caption{ \Pelectron and \APelectron production spectra in \ac{ism} per unit of hydrogen density $n_H$, as estimated in~\cite{stephens-originCRE}}

	\label{secondari}

\end{center}
\end{figure}

Secondary electrons and positrons are produced by protons via \Ppipm or \PKpm production in inelastic scattering on \ac{ism}: 

\begin{align*}
\HepProcess{\Pproton,\alpha + \mathrm{H,He} \HepTo \PKpm,\Ppipm}&+ \mathrm{X} \\
&\mspace{-10.0mu}\hookrightarrow \Pmupm + \Pnum \\
&\mspace{32.0mu} \hookrightarrow \Pepm + \Pnum + \Pnue
\end{align*}
The production of secondary electrons and positrons is determined by the flux of hadronic \acp{cr} and by the quantity of \ac{ism} that they cross, and the production process is regulated by reasonably well know cross sections.
As stated in section~\ref{CR-propag}, the production of secondary particles changes the composition of the \ac{cre} flux, because secondary \Pelectron and \APelectron are produced in similar quantities. For reference, the  rate of secondary \Pelectron and \APelectron calculated by~\citet{stephens-originCRE} is shown in figure~\ref{secondari}; the larger \APelectron rate is due to the charge asymmetry of primary particles (\Pproton and $\alpha$), which are all positively charged. At low energy, where the main process involves the $\Delta^+$ resonance, that can decay in \Pproton~\Ppizero or in \Pneutron~\Ppiplus, the charge asymmetry is larger~\cite{asimm-in-secondari-ism}.

Because of the Feynman scaling hypothesis, the  spectrum of produced secondary \APelectron and \Pelectron  will have the same energy dependence  of that of parent primary \acp{cr}, while we can assume that secondaries observed on Earth will have undergone the same diffusion processes of primary electrons. Therefore, if a source of primary electrons with injection spectrum $\propto E^{\gamma_0}$ will result in an observed electron flux at Earth $\propto E^{-(\gamma_0+\Delta)}$, where $\Delta$ describes the effect of diffusion and energy losses,  secondary electrons originated by a \ac{cr} population with a $\propto E^{\gamma_p}$ spectrum will be observed at Earth with a  $\propto E^{-(\gamma_p+\Delta)}$ spectrum. Under these assumptions, a model assuming that primary \ac{cre} sources produce only \Pelectron will lead to a $\APelectron + \Pelectron$ flux:

\begin{equation}
N_{\Pelectron+\APelectron}=N_0E^{-(\gamma_0+\Delta)}+N_{-}E^{-(\gamma_p+\Delta)}+
N_{+}E^{-(\gamma_p+\Delta)}
\end{equation}
 Defining $N_0$ as the normalization coefficient of the primary \Pelectron spectrum, $N_{+}$ as the normalization of secondary \APelectron and  $N_{-}$ as the normalization of secondary \Pelectron, the positron fraction will result:

\begin{equation}\label{rapporto-positroni}
R_{\APelectron/\APelectron+\Pelectron}=\frac{N_{+}}{N_{-}+N_{0}E^{(\gamma_p-\gamma_0)}}\stackrel{N_{0} \gg N_{-}}{\propto} E^{(\gamma_0-\gamma_p)}
\end{equation}
which is a decreasing function of energy when making the commonly (but not univocally, see~\citet{ams-pro-secondary-positron}) accepted assumption that primary \ac{cre} are produced with a spectral index  similar to that of other \ac{cr} species, and therefore $\gamma_p = \gamma_{inj}+ \Delta_{cr} \approx \gamma_0+\Delta_{cr}>\gamma_0$

A very important quantity related to the positron fraction is   the \APproton/\Pproton ratio: secondary \APproton are produced in a similar way to \APelectron, via hadronic interaction

\begin{align*}
\HepProcess{\Pproton + \mathrm{H,He} \HepTo \Pproton + \Pproton + \APproton}& + \mathrm{X}\\
\HepProcess{\Pproton + \mathrm{H,He} \HepTo \Pproton + \Pneutron + \APneutron}& + \mathrm{X}\\
&\mspace{-10.0mu}\hookrightarrow \APproton + \APelectron +\Pnue
\end{align*}

On the other hand, primary sources of \APelectron are not necessary also sources of \APproton, because in sources dominated by electro-magnetic interactions (like pulsar and \acp{pwn}, see \S~\cref{section-pulsar}) only leptons are produced. Therefore differences between \Pelectron/\APelectron and \Pproton/\APproton ratios put strong constraints on  the existence of sources of primary \APelectron, that will be treated in the next sections.

\subsection{Diffusive shock acceleration}\label{dsa-section}
\ac{dsa} (\citet{dsa-drury})  is a very efficient mechanism to accelerate particles to very high energies: it is thought that almost all the sources of primary \acp{cr}, either galactic or extra-galactic, accelerate particles through this process.

\ac{dsa} is also known as first order Fermi acceleration, to distinguish it from the original proposed mechanism by Fermi, the second order Fermi acceleration. The original mechanism was based on stochastic acceleration during scattering of particles on fast moving gas clouds: the energy gain, proportional to $(v/c)^2$ where v is the speed of clouds and c the speed of light, is the reason for the name 'second order mechanism'. This mechanism was insufficient to explain the observed acceleration of \acp{cr}, principally because the strong suppression in acceleration introduced by the factor  $(v/c)^2$: a new mechanism was therefore introduced, involving particle acceleration at strong shocks, where the energy gain is proportional to v/c (first order), where v is the speed of the shock~\cite{longair}.

\ac{dsa} takes places when a strong supersonic shock is moving  through matter: this is what typically happens in \acp{snr}, around fast pulsars, and probably also in Active Galactic Nuclei. It should be noticed that \ac{dsa} is able to accelerate only particles that are already non-thermal (Lorentz factor $\gamma\sim 8$): the origin of these particles is still unclear (injection problem), but is commonly accepted that they come from the high energy tail of the Maxwellian thermal distribution~\cite[page.~160]{vietri}. Because the cross section  is $\sigma \propto 1/E^2$, when an already non-thermal particle is accelerated to higher energy is not able anymore to quickly thermalize.

 When a non-thermal particle in the matter before the shock is reached by it, the interaction with the strong magnetic fields present  in the shock results in an increase of the particle energy. If the accelerated particle diffuses back to the shock it will suffer another process of acceleration. The repetition of this process can accelerate particles to very high energies, before they are released in the space. The time required to accelerate a particle to a certain energy is~(\citet{snr-review}): 
 
 \begin{equation} 
t_{acc}\propto \left(\frac{\lambda_{fmp}}{c}\right)\left(\frac{u_{shock}}{c}\right)^{-2}\propto \mathcal{M}^{-2} 
\end{equation}
where  $\lambda_{fmp}$ is the mean free path of the accelerated particles in the shock, u is the velocity of the shock and $\mathcal{M}$ is the Mach number of the shock, u/$c_{sound}$. For strong shocks (for example in \acp{snr}), $\mathcal{M}$ is of the order of $10^2-10^3$, and acceleration of particles can become very efficient.
 
 In \ac{dsa}, particles gain energy every time they cross the shock while a fraction of them escapes the acceleration site. This process leads to a resulting power-law particle spectrum $\frac{dN}{dE}\sim E^{-p}$. In the approximation in which the energy losses due by energetic particles escapes is negligible, the spectral index p should be $\approx 2$ for all accelerated particles (p, He, \Pelectron): this is an important result, because the observed spectral index of \acp{cr} of $p \sim 2.7$  and the esteems of the diffusion coefficient in the galaxy lead to value of the source injection index near to 2. However, the approximation of negligible  energetic loss due to particles escape is often not satisfied, while the presence of highly-relativistic particles can modify the very behaviour of the shock, therefore real source indexes are typically larger than 2, with the emitting spectrum becoming flatter at high-energy~(\citet{amato-cr}).
 
\subsection{Super Nova Remnants}\label{SNR}
The explosion of a \ac{sn}, either ignited by a thermonuclear explosion in a degenerate star  (type Ia) or by the collapse of a massive star (core-collapse, Type Ib,Ic, II), ejects a large quantity of matter in the surrounding space, with a kinetic energy of the order of $10^{51}~\erg$. Typically, the ejected matter expands almost spherically with velocities of the order of $5$--$10\times 10^3~\kilo\meter~\reciprocal\second$ that are many orders of magnitude larger than typical sound speed in surrounding space ($\sim $1--10$~\kilo\meter~\reciprocal\second$): this leads to the formation of a shock surface between the ejected matter and the surrounding matter~\cite{snr-review}.

At the beginning of the \acf{snr} life, during the \textit{eject-driven} phase, the shock expands almost freely, with the formation of an additional reverse shock that re-heats all the ejected matter. The density profile of the ejected matter is described by a power law $\rho \propto r^{-n}$, with n slightly depending   from the progenitor star (white dwarf or massive star) and, in case of a core-collapse supernova, from the behaviour of the \ac{csm}, that has been modified by the strong stellar wind emitted by  massive stars during giant phases. After a time varying from $10^2$ to $10^3~\yr$, an amount of matter of the order of several times the ejected mass has been shocked by the expanding wave, that starts to significantly slow down. From this point, the shock expansion is well described by the Sedov self-similar solution (\citet{sedov}, from which the name \textit{Sedov} phase) 

\begin{equation}\label{sedov}r = 1.15\left(\frac{\mathcal{E}_{SN}}{\rho_0}\right)^{\frac{1}{5}}t^{\frac{2}{5}}\end{equation}
where $\rho_0$ is the density of the surrounding \ac{ism} or \ac{csm} and $\mathcal{E}_{SN}$ is the initial energy of the ejected matter.

Finally, after $\sim 10^5~\yr$, the shock has slow down to velocities of the order of 100--300~\kilo\meter~\reciprocal\second\ and reached temperatures where elements like oxygen and hydrogen start to recombine: at this point radiative energy losses become dominant and the adiabatic expansion approximation breaks down, leading to the end of the Sedov phase; the shock keeps cooling for  $\sim 10^5~\yr$ and finally merges with the \ac{ism}.

During the ejecta-driven and the Sedov-expansion  phase, strong magnetic inhomogeneities are present at the shock, and its Mach number $\mathcal{M}$ is very high, therefore very efficient \acf{dsa} can take place (see \S~\ref{dsa-section}) until the end of the Sedov phase, when the speed of the shock becomes  low and particle acceleration  very inefficient. At $\mathcal{M}\sim\sqrt{5}$ (\citet{Mach-critico}) the shock is no longer able to accelerate particles via \ac{dsa}, even if the production of low-energy electrons (tens of MeV) should still be possible, as affirmed in~\citet{acc-post-dsa}.

  A strong support to the hypothesis that \acp{snr} are the main source of galactic \acp{cr} is the esteem of the energy transferred by these objects to the \ac{cr} population. Equation \eqref{LumCr} gives an esteem of the luminosity of the galactic \acp{cr} sources: if we compare this luminosity to the typical energy $\mathcal{E}_{SN}$ released by a \ac{sn} in the surrounding matter (excluding neutrinos) and to the measured rate  $\nu_{SN}$ of \ac{sn} events in the galaxy (of the order of 3 per century), the esteemed efficiency of \ac{cr} acceleration is  
  
  \begin{equation} \eta_{CR}\simeq 0.1\left(\frac{L_{CR}}{10^{41}~\erg~\reciprocal\second}\right)\left(\frac{0.03~\reciprocal\yr}{\nu_{SN}}\right)\left(\frac{10^{51}~\erg}{\mathcal{E}_{SN}}\right)
\end{equation}   
  The resulting  10\%--30\% efficiency is a reasonable value for shocks in \acp{snr}, even if the energy loss caused by \ac{cr} escape cannot be considered as negligible, and therefore the emission spectrum of \acp{snr} has a spectral index greater than 2. Recently, \citet{eff-dsa} gave a first demonstration that an efficiency of 10\%--20\% is perfectly achievable.
  
 An interesting way to estimate  the energy transferred by an \ac{snr}  to relativistic particles is presented in \citet{vink13}: escaping relativistic particles heat and compress the un-shocked matter close to the shock. When the escape of relativistic particles becomes significant, this results in the formation of a shock precursor, that significantly compresses the matter before it is reached by the subsequent shock (called sub-shock). Therefore, the sub-shock has a lower Mach number with respect to the pre-compressed matter, and the temperature of the shocked matter is lower than it would have been in absence of the shock precursor. Shocked matter temperature  can be univocally   related to the escaping flux of \acp{cr}: if  the velocity of the shock for a \ac{snr} is known, and the temperature of shocked protons (that is strongly related to the temperature of the other particle species)  can be measured using the thermal Doppler broadening of the H$\alpha$ line, the escaping cosmic-ray energy flux $\epsilon_{esc}$ can be estimated. Two \acp{snr} are analyzed in \cite{vink13}, with resulting $\epsilon_{esc}$ comprised between 6\% and 70\%. \citet{morlino,morlino2} present an in depth analysis of the \acp{snr} RCW 86, SNR 0509-67.5 and Tycho (the first two were analyzed also in~\cite{vink13}), concluding that, even if a value of $\epsilon_{esc}\sim0$ cannot be excluded because of uncertainties in the source distance, $\epsilon_{esc}$ is probably in the range 10--20\%.
  
Accelerated particles in \acp{snr}  are principally \Pproton, He and \Pelectron from the shocked matter: however, collisions between high energy particles and the \ac{snr} matter can result in the production of secondary particles like \Pproton \APproton and \Pelectron \APelectron pairs. The so produced antiparticles are accelerated in the same way of ordinary particles~\cite{Blasi}. Recently, \citet{positroni-decad-snr} suggested that supra-thermal \APelectron could be produced via $\beta^+$ decays of radioactive nuclei produced in the \ac{sn} explosion and subsequently accelerated by the reverse-shock.

There is a limit to the maximum energy reached by particles accelerated in a \ac{snr}, therefore the spectral law of emitted particles is 

$$N(E) \propto E^{-p}\exp[-(E/E_{max})^a]$$ 
$a$ depends from the diffusion coefficient of particles in the \ac{snr} matter that typically is described as $E^\beta$, and from the dominant process of energy loss, and is comprised between 0.5 and 4~(\citet{snr-sincro}).

The  processes that can limit the maximum energy that a particle accelerated in a \ac{snr} can reach are three, and they affect electrons and protons in a different way  (in following equations, u is the shock speed and B is the magnetic field downstream of the shock, usually of the order of 100~\micro\gauss):
\begin{itemize}
\item the time required for accelerating a particle to a certain energy can be longer than the life T of the \ac{snr}: 

$$E_{max}(age)\propto u^2 T  B$$
\item the energy loss by synchrotron processes is equal to that given by the acceleration process (this is typically the limiting factor for \Pelectron, while for protons synchrotron losses are negligible) 

$$E_{max}(loss)\propto u B^{1/2}$$

\item the acceleration process requires that particles are confined inside the \ac{snr} by the interaction  with \acp{mhw}; the existence of a maximum wavelength ($\lambda_{max} \sim 10^{17}~\centi\meter$) results in an abrupt increase of the diffusion coefficient for particles of the corresponding energy, that so escapes upstream of the shock and leave the \ac{snr} 

$$E_{max}(escape)\propto B\lambda_{max}$$

\end{itemize}
Expected $E_{max}$ can vary from tens of \TeV\ to (maybe) $\sim$ PeV from young \ac{snr}.
In addiction to this limit, electrons experience also a spectral break at high energy when their radiative loss time becomes smaller than the age of the \ac{snr} (ie, when the impact of radiative losses becomes significant for the electron population): above this energy the electron spectrum becomes softer, with an increase of the spectral index with respect to that of the protons, while below $E_{break}$ spectral index are expected to be similar~(\citet{snr-gabici}).

Highly-energetic particles in \ac{snr} produce photons in the X and gamma wavelengths: the study of high energy photons from \ac{snr} is a really important way to probe the effective presence of particle acceleration.
A population of electrons with spectral distribution $N(E) \propto E^{-p}$ emits photons via the synchrotron process with a spectrum $N(E) \propto E^{\frac{1-s}{2}}$ therefore the resultant gamma spectrum should have an index  $\geq 0.5$. Also, electrons can produce high energetic photons with spectrum $N(E) \propto E^{\frac{1-s}{2}}$ by \ac{ic} on \ac{cmb} or thermal photons, or also on the photon produced via synchrotron emission (Self Compton Scatter).
Highly energetic protons produce photons via \Ppizero production: the resulting photon spectrum peaks at 68~\MeV\  (half of the \Ppizero mass) and then decrease with the same spectral index  of the original protons population.

The detection of many \ac{snr} in the X-ray band (many with index near 0.5) is a strong confirmation of the presence of electron population of energy of the order of 100~\TeV, while an observation from \fermi\ (\citet{fermipi0}) has recently lead to the detection of the 68 \MeV\ signal in the gamma emission from a couple of  \acp{snr}, furnishing strong support to the hypothesis that also protons are accelerated in these sources. An updated list of  \acp{snr} detected in X or $\gamma$ wavelengths is presented in~\citet{catalogo-snr}.

An indirect proof of the production of high-energy particles could be found by precise measurement of the ionization rate of molecular clouds close to \ac{snr} and therefore subjected to a higher rate of \acp{cr}: \citet{snr-clouds} claim that  measurements made on clouds close to the \ac{snr} W28 returned a ionization rate more than 100 times the rate expected by the observation of far clouds.

\subsection{Pulsars and Pulsar Wind Nebulae}\label{section-pulsar}
During a \ac{sn} core-collapse explosion, the progenitor star can collapse in a neutron star, a very compact object with a mass comprised between 1.4 and 5 $\mathrm{M_\odot}$ and a radius of the order of 10~\kilo\meter, with density $\sim 10^{15}$~\gram~\centi\meter\rpcubed. Because the magnetic field energy is conserved during the star collapse, it is concentrated in the compact object and the magnetic field can reach  intensities of the order of $10^{10} - 10^{13}$~\gauss. The so formed neutron star is a very fast rotator, with period comprised between seconds and tens of milliseconds. Initially these objects were detected in the radio wavelength, where they exhibit a periodic signal: for this reason they were named pulsars~\cite{vietri}.

Pulsars are strong astrophysical sources: in the model developed by Pacini (see~\citet[chapter 8]{vietri}), where the magnetic field of a pulsar is modeled keeping only into account the dominant dipole term, the power $\dot E=\frac{dE}{dt}$ emitted by a pulsar is:

\begin{equation}
\dot E = \frac{1}{6}\sin^2(\beta) cB^2_p R^2_{ns}\left(\frac{\varOmega R_{ns}}{c}\right)^4
\end{equation}
where $\beta$ is the angle between the magnetic dipole and the rotational axis, $B_p$ is the strength of the magnetic field at the magnetic pole, $ R_{ns}$ is the radius of the pulsar and $\varOmega$ its angular velocity. This is a reasonable simplification of the more general model, presented in~\citet{pacini-bis}, where the emitted  energy, which  comes from the rotational energy of the star, is calculated using the relations:

\begin{equation}
\begin{array}{l}
\dot \varOmega = -\alpha \varOmega^n\\
\dot E = I |\dot \varOmega| \varOmega=\alpha I \varOmega^{n+1}\\
\end{array}
\end{equation}
where $I$ is the inertial moment of the pulsar, $n$ the braking index and $\alpha$ a constant that  in the case of simply magnetic dipole ($n=3$) becomes $\alpha=\frac{5}{8}B_p^2R_{ns}^4/M_{ns}c^3$. From previous relations it came straightforward that 

\begin{equation}
n=\frac{\ddot{\varOmega}\varOmega}{\dot{\varOmega}^2}
\end{equation}
and $n$ can be calculated if period parameters $(\varOmega, \dot \varOmega , \ddot{\varOmega})$ are measured.
Therefore it is possible to calculate the emitted energy with a precision depending on the time spanning of the observations . For observed young pulsars, the emitted power  is comprised between $4\times10^{36}~\erg~\reciprocal\second$ and $5\times10^{38}~\erg~\reciprocal\second$, while $n$ was measured for only 4 pulsars, spanning from 1.4 (Vela) to 2.8 (B1509-58).

Pulsars are a known source of high energy electrons and positrons: these particles are extracted from the pulsar surface and subsequently  accelerated by strong electric fields in the pulsar magnetosphere, generating a cascade of \Pelectron \APelectron pairs whose multiplicity (number of generated pairs) can be larger than 100 (\citet{blasi-amato-pwn}). The produced  \Pelectron \APelectron pairs are then ejected into the surrounding space as a relativistic wind with Lorenz factor $\sim 10^4$--$10^7$~\cite{aharonian,blasi-amato-pwn}, forming a \ac{pwn} of relativistic hot, magnetized fluid~(\citet{aharonian,blasi-amato-pwn}), where electron energy  is of the order of a few GeV. \citet{lista-pwn} report more than 50 sources identified as \acp{pwn} in the Milky Way or in the Large Magellanic Cloud.  An interesting test to verify the effective composition of \acp{pwn} was suggested in \citet{pwn-polariz}, that proposes the measurement of the circular polarization of the synchrotron radio emission of these sources: if the number of \Pelectron and \APelectron is equal, their contributes should cancel, and no circular polarization should be observed.

The release of the accelerated particles from the confinement in the magnetic field of the pulsar is possible because, at a distance $\sim \frac{c}{\varOmega}$ (the \textit{light cylinder}) from the star rotation axis, co-rotating particles would move to velocities near to the speed of light: relativistic effects cause the magnetic field lines to open, therefore allowing the release of the accelerated particles.

\begin{figure}[htb!]
\begin{center}
\subfigure[The polar cap model, where acceleration takes place near the poles (grey cylinder); also marked the zone of additional acceleration proposed in the  slot gap model. (figure from~\cite{slotgap})]{\includegraphics[width=.43\textwidth]{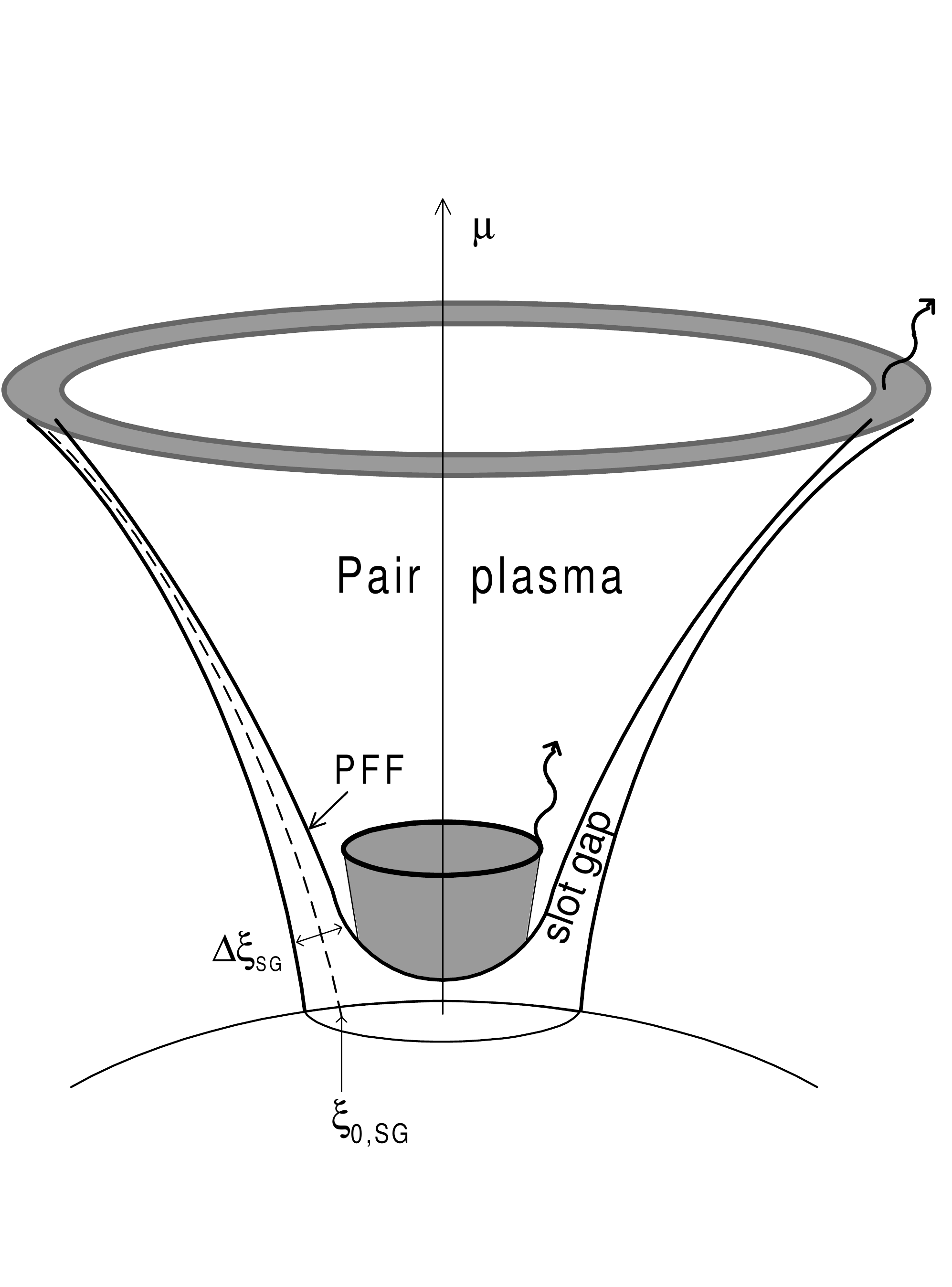}}\hspace{4mm}
\subfigure[Acceleration zones in the two-pole caustic (dotted lines) and outer gap (grey zone) models (figure from~\cite{tpc}). ]{\includegraphics[width=.43\textwidth]{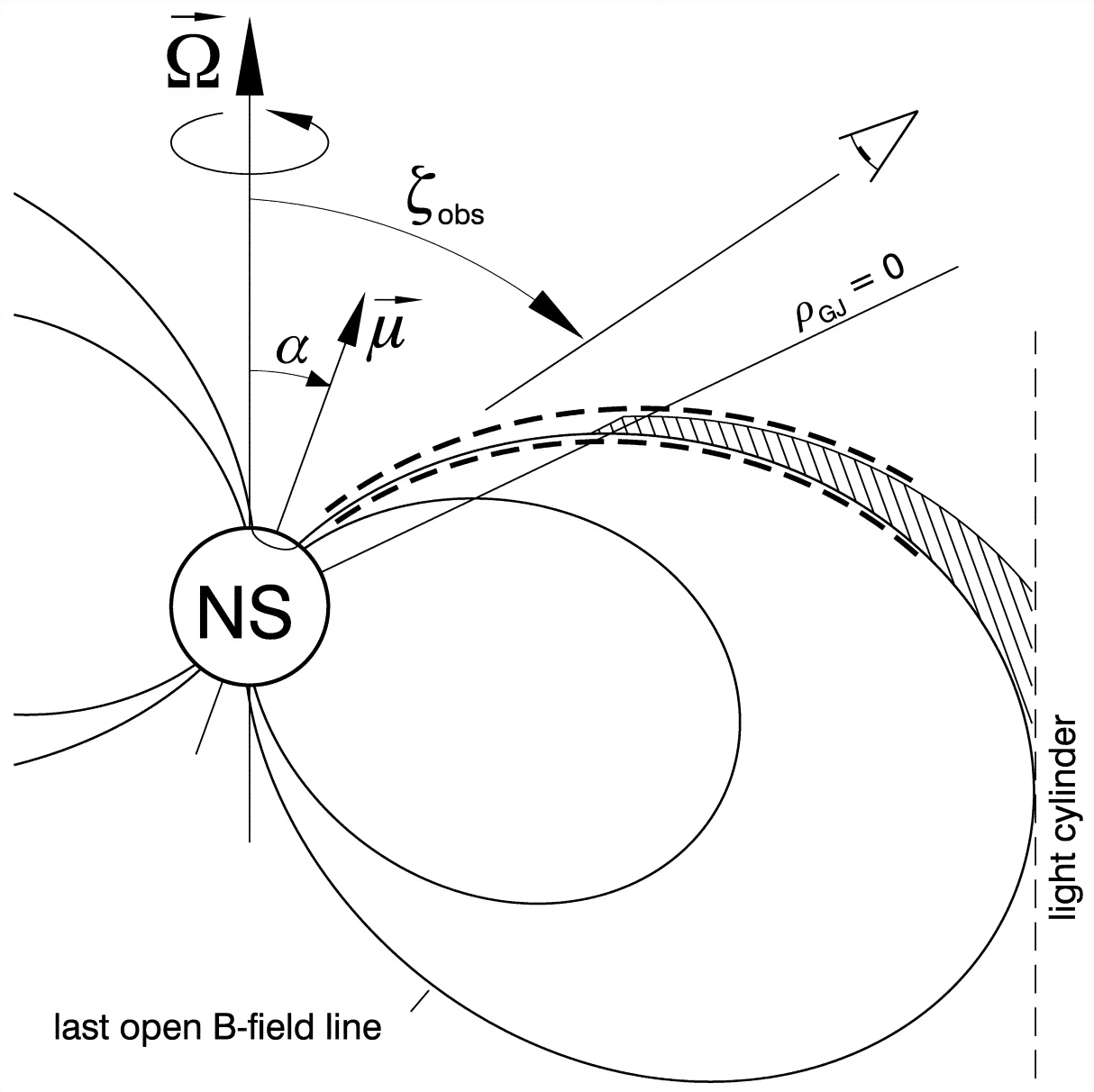}}
\caption{Different models for electron acceleration in pulsars.}
	\label{modelli-pulsar}

\end{center}
\end{figure}

The acceleration of electrons in the pulsar magnetosphere requires that the strong electric fields generated by the rotation of the magnetic field (in pulsars the magnetic axis is not aligned with the rotational axis) are not compensated by a corresponding charge flow, therefore a charge-depleted zone is required. There are different zones that can satisfy this request, consequently different acceleration models were elaborated:
\begin{description}
\item[Polar Cap: ]the acceleration takes place at the magnetic poles of the pulsar, near the surface~(\citet{polarcap,polarcapbis});
\item[Outer Gap: ]electrons are accelerated in zones (gaps) between the charge depleted zone that forms at $\Omega \cdot B\,=\, 0$ and the light cylinder, along the last closed lines of the magnetic field~(\citet{outergap});
\item[Slot Gap: ]In a Polar cap acceleration model, additional acceleration takes place in the space between the last open field line and the \Pelectron\APelectron plasma column arising from the pole~(\citet{slotgap});
\item[Two-pole caustic: ]particles are accelerated in a thin zone along the last open field line, from the pole until the light cylinder~(\citet{tpc}).
\end{description}
All the models are presented in figure~\ref{modelli-pulsar}.

Recent observations by \fermi~\cite{vela,vela2} have disfavored the Polar cap model: in all the models, high energy photons emitted from the high energy electrons interact with photons in the magnetosphere via \HepProcess{\Pphoton + \Pphoton \HepTo \Pelectron + \APelectron } processes, creating a cut-off in the gamma spectrum. Because in the Polar Cap model acceleration takes place at a lower altitude with respect to Outer gap or Two-pole caustic models, the expected cut-off is super-exponential $e^{-{(E/E_c)}^a}, \,a>1$, but \fermi\ observations have measured simple exponential cut-off ($a\sim 1$). Also, because  the radio emission of the pulsar is expected to come from the poles, radio and gamma peaks should be aligned in PC scenarios, while this was not observed by \fermi.

\begin{figure}[htb!]
\begin{center}
\includegraphics[width=.9\textwidth]{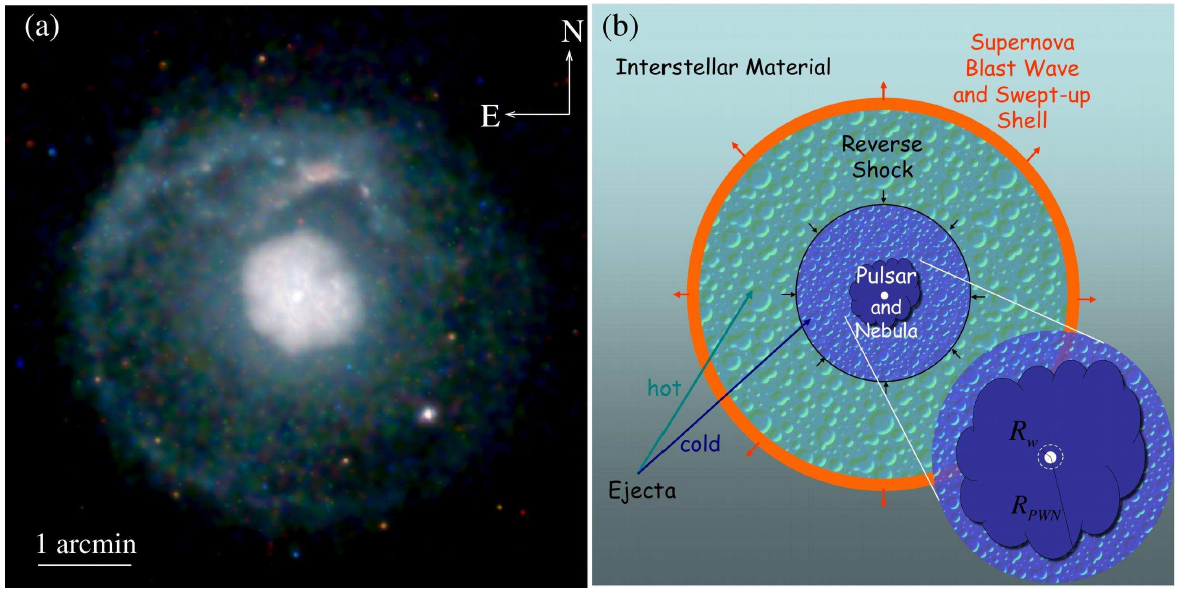}
\caption{A pulsar-powered \ac{pwn} surrounded by its \ac{snr}, from~\cite{PWNe}; on the left, \textit{Chandra} X-ray image of \ac{snr} G21.5-09~\cite{Matheson}, showing a circular \ac{snr} with the young pulsar J1833-1034 at the center, surrounded by a symmetric \ac{pwn}; on the right, a model of a composite \ac{snr} showing the   shell expanding in \ac{ism}, the reverse shock that is heating the ejecta and the central pulsar with the surrounding nebula, in which  the termination shock is shown. }
	\label{PWNe}

\end{center}
\end{figure}

 \acp{pwn} can assume very different behaviours depending on if the pulsar is surrounded by its \ac{snr}. For pulsars surrounded by their \ac{snr} (typical case for young pulsars), the \ac{pwn} expands almost freely in the un-shocked ejected matter~(\citet{PWNe}) until it meets with the reverse shock that, starting from the \ac{snr} front, is shocking the ejected material~(\citet{snr-review}): the reverse shock keeps the \ac{pwn} compressed in the \ac{snr} and, where the relativistic wind is stopped by the reverse shock, a termination shock is formed. The termination shock accelerates the electrons of the pulsar wind to very high energy (up to tens or hundreds of \TeV), even if \citet{blasi-amato-pwn} suggest that this is not because of  \ac{dsa}  (section~\ref{dsa-section}), as this process is inefficient in the case of quasi-perpendicular highly relativistic shocks, and therefore the acceleration mechanism is not clear. The presence of high energy electrons has however been confirmed by the detection of X and gamma emission from pulsars and from \acp{pwn}, with the characteristic signature of the synchrotron and \ac{ic} emission (for  example, in~\citet{velax} the detection  by \fermi\ of the nebula of the Vela pulsar is reported). Many \acp{pwn} were also detected at higher energies by ground-based gamma-telescopes: as of April 2013, the TeVCat~\cite{tevcat}, a catalog of \ac{vhe} sources compiled by the University of Chicago, classifies 31 very-high-energy sources as \acp{pwn}. A \ac{pwn} surrounded by a \ac{snr} is described in figure~\ref{PWNe}, even if the circular symmetry is maintained only if the pulsar remains close to the \ac{snr} center, with this expanding in a reasonably uniform medium (\citet{slane-pwn}) .

\begin{figure}[htb!]
\begin{center}
\includegraphics[width=.9\textwidth]{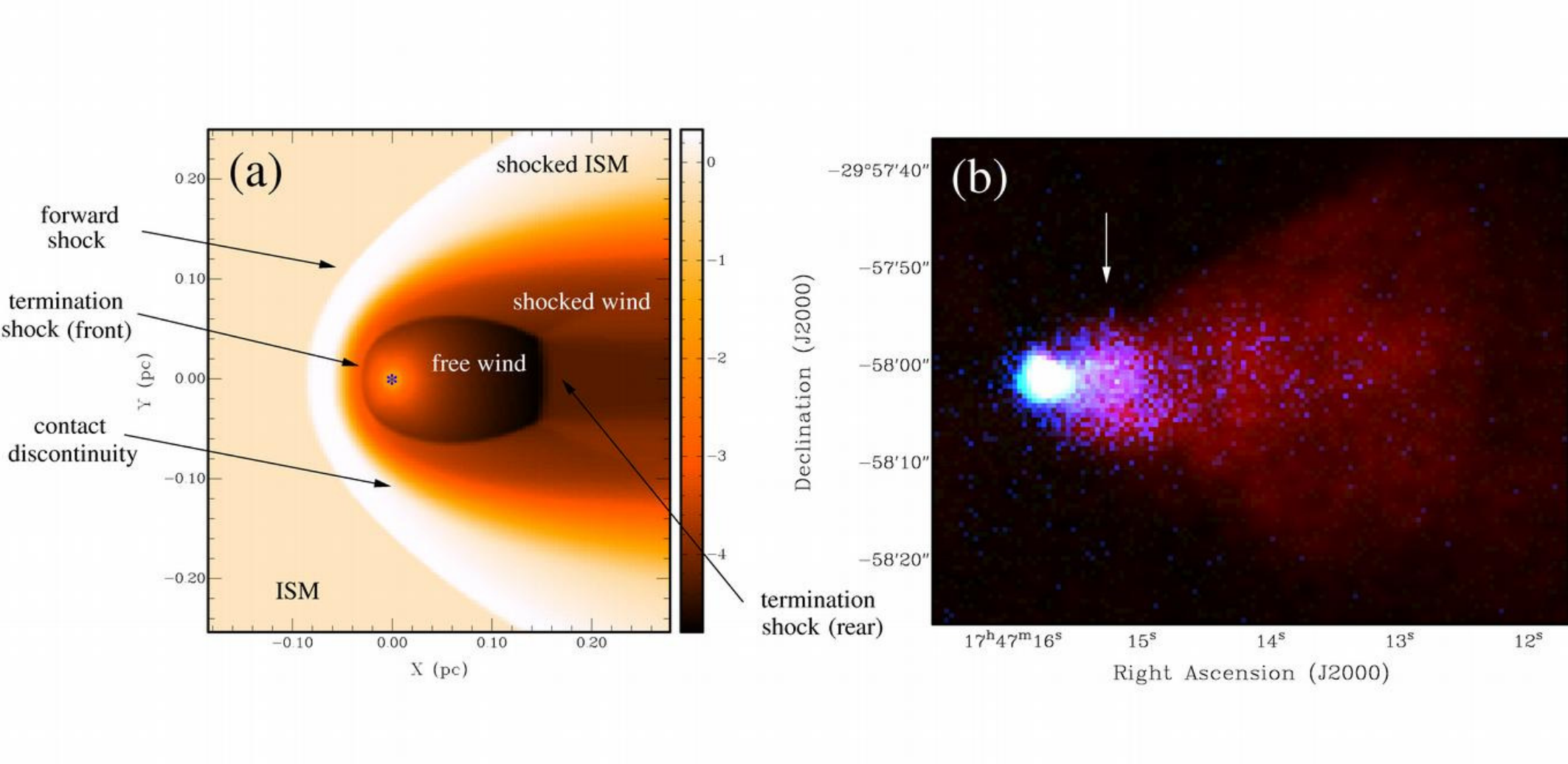}
\caption{A pulsar bow shock, from~\cite{PWNe}; on the left, a pulsar (marked by $\star$) moving leftwards with Mach number $\mathcal{M}=60$. The shock  caused by the pulsar motion in the \ac{ism} and the termination shock of the pulsar wind are marked. On the right, \textit{Chandra} X-ray (blue) and VLA radio (red) images of G359.23-0.82 (``the Mouse``),  the bow shock associated with pulsar J1747-2958~(\cite{Gaensler}). The white arrow marks a bright compact region of X-ray emission behind the apex, which possibly corresponds to the surface of the termination shock.
}
	\label{bow-shock}

\end{center}
\end{figure}

\acp{pwn} are not necessarily  surrounded by a \ac{snr}: during the \ac{sn} explosion, pulsars can gain high translation velocities (up to 400 or 1000~\kilo\meter~\reciprocal\second) and they can leave the expanding shell. If this happens when the pulsar  has still sufficient spin-down power $\dot E$, a \ac{pwn} is formed in the \ac{ism}. Because the sound speed in the \ac{ism} is comprised between 1~\kilo\meter~\reciprocal\second\ (cold matter) and 100~\kilo\meter~\reciprocal\second\ (hot matter), the pulsar and the surrounding nebula move at supersonic velocity, therefore  a  shock is formed between the \ac{ism} and the pulsar wind. Also in this case  a termination shock forms where the wind is stopped and at this shock particles are accelerated. This system, called bow shock,  assumes a highly asymmetric form, that is represented in figure~\ref{bow-shock}.

Accelerated  electrons assume a spectrum $\frac{dN}{dE}\sim E^{-s}e^{-E/E_c}$ with $s \sim 1.3\pm 0.3$~\cite{positroni-decad-snr} for simple \acp{pwn} and $\sim 2$ for \ac{pwn}+\ac{snr}  systems and $E_c \sim 10~\TeV$ or more (H.E.S.S estimated that the Vela pulsar has an  electron spectral break at 67~\TeV, see~\citet{elettroni-hess-vela}). If the radiative loss time due to synchrotron and \ac{ic} losses is smaller than the confinement time in the nebula, electrons experience strong energy losses and their spectrum at high energies becomes steeper, $\frac{dN}{dE} \sim E^{-s-1}$. 

The release of accelerated  \Pelectron and \APelectron in the \ac{ism} as \acp{cre} is very different between normal and bow-shock \acp{pwn}: in the first case, electrons are typically confined in the \ac{pwn} for a time of the order of $10^4$--$10^5$~\yr, until the pulsar exits from the \ac{snr}.  Because of the significant energy losses caused by radiative and adiabatic processes, the fraction of the accelerated electrons that are effectively released in \ac{ism} is not known,  representing the major uncertainty in the model.

On the other hand, bow-shock \acp{pwn} are not surrounded by anything, and accelerated particles quickly exit from the tail of the nebula, making these sources an interesting alternative as sources of \ac{cre}, the main uncertainty being about how much energy the pulsar still has when it leaves the \ac{snr}. The calculation is not easy: comparing the traveled distance $d=v_{ns}t$ with the radius of the \ac{snr} described by eq~\eqref{sedov} we obtain escape times $T_e$ of the order of 40--50~\kilo\yr; however, the residual pulsar energy strongly depends on the braking index $n$~\cite{blasi-amato-pwn}:

\begin{equation}
E(t>T_e)=\frac{1}{2}I\varOmega_0^2\left(1+\frac{T_e}{\tau_0}\right)^{\frac{2}{1-n}}
\end{equation}
where $\tau_0$ is the characteristic time of the pulsar and can be measured if the period evolution and age of the pulsars are known with sufficient precision:

\begin{equation}
\tau_0 + age=\frac{\varOmega}{\dot{\varOmega}(1-n)}
\end{equation}
In a Crab-like pulsar, the available energy would be $\sim10^{49}\erg$ for $n=3$ and $\sim10^{47}\erg$ for $n=2.5$, (similar to that of the Crab).

A rough estimation of the maximum luminosity of pulsar-injected \acp{cre}, $\mathcal{L}_{max}$, can  be calculated using the average lifetime of a \ac{pwn},  the total energy $E_{tot}$ transferred to \acp{cre} by a pulsar during this time and the estimated frequency $R_{cc}$ of core-collapse \acp{sn} in our galaxy, that is of the order of 2/century~(\citet{serpico}):

\begin{equation}
\mathcal{L}_{max}=6.3\times 10^{39}\erg\,\reciprocal\second \frac{R_{cc}}{2~\reciprocal\century}\frac{E_{tot}}{10^{49}~\erg}
\end{equation}

\subsection{Micro-quasar}\label{microquasar}
\begin{figure}[htb!]
\begin{center}
\includegraphics[width=.9\textwidth]{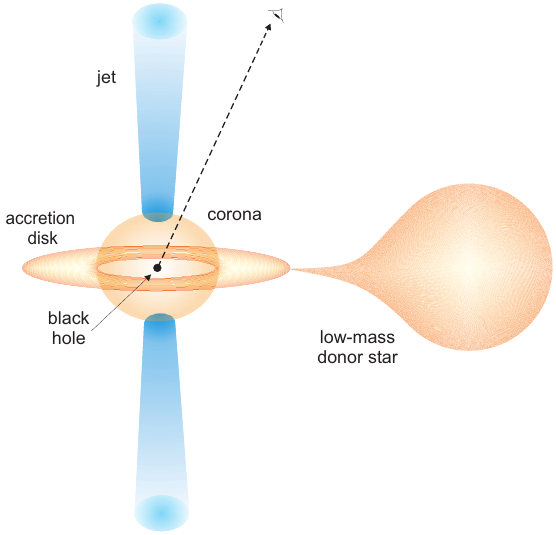}
\caption{Scheme of a low-mass micro-quasar from~\cite{vila-jets}}
	\label{microquasar-fig}

\end{center}
\end{figure}

Micro-quasars (see for example~\citet{rosswog} and figure~\ref{microquasar-fig}) are a special sub-class of X-ray binaries. Typically, these sources are composed by a star that is transferring matter to the compact companion star, either because it has expanded beyond its Roche lobe, therefore the matter that passes beyond  the Lagrangian point \textbf{L1} falls on the companion (Low mass X-ray binary), or because the star is highly massive, and its strong stellar wind furnishes a significant mass flow to the companion  (High mass X-ray binary). The estimated mass of the companion is typically far beyond 3~$M_\odot$, that is the probable mass limit of neutron stars, therefore the compact stars in micro-quasar are typically (but not always) supposed to be  stellar-mass black holes. The name micro-quasar stresses the strong similarity between these sources and the quasars, Active Galactic Nuclei where a super-massive black hole (millions of $M_\odot$) is, as the stellar-mass black holes of micro-quasar, surrounded by  an accretion disk and ejecting very powerful jets of relativistic matter.

Micro-quasars are transient source, with a luminosity of $10^{35}$ -- $10^{37}$~\erg~\reciprocal\second and a spectrum that extends from radio to gamma wavelengths (for the first gamma detection of a micro-quasar by \fermi, see~\citet{fermi-microquasar}, for a multi-wavelength study of the same source during a transient state, see~\citet{cygnus-transition}). It is thought that our Galaxy should host from hundreds to a thousand of micro-quasar, approximately distributed in the disc, bulge, halo in a ratio 2:1:0.3~(\citet{microquasar1,microquasar2}). 

The typical spectrum of micro-quasars has successfully been explained (\citet{microquasar-corona}) by a jets+corona+accretion disc model. In this model, the accreting matter forms a disk around the black hole, where its angular momentum and its gravitational energy are  dissipated via viscose friction, therefore heating the disk and allowing the matter to move towards the inner of the disk. The  accretion disk emits black-body radiation from the infra-red to the ultra-violet or also X wavelengths. Close to the black hole the disk disrupt, and a nearly spherical corona of hot mildly-relativistic plasma is present, whose non-thermal emission extends to 100 -- 150~\keV\ or more. 

The most interesting feature of a micro-quasar is the couple of opposite jets of highly relativistic ($\Gamma \sim$ 1 -- 10) collimated plasma of matter, that are orthogonal to the accretion disc plane and can extend up to hundreds of AU ($10^{13}~\kilo\meter$ from the source. The origin  of these jets is still under debate (\citet{rosswog,vila-jets}), however it is believed that they are generated by powerful magnetic fields in the accretion disk or in the accretion flow of matter, while it is possible that the jets are collimated by auto-induced magnetic fields. 

The jets are site of production of high-energy particles, probably  via \ac{dsa} process: at the end of the jets, where the relativistic plasma encounters the \ac{ism} matter, a front shock is formed; furthermore the relativistic matter flow that impacts on this shock gives origin to a reverse shock, that travels reversely the jet; finally, observations suggest that other internal shocks are present in the jets, originated by changes in the external matter or in the jets flow. For this reason the acceleration  have to be considered as inhomogeneous, and its study is really a complex matter.

The composition of the jets is still under debate: while the presence of high-energy \Pelectron (up to hundreds of \GeV or more) is clearly demonstrated by the strong radio emission of the jets, whose origin is clearly synchrotron radiation, it is not clear if positive charged matter is composed by \APelectron or by \Pproton.  As a matter of fact, both models are able to explain the observed gamma spectra, that is a power-law with a high-energy cut-off. At now, the presence of barionic matter has been proved for two sources, SS~433 (\citet{ss433}) and 4U1630-47 (\citet{trigo-jets}).

Micro-quasar are not strong \ac{cre} sources, \citet{fan-eccesso-interp} suggested that they could account  to less than 10\% of the total \ac{cr} luminosity,  and the interest about them rises by the possibility that their production could be strongly unbalanced towards \APelectron, if it is dominated by the interaction of  \Pproton with \Pphoton  produced via synchrotron emission by the high-energy \Pelectron:

\begin{align*}
\HepProcess{\Pproton + \Pphoton \HepTo \Delta^+  \HepTo \begin{cases} \Ppizero + \Pproton & \simeq \mathrm{(50\%)} \\ \Ppiplus + \Pneutron   & \simeq \mathrm{(50\%)} \end{cases}} \\
&\mspace{-130.0mu}\hookrightarrow \APmuon + \Pnum\\
&\mspace{-90.0mu}\hookrightarrow \APelectron + \Pnue +\APnum
\end{align*}
with both branches that can, if energy is sufficient, produce one or more additional \Ppiplus~\Ppiminus pairs or \Ppizero. This process has a minimum threshold energy, $\sim 145~\MeV$ in the \Pproton rest frame, below which only photo-pair production $\HepProcess{\Pproton + \Pphoton \HepTo \Pelectron + \APelectron +\Pproton}$ is possible.

This situation would require that \Pproton are a significant component of the jets but that the ratio \Pproton/\Pelectron is not too high, otherwise the target photon field would be too weak to make \Pproton\Pphoton interaction relevant with respect to \Pproton\Pproton, even if the matter density is not too high, as shown for example in the different models presented in~\cite{vila-jets}. 

The resulting \APelectron spectrum would be $\propto E^{(\gamma -4)}$, with $\gamma$ index of the photon spectrum (typically observed in the 1.5 -- 1.8 range),  if the \APelectron energy-loss distance is much smaller than the typical dimension of the production region, as expected for the relativistic jets (otherwise, the injection spectrum of \APelectron would be $\propto E^{(\gamma -3)}$). Because the estimated energy of \Pproton in the jets could extend up to 1~PeV, the \APelectron spectrum would extend to energies significant higher than that of \Pelectron.

The eventual presence of a spectral break in the \Pphoton spectrum would lead to a related spectral break in the \APelectron spectrum. The threshold  of $\Delta^+$ production is $E_pE_\gamma>0.14\, \delta_D^2~\squaren\GeV$, where $\delta_D$ is the observed Doppler shift factor of the jets; on average, the energy of the produced \Ppiplus will be $0.2\,E_p$, and it can be considered as equally shared between the four produced leptons, so that $E_{\APelectron} = 0.05\,E_p$.  Therefore, if \Pphoton have a break at energy $\epsilon_b$,  the energy at which the production of \APelectron will become dominated by the interaction of \Pproton with the more numerous low-energy \Pphoton will be $\frac{0.007\,\delta_D^2}{\epsilon_b}~\GeV$. This would lead, for $\epsilon_b = 0.1~\MeV$ and $\delta_D=3$ (values proposed in~\citet{elettroni-microquasar}), to a break in the \APelectron spectrum at 630~\GeV. Indeed, all models proposed for example in~\citet{vila-jets} and many observation of such objects show a spectral break at sub-MeV energies.

\subsection{Dark Matter}\label{dm-section}
\ac{cre} can be produced also by annihilation or decay of \ac{dm} particles (see \citet{panov} or \citet{serpico} for a brief review of the topic).
 The most recent theories suggest that \ac{dm} can be composed of weakly interacting non-Standard Model particles, like for example neutralinos. These particles are expected to annihilate, or to decay, forming \Pphoton\Pphoton or \Pleptonplus\Pleptonminus or other pairs. In the simplest hypothesis, a uniform halo of \ac{dm} surrounding the galaxy where \Pelectron\APelectron pairs are produced with fixed energy ($\mathcal{Q}(E)=\mathcal{Q}_0\delta(E-E_0)$), the expected spectrum at Earth is:

\begin{equation}
N(E)=
\begin{cases}
\propto \frac{\mathcal{Q}_0}{E^2}  & E<E_0 \\
0 & E>E_0
\end{cases}
\end{equation}

If the source spectrum of \ac{dm}-produced  \acp{cre}  is not a  $\delta$ function (for example if the decay/annihilation produces \Pmuon\APmuon or \Ptauon/\APtauon pairs), the observed flux will have a more complex energy dependence, however it will maintain very specific features  both in the total particle flux and in the \APelectron fraction, in particular the flux will show a quick drop above the mass of \ac{dm} particles.

The main problem of all \ac{dm} models is the magnitude of the expected \ac{cre} flux, which is proportional to the square of the \ac{dm} density $n$ and to the thermally averaged cross section $<\sigma v>$ (for annihilating \ac{dm} models), or to the decay rate $\Gamma_{dec} =1/\tau$ and to $n$ (for decaying \ac{dm} models). The values of $n$ and $<\sigma v>$ can be calculated from cosmological observations, with the \ac{dm} being a thermal relic from early universe, and the resultant value of the \ac{cre} flux is at least three orders of magnitude lower than the observed \ac{cre} flux  and therefore negligible and  impossible to identify. 

Two solutions have been proposed to justify an observable flux of \ac{cre} produced by \ac{dm}: a boost in the cross section/decay rate and the presence of sub-halos of \ac{dm}.

The first solution involves  a physical mechanism (some are referenced in~\cite{panov}, the most popular is the Sommerfeld enhancement presented in \citet{sommerfeld}), active at the low energies of galactic \ac{dm} but not in the early universe, which  increases the annihilation cross section  or the decaying rate by a boost factor. Such a mechanism would produce a measurable flux of \ac{cre} without significantly affecting the cosmological calculation of the \ac{dm} abundance. However, an increase in the \ac{dm} annihilation cross section/decay  rate would result in an increase in the production of   $\gamma$-ray, either directly produced or produced via inverse Compoton scattering of produced \Pelectron \APelectron: searches for the \Pphoton emission in dwarf spheroidal satellite galaxies of the Milky way (\citet{fermi-dm2,fermi-dm3}), in the Milky way halo (\citet{fermi-dm,fermi-dm3})  have put strong constraints on this process, and the upper limits on  $<\sigma v>$ or on $\Gamma$ almost exclude that a diffuse \ac{dm} halo can be the main responsible of an observable flux of  \Pepm.

The second proposed solution is the presence of sub-halos in the \ac{dm} halo (see \citet{dm-clumps} for an extensive review of the topic). Numerical simulations of cold \ac{dm} scenarios, like~\citet{clump-dm}, show that primordial \ac{dm} structures that formed the halo merged only incompletely, therefore the halo is clumpy. \citet{clump-dm} calculate that in the halo there should be $\sim 10^4$ sub-halos with a mass of $\sim 10^6~M_\odot$, inside which the \ac{dm} density is from hundreds to thousands times the average density in the galaxy. Because the \Pepm production is $\propto n^2$, it would be detectable if the Sun would be sufficiently close to a \ac{dm} clump. \citet{panov} estimates that a feature in the \ac{cre} spectrum would be observable for clumps with distance from the Sun 100 -- 500~\pc, a value that has  to be confronted with the average distance between sub-halos, that is of the order of 10~\kilo\pc\ for the calculated number of sub-halos. However, \citet{no-dm-clump} sustain that this model is very unlikely; furthermore  this model, together with the similar assumption that the \ac{dm} abundance is not uniform in the Galaxy, would result in a \Pphoton flux  sufficiently high to be detected by \fermi~\cite{no-dm-clump}.

 \chapter{History of gamma-ray astronomy}\label{storia-gamma}

\section{Gamma-ray astronomy}
Gamma-ray astronomy is a branch of Astrophysics that experienced a large development starting from the second half of the last century, principally because of technical advancements that made it possible to solve the challenges posed by the observation of cosmic gamma radiation. Cosmic gamma-rays are a fundamental probe for the study of high-energy sources in the Universe because, unlike charged cosmic-rays, they are not deflected by the galactic magnetic field, therefore cosmic gamma-rays observation can be used to collect direct information on the source of production and on the traversed matter. Furthermore, as stated in \S~\ref{cr-chapter}, gamma rays produced in the interaction of \acp{cr} with the interstellar matter are an independent way to study the \ac{cr} population and its propagation in the galaxy.

The observation of gamma rays is fundamentally different from that of photons of lower energy, because high-energy photons can not be collected and concentrated  on a detector using reflective or refractive systems.  Therefore, gamma-rays observation  requires instruments of completely different concept.

The dominant interaction process for photons, above a few tens of \MeV s,  is the production of \Pelectron\APelectron pairs: current gamma-ray telescopes rely on the direct or indirect detection  of the produced particles  to measure energy and direction of the incident photon.

\begin{figure}[htb!]
\begin{center}
\includegraphics[width=.88\textwidth]{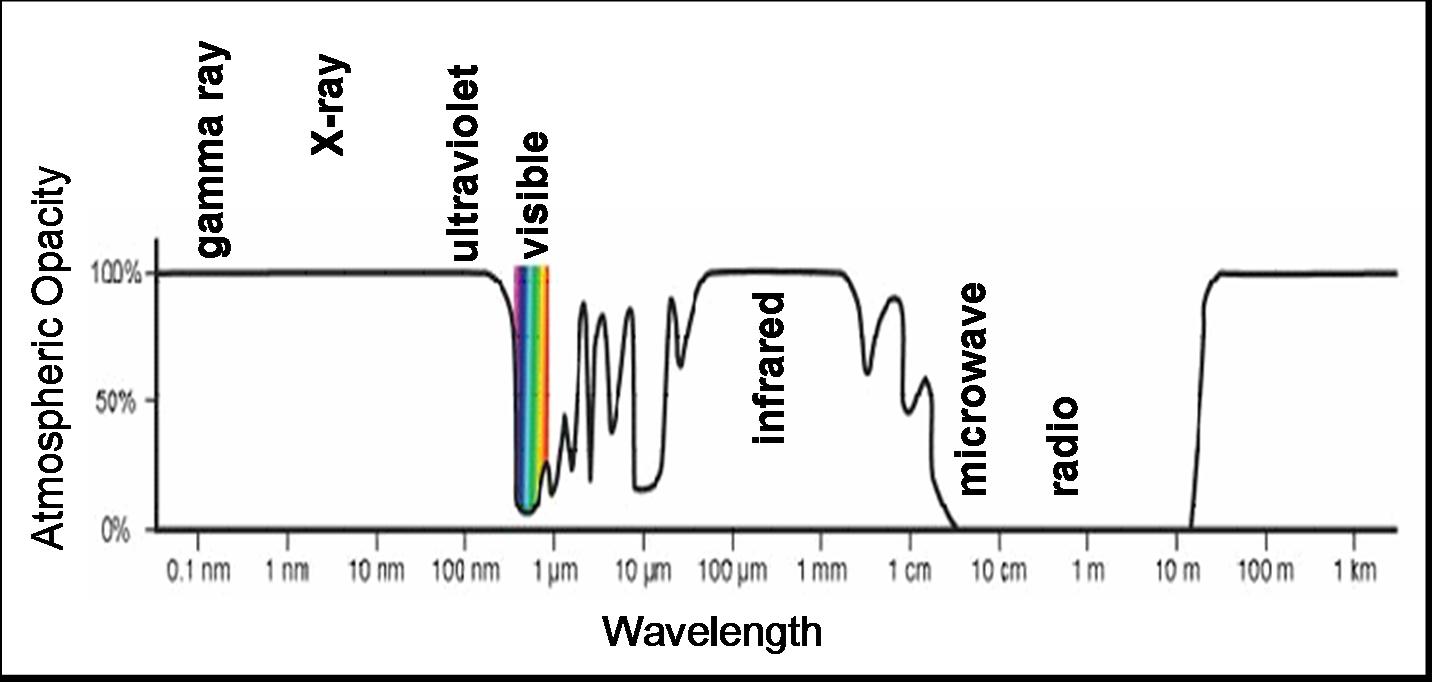}
\caption{Opacity of the atmosphere as a function of photons wavelength.~\cite{immagine-opacita}}
\label{atm-opac}
\end{center}
\end{figure}

Direct observation of gamma rays on  ground is not possible, because they are absorbed by the atmosphere (see figure~\ref{atm-opac}). This led to the development of two complementary approaches in the detection of high-energy photons, that are described in the following sections: space-based detectors directly collect gamma rays, while ground-based detectors reconstruct the shower developed by high-energy photons converting in the atmosphere. 

\subsection{Ground-based telescopes}\label{ground-telescopes}

When converted in the atmosphere, high-energy gamma rays produce an electromagnetic shower. If the shower is sufficiently extended (that is, if the energy of the incident photon is sufficiently high), the \v Cerenkov light emitted by charged particles in the shower can be detected by telescopes on ground. This type of instrument, first proposed in 1953 and named \acf{act}, is composed by a large mirror that collects the \v Cerenkov light  (principally emitted in the blue-UV wavelengths), concentrating it  on a collecting device, placed in the focus of the mirror. The first \acp{act} were totally unable to separate showers produced by gamma-rays  from the dominant background of showers originated by charged cosmic-rays, and only in recent years the development  of imaging detectors, that are able to reconstruct  the shower development in the atmosphere,  made these instruments fundamental in the study of gamma radiation. The second generation of \acp{act} (HESS, MAGIC, VERITAS, CANGAROO-III) are  all composed by two or more telescopes (stereoscopic configuration) and are therefore able to make a 3D-reconstruction of the shower, with great improvement in the ability of discriminating between electromagnetic and hadronic showers. These \acp{act} are a fundamental counterpart of space-based observations. 

The main advantage of \acp{act} with respect to space-based instruments is the very large effective area (of the order of $10^4~\squaren\meter$ with respect to $<1~\squaren\meter$ of LAT) obtained using  the atmosphere as converting medium. This huge effective area allows the detection of the really faint flux of photons emitted by very-high energy photon sources. Furthermore, the angular resolution in \acp{act} is  better than in space-based telescopes. 

The detection of \v Cerenkov light establishes intrinsic limitations to the use of these instruments. First, observations are possible only in  dark nights, therefore the duty cycle of an \ac{act} is of the order of 10\%;  on the contrary, space-based instruments have no intrinsic limitations and their duty cycle could virtually reach $100\%$. As an example, the \ac{lat} has an $\sim88\%$ duty cycle, principally because of the presence on the \fermi\ orbit of the \acf{saa}, a region where the intrument is keep active. (see \S~\ref{saa} for more information about this). 

Second, these instruments have a small acceptance cone (of the order of a few degrees) and can not perform a survey of the whole sky, but only  observations pointed to a very restricted area. This last feature  clearly marks the complementary role of ground-based and space-based detectors, where the former are (in most cases) pointed to sources detected by the latter in survey observations, to study their spectrum at energies that cannot be reached on space observatories, or to spatially resolve extended sources thanks to their better spatial resolution. 

Third, to be observable on ground a photon  must produce a sufficiently extended shower, therefore the minimum observable energy is limited. The lower boundary of the  energy range depends on the characteristic of the instrument,  from the brightness of the observed source and from its characteristics. The detection limit of current instruments is typically  above $\sim100~\GeV$, although MAGIC has performed specific observations down to  25~\GeV. Next generation instrument CTA (\v Cerenkov Telescope Array) is expected  to  detect photons down to 10~\GeV.
 
Finally,  space detectors are surrounded by an anti-coincidence shield that detects the incoming charged particles. This instrument gives  the main contribution  to the rejection power of space observatories, and makes it possible to obtain  samples of photons with reasonably small residual contamination. Because observed photons convert in the atmosphere, \acp{act} can not have an anti-coincidence shield, and particle discrimination can only be made using the different development of hadronic and electro-magnetic showers. Therefore,  signal in \acp{act} is strongly dominated by charged particle: even if hadronic showers are efficiently rejected, there is almost no way to know if an electro-magnetic shower is originated by a photon or by an electron, and electron flux is one or two order of magnitude greater  than photon flux. For this reason, \acp{act} can only observe  gamma sources that have limited spatial extension, for which  the photon signal can be estimated  via background subtraction, using  simultaneous observation of the source (on-source observation) and of close regions (off-source observation).

\subsection{Pair conversion telescopes}
\begin{figure}[htb!]
\begin{center}
\includegraphics[width=.75\textwidth]{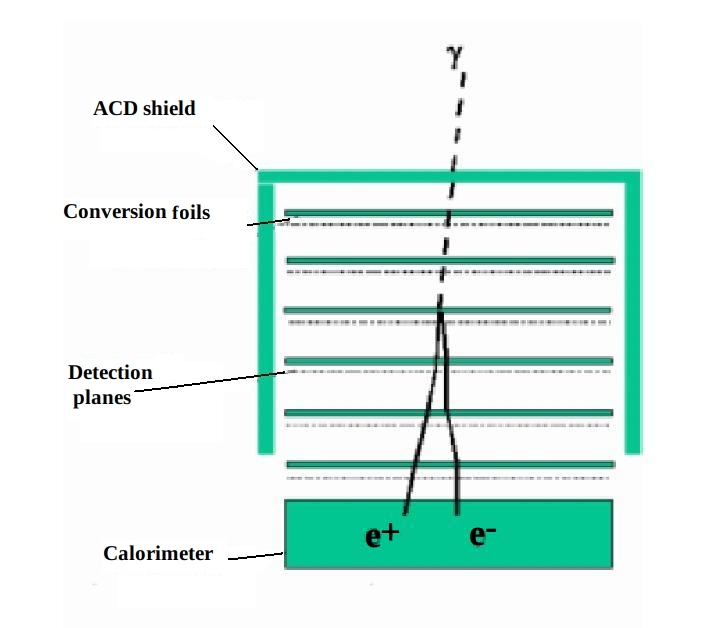}

\caption{Schematic description of the working principle of a pair-conversion telescope.~\cite{latdoc}}
\label{lat-schema}
\end{center}
\end{figure}

A typical space-based gamma telescope is an instrument where gamma photons are converted and absorbed. With the exception of early detectors, where only the energy of the incident photon was measured (or at least constrained in a known energy range) and the direction was only constrained  by the geometrical acceptance of the instrument, pair-conversion telescopes are equipped with  a tracker, where photons are converted in \Pelectron\APelectron pairs and the direction of produced particles is measured. Below the tracker  a calorimeter  absorbs the produced particles  and measures their energy. Typically, a high Z material is inserted in the tracker to increase the probability of gamma conversion. The instrument is surrounded by an anti-coincidence detector that detects incoming charged particles, that can therefore be rejected. Figure~\ref{lat-schema} shows a schematic gamma detector with an ``ideal'' gamma conversion.

Compared to \acp{act}, space telescopes clearly have better capability to reject charged particles; their \ac{fov} depends basically on the geometry of the instrument, that is, by  its capability to detect and reconstruct  events coming from a given direction, and can therefore be very large. The lower limit of detectable energy is set primarily  by the intrinsic difficulty of reconstructing an event that is quickly absorbed in the detector; for the \ac{lat} it is between 30 and 100~\MeV, depending on the scientific requirements of the observation. At high energy, problems arise because of the limit in weight imposed to a rocket-launched instrument. Even for a large instrument like the \ac{lat},  the electromagnetic shower is not fully contained in the instrument above a few \GeV\ for normally-incident photons  and  about half of the energy escapes from the back of the calorimeter at $\sim100~\GeV$~\cite{pass7-status}. For this reason the \ac{lat} uses a complex algorithm to reconstruct the initial energy of the incident photon: with this method, the \ac{lat} is able to reconstruct event with more than 500~\GeV\ of initial energy. Finally, the limited dimension of the instrument results in reduced effective area: because typical high energy sources have a power-law energy spectrum, the photon flux decreases as the energy increases, so that at high energy the statistic collected by a space-telescope can become really small.   

\section{History of space-based  gamma experiments}
\begin{figure}[htb!]
\begin{center}
\includegraphics[height=9cm]{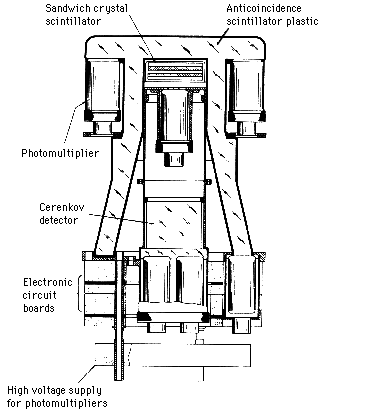}

\caption{Scheme of the gamma-detector placed  on Explorer 11.~\cite{explorer11}}
\label{explorer11-fig}
\end{center}
\end{figure}

The first space-based gamma detector was on-board of  \textbf{Explorer 11} satellite~\cite{explorer11}, launch\-ed on \nth{27} April  1961 on an eccentric orbit, with altitude comprised between 300 and 1100~miles. The gamma detector on-board  of Explorer 11 was a cylindrical sandwich of CsI and NaI crystal scintillators and a Lucite \v Cerenkov counter for total 20~in of height and 10~in of diameter. The entire instrument was surrounded by a plastic anti-coincidence scintillator (figure~\ref{explorer11-fig}) and had an effective area of the order of 7~\squaren\cm\ above 200~\MeV. The instrument was designed to identify photons of energy above 50~\MeV, and was not able to reconstruct their direction, which was only constrained by the solid angle determined by the two detectors, $\sim17^\circ$ half angle.  Explorer 11 was put on a tumble in order to scan the whole celestial sphere. The satellite operated until early September, when problems in the power supply shortly led to the end of data collection. Because of the failure of the on-board tape recorder during the launch, Explorer 11 was not able to store data on-board, therefore data could only be collected when the satellite was connected to a gro\-und receiving station. This, together with\- the problems in the power supply and with the crossing of the orbit with the radiations belt around the Earth, that periodically ``blinded" the instrument,  restricted the effective data-taking of the mission to only 141 hours, during which 1012 events were accepted as photons (\citet{explorer11result}). Among the collected photons, only 31 were of probable extra-terrestrial origin, that is, not produced by interaction of \acp{cr} with  Earth atmosphere. Because of the small amount of collected  extra-terrestrial photons, no definite clear anisotropy was detected.

\begin{figure}[htb!]
\begin{center}
\includegraphics[height=9cm]{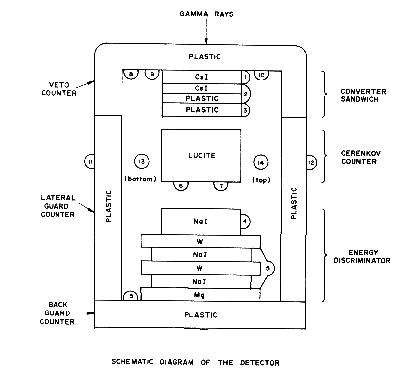}

\caption{Schematic view of the high-energy photons detector on-board of the satellite OSO-3.~\cite{oso3}}
\label{oso3}
\end{center}
\end{figure}

\vspace{2cm}
The \nth{3} Orbiting Solar Observer (\textbf{OSO-3}~\cite{oso3})  was launched on   \nth{8} March  1967 on an almost circular orbit at 550~\kilo\meter\ of altitude, inclined at $33^\circ$ with respect to the equatorial plane. It had a cylindrical symmetry, and rotated on his axis in 1.7~\second.  OSO-3 was principally  developed to study the Sun, therefore its  main instrument,   the hard X-ray detector developed by the University of California, San Diego, was always pointed to the Sun. This detector was composed by a NaI(Tl) crystal enclosed in an anti-coincidence CsI(Tl) crystal, mounted on the ``sail'' of the satellite, that was always exposed to the Sun. In addition to the hard X-ray detector, a gamma detector was mounted on the rotating wheel of OSO-3, therefore being able to scan the whole sky because of the rotation of the satellite. This detector (figure~\ref{oso3}), assembled at MIT, was a counting telescope designed to detect photons above 50~\MeV. The detector was composed by a sandwich of CsI and plastic scintillators, a \v Cerenkov directional counter for discriminating between forward and backward coming events, and a sandwich of Tungsten and NaI scintillators for energy discrimination. All the instrument was enclosed in a plastic scintillator dome. An event was accepted if a signal was registered by all  three detectors,  without any signal in  the front and lateral plastic scintillators. The direction of incoming gamma could only be restricted in the $\sim25^\circ$ acceptance cone of the instrument; its effective area was about 2.5~\squaren\cm\ at 100~\MeV\ for on-axis photons. OSO-3 operated continuously until \nth{27}  June   1968, when  only real-time data collection was possible because of the failure of the last tape recorder, and the gamma detector did not  work anymore. The last data were received on  \nth{10} November   1969. 

During its mission, OSO-3 collected 621 events~(\citet{oso3result}). The distribution of the events was clearly an-isotropic, with the photons concentrating around the galactic center and along the galactic plane.

\begin{figure}[htb!]
\begin{center}
\subfigure[The detector on-board of SAS-2..~\cite{sas2dati}]{\includegraphics[height=8cm]{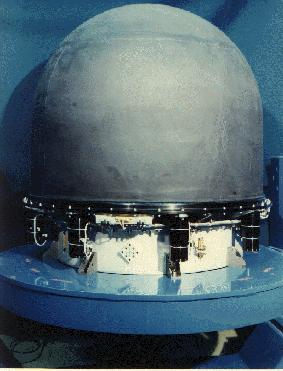}}
\subfigure[Scheme of the detector.~\cite{sas2dati}]{\includegraphics[height=8cm]{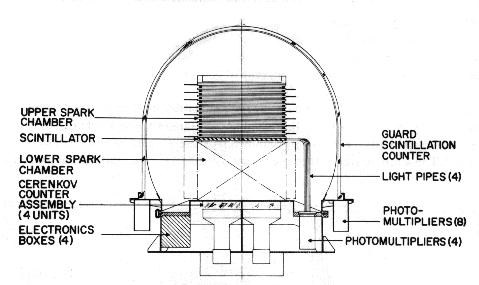}}

\caption{The SAS-2 observatory.}
\label{sas2}
\end{center}
\end{figure}

\vspace{2cm}
The second Small Astronomy Satellite (\textbf{SAS-2},~\cite{sas2dati}), also known as SAS-B and Explorer 48, was a dedicated mission to study the gamma emission in the sky. It was launched on  \nth{19} November  1972 on a near equatorial orbit, with an inclination of less than $2^\circ$. This orbit was chosen to minimize the background due to charged particles. Its altitude was comprised between 440 and 610~\kilo\meter, and the orbital period was of 95~minutes. The only on-board experiment was a gamma telescope designed to detect photons between 20~\MeV\ and 1~\GeV. With respect to previous experiments, SAS-2 was able to  reconstruct the energy of incident photons (roughly), and their direction. The detector (figure~\ref{sas2}) was composed by two 16-levels wire spark chambers, separated by a plane of four plastic scintillators. Below the second spark chamber there were four \v Cerenkov directional counters. Both the \v Cerenkov and the plastic scintillators had the purpose of triggering on forward-coming particles. The instrument was completely surrounded by a dome of plastic scintillator. The spark chamber modules were interleaved by 0.010~\cm\ tungsten foils: the purpose was to increase the conversion efficiency of photons and also to allow a rough measurement of the event energy through the Coulomb scattering of produced particles. The detector had an acceptance of $\sim30^\circ$ and operated in pointed mode; its effective area was 120~\squaren\cm\ at 300~\MeV, its \acf{psf} was $5^\circ$ at 20~\MeV\ and $1.5^\circ$ at 100~\MeV.

SAS-2 was expected to make a survey of the whole sky during 1~\yr\ of mission, but a failure in the low-voltage power supply ended data collection on \nth{8} June  1973. However, during the few months of operation, SAS-2 achieved important scientific results (\citet{sas2result}). Its angular resolution allowed the correlation of the gamma emission in the galactic plane with the galactic structural features, providing very strong evidences about high-energy photons production  through the interaction of cosmic ray with \acf{ism}. SAS-2 detected for the first time a high-energy component in the diffuse celestial background (above 35~\MeV). Finally, SAS-2 made the first detection of point sources of gamma rays, detecting the Crab and Vela pulsars, and an unknown source that was later identified as the radio-quiet Geminga pulsar.

\vspace{2cm}
The Cosmic-ray satellite, option ``B" (\textbf{Cos-B}~\cite{cosbdati,cosbdati-esa}), the first ESA mission devoted to the study of gamma rays, was launched on \nth{9} August 1975, on a highly eccentric orbit perpendicular to the equatorial plane. The orbit, with altitude comprised between 350 and 100000~\kilo\meter, was chosen to minimize the time spent by Cos-B in the radiation belts around the Earth, where the instrument can not be operated. This choice, that allowed to operate the instrument for 25~hours every 37~hours-orbit, precluded the observation of $\sim45\%$ of the celestial sphere. 

\begin{figure}[htbp]
\begin{center}
\includegraphics[height=9cm]{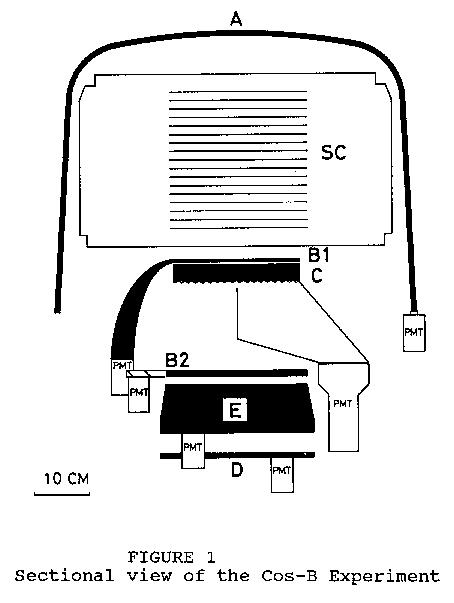}

\caption{Scheme of the \textit{Gamma-ray Telescope} on board  Cos-B.~\cite{cosbdati,cosbdati-esa}}
\label{cosb}
\end{center}
\end{figure}

The main instrument of Cos-B was the \textit{Gamma-ray Telescope} (see figure~\ref{cosb}), a pair-conversion  telescope composed by a magnetic-core, wire-matrix spark chamber, the trigger of which was provided by the coincidence of 2 scintillator counters and one directional \v Cerenkov counter  placed  below it~\cite{ric}. Every layer of the spark chamber was composed by two planes, each one formed by 192 wires spaced by 1.25~\milli\meter, and was able to reconstruct both the X and the Y position of the crossing particle. Between the layers a Tungsten plane was placed, to increase the gamma conversion efficiency, for total 0.4 radiation lengths. To minimize the effect of gas aging with time, Cos-B was provided with a mechanism to completely change the gas of the spark chamber, so that it was still active when the instrument was finally turned off. Below the spark chamber there were 4 layers of Mo, equivalent to 0.5 radiation lengths.  The device was entirely surrounded by a plastic scintillator counter for charged particles rejection. Cos-B was the first space instrument  to measure the energy of incident photons using an electromagnetic calorimeter, placed below the spark chamber and composed by CsI scintillator, for total 4.7 radiation lengths. On the bottom of the instrument, a plastic scintillator measured the rate of events that were not fully contained by the calorimeter and exited from the back (this meant an energy $\approx>300~\MeV$). 

In addition to the main instrument, an X-ray counter, with sensitivity 2-12~\keV, was placed on the side of the instrument, to provide synchronization on possible short-period pulsations of gamma-ray emission from X-ray pulsating sources. The pulsar synchronizer has also been used for monitoring the intensity of radiation from X-ray sources.

Cos-B gamma-telescope was sensitive to photons of energy comprised between $\sim30~\MeV$ and $\sim5~\GeV$, with an energy resolution of 40\% FWHM at about 100~\MeV\ and less than 100\% up to 5~\GeV. The effective area peaked at 50~\squaren\cm\ at several \MeV\ for on-axis photons, falling rapidly as the angle with respect to the axis of the instrument increases. Cos-B can not detect photons with angle greater than $30^\circ$; the resulting   \ac{fov} was $\sim2~\sr$. The small \ac{fov} leads Cos-B to be operated in pointed-mode, with subsequent pointed observation lasting for several weeks. The angular resolution was $\sim3.5^\circ$ at 100~\MeV\ and $\sim1^\circ$ at 300~\MeV.


Cos-B mission was initially planned to last for 2~years, but it was extended until \nth{4} April 1982, when the instrument was finally turned off because of the end of the fuel supply used for pointing the spacecraft. It had operated for 6~years and 8~months, and collected more than 600,000 photons. 

\begin{figure}[htbp]
\begin{center}
\includegraphics[width=1.\textwidth]{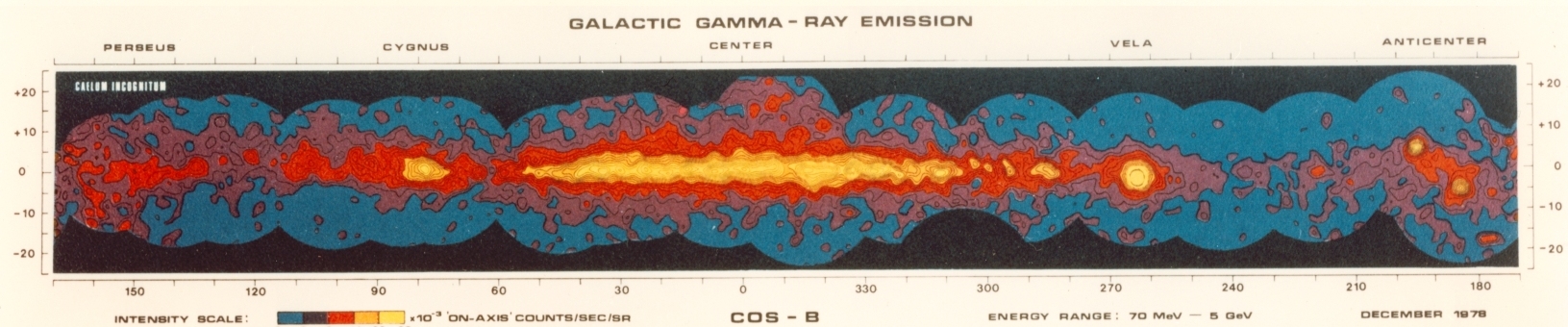}

\caption{Map of the galactic diffuse gamma emission, extracted by Cos-B data.~\cite{cosb-immagine-galassia}}
\label{galassia-cosb}
\end{center}
\end{figure}

Cos-B observed 25 gamma point-sources~(\citet{cosbresult},\citet{cosbresult2}), measuring  the spectrum of Vela, Crab and Geminga pulsars (the last was not known as a pulsar yet). These observations were collected in  the 2CG catalog~(\citet{cosb-catalogo}), the first published catalog of gamma-ray sources. Variability in the flux of Crab and Vela was observed, while a prolonged search (10\% of the total mission time) for variability in Cygnus-X-3, a binary pulsar which had an observed variability in the X-ray band, gave no result. Finally, Cos-B made the first complete map of the gamma emission from the galactic plane, reported in figure~\ref{galassia-cosb}.

\vspace{2cm}

\label{egretpage}

The success of the Cos-B mission persuaded NASA to develop the \textbf{\acf{egret}}~(\citet{egretdoc}), a gamma telescope with one order of magnitude greater sensitivity and better angular and energy resolution, that was placed on-board of the high-energy photon observatory \textbf{CGRO} (Compton Gamma Ray Observatory). 

\begin{figure}[htb!]
\begin{center}
\subfigure[Photo of CGRO taken from shuttle Atlantis during on-orbit placing.~\cite{cgro-orbita,cgro-strumenti}]{\includegraphics[width=.68\textwidth]{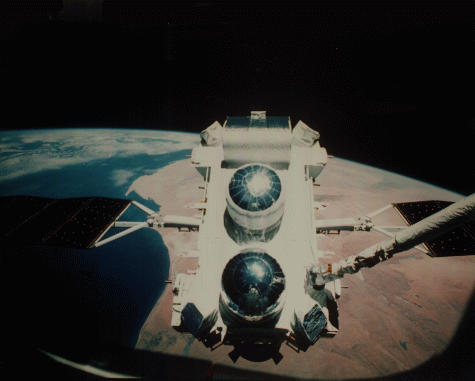}}
\subfigure[Main instruments of CGRO; only six on eight modules of BATSE (red) are visible.~\cite{cgro-orbita,cgro-strumenti}]{\includegraphics[width=.68\textwidth]{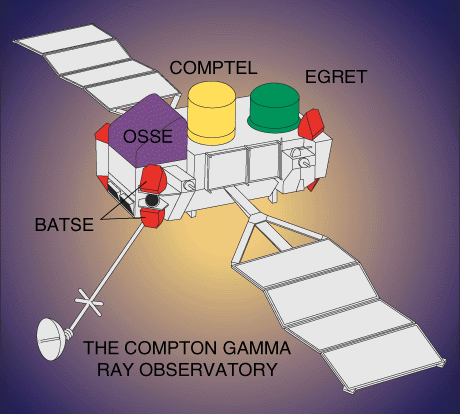}}
\caption{The CGRO observatory.}
\label{cgro}
\end{center}
\end{figure}

CGRO, placed in orbit by the shuttle Atlantis on  \nth{5} April   1991 and de-orbited on \nth{4} June   2000 following a failure in the gyroscopic stabilization system, was a gamma-ray observatory, equipped with four different instruments to cover a large energy range. These instruments, shown in figure ~\ref{cgro} were: the Oriented Scintillation Spectrometer Experiment (OSSE), designed to detect photons in the 60~\keV--10~\MeV\ range, the Imaging Compton Telescope for 1--30~\MeV\ photons, the Burst And Transient Source Experiment (BATSE) designed to detect gamma transient phenomena (like Gamma-ray bursts)  thorough the constant survey of the whole sky in the 20~\keV--100~\MeV\ range and \ac{egret}, sensitive to 30~\MeV--30~\GeV\ photons. 

\begin{figure}[htb!]
\begin{center}
\includegraphics[width=.8\textwidth]{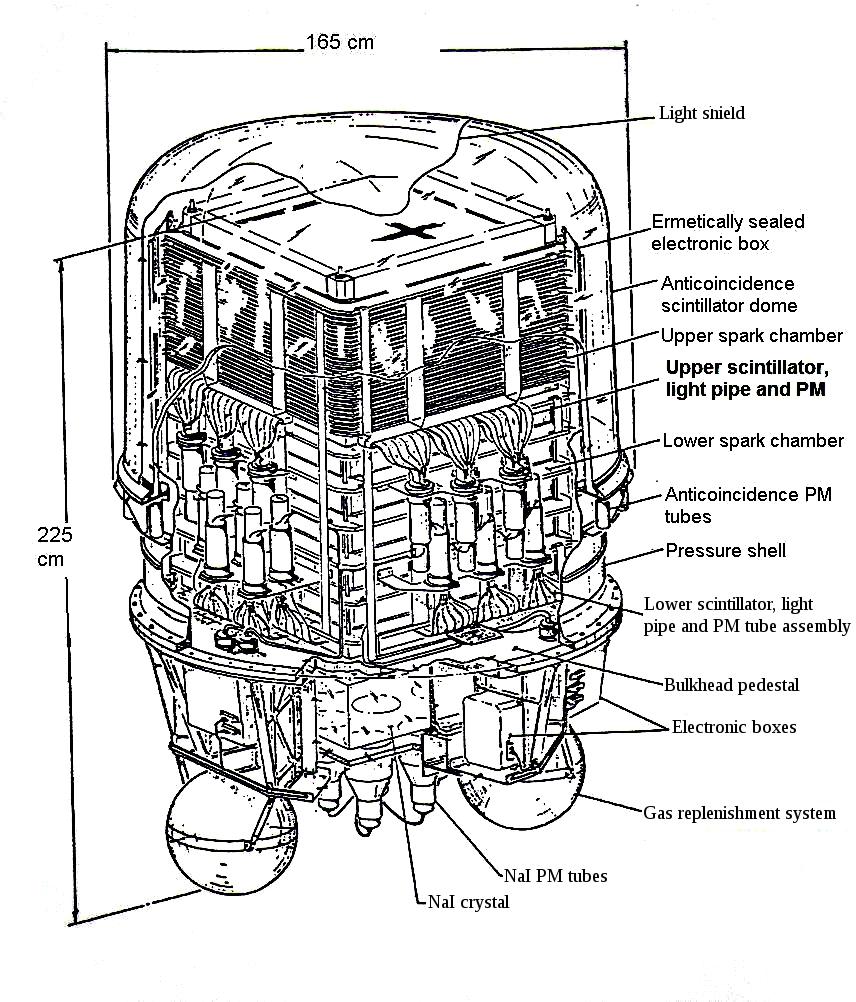}

\caption{Scheme of the EGRET detector (source:NASA)}
\label{egret-fig}
\end{center}
\end{figure}

\ac{egret} (figure~\ref{egret-fig}) had two spark chambers. The first, composed of 28 modules, detected the conversion point and the initial direction of generated particles. Conversion efficiency was increased by 27 Tantalum foils, 0.02 radiation lengths thick,  placed between the chamber plates. The lower chamber had 8 widely spaced planes, its purpose was to measure the angular separation and energy re-partition between \Pelectron and \APelectron, if particles could be spatially resolved. Furthermore, the chamber detected the entry point of particles in the underlying calorimeter. The trigger for the spark chambers was provided by a \ac{tof} system composed of two planes of 4x4 plastic scintillators, placed above and below the lower chamber, that detected downward-moving particles. The energy was mainly determined in the calorimeter placed below the lower spark chamber, with some aid from the measurement of the Coulomb scattering in the spark chambers. The calorimeter was a monolithic instrument of  $76\times76~\squaren\cm$, made  of  NaI(Tl) scintillators for total 7.7 radiation lengths. The active part of the detector was surrounded by a plastic scintillation dome. Because of gas aging in spark chambers, \ac{egret} was equipped with a mechanism to completely replace the gas in the chambers, and a gas provision for 4 complete replacements. After 1995, because of the end of the gas provision, \ac{egret} was used irregularly.

\ac{egret} energy resolution was 20\% (FWHM) over the central part of the energy range, degraded to about 25\% above several \GeV\ and below 100~\MeV, because of the incomplete absorption of the shower in the calorimeter and of relevant ionization losses in the spark chambers respectively. Effective area was about 1500~\squaren\cm\ at several hundred \MeV\ for on-axis photons, angular resolution improved from $10^\circ$  at 60~\MeV\ to $0.5^\circ$ at 10~\GeV.

\begin{figure}[htb!]
\begin{center}
\subfigure[The gamma sky seen by EGRET]{\includegraphics[width=.8\textwidth]{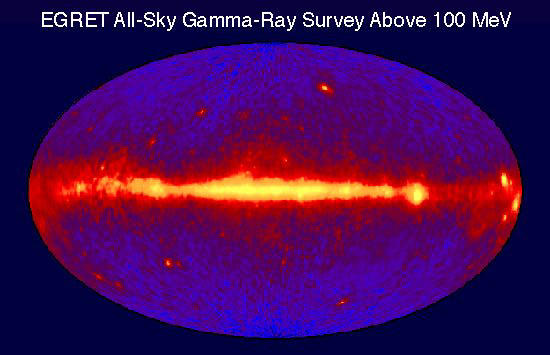}}
\subfigure[The \nth{3} EGRET source catalog]{\includegraphics[width=.8\textwidth]{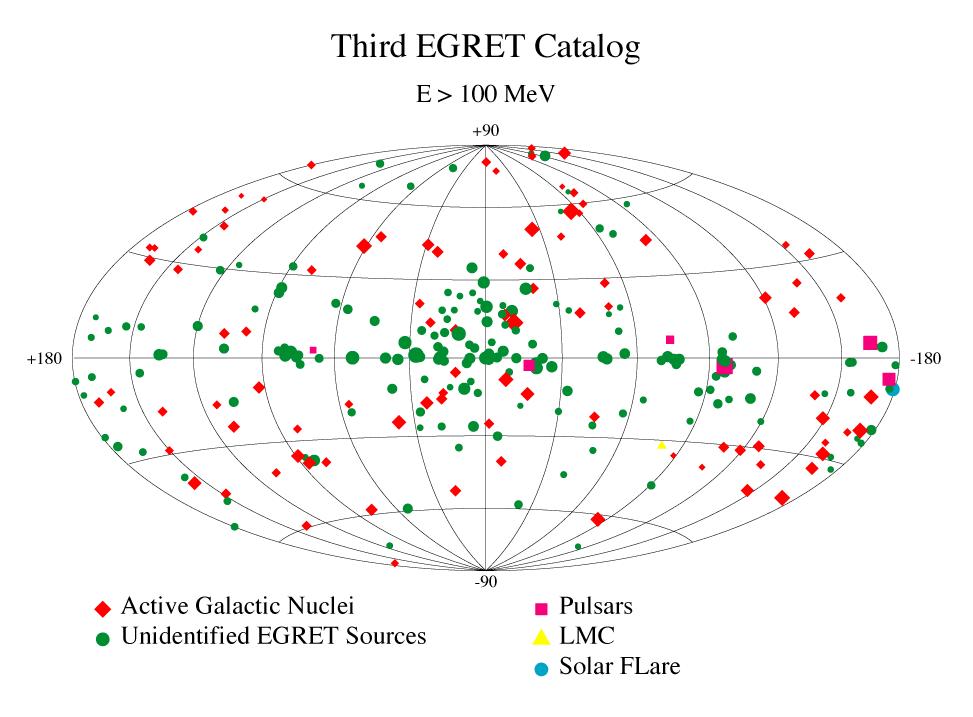}}
\caption{Results of the EGRET mission. Source: NASA.}
\label{egretbis}
\end{center}
\end{figure}

\ac{egret} was the first experiment to perform a complete survey of the sky in the gamma wavelength: the third \ac{egret} catalog (\citet{egretCat}) included 271 sources above 100~\MeV\ (see figure~\ref{egretbis}, bottom), 100 of which were identified with known sources in other wavelengths. A large part of \ac{egret} identified sources (94 out of 100) were \textit{blazars}, extra-galactic objects that were observed as gamma emitters for the first time by \ac{egret}. Moreover, \ac{egret} made a detailed map of the gamma emission from the Milky Way (figure~\ref{egretbis}, top), detected a strong solar flare and detected the emission of high energy $\gamma$ rays from a gamma ray burst for over an hour, with some gamma rays having energies over a \GeV\ and two having energies over 10~\GeV.

\clearpage
\vspace{2cm}
The first  \Pphoton-ray observatory   operated after the end of the \ac{egret} mission was the Italian mission \textbf{\acf{agile}} (\citet{agile-doc}), launched on an almost equatorial orbit (inclined by 2.5\degree) at an altitude of 535~km on \nth{23} April  2007, with the start of observation activities on \nth{1} December  2007. The \ac{agile} mission is still operative, even though the failure of one of the gyroscopes on November 2009 made it impossible to point the instrument, which is constantly spinning ever since and can only make surveys of a large portion of the whole sky;  a map of the sky seen by \ac{agile} after 4~\yr\ of observations is shown in \cref{cielo-agile}. The mission was recently prolongated at least to the first half of 2017. 

\ac{agile} is a light detector ($\sim 100~\kilo\gram$ on a total of $\sim 350~\kilo\gram$  satellite weight), designed to detect \Pphoton-rays in the energy range 30~MeV -- 50~GeV together with X-rays in the 18 - 60~\keV\ range. Notwithstanding its reduced mass, the use of a segmented anti-coincidence detector and of a tracker based on silicon detectors (both solutions being also at the basis of the \fermi-LAT impressive performances), makes it possible to significantly improve the spatial resolution and the Field of View of \ac{egret}.

\begin{figure}[htb!]
\begin{center}
\includegraphics[width=.8\textwidth]{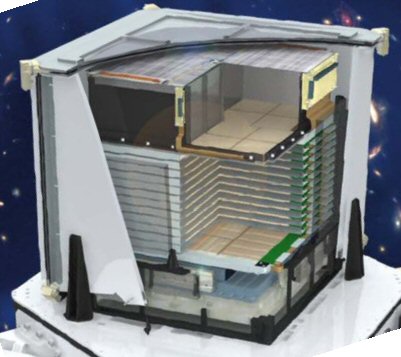}
\caption{The \ac{agile}  instrument showing the hard X-ray imager, the gamma-ray Tracker, and Calorimeter. The Anti-coincidence system is partially displayed, and no lateral electronic boards and harness are shown for simplicity.~\cite{agile-doc}}
\label{agile-figure}
\end{center}
\end{figure}

\begin{figure}[htb!]
\begin{center}
\includegraphics[width=.8\textwidth]{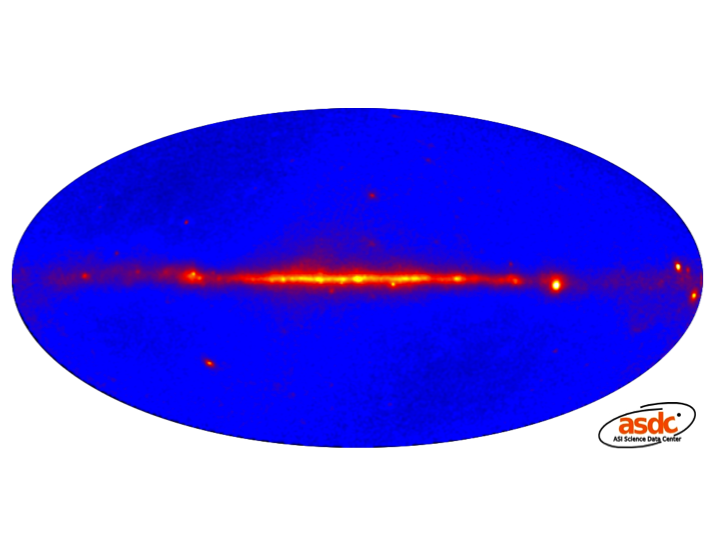}
\caption{The gamma sky seen by \ac{agile}, data collected before \nth{25} December  2012.~\cite{cielo-agile}}
\label{cielo-agile}
\end{center}
\end{figure}

\ac{agile}  is composed by three detectors~\cite{agile-doc}, a Silicon-Tungsten tracker, a MiniCalo\-rimeter (both form the GRID, Gamma-Ray Imaging Detector, the main instrument of \ac{agile}) and a Silicon-Tungsten X-ray detector (see \cref{agile-figure}), all enclosed by an anti-coinciden\-ce detector of plastic scintillator: to reduce the impact of false veto signals, generated by \acp{cr} in time with \Pphoton events or by charged particles of the electro-magnetic shower hitting the detector, the anti-coincidence detector is segmented, with 3 tiles for each side and a single tile for the top of the instrument. 

The GRID tracker is composed by 12 trays, each one supporting a couple of $38\times38~\squaren\cm$ silicon micro-strips detectors that  measure  the x and y position of particles crossing the plane; the first 10 trays also support a Tungsten foil to increase the probability of photon conversion; the total thickness of the tracker is $0.8~X_0$. The  choice of silicon detectors instead of gas detectors used in \ac{egret} is the same made for the \ac{lat}: the significant improvements introduced in terms  of detection efficiency, reduction of dead time, increase of the sensibility and several others are discussed in \S~\ref{ssd-section}.

The GRID MiniCalorimeter, placed below the tracker, is principally devoted to measure the energy of the shower generated by photon converting in the tracker; is composed by two planes of 15 CsI(Tl) crystals for a total of 1.5~$X_0$. The small thickness of the calorimeter is partially balanced by the read-out system, that reads every crystal separately and by both ends: this method collects more information than the simple value of the deposed energy, that can be used to perform a better calculation of the energy of the event; the MiniCalorimeter has a design very similar to that of the \ac{lat}, for a discussion about the advantages of a calorimeter able to make a spatial reconstruction of energy deposition see \S~\ref{cal}.
 
 The GRID is able to detect photons in the range 30~MeV -- 50~GeV, with an effective area of 200~\squaren\cm\ at 30 MeV, a \ac{fov} significantly larger than that of \ac{egret} (2.5~sr) and a better angular resolution (3.5\degree\ at 100~MeV and 0.2\degree\ at 10~GeV), while the reduced thickness of the calorimeter leads to a significantly worse energy resolution ($\frac{\Delta E}{E}\sim 1$); the dead time is significantly shorter than in \ac{egret}, 100 -- 200~\mycro\second.
 In addition to its role in the GRID, the calorimeter can be used as a self-standing detector of photon from burst events, with sensitivity in the range 350~keV -- 50~MeV.
 
 Above the GRID is placed the Hard X-ray Imager Detector (Super-Agile): it is a plane composed by a  $2\times2$ grid of $19\times19~\squaren\cm$ silicon detectors, with a thin tungsten plane placed 14~\cm\ above; it is sensitive to photons in the 18 -- 60~keV range. 

\newpage
The most important discoveries of \ac{agile} were:
\begin{itemize}
\item the  first detection of gamma-ray emission above 100~MeV from a colliding wind massive binary system in the Eta-Carinae region (\citet{agile1});
\item  the first detection of episodic transient gamma-ray flaring activity from microquasars Cygnus X-1 (\citet{agile2}) and Cygnus X-3 (\citet{agile3}) above 100~MeV;
\item  the first experimental confirmation of emission from a pulsar wind nebula, Vela-X, in the energy range from 100~MeV to 3~GeV (\citet{agile4});
\item  the     first experimental evidence of proton acceleration in a Supernova Remnant, W44, (\citet{agile-w44}), with the clear exclusion of leptonic emission models and a first evidence of the presence of the spectral bump caused by \Ppizero (the definitive detection of the \Ppizero bump was made by the LAT in \citet{fermipi0});
\item the   discovery of a spectral component up to 100~MeV in the cumulative energy spectrum of terrestrial gamma-ray flashes (\citet{agile-tgf});
\item  the  discovery, with the fundamental confirmation coming from the \fermi-LAT the following day,  of strong and rapid flares from the Crab Nebula above 100~MeV (\citet{agile-crab-flare}).
\end{itemize}

\chapter{The Large Area Telescope on-board of \fermi\ observatory}\label{capitolo-fermi}

\section{Fermi observatory}
The \fermi\ Gamma-ray Observatory~(\citet{latdoc}), formerly \acf{glast}, was launched on \nth{11} June  2008 on a circular orbit at 565 km of altitude, inclined at $25.6^\circ$ with respect to the equatorial plane. The orbit period is   96 minutes and the orbit precession time is  54.6 days. The  angle of \fermi\  with  respect to the local zenith (rocking angle), $35^\circ$ at the beginning of the mission,  $50^\circ$ starting from the second year of operation, is reverted at every orbit. This feature allows the main instrument of \fermi, the \acf{lat},  to make a complete scan of the sky every two orbits, with a duty cycle of $\sim88\%$, principally because of the presence on the \fermi\ orbit of the \acf{saa}~(\cite{saa1,saa2}). \label{saa} This extended region has  a particularly  high density  of charged particles  trapped by the geo-magnetic field, therefore  the high-voltage supply of the anti-coincidence shield of the \ac{lat} is lowered to prevent damages caused by high particle flux and no photon data is collected. Outside of the \ac{saa} the \ac{lat} data collection is almost continuous. 

The \ac{lat} and its sub-systems are described starting from \S~\ref{LAT-section}, while the secondary instrument on-board of \fermi, the \acf{gbm}, is described in \S~\ref{gbm-section}. To stress the large improvements introduced by \fermi\ in the observation of high-energy photons, comparisons will be made with the  previous similar  instrument  \ac{egret}~(\citet{egretdoc}), on board  the Compton Gamma Ray Observatory, described at page~\pageref{egretpage}. In table \ref{lattab} the main characteristic of the \ac{lat} are compared with \ac{egret}. Some of the \ac{lat} characteristics depends on the event reconstruction process, on the event selection adopted, and on the  simulation of the instrument and of the particles interaction. They are therefore subjected to changes during the mission.  Data in table \ref{lattab} are based on the \nth{3} version of the event reconstruction software, named Pass~8, that was developed after several years of \fermi\ mission, and is now used for processing the released photon data-sets in replacement of the original Pass 6 analysis. In particular, effective area, sensitivity, spatial and energetic resolutions are referred to the \acfp{irf} of Pass~8 Source class, that is designed to select a photon sample of sufficient purity to analyze faint point sources. It is important to note that  energy and angular resolutions similar to those in table \ref{lattab} can be achieved also for electrons, because they are principally related to the ability of the \ac{lat} to reconstruct electromagnetic showers. A description of the \ac{lat} \acp{irf} is given in \S~\ref{irf-sec}, while the meaning of  ``Pass'', the specific of Pass~8 and its difference from others ``Passes'' will be given in \S~\ref{passi}.

\subsection{The Gamma-ray Burst Monitor} \label{gbm-section}
\begin{figure}[htb!]
\begin{center}
\includegraphics[width=.55\textwidth]{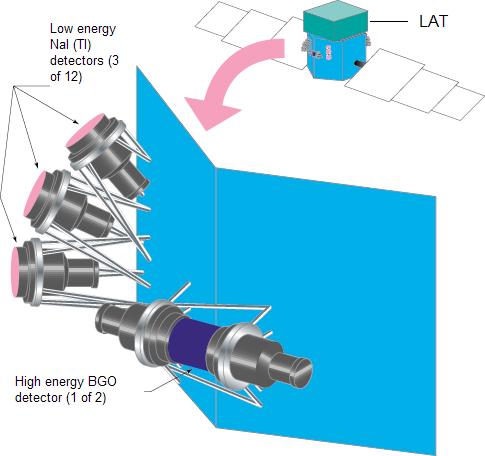}
\caption{\fermi\ with the arrangement of 3 NaI and 1 BGO devices. In the complete \fermi\ overview it is possible to see the \ac{lat} and another 3 NaI crystals (source: NASA~\cite{gbm-immagine}).}
\label{gbmschema}
\end{center}
\end{figure}

 The \acf{gbm}~(\citet{gbm-paper}) is the secondary   instrument on-board of \fermi, designed to continuously monitor all  the sky not occluded by the Earth, in the energy range 10~\keV--30~\MeV. Its purpose is the detection of transient phenomena. \ac{gbm} is composed by 12 \textit{Low Energy Detectors} (LED, 10~\keV--1~\MeV) and 2 \textit{High Energy Detectors} (HED, 150~\keV--25~\MeV). LEDs are grouped in 4 identical modules, one for each of the spacecraft's sides, while the HEDs are placed on opposite sides of \fermi\ (see figure~\ref{gbmschema}).  Every LED module is composed by a NaI crystal, a HED module is composed by a BGO crystal.
 
 The \ac{gbm} trigger for transient events is provided by the LEDs, the signal of which is compared with the average background, the latter being measured as the average event rate in the entire instrument within a given time span (typically 1024~s) and energetic range.  When  two or more detectors have an event rate that exceeds  the average background by more than a fixed number of $\sigma$, a trigger signal is issued.
 
  \ac{gbm} is able to (roughly) reconstruct on-board the direction of a detected transient by comparing the event rate in the various modules, so that a quick re-pointing of \fermi\ can be done, bringing the transient in  \ac{lat} \acs{fov}.

At the beginning of 2014, \ac{gbm} had detected more than 900 Gamma-Ray Burst, extra-galactic burst of gamma-rays that can emit more than  $10^{52}~\erg$ in a time ranging from hundreds of ms to some thousand of seconds (\citet{gbm-cat1,gbm-cat2}) and more than 400 Terrestrial Gamma Flares~\cite{gbm-tgf1} (for an example see \citet{gbm-tgf2}).

\section{The Large Area Telescope}
\label{LAT-section}
The \acf{lat} is the main instrument on-board of \fermi: it  is a pair conversion telescope, designed to study photons from 20~\MeV\ up to more than 300~\GeV.  Its commissioning,  after the launch of \fermi,  started on  \nth{11} June 2008 and the \ac{lat} began nominal science operation on \nth{4} August  2008. For a complete review of  the \ac{lat} design,  its construction and scientific goals, see~\citet{latdoc}.

The \ac{lat} was designed to reach  the performances required by the scientific goals stated at the beginning of the \ac{lat} project.
 One of the main capabilities requested to  the \ac{lat} with respect to \ac{egret} was to be able to continuously monitor  a large fraction of the sky. This capability, united with a strong increase in the effective area with respect to \ac{egret}, makes it possible  to study sources with a really faint  photon flux, leading to a strong increase in the number of known gamma sources. The request of a large \ac{fov} lead to the design of an instrument with a large base-to-height ratio, with the capability of triggering on events with large angle of incidence. 
 
 To study transients or variable sources, the \ac{lat} was required to have  a dead time many orders of magnitude smaller than the 100~\milli\second\ of \ac{egret}. To reach this performance, the spark chamber technology used in  previous spatial missions was replaced in the \ac{lat} with a tracker based on silicon strip detectors.  
 Furthermore, the  \ac{lat} was required to fill the energy gap between ground-based and space-based instruments, which implies it had to be able to reconstruct photons of energy greater than hundreds of \GeV. Because the maximum radiation length of the calorimeter is limited by the mass allowed by the launcher, it was necessary to design an imaging calorimeter, that through the reconstruction of the spatial development of the electro-magnetic shower made it possible to reconstruct the energy of an event that is not fully contained in the instrument. Moreover,  high-energy events produce significant back-splash of particles from the calorimeter, that can trigger a veto signal in the anti-coinciden\-ce detector that surrounds the instrument. For this reason, the anti-coinciden\-ce dome used in previous instruments was replaced with a segmented instrument that is  able to reconstruct the signal position, in this way allowing to discriminate between real veto signals and  self-vetoes. 
 
 Finally, the \ac{lat} was required to have an improved spatial and energetic resolution  with respect to \ac{egret} (in particular, the energy resolution was required to be below 20\% up to 300~\GeV). The final design of the  \ac{lat} is a trade-off between different and often conflicting  performance requirements, while respecting the constraints in mass and dimensions  posed by the choice of a Delta~5 launcher and the power consumption limits posed by a spacecraft-based instrument.

The \ac{lat} is composed by a $4\times4$ matrix of $\sim 37~\centi\meter\times37~\centi\meter
\times\sim 85~\centi\meter$ identical   towers. At the top of each tower  there is a module of the \acf{tkr} (\S~\ref{tkr}), composed of silicon micro-strip planes and Tungsten foils   which promotes pair conversion of photons and measures the direction of produced particles. Below every \ac{tkr} module there is   a module of the  \acf{cal} (\S~\ref{cal}) composed by  CsI(Tl) scintillation crystals for total 8.6 radiation lengths, that reconstructs the electromagnetic shower and measures its energy.  Every tower has a separate  \acf{tem} for  managing the trigger signals and the read-out of data. 
The towers are inserted in an aluminum grid  that is the main structure of the \ac{lat} and allows the dissipation of heat produced by the \ac{lat} subsystems. 

The choice to divide the \ac{tkr} and \ac{cal} in modules  instead of making two single instruments  reflects the difficulties, both technical and methodological,  that would have been encountered in producing instruments of this kind with the size of  \ac{lat} ($\approx 1.5~\meter$). Moreover, the modularity of the subsystems made the process of instrument manufacturing  quicker and more controllable. 

\begin{figure}[htb!]
\begin{center}
\includegraphics[height=15cm]{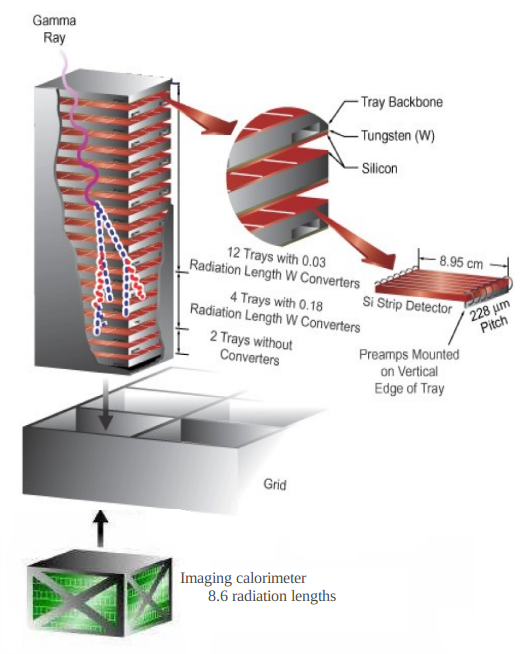}

\caption{A \ac{lat} tower, in which the main characteristics of the \ac{tkr} module, the calorimeter module and the placement of the tower in the \ac{lat} grid are displayed.~\cite{melissa}}
\label{lat-fig}
\end{center}
\end{figure}

\begin{figure}[htb!]
\begin{center}
\includegraphics[width=.55\textwidth]{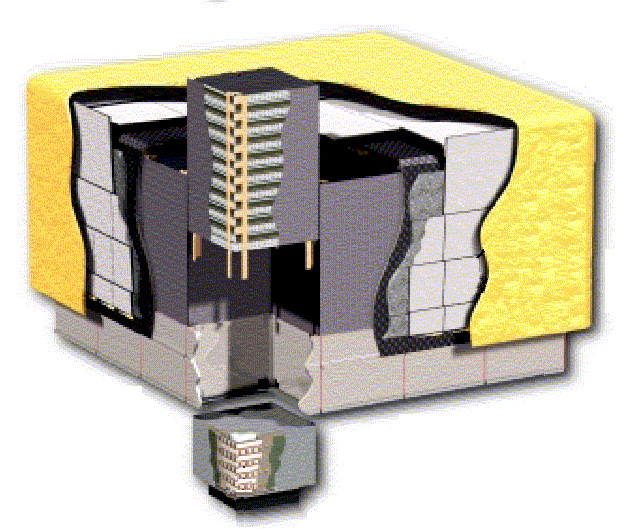}

\caption{Schematic view of the \ac{lat}, and of its basic components.~\cite{latdoc}}
\label{lat-fig2}
\end{center}
\end{figure}

\begin{figure}[htb!]
\begin{center}
\includegraphics[width=.95\textwidth]{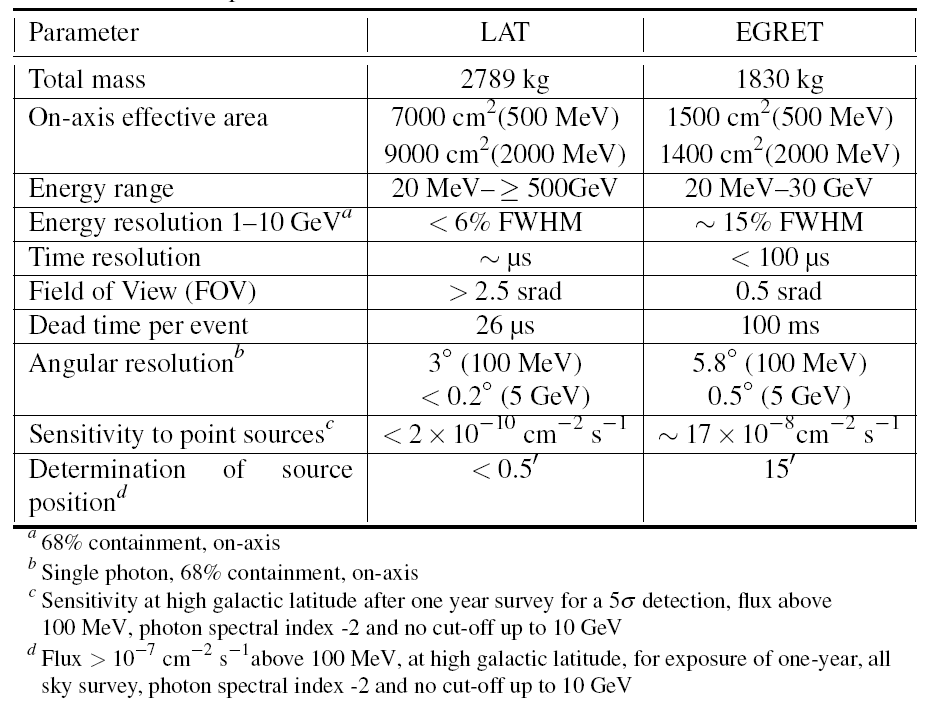}

\caption{Comparison of main characteristics of EGRET and \ac{lat}~\cite{lat-tabella}.}
\label{lattab}
\end{center}
\end{figure}

The \acf{acd} (\S~\ref{acd-section}) surrounding the \ac{lat} is composed by tiles of plastic scintillator.  The electronics for data acquisition, triggering and elaboration (DAQ) is located on the bottom of the instrument. The whole instrument is enclosed by a shield designed to protect it from thermal shocks and micro-meteoroids.  For a scheme of the instrument see figures~\ref{lat-fig} and~\ref{lat-fig2}. The main characteristics of the \ac{lat} are reported in table~\ref{lattab}, in comparison with those of \ac{egret}.

Each of the \ac{lat} sub-systems is able to produce one or more trigger primitives, that are then combined and evaluated to generate the global trigger signal and the read-out of the whole \ac{lat}. This feature allows great flexibility in the selection of acquired events. The \ac{lat} small dead time allows the acquisition of events at a rate of a few \kilo\hertz, that must then be reduced on-board to match the down-link bandwidth of 1~Mbps. The \ac{lat} trigger (\S~\ref{trigger}) and the subsequent on-board processing (\S~\ref{filtri}) are therefore optimized for maximizing the number of events triggered by gamma-rays which are transmitted to the ground. Moreover, the trigger design is also fundamental in the achievements of the \ac{lat}  performance in terms of effective area and \ac{fov}, because it removes the (geometrical) constraints caused by the presence of a single trigger device on the instrument (like the Time of Flight trigger on \ac{egret}).

The achievement of the requested performance requested a large simulation effort during the design process of the \ac{lat}.
All the \ac{lat} sub-systems (\ac{cal}, \ac{tkr},\ac{acd}) were designed making use of extensive Monte Carlo simulations to simulate and therefore optimize their responses  and to calculate their future performances.
After the assembly,  all modules  were tested  to verify their  capability of optimally  working in the typical environmental conditions of space (thermal and vacuum tests, \ac{tkr}~\citet{test-tkr}, \ac{cal}~\citet{test-cal}). Tests were also performed to verify the capability of all modules to tolerate  stress caused by the launch, the dissipation of produced heath and the sensitivity to electromagnetic interference.

The response of the devices and its correct reproduction by Monte Carlo simulations were verified in beam tests conduced at CERN and  GSI heavy ion accelerator laboratory,  described in~\citet{beam-test}, during which  was used a Calibration Unit~(CU) composed by two  \ac{tkr} modules, three \ac{cal} modules and several \ac{acd} tiles. Also, the performances of all the produced modules were tested with ground  muons.

\section{The Silicon-strip tracker}

\subsection{Balancing the characteristic of the Tracker}\label{tracker-effect}
The tracker main requirements, i.e. the high probability of  photon conversion, the large \ac{fov} and the good capability of reconstructing the direction of  the tracks produced by the generated \Pelectron\APelectron pair, depends on different, and often conflicting, characteristics of the instrument. This is particularly true when the instrument must respect the constraints in dimensions and power consumption posed by the space operation. The design of the \ac{lat} tracker is a balance between these different requirements. 

The \ac{fov} depends on the ratio between the instrument width and the minimum height that an event must cross to generate a trigger. Therefore the \ac{fov} does not depend only on the geometry of the instrument, but also on the triggering scheme. The use of silicon-strip detectors is critical to  achieve the \ac{lat} large \ac{fov}, as  discussed later in \S~\ref{ssd-section} and \S~\ref{trigger},  

Photon conversion probability is determined by the total thickness of the instrument in terms of radiation lengths: therefore foils of converting material (Tungsten in the \ac{lat}) are inserted between the detection planes of the instrument to increase its thickness.

The dependence of the angular resolution on the \ac{tkr} characteristics is different at low and at high energy. At low energy, the directional information of a track is quickly degraded by the \acf{mcs} in the traversed material.  Typically, the largest contribution to the total thickness of the  instrument comes from the foils of converting material: to minimize the effect of the multiple scattering on the tracks, the conversion foils are placed as near as possible to the detection planes, and the remaining part of the instrument  is constructed with materials that maximize the transparency to relativistic charged particles.

\begin{figure}[htb!]
\begin{center}
\includegraphics[width=.65\textwidth]{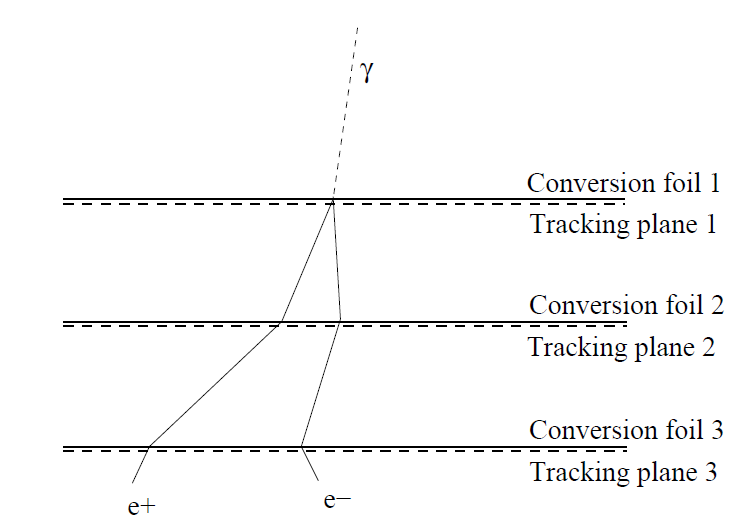}

\caption{Schematic view of a low energy photons converting in a tracker where the conversion foils are placed close to the detection planes and the structural material of the tracker has a thickness negligible with respect to the conversion foils. The conversion takes place in foil 1 and, because of \ac{mcs}, the original direction is quickly degraded, so that  only  planes 1 and 2 provide valuable information on it.~\cite{baldini}}
\label{tkr-LE}
\end{center}
\end{figure}

An approximated estimate of the angular resolution at low energy can be made from the over-simplified example of figure~\ref{tkr-LE}.
In this approximation, the conversion foils are sufficiently close to the detection planes so that we can consider negligible the effect of \ac{mcs} on the position measurement in the subsequent detection planes. The degradation of the track direction is severe, so that the direction is basically determined only by the first  two detection planes below the conversion point and therefore only the effective conversion foil contributes to the multiple scattering. Finally, the thickness \textit{t} of the conversion foils is small compared to the radiation length of the converting material $X_0$. This implies that the mean path traveled by the \Pelectron\APelectron pair in the converting foil is just half of its thickness (for vertical tracks), and the RMS of the deviation angle (in the x-z or y-z plane) can be approximated as Gaussian and is described by~\cite{pdg}:

\begin{equation}\label{risoluzione-LE}
\theta_{RMS}=\frac{13.6~\MeV}{\beta E}\sqrt{\left(\frac{t}{2X_0}\right)}\left[1+0.38\ln\left(\frac{t}{2X_0}\right)\right]
\end{equation}
where $X_0$ is the radiation length and $\beta=\frac{v}{c}\simeq 1$ for  \Pelectron and \APelectron produced by gamma-rays in the \ac{lat} energy range and $\left \langle E \right \rangle\simeq E_\gamma/2$ (on average the energy of the converting photon is equally divided between \Pelectron and \APelectron). Equation~\eqref{risoluzione-LE} shows that the low-energy angular resolution approximately scales as the inverse of the energy and the square root of the thickness of the converting foil. Therefore, an improvement in the angular resolution can be achieved by reducing the  thickness of the converter foils, with a consequent reduction of the conversion probability if the number of layers is kept constant.

\begin{figure}[htb!]
\begin{center}
\includegraphics[width=.75\textwidth]{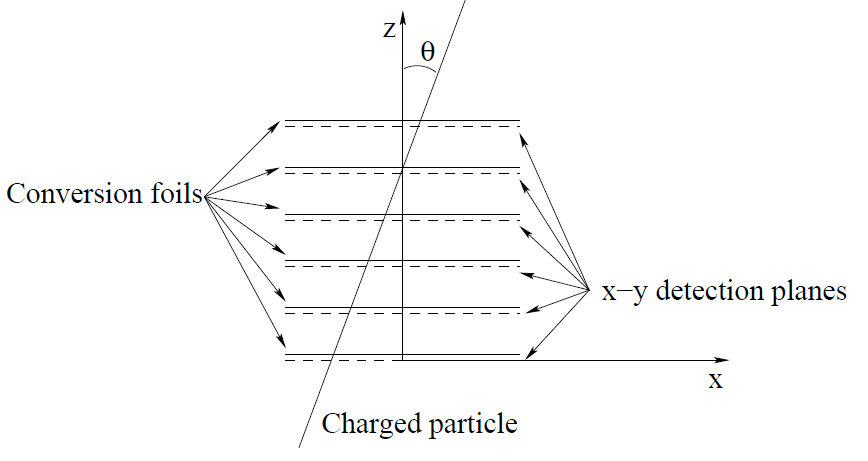}

\caption{A charged particles with sufficient energy crosses all the planes below the entering/conversion point with no significant deviation.~\cite{baldini}}
\label{tkr-HE}
\end{center}
\end{figure}

The angular resolution improves with energy until, at high energy,  the impact of the multiple scattering becomes negligible and the resolution reaches an asymptotic value that depends on the intrinsic design of the tracker.  At high energy, a track can be approximated as in figure~\ref{tkr-HE}, where a charged particle crosses all the planes below the entering or converting point with no significant deviation. 

Therefore the reconstruction of the direction of the track will depend on the precision with which we can reconstruct the position of the track on the traversed layer, that is, for a digitally-read micro-strip detector:

\begin{equation}
\sigma_{hit}=\frac{p}{\sqrt{12}}
\end{equation}
where $p$ is the pitch of the strips. From figure~\ref{tkr-HE} it is clear that if we increase the lever arm between layers the effect of the uncertainty on the hits position is decreased. Furthermore, an increase in the number of layers of the tracker will increase the number of hits on the track, and therefore the precision of the reconstruction. Finally, we can made a rough estimate of the dependence of the angular resolution on the characteristics of the tracker:

\begin{equation}\label{risoluzione-HE}
 \Delta \theta \propto \frac{p}{d N^\frac{3}{2}}
\end{equation}
where $d$ is the distance between planes and $N$ the number of planes. This is an estimate of the uncertainty on the reconstructed direction of a single track. The reconstruction of the original direction of a high energy photon is actually more complicated, because such events typically have more than one track, both because the electromagnetic shower produces high-energy particles in the tracker and because the  back-splash (page~\pageref{backsplash-page}) fills the tracker with charged particles.

However, from equation \eqref{risoluzione-HE} we can see that the asymptotic angular resolution can be improved by a finer segmentation of the detection planes or by increasing the distance between the planes. The first approach leads to an increase of the electronic channels of the tracker and therefore of its power consumption (that, as stated before, is constrained by the characteristics of the satellite); the second approach, if we keep constant the total number of detection planes, worsens the base-to-height ratio, therefore reducing its \ac{fov}.

\subsection{The tracker layout}
\begin{figure}[htb!]
\begin{center}
\includegraphics[width=.65\textwidth]{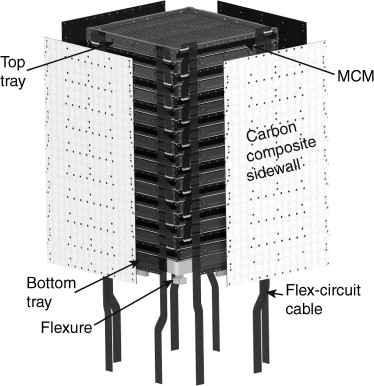}

\caption{A tracker tower module, showing the trays and the Multi-chip Modules (MCM) containing the electronics for the readout of the silicon device planes~\cite{trackerdoc}.}
\label{tkr-torre}
\end{center}
\end{figure}

Each  of the 16 tracker modules (see~\citet{trackerdoc} for a complete review of the instrument) is composed of a stack of 19 trays of carbon with a honey-comb core of aluminum, with density of 48 or 16~\kilo\gram~\meter\rpcubed\ depending on the load of the tray. Trays are supported by four carbon sidewalls. Carbon was chosen because of its high radiation length, that reduces the probability of undesired gamma conversions far from the silicon detectors. A tray is $\sim 3~\centi\meter$ thick. Figure~\ref{tkr-torre} shows one of the modules.

Each tray has a \acf{ssd}  plane on both faces, except for the top and the bottom trays that have only one detection plane. The silicon planes of subsequent trays are separated by 2~\milli\meter, and their strips are rotated by $90^\circ$. Detectors of neighbor layers combine to forms an x-y detection layer, for a total of 18 x-y detection layers. Each tracker tower is 66$\times37.6\times37.6$~\cubic\cm: the lateral size of the tower was determined by the maximum number of \acfp{ssd} that can be bonded while keeping  the noise, which is linearly dependent on the strip length,  $5\sigma$ lower than the hit signal threshold. The height of the tower was optimized to grant a reasonable sampling of the generated particle tracks, which is determinant for the high-energy angular resolution (see eq~\eqref{risoluzione-HE}) while keeping the \ac{lat} base/height ratio high, so to endow the instrument with a large \ac{fov}.

Read-out of the \acp{ssd} is provided by 36 \acfp{mcm}, placed on the sides of the trays. 8 cables connect the \acp{mcm} to the \acf{tem} placed at the bottom of the tower, below the \ac{cal} module.

The conversion material chosen for the \ac{lat} tracker is Tungsten (W), placed in foils  immediately above the lower silicon detector of a tray. This choice places the conversion foils immediately above a x-y detection layer, therefore minimizing the material traversed by the produced particles after a photon conversion and the lever arm between the conversion point and the first measured position.

From equation~\eqref{risoluzione-LE} we can see that the width of the conversion foils is fundamental in determining the angular resolution when the effect of the multiple scattering is important (in the \ac{lat}, below 10~\GeV). Thinner foils improve the angular resolution, but decrease the effective area. In the \ac{lat} the choice made was to distribute unevenly the Tungsten in the tracker, with the 18 x-y planes of the tracker divided in 3 different configurations. The  12 upper planes constitute the 'front' section (or \textit{thin} section) of the tracker, where each plane is preceded by  a thin ($0.095~\milli\meter = 0.027~X_0$) Tungsten layer, in order to reduce the impact of multiple scattering and obtain good angular resolution for events converting in this section. The lower planes constitute the 'back' section of the tracker, designed to almost double the conversion rate of photons and therefore the effective area. The first 4 layers of the back section (\textit{thick} section) have foils  $0.72~\milli\meter =0.18~X_0$ thick, therefore
containing more than 70\% of the total Tungsten thickness (0.72~$X_0$ with respect to 0.324~$X_0$ of the front section). As described in \S~\ref{trigger}, an event will generate a trigger primitive (a signal that can lead to the issuing of a trigger signal) if three adjacent x-y planes contemporary have one or more hits: for this reason a conversion in the last 2 planes is not desirable, and so they have no Tungsten foil (\textit{blank} section).
The total thickness of the Tungsten in \ac{tkr} is 1.08~$X_0$, with a conversion probability of 63\% for photons above 1~\GeV\ with normal incidence; the total width of the tracker is $\sim$ 1.5~$X_0$.

\subsection{The silicon-strip detectors}\label{ssd-section}
One of the most crucial improvements in the \ac{lat} with respect to  previous instruments is the use of silicon detectors in the tracker, which grant high hit efficiency, low noise occupancy and good spatial resolution.

Detection planes are assembled starting from  single-sided \acfp{ssd} of $8.95\times8.95$~\centi\squaren\meter, with a thickness of 400~\mycro\meter. This thickness, $\sim30\%$ greater than in typical particle-physics detectors, was chosen to optimize the signal-to-noise ratio: because of the large amount of Tungsten in the tracker, the thickness of the silicon detectors is however negligible. The choice of using two close single-sided detectors for each x-y detection plane instead of using double-sided detectors was dictated by the reduced cost, noise and complexity of the first solution~\cite{trackerdoc}. Each \ac{ssd} is obtained  from a 6~\inch\ wafer, leaving an inactive zone at the side of the device of $\sim 1~\milli\meter$, and has 384 strips. Four detectors are micro-bonded together head-to-head to form a ladder with strips that are effectively  $\sim 35~\centi\meter$ long. Four alongside ladders form a detection plane, with a sensitive surface of 95.5\% \cite{pass7-status} of the total plane surface. Every plane contains a total of 1536 strips, leading to a total of  more than 885000  strips for the entire \ac{lat} and a total power consumption of the entire tracker of 160 W (only 180~\mycro\watt\ per channel).

The micro-strips have a width of 56~\mycro\meter\ and are spaced by 228~\micro\meter.  The limiting factor in the number of strip per plane was the need to have a low power consumption and a low heat production. From equations~\eqref{risoluzione-LE} and~\eqref{risoluzione-HE} we see that the strip pitch is a limiting factor for the angular resolution only when energy is sufficiently high and the multiple scattering is negligible (for the \ac{lat}, above 10~\GeV).  

\begin{figure}[htb!]
\begin{center}
\subfigure[Average hit efficiency in \ac{tkr}]{\includegraphics[width=.7\textwidth]{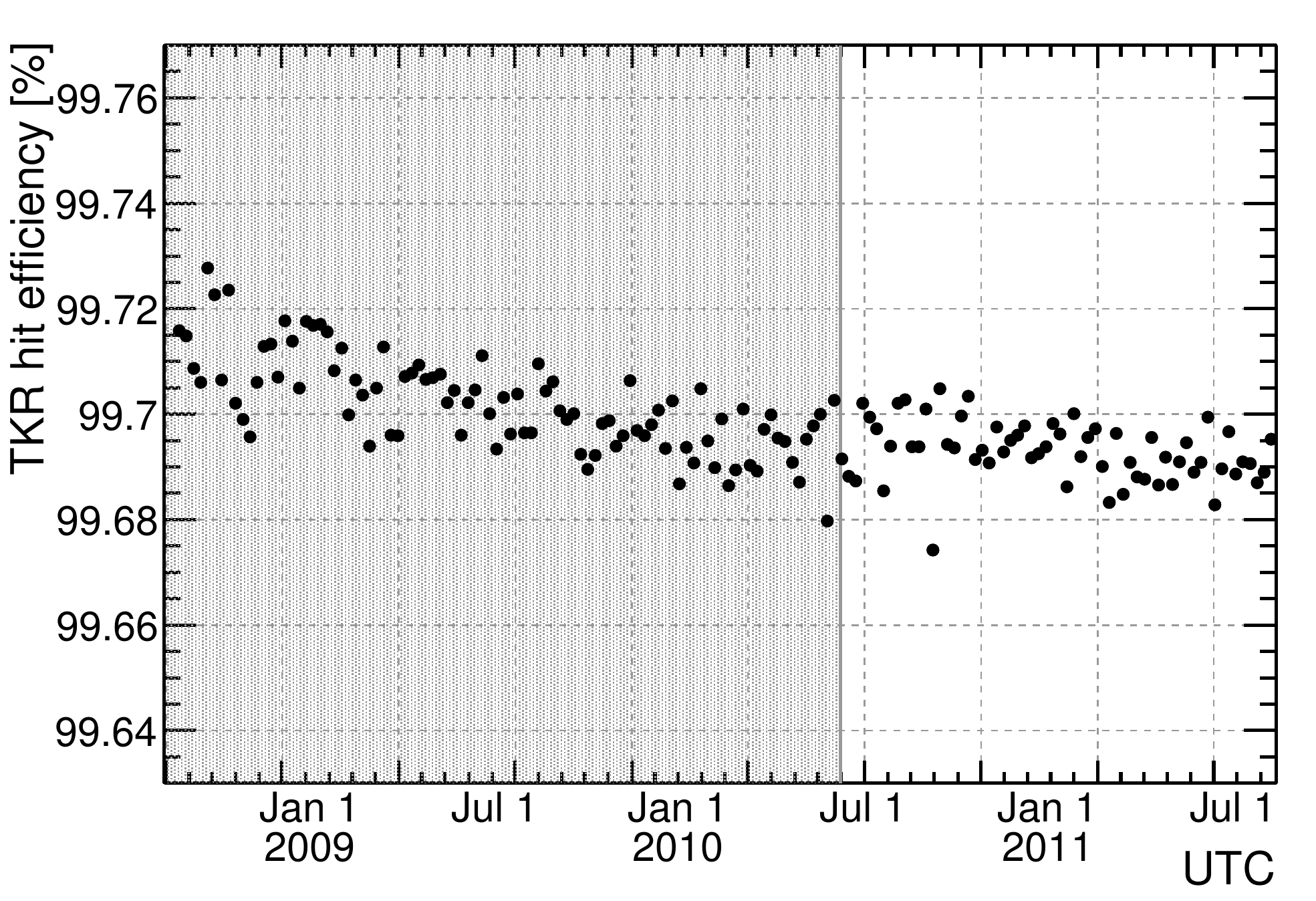}}
\subfigure[Single-strip noise occupancy]{\includegraphics[width =.7\textwidth]{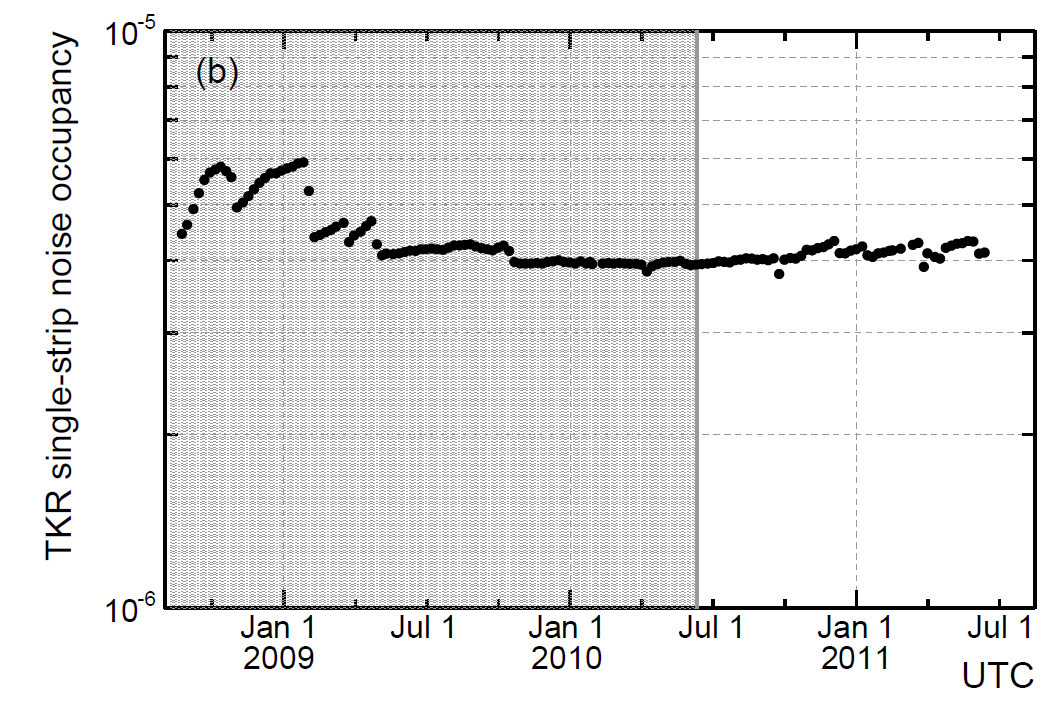}}
\caption{Data from the first three years of the mission. Each point is the average on a week of data taking~\cite{pass7-status}.}
\label{tkr-efficienza}
\end{center}
\end{figure}

\begin{figure}[htb!]
\begin{center}
\includegraphics[width=.75\textwidth]{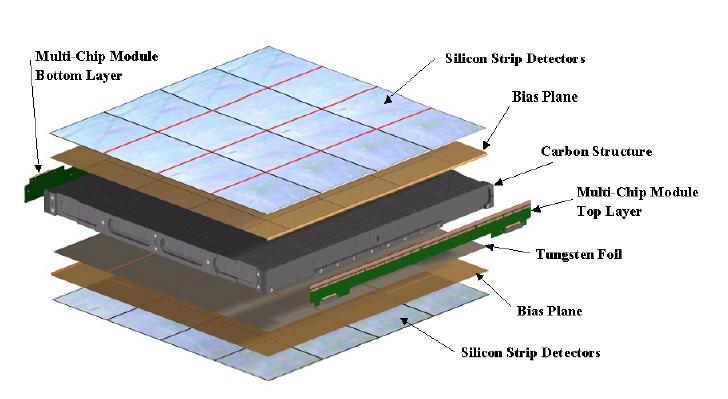}

\caption{A tracker tray, with  the carbon-aluminum base structure, the two electronic modules on the sides, the Tungsten plane, and the two micro-strip planes.~\cite{trackerdoc}}
\label{tkr}
\end{center}
\end{figure}

When compared to spark gas-chambers used in previous space-based experiments, silicon sensors  grant very high particle detection efficiency ($>99.9\%$)~\cite{pass7-status} and a small noise-occupancy: the averaged electronic noise of \ac{lat} micro-strip, measured as the probability of having a hit signal caused by noise during event read-out, is $5\times10^{-7}$. Moreover, silicon detectors are intrinsically fast detectors. This allows, even within the limits on the electronics imposed by power constraints, to strongly reduce the dead time with respect to spark chambers, from \milli\second\ to \mycro\second. The really low dead time allows the \ac{lat} to acquire  events at a rate of several~\kilo\hertz. At such a  rate, a large fraction of the events originated in the \ac{lat} can be acquired, and only after the acquisition they are analyzed to eliminate the dominant background. This is a complete reversal of the approach used with gas detectors, where the high dead time forced to perform an initial event selection at the trigger level, with strong impact on the instrument performance. Furthermore,  micro-strip technology makes it possible for  the \ac{tkr} to autonomously  detect the passage of charged particle in the detector, and therefore to self-trigger, while \ac{egret} needed a \ac{tof} system that detected only particle crossing the whole lower spark-chamber. This characteristic of the \ac{tkr}, that allows the acquisition  of events that cross only 3 detection planes, is the key of the large \ac{fov} of the \ac{lat}. Finally, silicon detectors don't have  consumable elements like gas, so the potential life of the instrument is significantly increased, and there is no need of a mechanism for gas recharging on board, and the degrading of the performances because of aging is  low. In figure~\ref{tkr-efficienza} (from~\citet{pass7-status})  the efficiency and the effective noise occupancy over the first three years of operation are shown. The changes in the noise occupancy are caused by the masking, at different times during the mission,  of noisy strips that dominate the average single-strip noise, while the baseline at $4\times10^{-6}$ is the effect of accidental coincidences between event readouts and charged particles tracks ($\sim 3$ hits per event over the full \ac{lat}).

\subsection{The tracker electronics}

The  read-out electronics is placed on the sides of the towers, with two \acp{mcm} on opposite sides of each  tray, forming an angle of $90^\circ$ with respect to the \acp{ssd}: this particular configuration is designed to reduce the insensitive gaps between towers to only 17.9~\milli\meter, with a resultant sensitive surface of $\sim 89\%$ of the total \ac{lat} surface~\cite{pass7-status}. Each \ac{mcm} reads an \ac{ssd} plane: the first and last trays, having only one detection plane, have only one \ac{mcm} module. Each  \ac{mcm} contains 24    64-channel amplifier-discriminator Application Specific Integrated Circuits (ASICs) for the read-out of  64 strips (GTFE chips), and 2 digital read-out-controller ASICs (GTRC). On every side of a tower, 2 cables connect the nine  \acp{mcm} with the \ac{tem} located below the calorimeter. GTRCs are redundant, and GTFEs can be assigned to a different controller via remote control, so that a failure in a cable or in one of the 24 chips of a plane would result only in the loss of at most 64 channels out of 1536~(\citet{latdoc,trackerdoc}).

Strip  read-out  is \textit{binary}, that is, the amplifier output is discriminated by a single threshold, that determines if a single strip has a \textit{hit} signal or not. Then,  only the addresses of the  hit strips are acquired for each plane, until a maximum of 64 for each GTRC chip (roughly corresponding to a half plane). The threshold can be remotely adjusted for every GTFE chip. The binary read-out and the zero suppression procedure greatly reduces the amount of data acquired (and downloaded) for each  event. The electronics also calculates for every  GTRC the logical OR of all the strips, issuing a signal   when a strip in the plane go over threshold. This signal is used to construct the trigger primitive related to the tracker (see \S~\ref{trigger} for all the \ac{lat} trigger primitives). The logical OR of all the strips is also used to calculate, for every acquired event, the \acf{tot}: \label{totref} at the opening of the trigger  window, each GTRC with OR$=1$ starts to measure the time it takes to return to 0.  The \ac{tot}, that is added to the downloaded information, provides very valuable information on the amount of ionization in the device. Finally, the electronics allows the masking, via  remote control, of single noisy strips, that can  be excluded both from the issuing of trigger primitives  or from the data read-out.

\section{The electro-magnetic Calorimeter}\label{cal}

The \acf{cal} (\citet{latdoc,pass7-status,caldoc}) is designed to absorb  the electro-magnetic shower produced by the initial \Pelectron\APelectron pair and to measure its energy. In addition, the \ac{cal} is requested to provide a good correlation between tracks in the \ac{tkr} and the position of the energy deposition in the instrument. This request, that is fundamental both for background rejection and for the high-energy extension of the \ac{lat} sensitivity, led to the design of an imaging calorimeter instead of a monolithic instrument similar to that on-board of \ac{egret}.

\begin{figure}[htbp]
\begin{center}
\includegraphics[height=8cm]{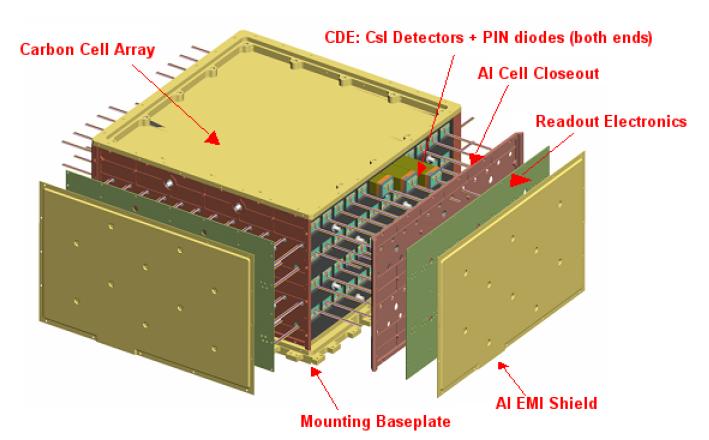}

\caption{A calorimeter module.~\cite{caldoc}}
\label{calebbasta}
\end{center}
\end{figure}

The \ac{cal} reproduces the \ac{tkr} segmentation, with a \ac{cal} module placed immediately below each one of the 16 \ac{tkr} modules.
Each module is composed of 8 layers of 12 CsI(Tl), $19.9\times26.7\times326$~\milli\cubic\meter\ crystal logs, optically isolated with reflective material, each layer rotated  by $90^\circ$ with respect to the preceding and following layers (\textit{hodoscopic configuration}). CsI was chosen because of its non-hygroscopic nature, high light yield and modest cost.  Non-hygroscopic crystals are easier to handle with respect to hygroscopic ones, and also there is no need to introduce hermetic housing for individual crystals in the instrument, which would cause a decrease in the instrument performances. The transverse dimensions of the crystal logs are of the same order of magnitude of CsI radiation length (height is 1.07~radiation lengths) and of its Moli\`ere radius\footnote{The Moli\`ere radius ($R_M$) is the transverse distance that a particle of an electromagnetic shower travels while traversing the last radiation length before it goes under the critical energy and is quickly stopped by ionization energy losses; therefore the $R_M$ of a material is a measure of the transverse dimension of an electro-magnetic shower, with more than 90\% of the shower contained in 2 $R_M$ from the longitudinal axis.~\citet{leo,green}} (width is 0.75~Moli\`ere radius), and is a compromise between the request of a fine segmentation in the instrument and the constraint on power consumption. The total width of the \ac{cal} is 8.6~$X_0$ (only $\approx~0.4$ nuclear interaction lengths~\cite{pdg}), therefore the total thickness of the \ac{lat} is 10.1~$X_0$. The mass of the \ac{cal} is about 1800~\kilo\gram. A schematic view of one module is given in figure~\ref{calebbasta}.

\begin{figure}[htbp]
\begin{center}
\includegraphics[height=11cm]{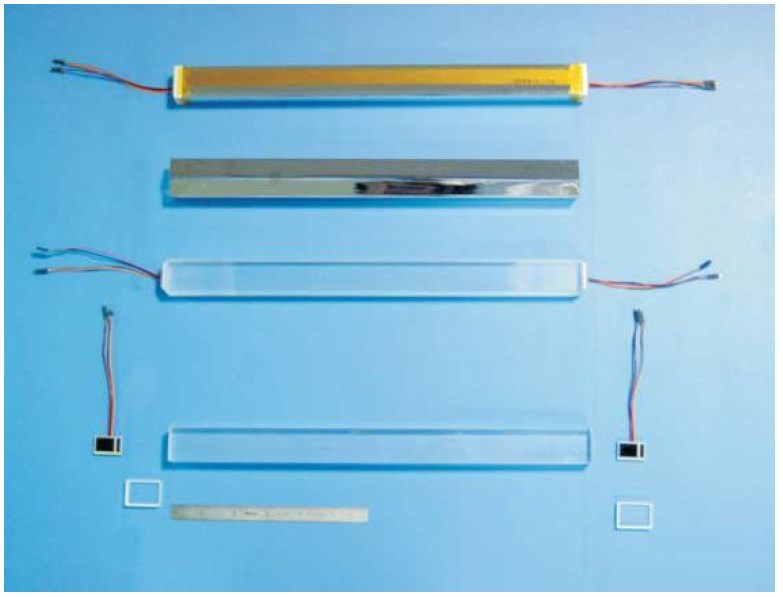}

\caption{One crystal detector element in various level of assembly. From top: a complete crystal detector element; the reflective optical wrapper used to isolate crystals; a CsI(Tl) crystal with bonded the two photo-diode assemblies; a crystal, the two photo-diode assemblies and two polymer end caps. At the bottom a 15~\cm\ scale.~\cite{caldoc}}
\label{calcrystal}
\end{center}
\end{figure}

Each CsI(Tl) crystal is read  at both sides by two photo-diodes, one with surface 25~\milli\squaren\meter\ (High Energy PIN Diode, HEPD), the other with surface 147~\milli\squaren\meter\ (Low Energy PIN Diode, LEPD). The two photo-diodes are enclosed in a single ceramic carrier for easier bonding to the crystal, forming a \acf{pda}; a complete calorimeter crystal detector and its components are shown in figure~\ref{calcrystal}. Photo-diodes operate with low bias voltage, therefore their power consumption is small, only 5~\watt\ for the 192 \acp{pda} of a \ac{cal} module.

The signals from both diodes of a \ac{pda} are sent to a single chip (GLAST Calorimeter Front End, GCFE) where they are amplified with  charge-sensitive pre-amplifiers. The output of the amplifiers is then analyzed by a slow shaping amplifier ($\sim$ 3.5~\mycro\second\ peaking time)  and by a fast one ($\sim$ 0.5~\mycro\second\ peaking time). The output of the fast shapers is used for trigger purpose: it is compared to a programmed threshold by a programmable discriminator and, if the threshold is exceeded, a trigger primitive is generated. In nominal flight operation two thresholds are set, to $\sim 100~\MeV$ and $\sim 1000~\MeV$ of deposed energy.  

The output of the slow shapers is used for the  measurement of the deposed energy in the crystals. Each output signal is split in two track-and-hold stage with nominal gains 1 and 8. Including the ratio of the diode responses, each GCFE provides energy measurement at a crystal end in four energy ranges, with effective gains x1, x8, x64 and x512~\cite{caldoc}. The highest not saturated signal is selected as the ``best'' signal and converted by digital-analogical converter. This particular configuration was designed to have a large energetic range in the calorimeter: the resulting energy ranges  are 2~\MeV--100~\MeV\ and 2~\MeV--1~\GeV\ for the two LEPD outputs  and 30~\MeV--7~\GeV\ and 30~\MeV--70~\GeV\ for HEPD outputs~\cite{on-orbit-calibration} (for comparison, the energy deposited in a crystal by on-axis \ac{mip} is $\sim 11.2~\MeV$). For every row on a \ac{cal} module side, a GLAST Calorimeter Readout Control chip (GCRC) collects the information from the 12 GCFE-ADC pairs and sends them to the \ac{tem}. As for the tracker, there is a zero suppression algorithm that prevents crystals below a specific threshold (2~\MeV) to be included in the data stream out of the \ac{tem} in nominal science operation. The read-out dead time for a \ac{cal} event in nominal science operation   is 22.3~\mycro\second , the power consumption of the whole \ac{cal} is $\sim46~\watt$, that is only 20~\milli\watt\ for a single channel.

\begin{figure}[htbp]
\begin{center}
\includegraphics[width=1.\textwidth]{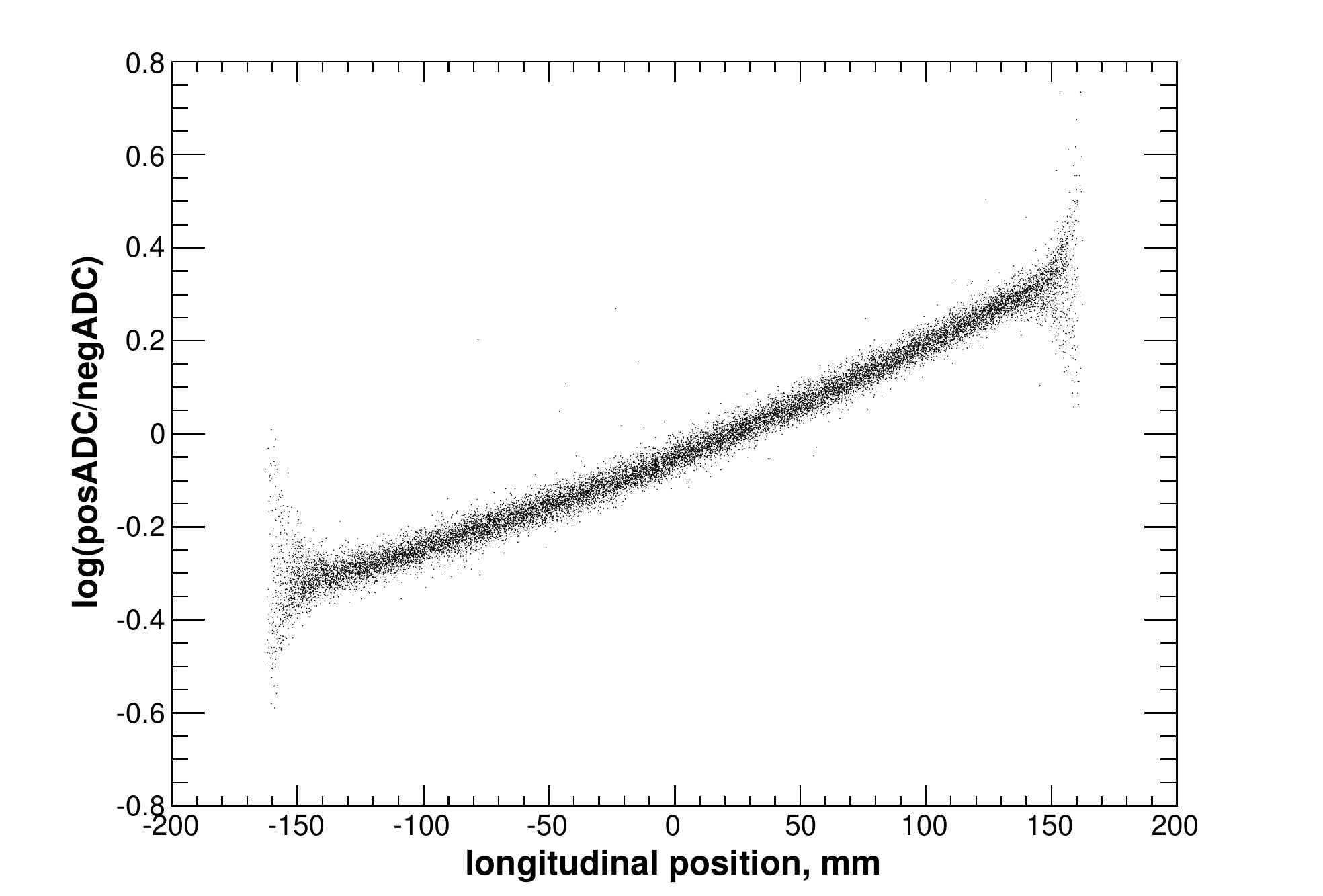}

\caption{Light asymmetry in collected light for a crystal exposed to the muon flux at sea level. Light asymmetry is defined as $\log(S_{left})/\log(S_{right})$. The width of the distribution is due to the statistical fluctuations  of collected light and has a dependence from the deposed energy of $E^{-1/2}$.~\cite{latdoc}}

\label{asimm}
\end{center}
\end{figure}

The read-out at both  ends of each crystal is not  simply a redundancy to prevent crystal losses: it permits the reconstruction of the longitudinal position of the energy release using the light yield asymmetry at the ends of the crystal, with a resolution that scales with the deposited energy, from  some \milli\meter\ at $\sim10~\MeV$ to a fraction of \milli\meter\ above 1~\GeV. Figure~\ref{asimm} shows  the relation between the asymmetry in light collection in one crystals and the position of the energy deposition.

\begin{figure}[htbp]
\begin{center}
\includegraphics[width=1.\textwidth]{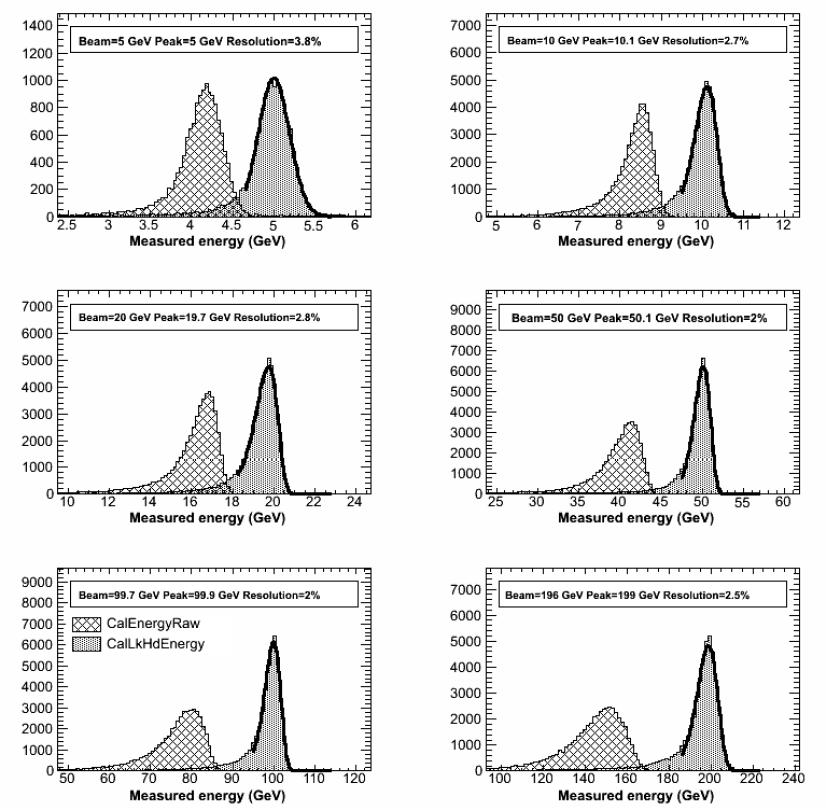}

\caption{Example of the results of the beam test at CERN: test of  the energy resolution of the \ac{cal} with mono-energetic electrons of energy 5, 10, 50, 99.7 or 196~\GeV.  The incidence angle is $45^\circ$. Every pad shows the energy collected  in the \ac{cal} (grid area) and the energy reconstructed by the shower profile method (filled area). The shower profile method is the best method for the reconstruction of energy above some \GeV, see \S~\ref{energy-reconstruction}). In every pad labels show the peak energy of electrons, the peak of the reconstructed energy distribution and the energy resolution  $\Delta E / E$. In these results the electro-magnetic shower is contained in a single \ac{cal} tower, therefore the energy resolution is better than that achieved in typical flight-data.~\cite{latdoc,beam-test}
}

\label{ricostr-energia}
\end{center}
\end{figure}

The hodoscopic arrangement of the CsI crystals  of the \ac{cal} and the reconstruction of the  longitudinal position of the energy deposit in crystals allow a full 3D reconstruction of the shower in the \ac{cal}. This  permits the reconstruction of the shower development in the \ac{cal},  making  the reconstruction of the energy possible also when the shower is not fully contained in the \ac{cal}:  for energies above 100~\GeV\ about half of the total energy of an on-axis electromagnetic  shower  escapes from the back of the \ac{cal}~\cite{pass7-status}. \S~\ref{energy-reconstruction} will give a description of the energy reconstruction methods used in the \ac{cal}, while figure~\ref{ricostr-energia} shows the precision of the energy reconstruction measured during beam test at CERN. Moreover, \S~\ref{energia-p8} will show the large improvement obtained at high energy by the new energy-reconstruction method.

Moreover, the 3D reconstruction of the shower allows the measurement of some important characteristics of the shower  like its transverse dimension, that is largely used in the identification of the nature (hadronic or electromagnetic) of the particle and is fundamental for the electron analysis described further in this work.  Furthermore, the reconstruction of the shower momenta gives a  robust estimation of the event direction (above some \GeV), which can be usefully compared with the one reconstructed in the tracker. Finally, the reconstruction of the event direction in the \ac{cal} can be used in the study of photons not converting in the \ac{tkr}, or without usable information in the \ac{tkr}. This class of photons can be very useful for the study of high-energy faint sources, where the photon flux is very low and the substantial increase in the effective area achievable with the recover of cal-only events would be very important. Indeed, a class of caloimeter-only events is under development and will be added to the \ac{lat} data products (\citet{calonly}).

\section{The Anti-Coincidence Detector}\label{acd-section}
The \ac{acd} (see~\citet{acddoc} for a complete review) is designed to identify charged particles entering the \ac{lat} field of view (FoV); it surrounds all the upper part of the \ac{lat} for a total of 8.3 \squaren\meter. When the \ac{lat} was designed, it was requested to be able to produce photon data-sets with residual contamination from charged particle $\leq 10\%$ of the diffuse gamma-ray background intensity, therefore having a suppression factor of $\sim10^6$ for protons and $\sim10^4$ for electrons: keeping in account the contribution of \ac{tkr} and \ac{cal} to the identification of charged particles, the \ac{acd} must have a suppression factor for charged particle  of $\sim 10^3$  for protons and  $3\times 10^3$ for electrons. Furthermore, it was requested that the photon absorbed by the \ac{acd},  including those converting  in the thermal blanket and in the micro-meteoroid shield,  were less than 6\% of the total incident gamma rays. Finally, the characteristics of the spacecraft requested that the \ac{acd}, including the thermal blanket and the micro-meteoroids shield, had a mass lower than 290~\kilo\gram\ and a total power consumption  less than 12~\watt.

\begin{figure}[htbp]
\begin{center}
\includegraphics[width=.7\textwidth]{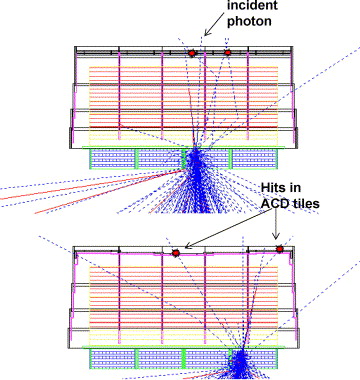}
\caption{Simulation of the back-splash in the \ac{lat} using Monte Carlo photons: charged particles are in red, photons are in blue, signals in \ac{acd} caused by back-splash are red dots.~\cite{acddoc,test-acd}}
\label{backsplash-fig}
\end{center}
\end{figure}

The  \ac{acd} was designed to minimize self-vetoes in the \ac{lat}. \label{backsplash-page} Self-vetoes can be originated by two different phenomena: the first is a random coincidence between a gamma conversion and the passage of a charged particle through the \ac{acd}, that has a surface of 83000~\squaren\cm\ ($170\times170~\squaren\cm$ on the top and $170\times80~\squaren\cm$ on each side). The second phenomena is the \textbf{back-splash}: low energy particles produced by the electromagnetic shower in the \ac{cal} (typically 100--1000~\keV\ photons) travel backward and hit the \ac{acd}, where they generate a veto signal (for low-energy photons, through electrons produced via Compton scatter). The impact of back-splash increases dramatically with energy: in \ac{egret}, whose \ac{acd} was a monolithic dome of plastic scintillator, the detection efficiency at 10~\GeV\ was almost halved with respect to the efficiency at 1~\GeV, while at 50~\GeV\ almost all the events were rejected because back-splash. The \ac{lat} was requested to reject less than 20\% of photons at 300~\GeV\ because of back-splash. The probability of having a back-splash signal in the \ac{lat} \ac{acd} was estimated to be, at 300~\GeV\, $\sim 10\%$ on a surface of 1000~\squaren\cm. Figure~\ref{backsplash-fig} shows the simulation of the back-splash effect on the \ac{acd}.

\begin{figure}[htbp]
\begin{center}
\includegraphics[width=.8\textwidth]{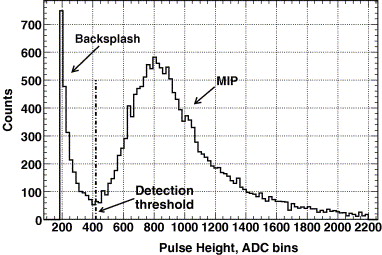}
\caption{Result of a beam-test performed on a \ac{acd} tile posed before a block of material mimicking the \ac{lat} calorimeter. The \ac{mip} peak is obtained with a beam of protons, the back-splash peak is obtained with a beam of electrons. Pulse-height distribution of signals after pedestals subtraction is shown: on the left the peak generated by back-splash photons through Compton scatter on electrons in the \ac{acd} material; on the right the peak generated by \acp{mip}, which follows a Landau distribution.~\cite{acddoc,test-acd}}
\label{soglia}
\end{center}
\end{figure}

In order to reduce self-vetoes (\citet{test-acd}), the \ac{acd} was therefore segmented, to make  the reconstruction of  the signal position possible and to check whether it was related to the event reconstructed in \ac{tkr} and \ac{cal}. Furthermore, the choice of the detection threshold of the \ac{acd} required a careful balancing between two requests:  high efficiency, requiring a low threshold because of statistical fluctuation of the energy loss in the \ac{acd} material (assumed to follow a Landau),  and low impact of back-splash, requiring a high threshold because electron signal  produced by  Compton scatter of low-energy photons is typically low. Figure~\ref{soglia} shows the pulse-height distribution in an \ac{acd} tile:  the peaks of \ac{mip} and back-splash particles are clear; also shown is the chosen detection threshold of 0.45~\ac{mip}~\cite{pass7-status}. Since the signal threshold is constant on a single tile, another fundamental requirement for the \ac{acd} is that also the light signal from a tile is almost constant on all its surface.


\begin{figure}[htbp]
\begin{center}
\subfigure[Scheme of the \ac{acd}.  BEA modules (Base Electronics Assembly) contain photo-tubes, high-voltage supply and read-out circuits.]{\includegraphics[width=.8\textwidth]{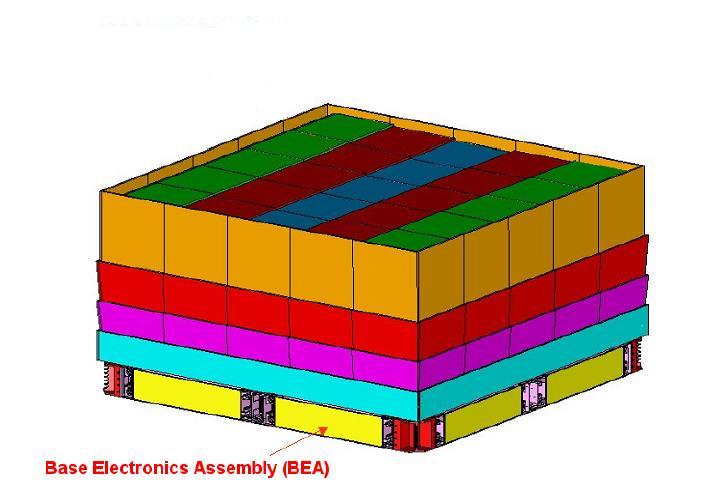}}
\subfigure[\ac{acd} tile shell assembly, clear fiber cables are seen in the cutout.]{\includegraphics[width=.42\textwidth]{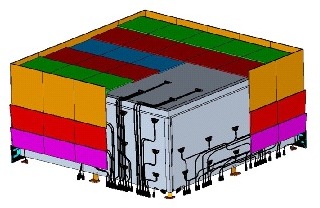}}
\subfigure[\ac{acd} Base Electronics assembly with photo-tubes  shown; below it, the \ac{lat} grid.]{\includegraphics[width=.42\textwidth]{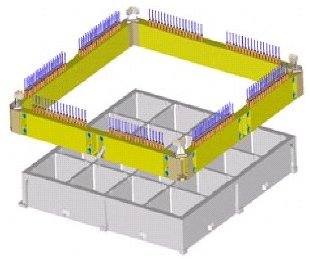}}

\caption{Design of the \ac{acd}.~\cite{acddoc}}
\label{acd-figure}
\end{center}
\end{figure}

The \ac{acd}  consists of 89 tiles of plastic scintillator and 8 ribbons to cover gaps between tiles. Figure~\ref{acd-figure} shows the \ac{acd} with  its components. Plastic scintillators were chosen because they are cheap, robust and largely used in previous space mission: all previous space-based gamma-detectors had an anti-coincidence detector of plastic scintillator. Of the 89 tiles, 25 compose the top of the instrument, while every side is composed by 16 tiles. All tiles are 10~\milli\meter\ thick, with the exception of the five composing the middle row of the top face of the \ac{acd}: because they have the longest light-path to the read-out photo-tubes, the light lost in transmission is larger than for other tiles, and they are 12~\milli\meter\ thick to increase their light yield. This thickness was chosen to obtain $\sim20$ photo-electrons generated by a \ac{mip} in each one of the photo-tubes reading a tile.

All tiles are optically isolated, to obtain independent signals and also to reduce the effect of  possible perforation caused by micro-meteoroids. Isolation is provided, for every tile, by two layers of white high-reflective material, followed by two layers of black absorbent material.

Tile dimension is not uniform:  tiles on the top and in the higher row on every face are $32\times 32~\squaren\cm$, while subsequent tiles are smaller, with tiles in the third row of every side $15\times 32~\squaren\cm$. The size of the side tiles decreases in order to maintain the solid angle seen from the calorimeter almost constant, and therefore to have the same probability of a back-splash signal in every tile. The estimated probability of a back-splash signal in a tile at a distance $x~\cm$ from the calorimeter face was estimated~\cite{acddoc} 

\begin{equation} 
P\propto \left(\frac{55}{x+15}\right)^2\times \sqrt{E}
\end{equation}
 where $E$ is the energy of the photon. The bottom row of each side is covered with a single $17\times170~\squaren\cm$ tile: since this area is outside of the \ac{lat} primary field of view, segmentation to prevent back-splash signal is not required, and an efficiency of 0.999 is sufficient.

The first row on every side is some \milli\meter\ higher than the top face of the \ac{acd}, therefore reaching the top of the thermal blanket and the micro-meteoroid shield. This protruding structure, the \textit{crown}, detects charged particles grazing the top face of the \ac{lat}. These particles can traverse  the  micro-meteoroid shield without interacting with the tiles of the top face. Because of their long path-length in the shield, these particles have a high probability of generating a gamma ray: if they were to go undetected, the produced gamma rays would constitute a background of about 5\% of the diffuse gamma ray radiation.

The coverage of the \ac{acd} must be total, therefore no insensitive gap between the tiles has to exist.  In one direction, consecutive tiles  overlap each other by 20~\milli\meter. Because overlap tiles in two directions would add excessive complexity, mass and volume, 2--4~\milli\meter\ gaps on the remaining side of the tiles, that are requested to allow thermal expansion of tiles,  are covered by flexible ribbons of plastic scintillator (8 ribbons in the \ac{acd}).

To protect the \ac{acd} from perforation caused by micro-meteoroids and space debris that could penetrate the light-tight wrapping of tiles making them useless, the whole instrument is surrounded by a \ac{mms}. Since interaction of charged particle with the \ac{mms} would produce an irreducible local background of gamma-rays, the \ac{mms} must be as thin as possible. The final design has an \ac{mms} with a total area density of 0.39~\gram~\cm\rpsquared\ that, based on actual orbital debris model, should allow no more than one penetration in five years. It has to be noted that, in case of a penetration, it should be however possible to identify events coming from the dead spot in the \ac{acd}~\cite{acddoc}. At present (June 2016), the \ac{mms} has not suffered any perforation.

\subsection{Light collection and electronics}

\begin{figure}[htbp]
\begin{center}
\includegraphics[width=.8\textwidth]{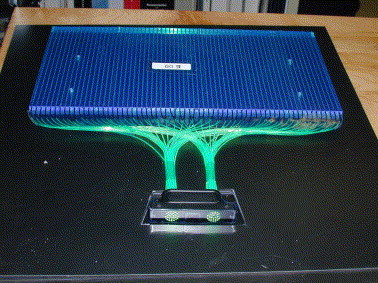}

\caption{An unwrapped tile of the \ac{acd}.  \acp{wls} inside the tile and  \acp{wls} that collect the light and transmit it to the optical collector in the foreground are clearly visible.~\cite{acddoc}}
\label{fibre}
\end{center}
\end{figure}

To achieve the \ac{acd} requested performances, light collection must be as uniform as possible inside the tiles. Therefore,  light is collected by \acp{wls} (absorption peak 425~\nano\meter, emission peak 490~\nano\meter) inserted inside the tiles at 2~\milli\meter\ from the surface with a pitch of 5~\milli\meter\ (see figure~\ref{fibre}). This read-out scheme grants  fluctuation in the light collection below 10\%, except for the tile edges, where light yield decreases to 70\% recovering to 100\% at $\sim3~\centi\meter$ from the edge. Light is transmitted to the read-out photo-tubes using \acp{wls} or, for long paths, by clear fiber cables coupled to \acp{wls} (top and upper two rows of sides). To prevent losses caused by failures of photo-tubes or by the high-voltage supply, every tile or ribbon is read by two photo-tubes,  with neighboring \acp{wls} connected to different photo-tubes. Photo-tubes of the same tile are powered and read by independent electronic circuits.

The electronics of the \ac{acd} is situated in eight chassis at the base of the detector. It is composed by 12 Front End Electronics (FREE) circuits, that handle the read-out and high-voltage power supply of the photo-tubes, and the communication between \ac{acd} and the \ac{lat}. Each FREE contains 18 GLAST ACD Front End Electronic chips (GAFE), each one reading a photo-tube (only 194 on 216 available channels are used). Since the signal in an \ac{acd} tile  spans from 0.1~\ac{mip} to hundreds of \ac{mip}, the signal is split between two amplification circuits with different gains: a specific discriminator selects which  output is sent to the digital-analog converter associated with each GAFE. Furthermore, two different discriminators are used to generate two different trigger primitives: the first, with a low threshold, detects the passage of any charged particle while the second, with a high threshold, detects the passage of heavy nuclei.

 Each FREE module contains a GLAST ACD Readout Control (GARC) chip: this chip handles the transmission of trigger primitives and read-out signals, the regulation of the threshold of the discriminators (which is dependent from the instrument temperature) and the power supply of the photo-tubes of the FREE.

The \ac{acd} electronics is required to handle a rate of charged particles through a single tile up to 3~\kilo\hertz\ without performance degradation: the expected rate should not exceed 1.5~\kilo\hertz.
\section{Event triggering}
\label{trigger}
The \ac{lat} trigger is designed to collect the largest possible fraction of the events that are potentially  originated by gamma rays. Only after acquisition events are analyzed at different stages to reject the dominant background of charged particles. The read-out of the instrument takes at least 26.5~\mycro\second, therefore, in order to limit the dead time fraction to less than 10\% the trigger is designed to keep the total trigger rate below 3.8~\kilo\hertz. The trigger is also designed to acquire special samples of events for monitoring or calibration purposes. 

Each  of the \ac{lat} subsystems is able to detect the passage of particles through itself, generating different trigger primitives that are collected by the Central Trigger Unit in the Global Electronic Module (GEM) of the \ac{lat}. The trigger primitives generated by the \ac{lat} subsystems are:
\begin{description}
 \item[TKR:] issued when a signal above threshold (nominally 0.25~\acp{mip}) is detected in three consecutive x-y layers of the \ac{tkr};
 \item[CAL\_LO: ]issued when the signal in one end of a crystal in the \ac{cal} is above the low-energy threshold (nominally 100~\MeV);
 \item[CAL\_HI: ]issued when the signal in one end of a crystal in the \ac{cal} is above the high-energy threshold (nominally 1~\GeV);
 \item[ROI: ]when the signal in an \ac{acd} tile is above the low threshold (nominally 0.45~\acp{mip}), a VETO primitive is issued. Towers of the \ac{tkr} have a programmable list   of associated \ac{acd} tiles that ``shadow'' them: if a VETO signal is present in a tile together with a \ac{tkr} primitive in the associated tower, a ROI primitive is issued;
 \item[CNO: ]issued when the signal in one of the \ac{acd} tiles is above the high threshold (nominally 25~\acp{mip}). This primitive signals the passage of a highly ionizing nucleus (CNO stands for Carbon, Nitrogen, Oxygen).
\end{description}
In addition, a special \textbf{PERIODIC} primitive is generated at a nominal rate of 2~\hertz.

Some of the trigger primitives are enabled to open a trigger window  (nominally\- 700~\nano\second) during which trigger primitives from all the instrument are collected. The combination of the trigger primitives collected during the trigger window is compared to a table of allowed trigger combinations: when a trigger condition is satisfied, a read-out of the entire instrument starts. All the primitive combina\-tions are mapped into so-called trigger engines, that define the \ac{lat} readout mode (for example the zero suppression in \ac{cal} and \ac{acd} and the single or four-range read-out of \ac{cal} crystals) and in some case set a pre-scale on the trigger signal, that is the number of valid trigger requests necessary to start a single read-out of the \ac{lat}.

The trigger engines used by the \ac{lat} on orbit are defined in table~\ref{trigger-table}. In the usual science operation mode, the principal source of \Pphoton triggers is when three consecutive x-y layers of the tracker have a signal above threshold without vetoing from the \ac{acd} (engine 7), but other engines are designed to handle  high energy events or for  calibration purposes.
\ac{acd} signals are used on-board as a throttle on the first-level  trigger, using the \textbf{ROI} primitive as a veto for some of the engines (see \cref{rate_eventi}). Again, it has to be remarked that, also for engines designed to acquire \Pphoton events, the vast majority of acquired events are charged particles.

\begin{table}[htbp]
 \begin{center}
  \begin{tabular}{|>{\footnotesize}l|ssssss|s|>{\footnotesize\raggedleft}p{.092\textwidth}|>{\footnotesize}p{.099\textwidth}|}
  
\toprule
\footnotesize Engine&PERIODIC&CAL\_HI&CAL\_LO&TKR&ROI&CNO&Prescale&\raggedright Average Rate (Hz)& Note\\
\midrule
3 & 1 & $\times$ & $\times$ & $\times$ & $\times$ & $\times$ & NO & 2 &monitoring\\

4 & 0 & $\times$ & 1 & 1 & 1 & 1 & NO & 200 &heavy ions\\

5 & 0 & $\times$ & $\times$ & $\times$ & $\times$ & 1 & 250 & 5 &calibration\\

6 & 0 & 1 & $\times$ & $\times$ & $\times$ & 0 & NO & 100 &   HE \Pphoton \\

7 & 0 & 0 & $\times$ & 1 & 0 & 0 & NO & 1500 & \Pphoton\\
9 & 0 & 0 & 1 & 1 & 1 & 0 & NO & 700 & \Pphoton \\
10 & 0 & 0 & 0 & 1 & 1 & 0 & 50 & 100 &calibration\\
\bottomrule
  \end{tabular}

 \end{center}
\caption{Definition of the trigger engines used by the \ac{lat} during on-orbit operation and their typical trigger rates. Trigger primitives can be 1: required, 0: excluded, $\times$: not relevant for the engine. Trigger combination are assigned to the first engine matching them. There are trigger engines map conditions that should not happen on-orbit (0,1,2) or are disabled (8).}
\label{trigger-table}
\end{table}

\section{On-board filter}
\label{filtri}

\begin{figure}[htp!]
\begin{center}

\includegraphics[width=.75\textwidth]{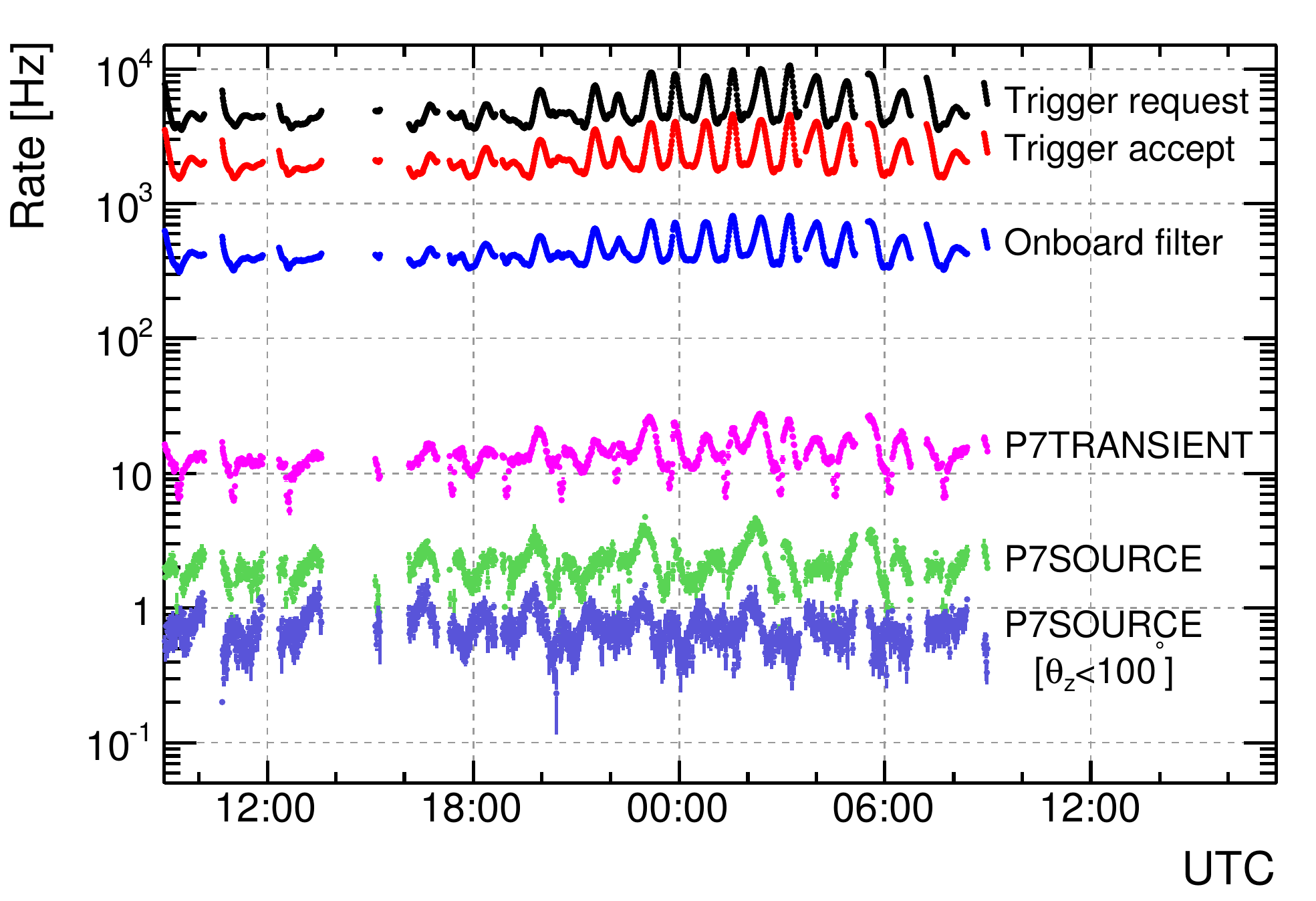}
\caption{Event rate at different stages of  acquisition and processing, on a typical day of operation of the \ac{lat},  \nth{17} August 2011. From top: event generating a trigger request, event accepted by the trigger system, event downloaded to ground by the on-board filter, loose and stricter gamma selections after ground processing.~\cite{pass7-status}}
\label{rate_eventi}
 
\end{center}
\end{figure}

The rate of events acquired by the \ac{lat} trigger   is $\sim 2.5~\kilo\hertz$, and must be reduced to match the available download bandwidth of the spacecraft to $\leq 400~\hertz$ (see \cref{rate_eventi}), a rate however greatly larger than the typical \Pphoton rate, that is of the order of a few \hertz.

The events are therefore processed on-board by three parallel and independent filters, each one  designed to select different categories of events. 

The \textit{Gamma filter} is optimized to reject events clearly generated by charged particles and events where no reconstruction is possible, while it accepts the large majority of events generated by photons (if there is a chance to reconstruct them). It is structured in a hierarchical sequence of tests, starting with least CPU-consuming ones. When an event fails one of the test, is rejected and does not proceed to further processing. Last tests include a very rudimentary track reconstruction, to check for correlation with hits in \ac{acd}. All events with more than 20~\GeV\ of deposed energy are downloaded (High-pass): this is done because high energy events are a small but often very interesting fraction of total events. This last feature is fundamental for the analysis of \ac{cre}, because almost all the  electrons with more than 40~\GeV\ of initial energy are downloaded to ground and can therefore be studied. The average rate of events passing the Gamma filter is $\sim 350~\hertz$.

The \textit{Diagnostic filter} sends to ground the events acquired through the periodic trigger and an  unbiased sample of all the \ac{lat} triggers, pre-scaled by a factor of 250. The average rate of events selected by the Diagnostic filter is $\sim 20~\hertz$. This filter is fundamental for the analysis of \ac{cre} below some tens of \GeV: at this energy, a large fraction of the events generated by electrons is rejected by the Gamma filter and not downloaded; furthermore, the Gamma filter introduces a strong bias in the electron sample that is very hard to correct. The low energy part of the spectrum published in~\citet{electronpaper} is obtained using data selected by the Diagnostic filter.

The \textit{Highly Ionizing Particle filter} selects heavy ion candidates for calibration purposes, with an average rate of $\sim 10~\hertz$.

\section{Event reconstruction}
Event reconstruction translates the \ac{digi} (provided in ROOT format, see \citet{root}) in physical quantities and uses them to give a high-level description of the event. Event reconstruction of the \ac{lat} is summarized in~\citet{pass7-status} and fully described in~\citet{latdoc}.

The event reconstruction described in this section is that used until 2015 in \ac{lat} data processing: the changes introduced with the new event reconstrution process (Pass~8) are described in  \S~\ref{pass8}.
\paragraph{Reconstruction of the energy deposition in Calorimeter}
The first step of the event reconstruction is the analysis of the energy deposition (if present) in the \ac{cal}: pedestal values are subtracted from the output of the photo-diodes, and the signals from both ends of each crystal are combined to obtain the energy deposition and  longitudinal position, therefore obtaining a   3D matrix of the energy deposition in the \ac{cal}. Through moment analysis, the centroid of the energy deposition and its main axis (that will be used as \ac{cal} event direction) are calculated. The energy deposition is treated as a whole, with no attempt to divide it in separate clusters in order to identify signals coming from different particles. The new event reconstruction process described in \S~\ref{pass8} will include a \ac{cal} clusters identification.
\paragraph{Track finding}\label{track-rec}
The following step in the event reconstruction is the analysis of the signals in the \ac{tkr} and the track finding and fitting process. First, adjacent hit strips are merged in clusters. The track reconstruction algorithm  generates various track hypothesis and uses a Kalman filter (\S~\ref{kalman-sez}) to populate the tracks with hits, fit them and calculate their quality, keeping into account the effect of multiple scatter in the \ac{tkr} material and the uncertainties in the measured positions. Tracks are  requested to have at least 5 hits and to be classified as ``good'' in terms of quality parameters: tracks failing one of these requests are  discarded.   When the algorithm terminates with more than one valid track, the ``best'' track is selected discarding all the other hypothesis, its hits are marked as used and the algorithm is run again on the remaining hits.  Only the first hit of  reconstructed tracks and  clusters with number of strips  too large to be reasonably generated by a single charged particle are allowed to be shared between two tracks. The track reconstruction process  ends when no more tracks can be reconstructed or when 10 tracks are reconstructed.

The track reconstruction process uses two  methods to generate track hypothesis: the first invoked is the \ac{cspr}, that uses information from the \ac{cal} moment analysis to generate track hypothesis and to evaluate their quality, and the \ac{cal} deposed energy as an initial estimate of the tracks energy. To generate a trial track, the centroid of the energy deposition in the \ac{cal} is connected  with a random x-y pair in the furthest \ac{tkr} plane from the \ac{cal}. If a subsequent hit is found near this line a track hypothesis is generated. Above 1~\GeV\ the search for the initial x-y pair is restricted to a cone around the \ac{cal} reconstructed axis. The cone opening angle is narrowed with increasing \ac{cal} energy. Trial tracks are generated and populated until a sufficiently good track has been found and at least all the x-y pairs on two layers have been looped over. To evaluate the quality of the tracks,  the quality parameter of the Kalman fit, that tends to better evaluate longer and straighter tracks,  is combined with the distance of the track from the energy centroid  and, above 1~\GeV\ of deposed energy, also with the distance of the tracks from the shower axis, with a weight that increases with the deposed energy. When \ac{cspr} is not able to find more tracks, or when the moment analysis of the energy deposition in the \ac{cal} has failed, \ac{bspr}, that does not use any information from the \ac{cal}, is invoked. To generate a trial track, \ac{bspr} connects at random all x-y pairs from the two \ac{tkr} plane furthest from the \ac{cal}. If a third x-y pair is found in the subsequent layer along this direction, a trial track is generated and tested. Because \ac{bspr} does not use any information from the \ac{cal}, the energy of the tracks is initialized at 30~\MeV\ by default.

\paragraph{Energy reconstruction}\label{energy-reconstruction}
When the track reconstruction process ends, tracks are ordered on the basis of their quality, and the direction provided by the ``best'' track is used to  correct the deposed energy in the \ac{cal}, for the partial containment of the shower in the instrument. Two different methods are used: a parametric correction and a 3D fit of the shower profile. The parametric correction covers the entire phase space of the \ac{lat}, while the shower profile method often fails at low energies, but above some \GeV\  returns a better energy estimate in almost all cases.  A third method, based on a maximum likelihood approach comparing the energy deposed in the first layer of the \ac{cal} with the total deposed energy, was used in Pass 6 reconstruction but is not used in  Pass 7  (see \S~\ref{passi} for the definition of Pass 6 and Pass 7), because it creates artifacts in the energy distribution caused by the binning used in the parametrization of the energy ratio and because it was designed  not to return values above 300~\GeV, therefore it tends to concentrate values just below this energy. In the Pass 7 reconstruction, the energy returned by all these methods is corrected subtracting an energy-dependent bias estimated using Monte Carlo simulations.

At low energy a large fraction of the event energy ($\sim 50\%$ at 100~\MeV) can be  deposed in the \ac{tkr}, and must be estimated and  added to the reconstructed energy in the \ac{cal}. Considering the \ac{tkr} as a sampling calorimeter, where the deposed energy is proportional to the strips over threshold, we can estimate this energy as  

\begin{equation}\label{energia-tracker}
 E_{Tkr}=\sum_{sec}{N_{sec}\overline{E}_{sec}}
\end{equation}
where $\overline{E}_{sec}$ is the average deposed energy in all the \ac{tkr} material for each hit strip, calculated for each  \ac{tkr} section (thick, thin, blank), $N_{sec}$ is the number of hit strips in a cone whose opening angle decreases as $E^{0.5}$, where $E$ is the apparent energy in \ac{cal}.
\paragraph{Track refitting}
The sum of the energy estimate in the \ac{tkr} and in the \ac{cal} is used to improve the reconstructed tracks, with \ac{cal} energy calculated using parametric correction, because it is the only method which always returns a value. The energy is apportioned between the best two tracks (if more than one track is present), according to the amount of multiple scatter observed for each. Then, the Kalman fit is run again on these tracks using the assigned energy as starting value, without repopulating the tracks. 
\paragraph{Vertex finding}
After  tracks are reconstructed, vertexes are searched between tracks. The presence of a vertex between two tracks is an important signature of a gamma conversion, although it is infrequent: at low energy one particle of the \Pelectron \APelectron pair often has so little energy that is not reconstructed by the \ac{lat}, while at high energy the angle between the pair is too small to  resolve it. Vertex search starts from the best tracks and loops over all the other tracks searching for the one with the closest approach distance. If the approach distance is below 6~\milli\meter, a vertex solution is generated, with a reconstructed position calculated using the first hits and a quality parameter calculated using the distance of approach and the $\chi^2$ of the combination of the tracks. If a vertex is found, the tracks are marked as used, otherwise only the best track is marked as used: in both cases, the process is repeated on the next best unused track. The direction given by the ``best'' track and that given by the first vertex are combined co-variantly to the direction given by the straight line linking the energy centroid in the \ac{cal} with the hypothetical conversion point (vertex or beginning of the best track) to obtain two ``neutral energy'' directions: in events where an important fraction of the  energy is carried by \Pphoton-rays produced during the conversion or immediately after, the neutral energy solution can be a better estimate of the initial photon direction.
\paragraph{ACD reconstruction}
Pedestal values are subtracted from the output of the photo-tubes and the energy deposed in each tile or ribbon is reconstructed. After the conclusion of the track reconstruction process, all tracks are propagated to the \ac{acd}, searching for an intersection with a tile or a ribbon with an over-threshold energy deposition: if there is no such an intercept, the distance between the track projection and the nearest hit \ac{acd} element is calculated.

\paragraph{Extraction of the ``figures of merit''}\label{figures-of-merit}
The last step is the extraction from the \ac{recon} of a few hundreds of \ac{merit}\footnote{\ac{recon} and \ac{merit} are provided in ROOT format~\cite{root}}. These quantities give a high-level description of the whole event. This step includes the selection of the ``best'' event reconstructed direction and of the ``best'' reconstructed energy, choosing between the different options produced during the reconstruction, and the calculation of a parameter that estimates the quality of the direction or energy reconstruction. Both the selection of the best energy and  direction and the estimate of their quality are made using \acfp{ct}: for a brief description of this statistical tool, see \S~\ref{CT}.

The best energy is selected among the energies provided by the different energy reconstruction methods using a \ac{ct} that selects the method that most likely would provide the best energy estimate. Other \acp{ct} estimate the probability that the reconstructed energy is within $2\sigma$ or $3\sigma$ from the core of the energy dispersion of the instrument (that depends on the energy and reconstructed direction). These two quantities are used to define a quality estimator of the energy reconstruction:

\begin{equation}
 P_E=\sqrt{P_{2\sigma}P_{3\sigma}}
\end{equation}

The best direction is selected by a \ac{ct} between the four possible reconstruction (when available): best track, best vertex and the related neutral energy solutions. A second \ac{ct} estimates the probability that the chosen direction falls within the 68\% containment angle, defined as

\begin{equation}
 C_{68}(E)=\sqrt{\left[c_0\left(\frac{E}{100~\MeV}\right)^{-\beta}\right]^2+c_1^2}
\end{equation}
where the coefficient $c_0$ and $c_1$ are estimated by Monte Carlo and  are different for events converting in the front and back section.

\subsection{Classification Trees}\label{CT}

\acp{cart}, first described in~\citet{breiman}, are learning machines, the purpose of which is to predict the output of a dependent variable $y$ starting from a set of measured quantities $\mathbf{\bar{x}}$. The values of $y$ are discretized in Classification Trees (an example is the choice of the best energy correction method in the energy reconstruction) and continuous in Regression Trees (the estimated probability of being in the core of the energy distribution).
 
 \acp{cart} are ``grown'' using a \textit{learning sample}, that is a set of events where the dependent quantity $y$ is known and its correlation with $\mathbf{\bar{x}}$ can therefore be studied. The learning sample can be  a Monte Carlo generated data-set or the result of previous measurement ($\mathbf{\bar{x}}$,$y$) on events from the same population. \acp{cart} map the learning sample on a d-dimensional space, where d is the dimension of the vector of measured quantities $\mathbf{\bar{x}}$ and therefore each event is represented by a point in the space. This space is then divided  using a series of subsequent splits, creating  sub-spaces where mixing between events with different $y$  is  reduced by each split. The measure of the event
mixing is not unique, typically used indexes are the Gini index of heterogeneity $\mathit{I}$, introduced in~\cite{gini}, and the information entropy $\mathbb{H}$, defined by Shannon in~\cite{entropia-shannon}
 
\begin{eqnarray}
 \mathit{I}(\mathbf{X})&=&1 - \sum_{j}p(j\mid \mathbf{X})^2  \\
 \mathbb{H}(\mathbf{X})&=&-\sum_{j} p(j\mid \mathbf{X})\log{p(j\mid \mathbf{X})}
\end{eqnarray}  
 where $p(j\mid \mathbf{X})$ is the probability that an event of  $\mathbf{X}$ belongs to the $j^{th}$ class in which the population is divided, that is, the fraction of events of $\mathbf{X}$ for which $y\in \bar{y}_j$. Both indexes are 0 if all the events in $\mathbf{X}$ belong to the same class and reach a maximum when events are equally distributed in all classes.
 
 The  series of subsequent splits constitutes  a tree, where each node is a cut on a combination of quantities from $\mathbf{\bar{x}}$ and where each terminal node is assigned to a specific class $j$. Therefore, any event for which $\mathbf{\bar{x}}$ is measured can be assigned to a class simply passing it through the tree. The training process is not univocal and can be technically very complex; \citet{CT-bagging} describes a methods commonly used in the \ac{lat} event reconstruction and selection, where the training sample is divided in sub-sets and trees are grown on each subset and then averaged.
 
 \acp{cart} are a valuable alternative to other selection methods, particularly when the dependence of $y$ on $\mathbf{\bar{x}}$ is scarcely known or very complex.
 
\begin{figure}[htbp]
\begin{center}
\includegraphics[width=1.\textwidth]{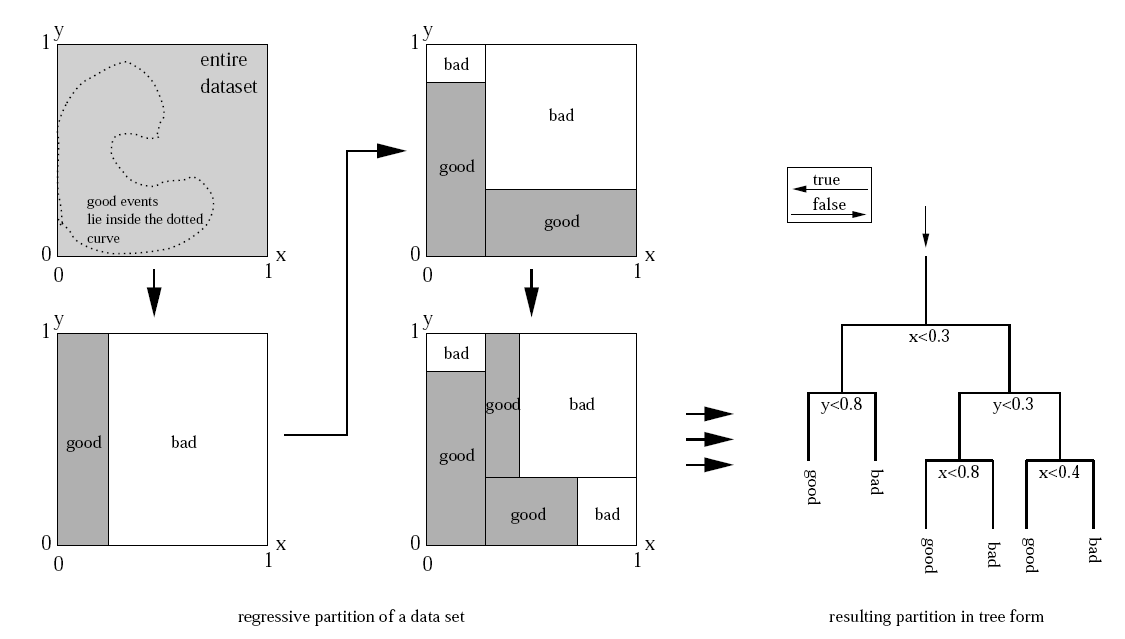}

\caption{Growing of a \ac{ct} on a learning sample  of events from two different classes, ``good'' and ``bad''. Events are described by two normalized quantities, x and y. 
On the left there is the x-y plane containing all the events, that is repeatedly split in sub-spaces of increasing purity. On the right the resulting tree, its terminal nodes classifying events as ``good'' or ``bad''.\cite{ric}}
\label{ct-fig}
\end{center}
\end{figure}

 An example of the training of a \acf{ct} on a 2-d space ($\mathbf{\bar{x}}=(x,y)$) with two classes of events (``good'' and ``bad'') is shown in figure~\ref{ct-fig}, together with the resulting tree.

\subsection{The Kalman filter in track reconstruction}\label{kalman-sez}

The track reconstruction algorithm described in here was used in Pass~6/Pass~7 analysis; the new Pass~8  track reconstruction algorithm is treated in \S~\ref{tree-section}

The track reconstruction process described at page~\pageref{track-rec} requires an algorithm that is able to simultaneously find out which hits in the \ac{tkr} belong to the hypothetical track analyzed and to find the best fit of the track direction, therefore being able to find subsequent hits. While doing so, the mechanism must take into account both the impact of the multiple scattering and the uncertainty in the measurement of the hits position. The chosen algorithm in the \ac{lat} reconstruction is the Kalman filter, initially introduced  in~\citet{kalman1} and the application of which to track finding in physics can be found in~\citet{fruhwirth} or in~\citet{regler}. The Kalman filter is a recursive process that uses   information  about the track in a layer of the detector to propagate it in the following layer, using a linear approximation of the track propagation (this means that the information about the track in each layer are represented by a state vector $\mathbf{\overrightarrow{x}_k}$ that is linearly propagated to the next layer using a matrix). This step-by-step approach  strongly reduces the  computational power needed, making the process affordable. It has to be noted that in the case of fitting-only problem, where all the hits are known to belong to the same track, if the uncertainties on the hits location and the multiple scattering can be approximated as Gaussian, the Kalman filter is an optimal fit method,  equivalent to a $\chi^2$ or a maximum likelihood fit. In the case that not all the hits belong to the track, so that the problem is track finding-and-fitting, an analytic solution cannot be found; however, the Kalman filter has proved to be a very valuable method. In this section a general description of how the Kalman filter works is given, a more detailed description of the application of the Kalman filter to the \ac{lat} reconstruction is given in \S~\ref{kalman-annotato}.

\begin{figure}[htbp]
\begin{center}
\includegraphics[height=110mm]{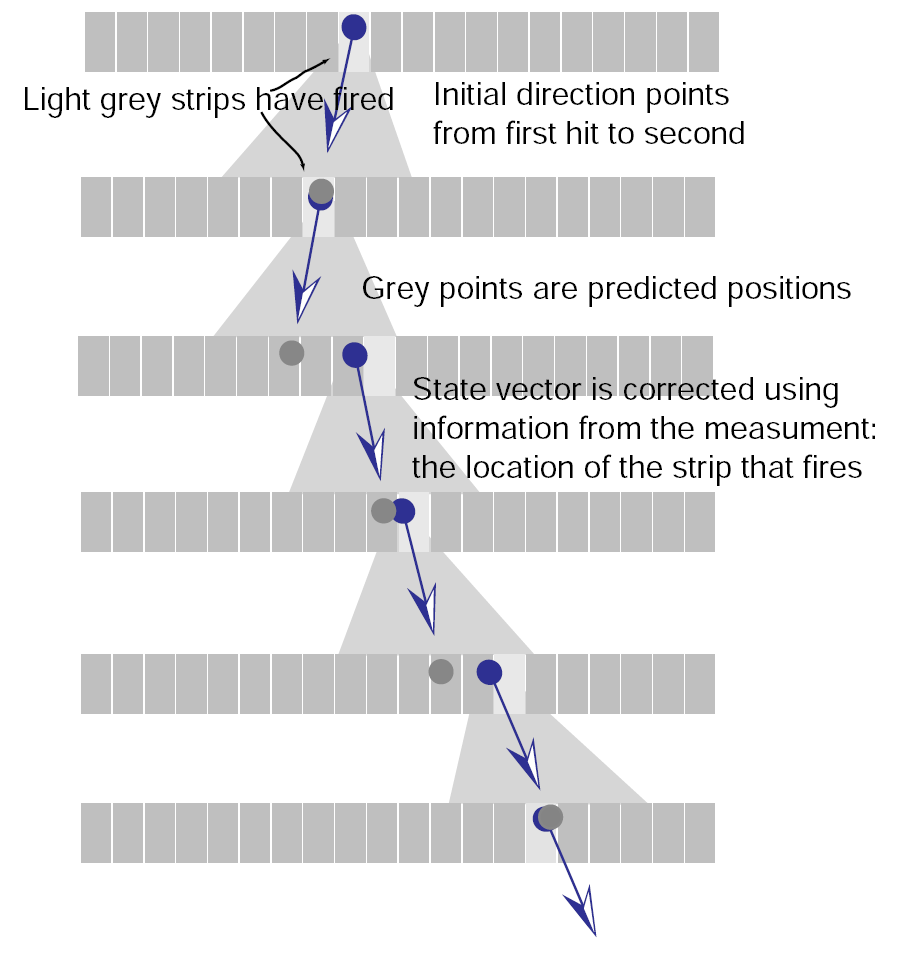}

\caption{Kalman filtering in a silicon strip tracker. The Kalman filter uses the current track direction  to predict the position of the hit on the next layer (grey circles). The spread on the position caused by multiple scatter is also calculated and creates a cone around which a new hit is searched (grey area). The effective position of the hit on the plane (blue circles) is used to correct the track direction, that is used to predict the hit position in the subsequent layer.~\cite{kalman2}}
\label{kalman-filter}
\end{center}
\end{figure}

The Kalman filter consists of two subsequent steps: filter and smoother. The filter  starts from the first hit of the hypothetical track and uses the initial state vector  to propagate the track in the next detection layer. The impact of the multiple scatter is used to calculate the covariance matrix of the state vector, that gives a region on the layer where a new hit is searched. If a hit is found near this ellipse, the state vector is updated using its position. If, as in the case of the \ac{lat}, the measured quantities are different from the quantities used to describe the track, also the vector of measured quantity $\mathbf{\overrightarrow{m}_k}$  is introduced, together with measurement matrix that relates it to $\mathbf{\overrightarrow{x}_k}$. When updating $\mathbf{\overrightarrow{x}_k}$ using $\mathbf{\overrightarrow{m}_k}$, the uncertainties on the measured quantities are used to update the covariance matrix of $\mathbf{\overrightarrow{x}_k}$. Then the procedure is iterated to the next layer until the track is terminated. The filter is described in figure~\ref{kalman-filter}. The filter weights hits in function  of their associated covariance matrix: if this matrix is large compared with that associated with the multiple scattering, single hits will have a small weight and therefore their impact on the track direction will be small, so that the track will tend to a straight line. Instead, if multiple scattering dominates on the uncertainty on hits position, each measure will have a strong impact on the track direction, and the filter will simply connect the hits to fit the track.

\begin{figure}[htbp]
\begin{center}
\includegraphics[height=110mm]{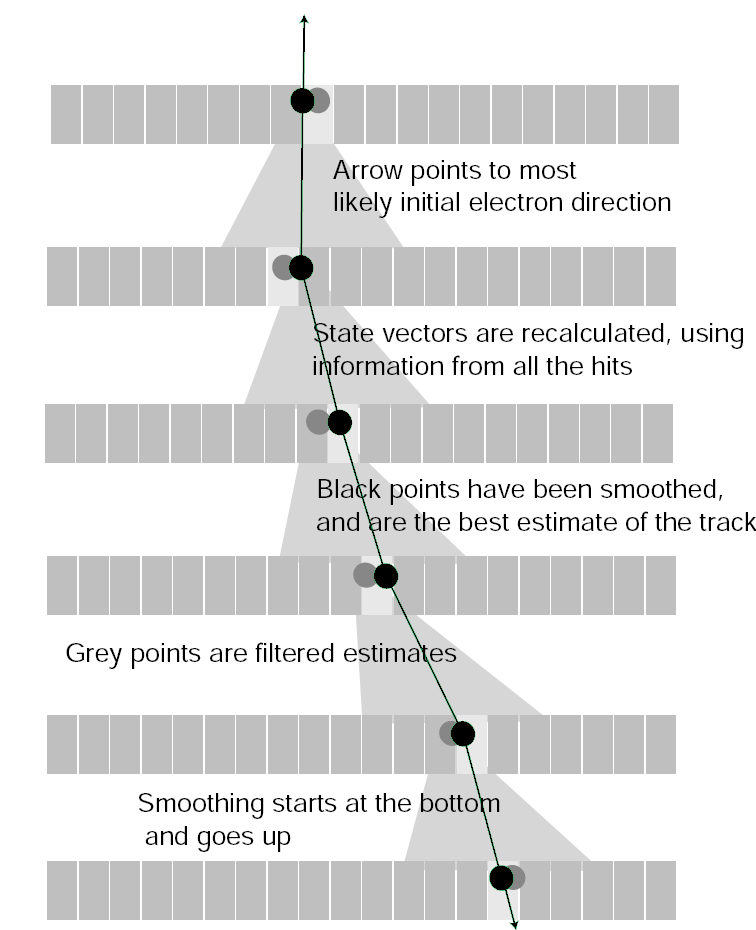}

\caption{Kalman smoothing in a silicon strips detector. Starting from the bottom plane, the smoothing uses the hits calculated during the previous filtering (grey points) to back-propagate the track into the previous layer (grey area) and improve the position estimation in the layer (green points).~\cite{kalman2}}
\label{kalman-smoother}
\end{center}
\end{figure}

When the filtering process ends, the state vector $\mathbf{\overrightarrow{x}_k}$ at each layer depends only on previous layers, therefore it does not contain all the available information: this is achieved with the smoother. Smoothing steps up the track from the bottom, starting from the last hit and propagating backward along the track,  keeping into account the effect of multiple scatter and the uncertainty on measured quantities, updating the state vector and the covariance matrix at every plane. At the end of the smoothing process, the state vector in the first plane contains the initial track direction, and the covariance matrix in the first plane gives its uncertainty. The smoothing process is described in figure~\ref{kalman-smoother}. At each layer the residual vector with respect to measured hit is calculated, together with its covariance matrix. From these two quantities it is possible to extract a quantity distributed as a $\chi^2$, that can be used to test hits in order to identify outliers. The sum of this quantity  on all the layers (that is not necessarily distributed as a $\chi^2$, but in the \ac{lat} the difference is not too large) gives us an estimation of the track quality.

\subsubsection{Track termination in the LAT}
In the \ac{lat} event reconstruction, track populating can be terminated before the track reaches the edge of the \ac{tkr}. When the filter does not find a hit around the region described by the covariance matrix on the layer, the track is projected into the subsequent layer but, if the track doesn't cross an un-instrumented region (like the space between towers), a ``gap'' is added to the track. The track is terminated when more than two gaps have accumulated.


\subsection{The annotated Kalman filter in LAT track reconstruction}\label{kalman-annotato}
This section will briefly describe the implementation of the Kalman filter used in the \ac{lat} track reconstruction. For a complete review of this topic, see~\citet{kalman2,kalman-glast1,kalman_guida,kalman-last}, the last reference contains the last updates made on the Kalman filter implementation to the \ac{lat} in 2002.  The complete description of the used notations and the demonstrations of the general formulas are given in~\citet{fruhwirth}.

The track is described at each layer $\mathbf{k}$ by a state vector that contains all the relevant information about it:

\begin{equation}
 \mathbf{\overrightarrow{x}_k}=\left(
 \begin{array}{c}
  \mathrm{horizontal~position~(x,y)}\\
  \mathrm{track~slope~(\theta,\phi)}\\
  \mathrm{current~energy}\\  
 \end{array}
\right)
\end{equation}
The state vector is linearly propagated to the next layer by the equation

\begin{equation} 
\mathbf{\overrightarrow{x}_{k+1}}=\mathbf{\overrightarrow{x}_k F_k}+\mathbf{\overrightarrow{q}_k(\theta)}
\end{equation}
where $\mathbf{\overrightarrow{q}_k(E)}$ describe the impact of the multiple scatter on the track and the matrix $\mathbf{F_k}$ simply projects the track from a layer into the subsequent one, correcting the energy by the losses in the infra-layers material, calculated with the Bethe-Block formula~\cite{pdg} 

\begin{equation}
\mathbf{\overrightarrow{x}_k F_k}= \left(
 \begin{array}{c}
  \mathrm{x+d_k \sin{\theta}\cos{\phi}}\\
  \mathrm{y+d_k \sin{\theta}\sin{\phi}}\\
  \mathrm{E -\Delta E(k,E,\theta)}\\
 \end{array}
 \right)
\end{equation}

To compare the state vector with the quantities measured in the \ac{tkr}, the vector of measured quantity $\mathbf{\overrightarrow{m}_k}$ is introduced together with the measurement matrix that relates it to the state vector

\begin{equation}
 \mathbf{\overrightarrow{m}_k}=\left(
 \begin{array}{c}
  \mathrm{strip~number~(x,y)}\\
  \mathrm{current~energy}\\  
 \end{array}
\right)
\end{equation}
\begin{equation} 
\mathbf{\overrightarrow{m}_k}=\mathbf{\overrightarrow{x}_k H_k}+\mathbf{\overrightarrow{\epsilon}_k}
\end{equation}
$\mathbf{\overrightarrow{\epsilon}_k}$  introduces the uncertainties on the measured position, while $\mathbf{H_k}$, because the \ac{tkr} provides no energy information, is  

\begin{equation}
\mathbf{H_k}= 
\begin{bmatrix}
  1 & 0 & 0 & 0 & 0 \\
  0 & 1 & 0 & 0 & 0 \\
  0 & 0 & 0 & 0 & 0\\

\end{bmatrix}
\end{equation}
$\mathbf{\overrightarrow{q}_k}$ and $\mathbf{\overrightarrow{\epsilon}_k}$ are both random quantities. Because their expectation value is 0, the only relevant quantities are their covariance matrices. Defining $\mathbf{V_k}=cov\{\mathbf{\overrightarrow{\epsilon}_k}\}$, $\mathbf{Q_k}=cov\{\mathbf{\overrightarrow{q}_k}\}$ and $\mathbf{C_k}=cov\{\mathbf{\overrightarrow{x}_k-\overrightarrow{x}_{k,true}}\}$, it is possible to demonstrate~\cite{fruhwirth}:

\begin{equation}
 \mathbf{C_k=\left[\left(F_{k-1}C_{k-1}F_{k-1}^T+Q_{k-1}\right)^{-1}+H^T_kV^{-1}_kH_k\right]^{-1}}
\end{equation}

It becomes evident what already stated in \S~\ref{kalman-sez}: if the uncertainty on the measure (and therefore its covariance matrix $\mathbf{V}$) is large with respect to the impact of the multiple scatter, the contribution of each measure to the track will be small.

Finally, at each layer, the residual vector  is defined as 

\begin{equation}
\mathbf{\overrightarrow{r}_k}=\mathbf{\overrightarrow{m}_k}-\mathbf{H_k\overrightarrow{x}_k}
\end{equation}
and its covariance matrix results

\begin{equation}
 \mathbf{R_k=V_k-H_kC_kH^T_k}
\end{equation}
The $\chi^2$ of the track in each layer is 

\begin{equation}
 \chi^2_k=\mathbf{\overrightarrow{r}^T_kR^{-1}_k\overrightarrow{r}_k}
\end{equation}
and it's distributed as a $\chi^2$ with dim($\overrightarrow{m}_k$) degrees of freedom. The sum $\sum_{k}\chi^2_k$, that is not distributed exactly like a $\chi^2$, gives an estimator of the quality of the track.

\section{Event classification}
The main purpose of the \ac{lat} is to produce photon data-sets to be released to the whole scientific community. The process of \Pphoton identification in the dominant background of charged particles is very complex and  will not be  described in this work; a full review can be found in \citet{pass7-status}. 

However, photons are not the only particle type that can be identified with the \ac{lat}: different selections have been developed to identify electrons, protons or ions. While each selection has very specific characteristics that depend on the desired signal/background ratio and on the  information available in the \ac{lat} output (the \ac{lat} is a detector designed to optimally reconstruct and identify electro-magnetic showers and which trigger is programmed principally to collect events  potentially originated by photons and to reject clear charged-particle events, see \S~\ref{trigger}), the general scheme of event selection is similar for different event classes. This section will give a very general overview of the scheme of \ac{lat} event classification.

Event classification is made  using the \ac{merit} quantities. Typically, events undergo a series of cuts, designed to remove from the data sample specific types of events, that can either be  difficult to reconstruct or to classify, or that are clearly background. Events passing the cuts are passed to one or more \acp{ct}, the output of which are continuous quantities that classify the events as background-like or signal-like. These, and other significant quantities, are used to set one or more cut, each one defining a class of events (see \cref{rate_eventi}). In the photon analysis, events are classified in a series of inclusive classes with decreasing residual background contamination, while the electron analysis described in \citet{electronpaper} and discussed in \S~\ref{CRE-fermi} is based on two independent electron classes, selecting high-energy or low-energy electrons.   

Events passing one of the photon selection are then released to public through the \acf{fssc}~\cite{lat-dati} in FITS format~\cite{fits}, providing for each event all the information that are relevant for astrophysical analysis.

\section{Updates in the LAT analysis and different data-sets }
\label{passi}
This section  describe the principal changes introduced in \ac{lat} event reconstruction and classification, while introducing the corresponding terminology.

\paragraph{Pass}

The data analysis of the \ac{lat} is constantly improved through   continue updates of the event reconstruction, simulation, classification, of the \acp{irf} and of the instrument calibrations: this section  briefly describe the principal changes introduced from the start of the mission.

For historical reasons, the complete process of reconstruction and classification, that starting from the \acf{digi}  produces the \acf{merit} and the photon classes FITS files  is named \textbf{Pass}: \textbf{Pass 6} is the first Pass used to analyze \ac{lat} data; it was developed before  launch, using exclusively Monte Carlo simulations and beam test results. The first \ac{lat} electron spectrum was measured using Pass 6 reconstructed data.

After launch, building on the knowledge acquired with flight data, important Monte Carlo simulation changes were implemented in order to correct for unanticipated effects, the most important of which is the presence in data of the \textit{ghost events}. \label{overlay}

\begin{figure}[htp!]
\begin{center}

\includegraphics[width=.95\textwidth]{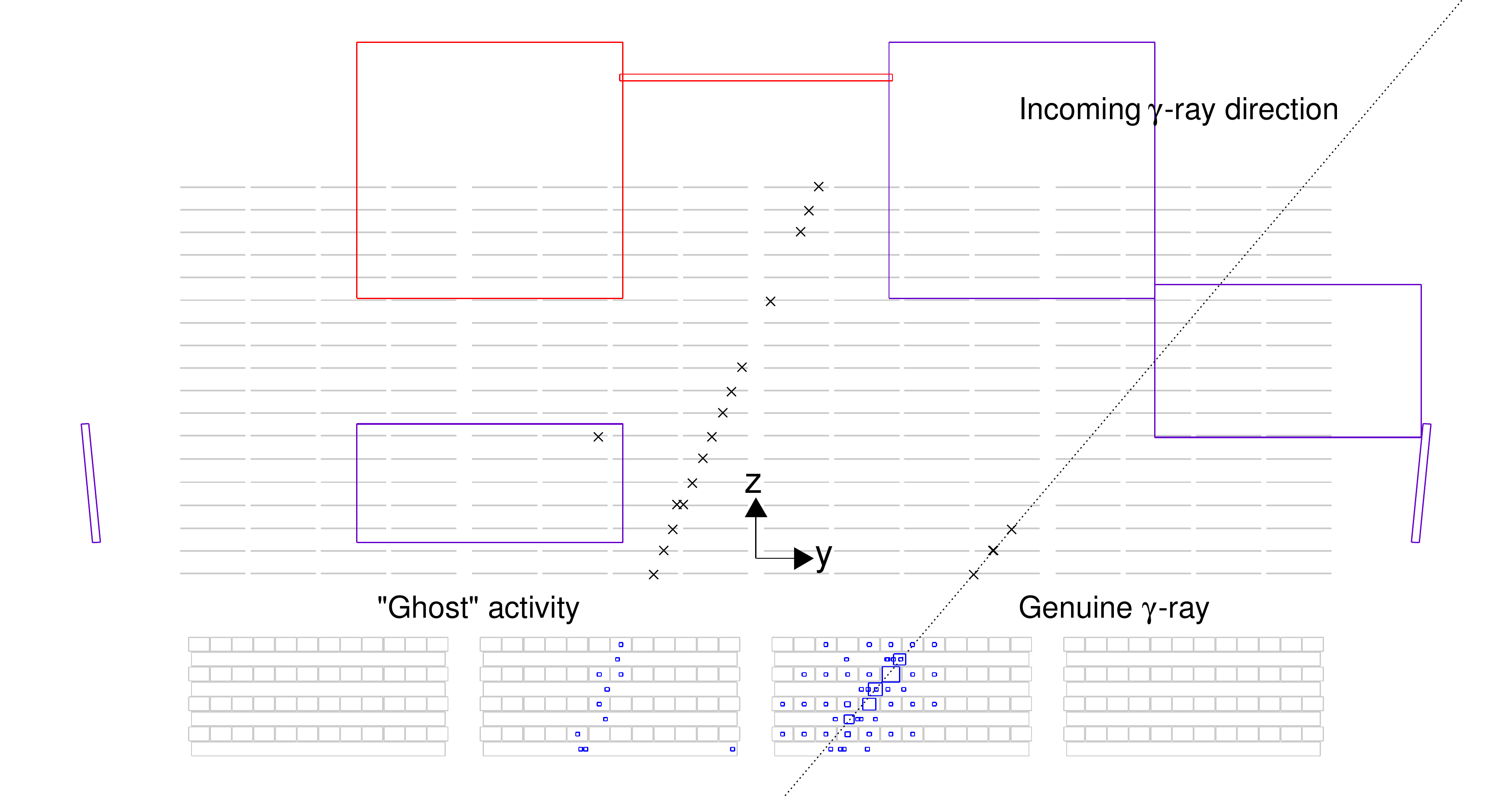}
\caption{A 8.5~GeV $\gamma$ event with a significant overlay caused by a non in-time charged particle; crosses are cluster of hit strips in \ac{tkr}, hit crystals in \ac{cal} are represented with squares with side proportional to the deposed energy, \ac{acd} tiles are shown only if they are over threshold; dashed line indicates the \Pphoton direction.~\cite{pass7-status}}
\label{overlay-fig}
 
\end{center}
\end{figure}

A ghost (see for example~\citet{overlays})  is the pile-up (or \textit{overlay}) of a triggering gamma event with an out of time background event. The signal generated by a charged particle typically lasts in the \ac{tkr} for 8~\mycro\second, a time that can increase to 150~\mycro\second\ for heavy ions. Signals in \ac{cal} and \ac{acd} last for shorter but not negligible time. If during this time a \Pphoton triggers the \ac{lat}, its signal will pile-up with the remains of the charged particle signal, so creating ``ghost'' signals in the event. Figure~\ref{overlay-fig} shows a $\gamma$ event with a significant overlay.

Shortly after the launch of \fermi, the Monte Carlo simulation was updated to keep into account this effect, by superimposing to each simulated  event an event acquired with the periodic trigger of the LAT. The periodic trigger (see \S~\ref{trigger}) is a signal with 2~\hertz\ rate that causes the read-out of the LAT independently of other on-board trigger primitives. The acquired read-out is then sent to ground without any  filtering. Because the rate of charged particles, and therefore the expected rate of overlays, depends on the geomagnetic latitude of the spacecraft, signals from the periodic trigger are mapped in function of the geomagnetic latitude of acquisition, defined using the parameters  developed in~\citet{mcilwain}. The Monte Carlo simulation includes a detailed simulation of the \fermi\ orbit, and the events of the periodic trigger are randomly added to Monte Carlo events simulated at the same geomagnetic latitude. 

Based on the new simulation and on real flight-data, new background rejection and data classification were developed; these improvements formed the new \textbf{Pass 7}, that was used for data analysis until 2015.

\begin{figure}[htp!]
\begin{center}

\subfigure[Difference in the best reconstructed direction, the angular separation is measured in units of the nominal 68~\% containment radius, therefore factorizing out the energy dependence of the angular resolution.]{\includegraphics[width=.55\textwidth]{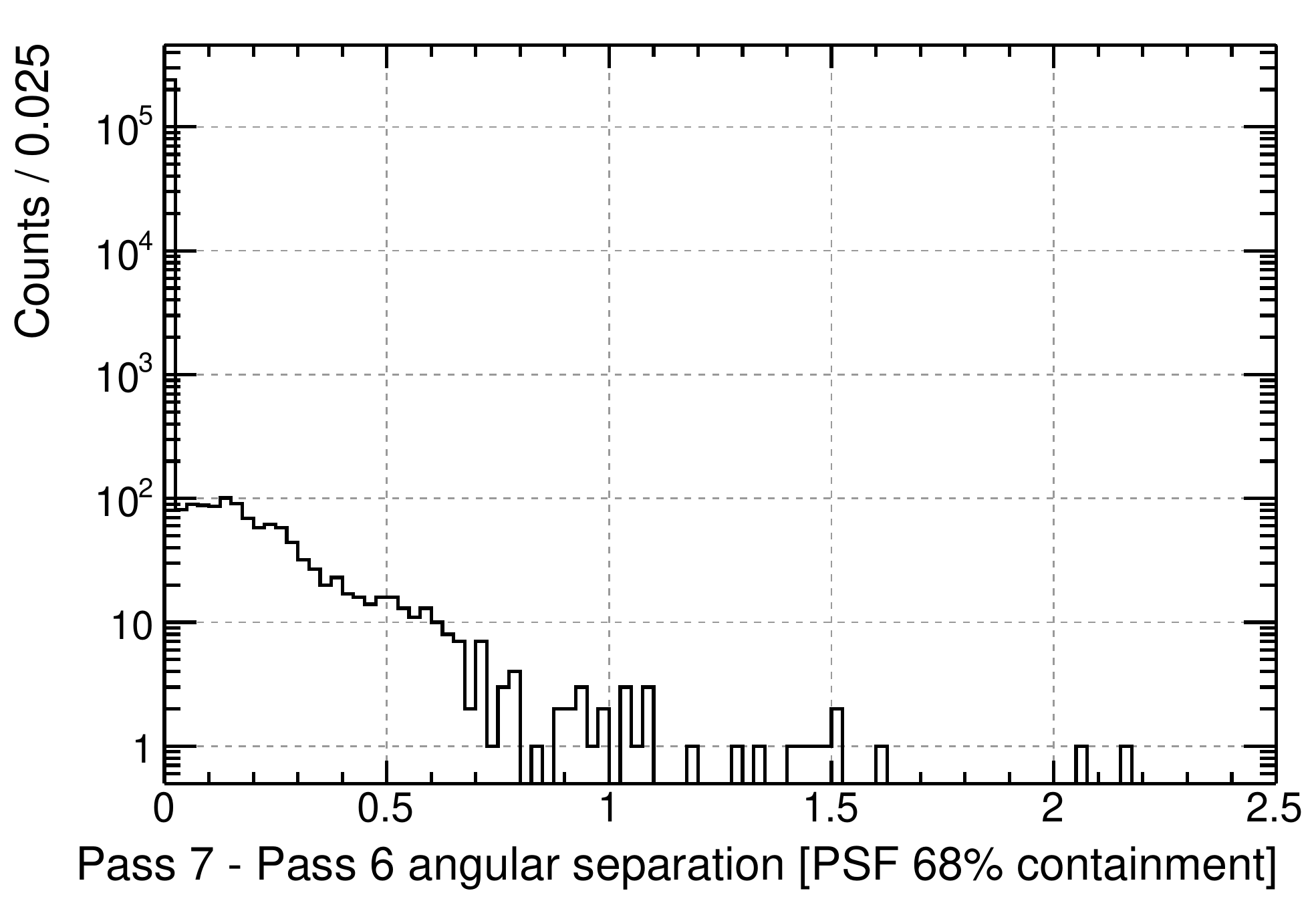}}
\subfigure[Difference in the best reconstructed energy. This plot shows a clear shift introduced by the Pass~7 energy un-biasing.]{\includegraphics[width=.55\textwidth]{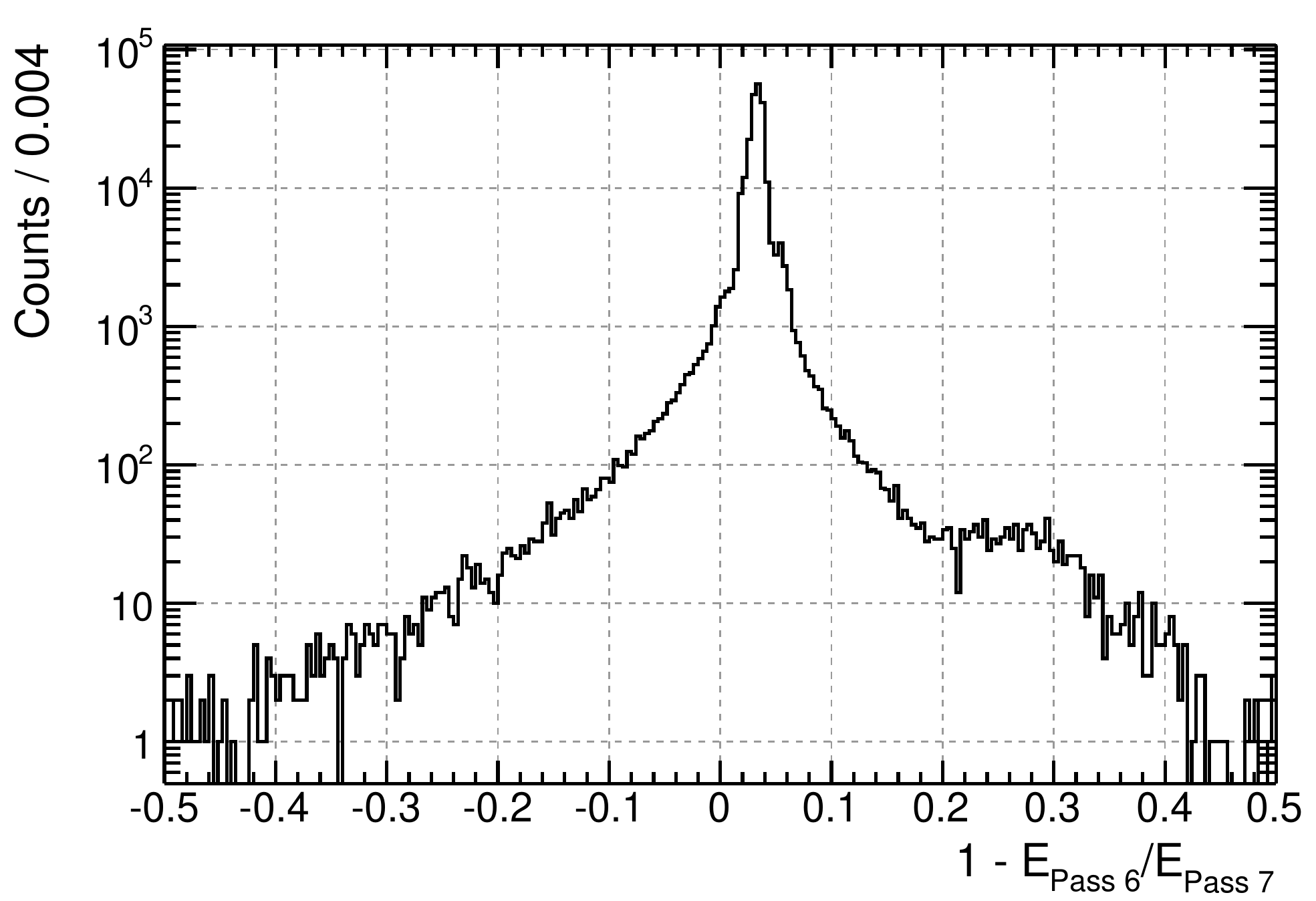}}
\caption{Event by event difference  between Pass~6 and Pass~7 events.  Histograms show events from a sample of P7 SOURCE events above 100~\MeV, with a zenith angle cut at $100^\circ$
to  remove photons from the Earth Limb.~\cite{pass7-status}}
\label{diff-p6p7-Edir}
 
\end{center}
\end{figure}

Pass~7 event reconstruction is substantially identical to that of Pass~6, the only differences being the introduction of the Monte Carlo-based un-biasing of the reconstructed energy and the exclusion of the Likelihood energy reconstruction method (see page~\pageref{energy-reconstruction}). The main differences is the re-training of the \acp{ct} described in \S~\ref{figures-of-merit} using the new Monte Carlo, that can result in different energy and direction assignment to the same events (see figure~\ref{diff-p6p7-Edir}). The main difference between Pass~6 and Pass~7 reconstruction is in the event classification: photon classes were re-defined, and all the \acp{ct} that classify  events as gamma-like or background-like were re-trained using the new Monte Carlo.

The \ac{lat} collaboration has developed a largely new data-analysis process, named \textbf{Pass 8}, that is described in \S~\ref{pass8}. Pass~8 substantially improves the performance of the \ac{lat}, and is used for the data analysis since 2015.

\paragraph{Glast Release}

A \acf{gr} is the software suite, developed by the \ac{lat} collaboration, that contains all the  software necessary for running the instrument simulation, event reconstruction  and classification. A \ac{gr} includes the GEANT~4-based (\citet{geant}) event-simulation tool, the modelization of \fermi\ and the orbit description required to correctly run the simulations, the complete event reconstruction program including \ac{lat} calibrations, all the \acp{ct} classifiers and the external software needed to run it. 

Pass 6 analysis runs on \ac{gr} version~15 (\ac{gr}~v15), Pass~7 was fully introduced with \ac{gr}~v17, and Pass~8 was first introduced in \ac{gr}~v19, while its definitive version runs on \ac{gr}~v20. 

\paragraph{Reprocess}

When significant improvements have been introduced either in the estimation of calibration constants, in the event reconstruction or in the event classification, all the collected data may be either fully reprocessed, starting from the \ac{digi} level, or just reclassified, starting from the \ac{recon} level. \textbf{P120} is a first reclassification of all events, launched after the first 3 years of the mission, using \textbf{Pass 7} classifiers. P120 reprocessed all the existent data, while  the \ac{l1}, that processes the newly acquired data,  was updated to process further acquired data using  Pass~7 classifiers. During 2011 the \ac{l1} was updated to GR v17 and  data also benefited of a slightly better event reconstruction.

\begin{figure}[htp!]
\begin{center}

\includegraphics[width=.75\textwidth]{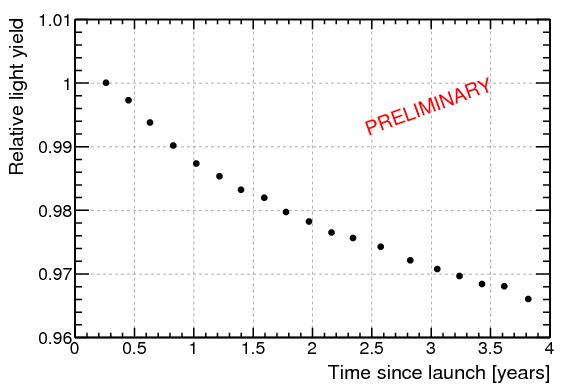}
\caption{Relative variation of the light yield in  \ac{cal} crystals throughout the first four years of mission. Light yield is measured using minimum ionizing protons selected by the \ac{lat} trigger.~\cite{p202}}
\label{light-decrease-cal}
 
\end{center}
\end{figure}

\begin{figure}[htp!]
\begin{center}

\includegraphics[width=.75\textwidth]{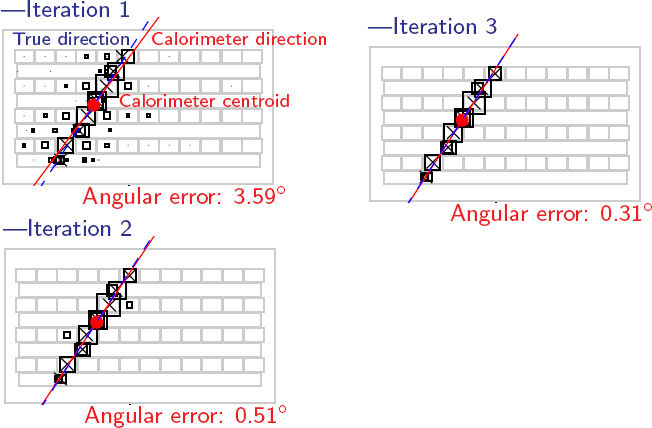}
\caption{Iteration in the calculation of the event direction in \ac{cal} (red line), blue line is the true (Monte Carlo) event direction. The reconstructed direction is the main axis of the event distribution, calculated through moment analysis. The algorithm iteratively removes from the calculation \ac{cal} crystals far from the calculated axis: deposed energy in crystals is represented by a black square of proportional dimension. When a crystal is removed from the calculation it is not marked anymore.~\cite{moment-analysis-symposium}}
\label{moment-fig}
 
\end{center}
\end{figure}

P120 presents some features which affect data quality: for instance, \ac{cal} calibrations have not been updated to correct for the decrease in light-yield of the crystals due to radiation damages (which effect was estimated to be $\sim 1\%/\yr$, see figure~\ref{light-decrease-cal}), therefore the measured energy of events artificially decreases with time. Furthermore, the change in 2011 of the \ac{l1} software (updated from \ac{gr}~v15 to \ac{gr}~v17) introduced a discontinuity in the data. \ac{gr}~v17 introduced a change in the calculation of the \ac{tot} (see page~\pageref{totref}), while the moment analysis  of the energy distribution in the \ac{cal} has been updated with a recursive algorithm that removes outliers crystals from the moments calculation, as shown in figure~\ref{moment-fig}. For these reasons, all \fermi\ data were later reprocessed starting from the \textit{digi} level, introducing periodic updates of the \ac{lat} calibrations and making use of \ac{gr} v17: this data reprocessing is named \textbf{P202} or \textbf{P7REP} (\citet{p202}).

The reprocess of the data from the \ac{digi} level using the definitive Pass~8 reconstruction started at the end of summer 2013 and it is named \textbf{P300} for the production of \ac{recon} files, \textbf{P301} for the production of \ac{merit} files and \textbf{P302} for the production of FITS files. \textbf{P302} was completed at the beginning of 2015.

\section{LAT Instrumental Response Functions (IRFs)}
\label{irf-sec}

The \acfp{irf} describe the response of the instrument to the observed particles in terms of the transformation probability from the true physical quantities (in the \ac{lat}, the energy $E$ and the solid angle direction $\varOmega$) to the corresponding measured quantities ($E',\varOmega'$), keeping into account either the detector characteristics and the whole process of event reconstruction and selection. Therefore, when the detector is exposed to a flux of particles $F(E,\varOmega ,t)$, the expected signal rate is 

\begin{equation}
 \frac{dN}{dt}(E',\varOmega',t)=\int_E \int_\varOmega A_{eff}(E,\varOmega)\Delta E(E'\mid E,\varOmega)PSF(\varOmega' \mid E,\varOmega)F(E,\varOmega ,t)dEd\varOmega
\end{equation}
In this equation integration is performed on the \ac{lat} phase space, while the defined response functions are the effective area $A_{eff}(E,\varOmega)$, the \acf{psf} $PSF(\varOmega' \mid E,\varOmega)$ and the energy dispersion $\Delta E(E'\mid E,\varOmega)$, which are supposed to  depend on the true energy and direction of the events, but not on the corresponding reconstructed quantities.

In the \ac{lat}, \acp{irf} are calculated using large data-sets of simulated photons, generated isotropically and with an energy spectrum of $E^{-1}$, then some corrections based on flight-data are applied, see \citet{pass7-status} for an exhaustive description of the calculation of the \acp{irf}. Where not specified, all the plots shown in this section are published at~\cite{irf} and refer to the P8R2\_V6 \acp{irf}, which describe the performances of the instrument with Pass~8 analysis.

\subsection{Effective area, Geometric factor and Field of view}
\begin{figure}[htp!]
\begin{center}

\subfigure[The effective area as a function of energy for normal incidence photons.]{\includegraphics[width=.8\textwidth]{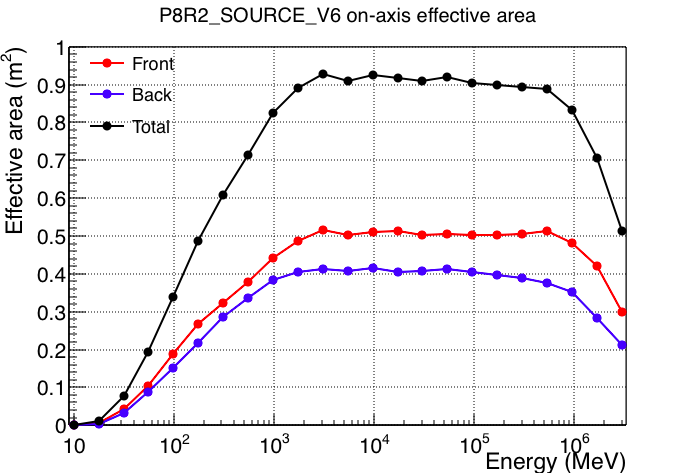}}
\subfigure[The effective area as a function of incidence angle $\theta$ for 10~\GeV\ photons.]{\includegraphics[width=.8\textwidth]{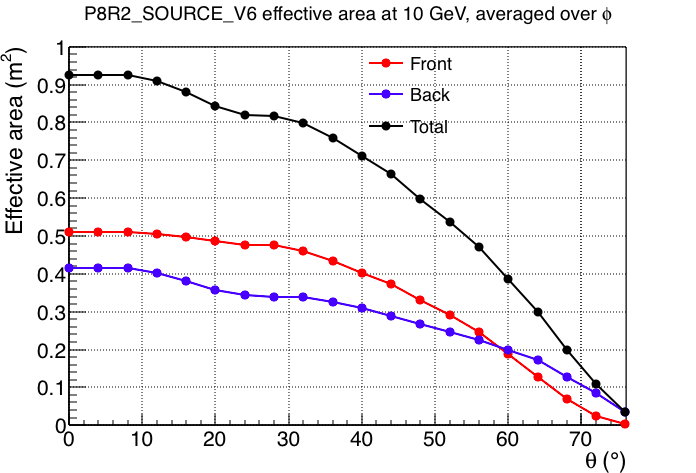}}

\caption{The effective area for the P8\_SOURCE photon class.}
\label{aeff1}
 
\end{center}
\end{figure}

\begin{figure}[htp!]
\begin{center}
\subfigure[The effective area as a function of the azimuth angle phi for 10~\GeV\ photons, $30^\circ$ off-axis]{\includegraphics[width=.7\textwidth]{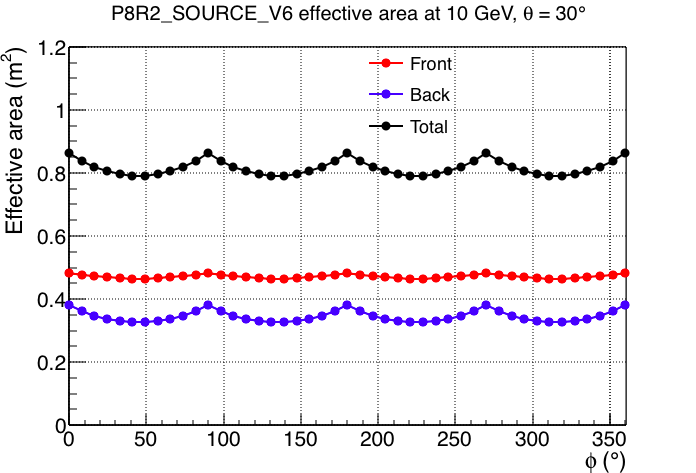}}
\subfigure[The effective area for 10~\GeV\ photons as a function of $\theta$ and $\phi$, figure from~\cite{pass7-status} (therefore referring to Pass~7 \acp{irf}, but behavior is similar for Pass~8). The plot is shown in a zenith equal-area projection, with the \ac{lat} boresight at the center of the image; concentric rings correspond to 0.2 increments in $\cos(\theta)$.]{\includegraphics[width=.7\textwidth]{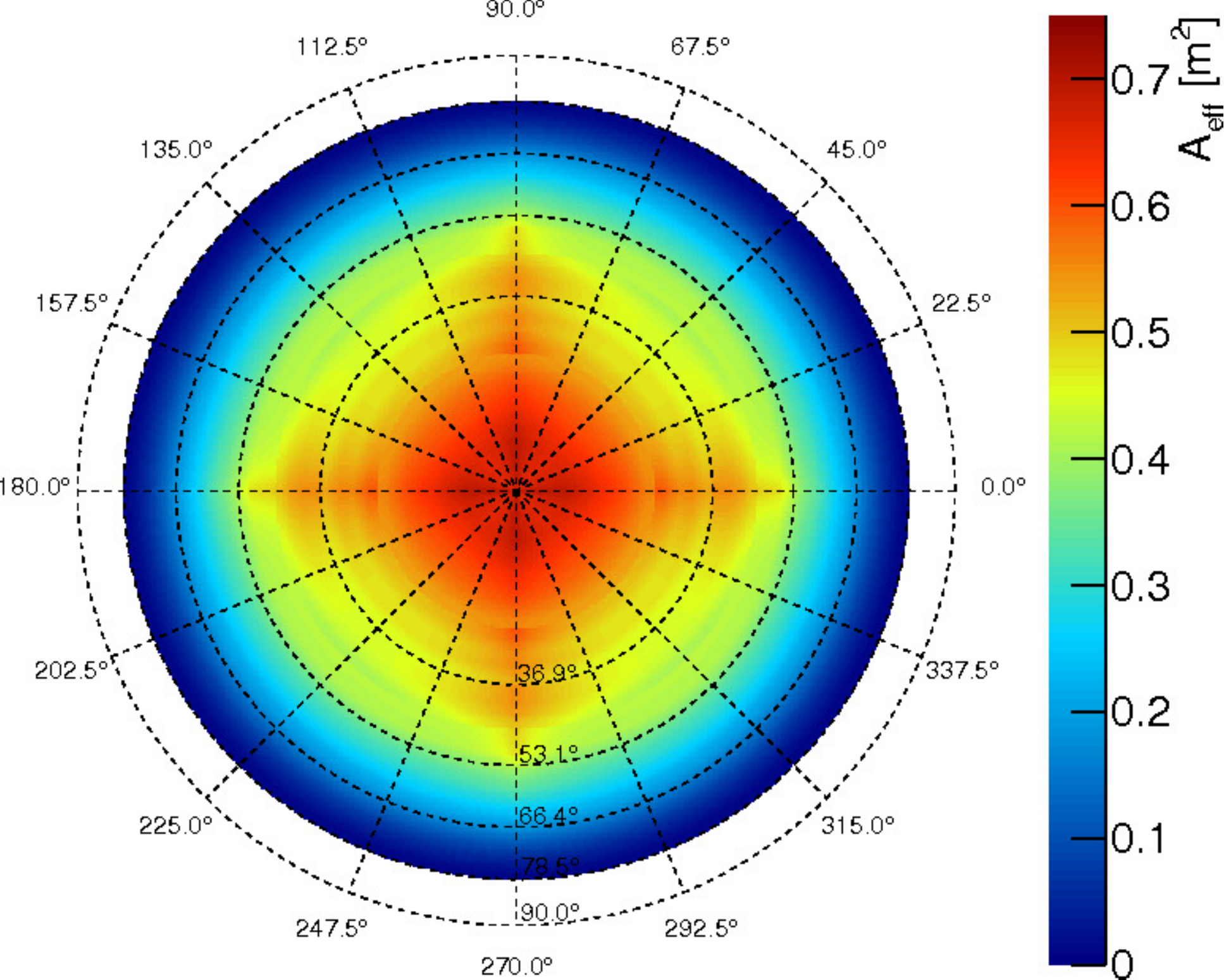}}
\caption{Dependence of the effective area on the azimuth angle.}
\label{aeff-phi}
\end{center}
\end{figure}

\begin{figure}[htp!]
\begin{center}

\subfigure[Event display of a simulated 100~\MeV\ photon]{\includegraphics[width=.8\textwidth]{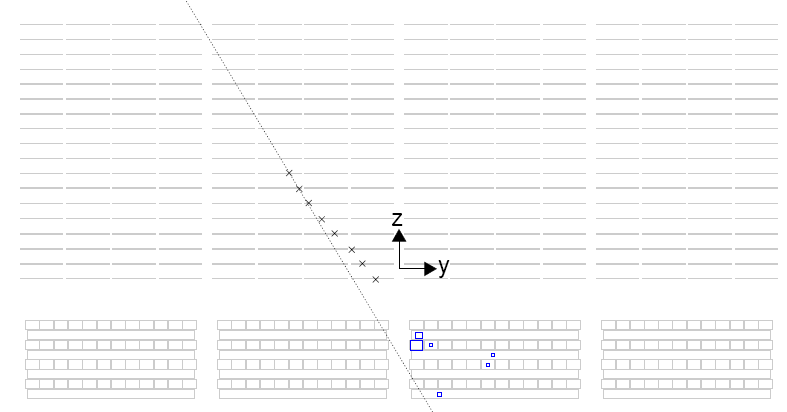}}
\subfigure[Event display of a simulated 540~\GeV\ photon]{\includegraphics[width=.8\textwidth]{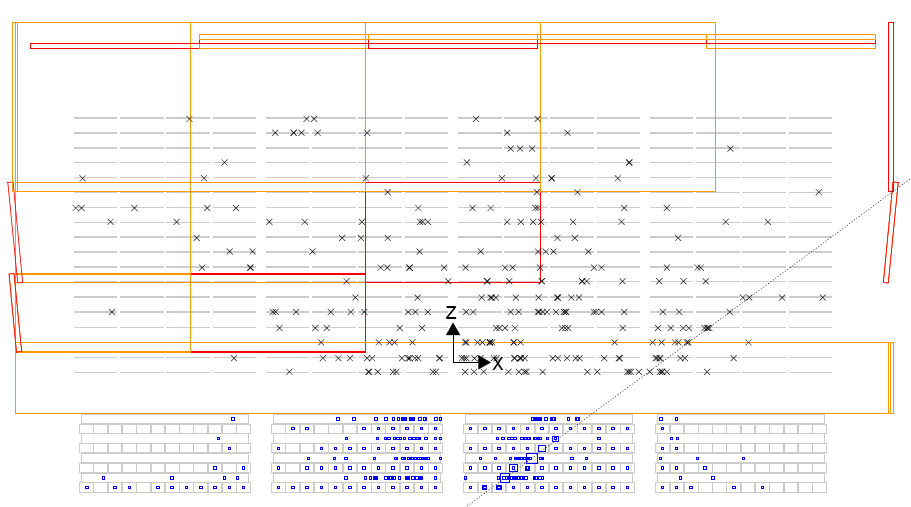}}
\caption{Event display of simulated  photons. Hit strips in \ac{tkr} are marked with x, blue square are \ac{cal} crystals with deposed energy and hit \ac{acd} tiles are marked in red or orange.}
\label{fotoni-aeff}
 
\end{center}
\end{figure}

The effective area describes, in function of the true energy and direction, the rate of detected events obtained exposing the instrument to a flux of particles $F(E,\varOmega ,t)$:

\begin{equation}
 \frac{dN}{dt}(E,\varOmega,t)=\int_E \int_\varOmega A_{eff}(E,\varOmega)F(E,\varOmega)dEd\varOmega
\end{equation}
To be detected, an event must cross the active area of the detector, interact with the detector (for $\gamma$ it must convert in an \Pelectron \APelectron pair), generate a trigger signal in the detector and finally must be reconstructed as a signal event and not discarded as background. Therefore, the effective area can be factorized in four terms describing these stages of the detection process:

\begin{equation}
 A_{eff}(E,\varOmega)=A_{geo}(E,\varOmega)P_{conv}(E,\varOmega)\epsilon_{trig}(E,\varOmega)\epsilon_{rec}(E,\varOmega)
\end{equation}
Figure~\ref{aeff1} shows the dependence of the effective area on the energy and on the incidence angle, figure~\ref{aeff-phi} shows the dependence on the azimuth angle, for the photon class P8R2\_V6\_SOURCE, a class designed to study point sources. The effective area is calculated separately for the front (red line) and back (blue line) sections of the \ac{tkr}, and the total effective area (black line) is just the sum of the two. It can be seen that the contribution to the effective area of the four thick layers of the back section is almost equivalent to that of the 12 layers of the front section. The dependence of the effective area on energy shows that the function rapidly decreases below 100~MeV, because  of the decrease of the pair-conversion cross-section and because low energy events are quickly absorbed in matter, therefore  the number of hits in \ac{tkr}  and deposed energy in \ac{cal} can be so small that the event will not trigger or will be impossible to reconstruct. The effective area reaches a substantial plateau at $\approx 1~\GeV$, with the subsequent increase  principally due to the decrease of the incidence of overlays (see page~\pageref{overlay}), because at high energy the impact of the overlay  on the whole event becomes less important and the event is more easily accepted as signal. Finally, at high energy the impact of back-splash is so large that events start to be rejected as charged particles. An example of these effects is shown in figure~\ref{fotoni-aeff}, where a low-energy and an high-energy simulated photon are shown.

The dependence on the incidence angle shows that the effective area slowly decreases with increasing angle until, at high incidence angles, the reconstruction becomes very hard. Above $70^\circ$ the \ac{tkr} is typically unable to reconstruct any track, because typically less than three planes are crosses, and the effective area rapidly  falls to zero.

The dependence on the azimuth angle shows the existence of privileged directions along with events are more easily reconstructed: these  are the directions of the silicon micro-strips in the detection planes, that define the x and y axis of the \ac{lat}.

\begin{figure}[htp!]
\begin{center}

\includegraphics[width=.8\textwidth]{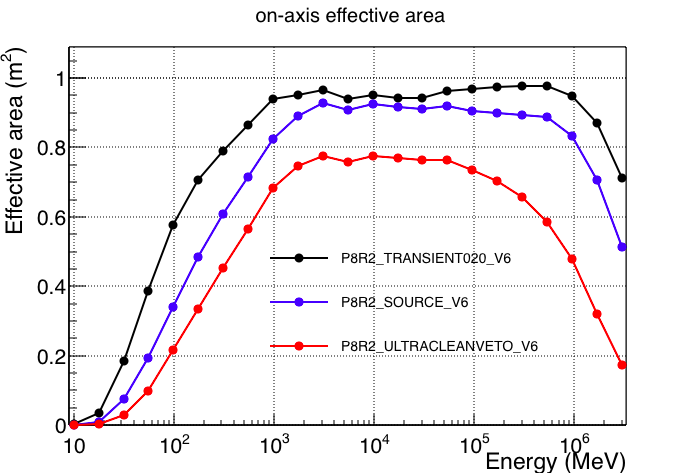}
\caption{A comparison of the on-axis effective area (as a function of the energy) for different P8R2\_V6 photon classes}
\label{aeff2}
 
\end{center}
\end{figure}

Figure~\ref{aeff2} shows the dependence of the (total) effective area on the applied event selection: the TRANSIENT class is designed to study transient events, for which the signal to background ratio is smaller, therefore the event selection can be made less stringent and more photons pass the cuts. On the other hand, the ULTACLEANVETO class is designed for analysis, for example  the study of the  diffuse \Pphoton radiation, for which the residual background contamination must be very small: therefore the cut are more stringent than in SOURCE class, and the effective area is smaller.

In some analysis, for example that on \acp{cre} described in \S~\ref{CRE-fermi}, the direction of incoming events is not important, because for example the incident flux can be supposed isotropic and therefore independent from $\varOmega$. In these analysis, the effective area can be integrated in $\varOmega$, obtaining a function that describes more immediately the instrument response to such a flux, named \textit{Geometric Factor} or \textit{Acceptance}:\label{geometric-factor}

\begin{equation}
 G(E)=\int_{\varOmega}A_{eff}(E,\varOmega)d\varOmega
\end{equation}

Finally, to express the portion of the solid angle that is effectively observed by the instrument, the \acf{fov} is defined

\begin{equation}
 FOV=\frac{\int_{\varOmega}A_{eff}(\varOmega)d\varOmega}{A_{peak}}
\end{equation}

\subsection{Point Spread Function}
\begin{figure}[htp!]
\begin{center}
\includegraphics[width=.8\textwidth]{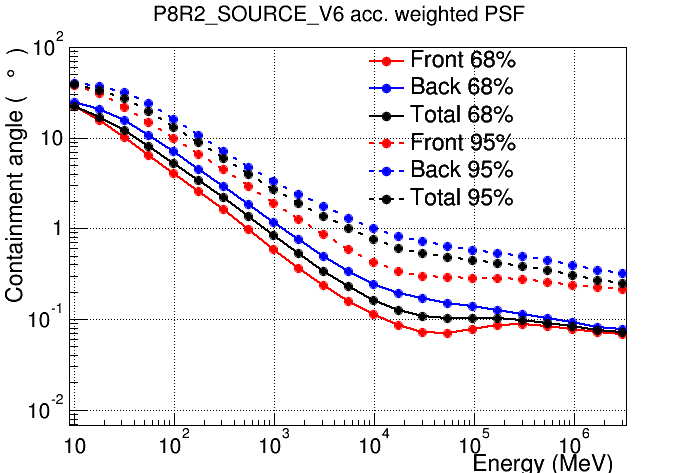}
\caption{68\% and 95\% containment angles of the reconstructed incoming photon direction, weighted on the \ac{lat} acceptance, as a function of photon energy.}
\label{psf} 
\end{center}
\end{figure}

The \acf{psf} describes the effective angular resolution of the instrument on the detected signal events, and is defined as the probability that an event of incoming energy $E$ and direction $\varOmega$ is reconstructed with a direction $\varOmega'$: $PSF(\varOmega' \mid E,\varOmega)$. It is expressed in terms of 68\% and 95\% 2-D containment angles, that are the angles containing 68\% or 95\% of the events produced by a point source. As discussed in \S~\ref{tracker-effect}, at low energy the \ac{psf} is dominated by the multiple scattering, the impact of which decreases with energy. Therefore the angular resolution improves with energy from $\sim 4^\circ$ at 100~\MeV\ to $\approx 0.1^\circ$ above 20~\GeV, where the dominating factor in the angular resolution is the strips pitch, and the \ac{psf} becomes almost constant. As the other \acp{irf}, the \ac{psf} is calculated separately for the front and back sections of the \ac{tkr}, the former having (by design) a significantly better resolution than the latter. The dependence of the \ac{psf} on the energy, weighted on the \ac{lat} acceptance, is shown in figure~\ref{psf}.

The  distribution of angles typically is not Gaussian, because the multiple scatter has a significant tail at high values, for this reasons the 68\% containment is not equivalent to a $\sqrt{2}\sigma$ containment. The ratio between the 68\% and the 95\% containment radii can be used as an estimator of the non-gaussianity of the distribution.

\clearpage
\subsection{Energy dispersion and energy resolution}\label{sec-energy-dispersion}
The energy dispersion describes the spread of the reconstructed energy around the true energy of the event as the probability of an event of energy $E$ and direction $\varOmega$ of being reconstructed with energy $E'$.  \ac{lat} energy distribution function can not include the presence of a systematic shift between  reconstructed and true energy. This possible shift was estimated as energy-independent during the beam test campaign described in~\citet{beam-test}, so that an observed spectrum could be only rigidly shifted but not deformed; for this reason this shift  is essentially irrelevant for almost all \ac{lat} spectral analysis (see for example~\citet[\S~7.3]{pass7-status}), and is treated at page~\pageref{absolute-energy}.

 The reconstructed energy distribution around the true energy is not symmetrical, because of the incomplete containment of the electro-magnetic shower in the \ac{cal}. Only a fraction of the event energy is collected and then a correction is applied (see \S~\ref{energy-reconstruction}): when the event is poorly reconstructed, the correction is typically underestimated, therefore   the low-energy tail is higher. Furthermore, the typical \ac{lat} observed sources have a power-law spectrum (sometimes with a cut-off), therefore while the under-estimation of the energy causes a small increase in the number of lower energy events, the over-estimation of the energy can cause a significant increase of the number of higher-energy events, with potential impact on the measured spectral index or cut-off. For this reason, the energy reconstruction and the event selection are optimized to not over-estimate the energy.

\begin{figure}[htp!]
\begin{center}
\subfigure[Energy resolution as a function of energy, weighted on the \ac{lat} acceptance.]{\label{Edisp-a}\includegraphics[width=.8\textwidth]{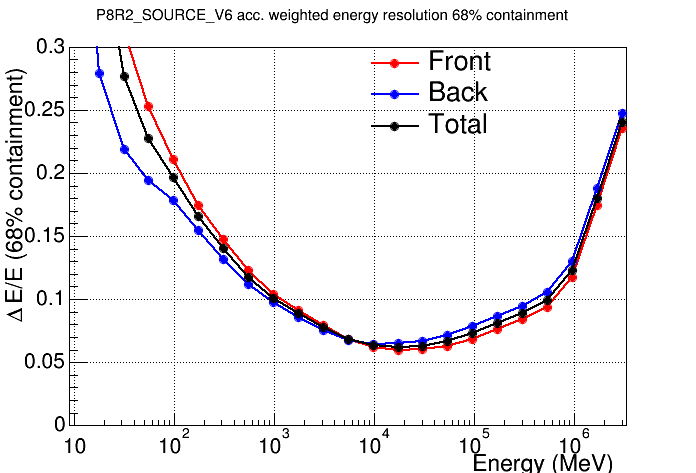}}
\subfigure[Energy resolution  as a function of incidence angle for 10~\GeV\ photon]{\label{Edisp-b}\includegraphics[width=.8\textwidth]{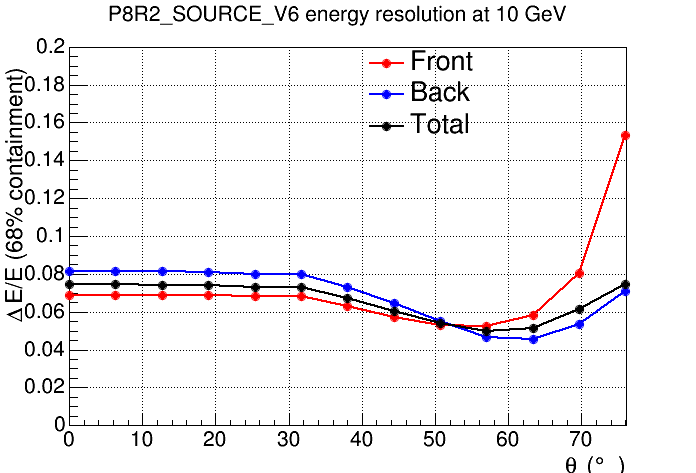}}

\caption{Energy resolution, defined as 68\% containment of the reconstructed incoming photon energy}

\end{center}
\end{figure}

Starting from the energy dispersion, the \textit{energy resolution} is defined as  the half-width of the energy window containing 34\% + 34\% (i.e., 68\%) of the energy dispersion on both sides of its most probable value, divided by the most probable value itself. Figures~\ref{Edisp-a} and~\ref{Edisp-b} show the dependence of the energy resolution on energy and incidence angle. Because the \ac{tkr} has a significantly worse energy reconstruction than the \ac{cal} (only the number of hit strips is used as an estimation of the deposed energy), the energy dispersion is larger at low energy, where a significant fraction of the event energy is absorbed in the \ac{tkr}. As a consequence, front-converting low-energy events have a worse energy resolution, because the charged particles have to cross a larger fraction of the \ac{tkr} before reaching the \ac{cal}. The energy resolution reaches its minimum between 1 and 20~\GeV. As the energy increases and the contained fraction of the shower decreases, the energy resolution  slowly worsen and the resolution of the front section becomes slightly better because of the larger crossed path-length of these events. Finally, the energy resolution has a small improvement at  angles $>30^\circ$, because sloping events cross a larger amount of material and a larger fraction of energy is absorbed by the \ac{cal}. At very high angles ($>60^\circ$) the shower starts to exit from the sides of  the \ac{cal} (or even to significantly miss it) and the energy resolution worsen (this is particularly important for front-converting events, whose conversion point is more distant from the \ac{cal}).

\subsubsection{Absolute energy scale}\label{absolute-energy}
The absolute energy scale of the \ac{lat}  is related to the quantitative knowledge of the response of the \ac{cal} modules to the deposition of a certain quantity of energy. The uncertainty on this  quantity was tested during the beam-test described in~\citet{beam-test}, and evidenced no significant energy dependence. Therefore the only impact of the uncertainty on the absolute energy scale  in spectral measurements is a rigid shift of the measured spectrum, that is essentially irrelevant for almost all \ac{lat} spectral analysis~\cite{pass7-status}.

\begin{figure}[htp!]
\begin{center}

\includegraphics[width=.9\textwidth]{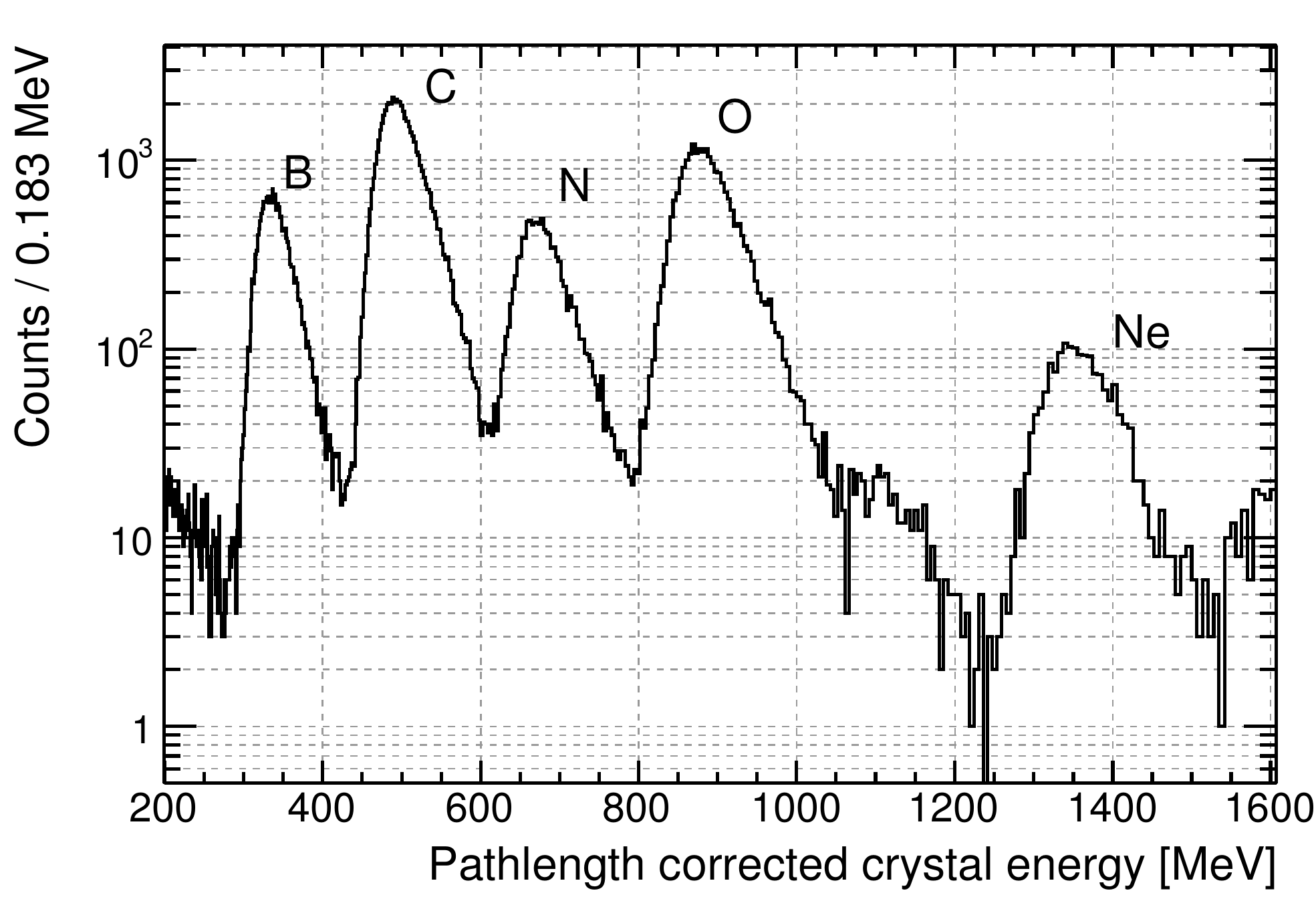}

\caption{Distribution of the path length-corrected crystal energy deposition for a sample of calibration events, acquired by the Heavy-ion trigger. The peaks corresponding to different \ac{cr} species are clearly visible.~\cite{pass7-status}}
\label{picchi-ioni}
 
\end{center}
\end{figure}

This feature was also confirmed by the observation of the energy deposition of different \ac{cr} species, from \Pproton to Ne~\cite{pass7-status}. Because the deposed energy by non-interacting particles is linearly related to the crossed path-length, and therefore to the incidence angle (limited in the 0--$60^\circ$ range), the linearity of the crystals response can be checked if the peaks generated by different nucleus can be resolved, as figure~\ref{picchi-ioni} proves. These measurements confirmed the linearity of the crystal responses, and as a consequence the absence of an energy dependence of the absolute energy scale, in the ranges 11 -- 22~MeV (\Pproton), 45 -- 90~MeV (He) and 180 -- 1500~MeV (heavier ions) of deposed energy per crystal. 

The  uncertainty on the absolute energy scale, measured with data from the beam-test and described in~\cite{elettroni,electronpaper}, is -10\%/+5\%.

After the launch of \fermi, the \ac{lat} undergone an extensive calibration campaign, described in~\citet{on-orbit-calibration}, to check the effective behavior of the \ac{lat} during science operation, which also interested the measurement of the absolute energy scale.

 Because the uncertainty on the absolute energy scale is caused both by the non-linear response of the \ac{cal} front-end electronic and by the uncertainty on the response of \ac{cal} crystals to energy deposition, these quantities were separately checked. The response of the electronic was checked by measuring the response to controlled charge injections, and  is known with a precision $\leq 1\%$. The response of single crystals to energy deposition was checked using the position of the \ac{mip} peak of protons, for which the $\frac{dE}{dx}$ is well known. A similar approach at higher energies using heavier species can not be used because the scintillation efficiency of nucleus in CsI(Tl) is not known with sufficient precision. Using \ac{mip} protons,  only the lowest two energy ranges of the \ac{cal}  electronics (see \S~\ref{cal}) can be calibrated directly, the others are calibrated using inter-range events. These events, acquired by the Heavy ions trigger (see \S~\ref{trigger}), are generated by heavy ions or interacting high-energy protons, and the signals from all the four \ac{cal} crystal ranges are acquired. Events for which the energy deposition is in the overlap between ranges calibrated using \acp{mip} and other ranges can therefore be used for calibration purposes.  

 An independent method to calibrate the absolute energy scale  was presented in~\citet{energy-scale}. This method, that uses the cut-off in the \ac{cre} spectrum caused by the geomagnetic field (see page~\pageref{geomagnetic-cutoff}) at different geomagnetic latitudes, suggests that the uncertainty could be lowered by 5 -- 7\%.

\begin{figure}[htp!]
\begin{center}

\includegraphics[width=.9\textwidth]{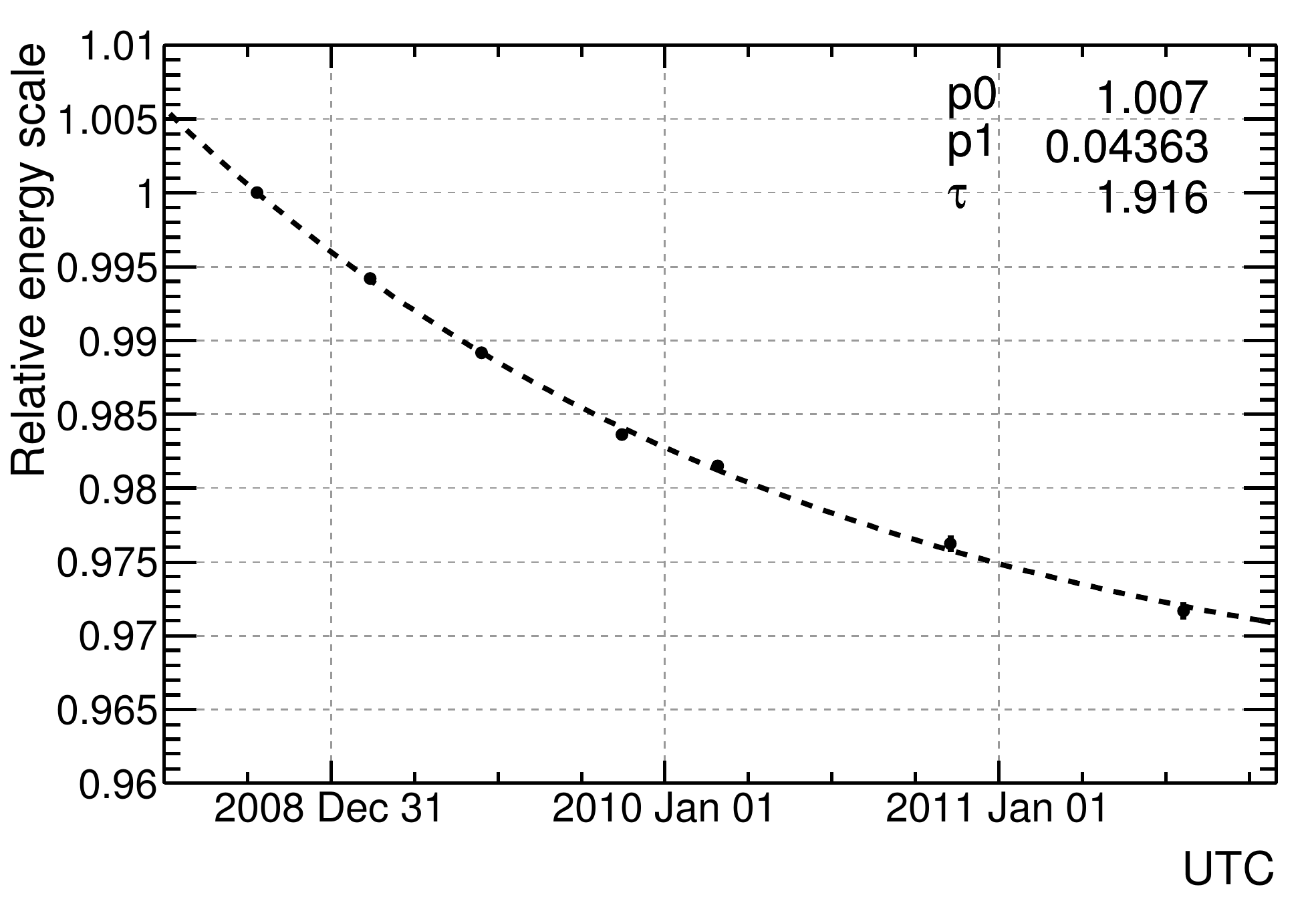}

\caption{Relative variation of the absolute energy scale, as measured from the position of the proton peak position, throughout the first three years of the mission.~\cite{pass7-status}}
\label{variazione-scala-energia}
 
\end{center}
\end{figure}

Calibrations of the \ac{lat} subsystems are repeated every few months to correct for changes caused by the aging of the instrument. The measured shifts of the \ac{mip} proton and nucleus peaks give a very good measurement of the changes in the \ac{cal} energy scale.
The time evolution of the absolute energy scale  during the mission is shown in figure~\ref{variazione-scala-energia}, and is completely compatible with the expected decrease of the crystals response caused by radiation damage.

\chapter{Measurement of the CRE spectrum using the Large Area Telescope}

\label{CRE-fermi}
The \fermi-LAT collaboration  published in 2009 a measurement of the \ac{cre} spectrum, from 20~\GeV\ to 1~TeV using 6~months of \ac{lat} data. This measurement is described in~\citet{elettroni}. In 2010, an independent selection for low energy electrons was developed, and the spectrum was extended down to 7~\GeV,  see~\citet{electronpaper}, with the statistics extended to 1~\yr.  Both measurements are  made using  data processed with Pass~6 (see \S~\ref{passi}). The full analysis is described  in \cite{electronpaper}. In 2011, the \fermi-LAT collaboration has published a separate measurement of the \Pelectron and \APelectron spectra in the 20--200~GeV range, discriminated using the Earth magnetic field. This analysis, briefly described  in \S~\ref{lat-positroni}, uses 39~days of Pass~7 data (because only runs with the Earth in the \ac{lat} \ac{fov} can be used) and a completely different approach with respect to~\cite{electronpaper}, and  the resultant $\Pelectron + \APelectron$ is compatible with that from~\cite{electronpaper}.

This chapter  gives a review of the methodology used in the analysis (\S~\ref{cre-analysis}) and of the obtained \ac{cre} spectrum (\S~\ref{misura-spettro})

\section{Analysis} \label{cre-analysis}
\subsection{CRE selection}\label{cut-CRE}
\begin{figure}[htp!]
\begin{center}

\includegraphics[width=.85\textwidth]{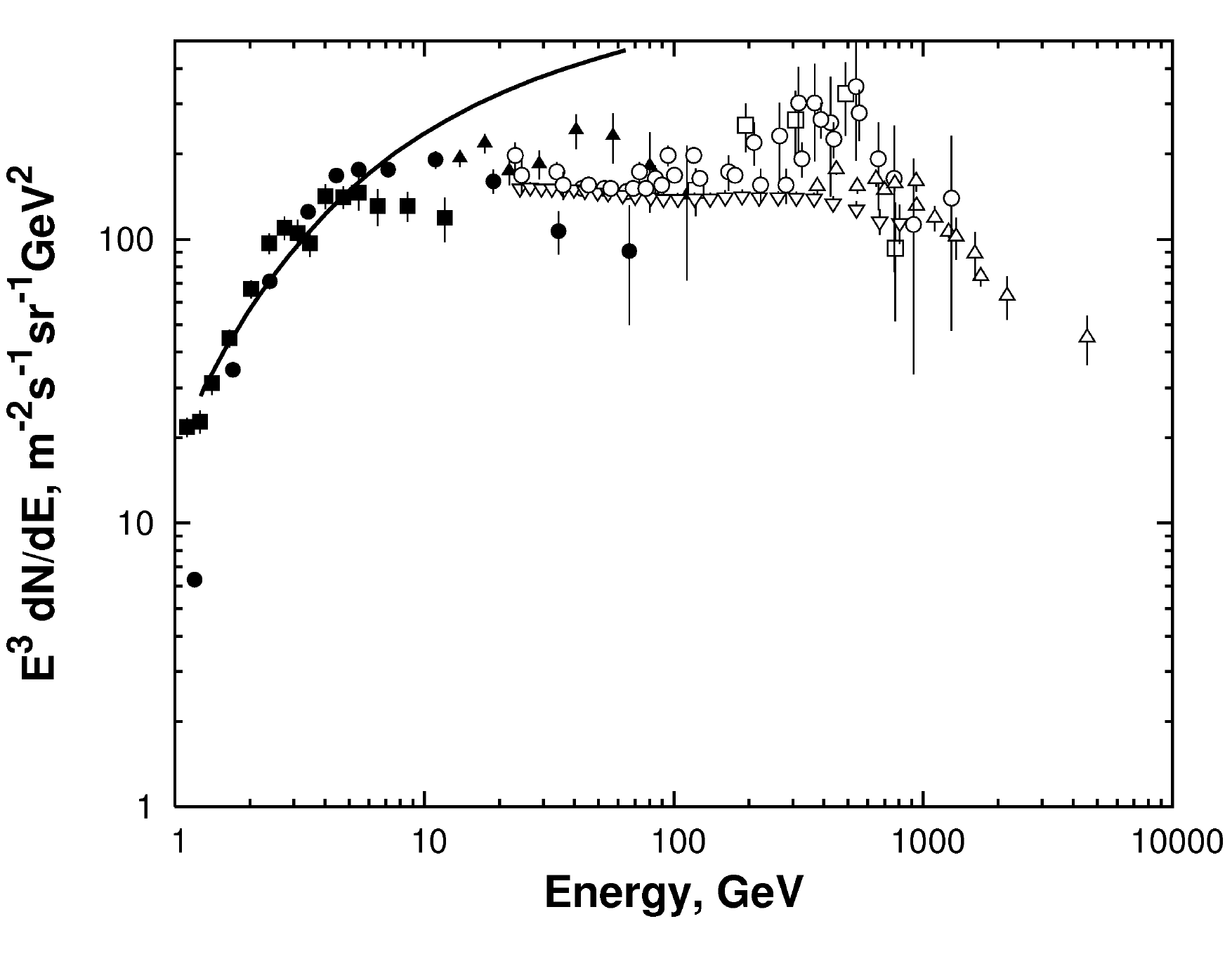}
\caption{\Pelectron$+$\APelectron spectrum measured by recent experiments (points), multiplied by $E^3$. Black solid line shows the \Pproton spectrum multiplied by 0.01. \cite{pdg}}
\label{elet-segn-rumore}
 
\end{center}
\end{figure}

Electrons of the analysis described in~\cite{electronpaper} are identified from the dominant background using two independent selections: a low-energy selection (\textbf{LE}, $\sim~100~\MeV$ -- 100~\GeV) and an high-energy (\textbf{HE}, 20~\GeV -- 1~TeV) one. The development of a preliminary  HE selection is analyzed in~ \cite{carmelo}, while the development of the LE selection is described in~\cite{melissa}. Both selections were developed using Monte Carlo simulation of electrons following an energy power-law with -1 index and the LAT standard background simulation, reproducing the whole particle population on \fermi\ orbit.  Both simulations were generated using  the \ac{gr} version~15 (see \S~\ref{passi}). The amount of the background with respect to the signal  is shown in figure~\ref{elet-segn-rumore}, where recent measurement of the \Pelectron$+$\APelectron spectrum are compared with the spectrum of the protons times 0.01.

Two independent selections were developed because of the different issues that arise when extracting  low and high-energy \ac{cre} populations  from LAT ground data. 

The first difference arises from the low altitude of the \fermi\ orbit (565~km), where the effect of the Earth magnetic field  is not negligible: below 20~\GeV\  the detectable primary \ac{cre} flux (without secondary electrons generated in the Earth Limb) is limited in energy by  the strength of the geomagnetic field, that varies along the \fermi\ orbit,  therefore the measured \ac{cre} flux is dependent from the LAT position.

The second and most important difference is caused by the on-board filtering described in \S~\ref{filtri}: above  20~\GeV\ of deposited energy, all events are accepted by the \textit{gamma} filter. Therefore, above $\sim$ 20~\GeV\ of reconstructed energy the electron analysis can be performed using events downloaded through the gamma filter.

Below 20~\GeV\ of deposited energy, events are selected by the gamma filter, that is designed to reject events that have very high probability of being originated by charged particles. Therefore, a large fraction of low-energy triggering electrons are discarded, and the downloaded sample is strongly biased. Furthermore, the on-board filter processing is not easy to reproduce in Monte Carlo simulations, therefore a \ac{cre} analysis that uses events processed by the gamma filter below 20~\GeV\ could be affected by strong systematic effects. To prevent this, the LE electron sample is extracted by the output of the \textit{diagnostic} filter, that sends to ground an unbiased sample of all the LAT triggering events, pre-scaled with a factor of 250.

The starting point of both selections is that the event has at least one reconstructed track in the \ac{tkr} and a minimum of deposited energy in \ac{cal} (5~\MeV\ for LE selection, 1~\GeV\ and 7 radiation lengths for HE). This request removes events that are poorly reconstructed or that have insufficient information in one of the LAT subsystems, and that consequently would be difficult to correctly identify. Furthermore, it is necessary to remove events coming from the Earth Limb, because it is a very powerful source of photons that could be misidentified as electrons. To do so, a selection is applied to the reconstructed event angle with respect to the local zenith. The Earth atmosphere enters in the \fermi\ \ac{fov} at angles greater than $\sim 112^\circ$, and the \ac{psf} is $\sim 7^\circ$ at 100~\MeV\ for 95\% containment radius (see \S~\ref{irf-sec}). Therefore, events with angle $>105^\circ$ with respect to the local zenith are rejected.

The \ac{acd} is designed to detect charged particles, and therefore its utility in discriminating between electrons and protons is limited. It can be however used to identify \Pphoton  and nuclei. The photon rejection exploits the first stage of the on-ground \Pphoton selection process, named charged particles in the \ac{fov} (\textbf{CPF}). This stage mainly uses  \ac{acd} quantities to reject charged particles (for example particles with best track pointing to a hit tile in \ac{acd}): both LE and HE selections require that CPF selection is not passed. 

\begin{figure}[htp!]
\begin{center}

\subfigure[Average energy per \ac{acd} tile, used in HE selection. 615~--~772~\GeV.]{\label{fig-HE-acd}\includegraphics[width=.7\textwidth]{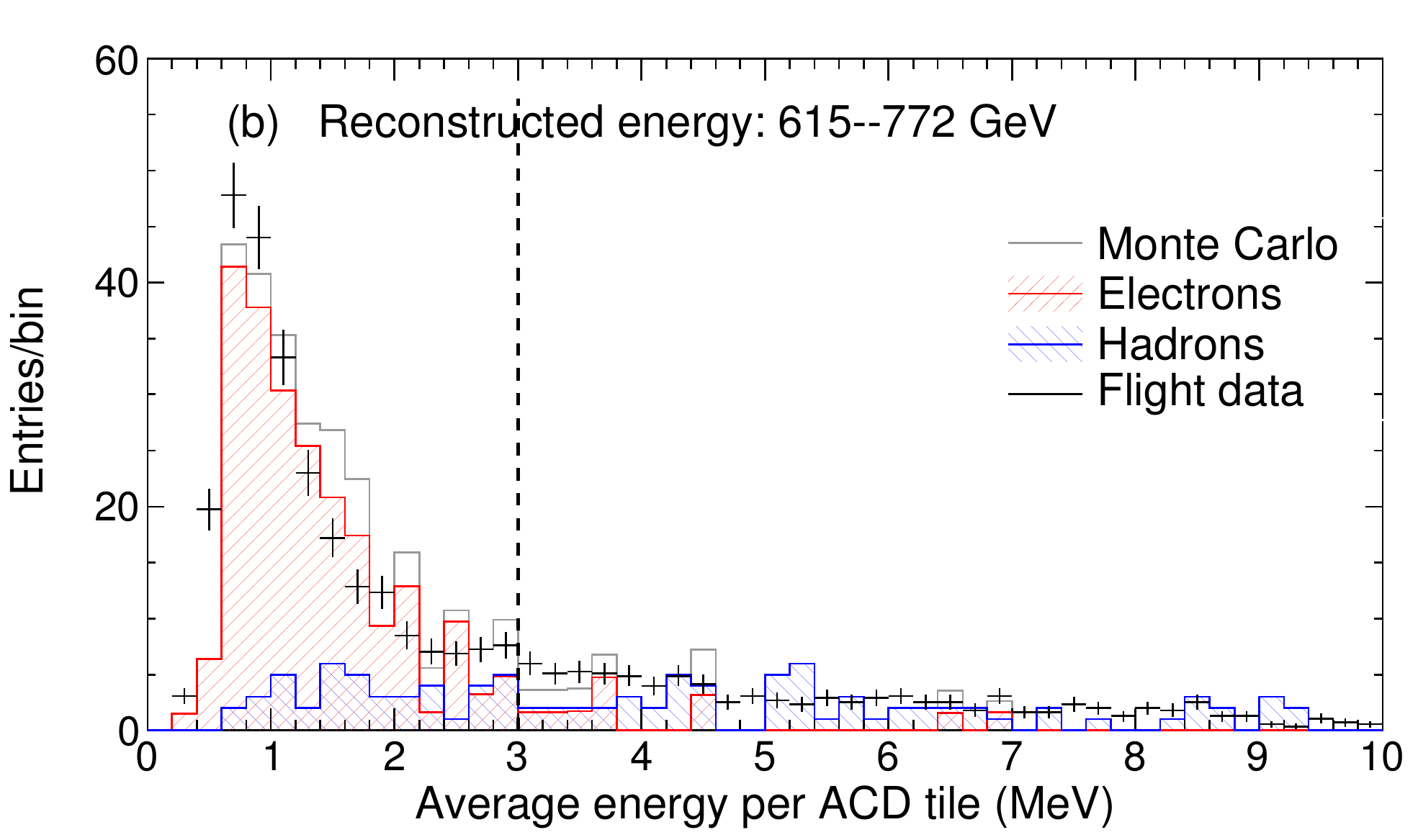}}
\subfigure[Total energy deposition in \ac{acd}, used in LE selection. 5~--~10~GeV.]{\includegraphics[width=.7\textwidth]{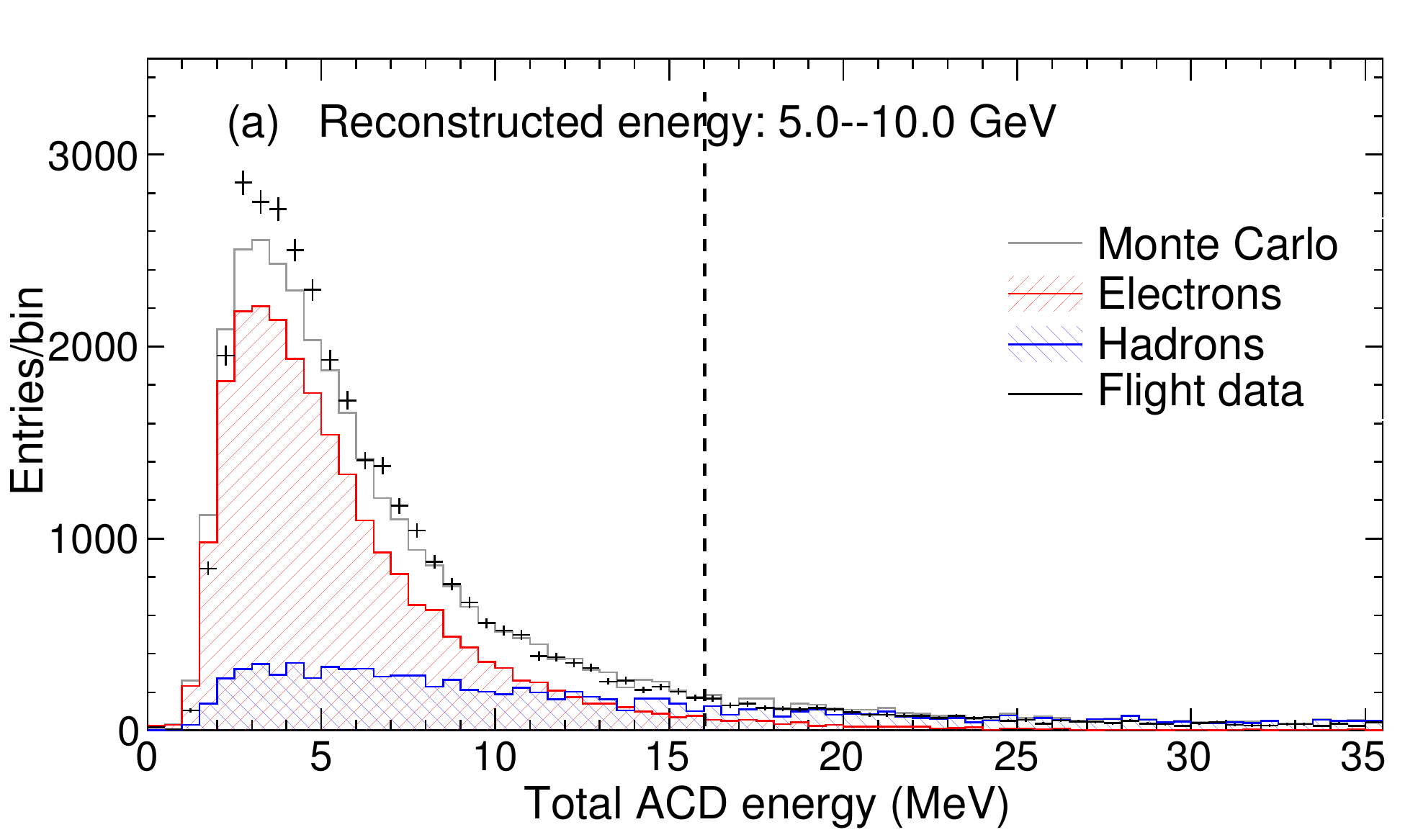}}
\caption{Selections on \ac{acd}. All  cuts applied with the exception of the cut on the shown quantity. Monte Carlo electrons are in red, background in blue, sum of the two Monte Carlos is in grey and flight data are black points. Black dashed lines are the cut value.~\cite{electronpaper}}
 
\end{center}
\end{figure}

$\alpha$ and heavier nuclei can be identified using their larger energy deposition in material, predicted by the Bethe-Bloch law $\frac{dE}{dx}\propto z^2$. Furthermore, in events with energy larger than hundreds of \GeV, the back-splash from the \ac{cal} is larger in hadronic showers than in electromagnetic ones, therefore the energy deposition in the \ac{acd} is larger. Figure~\ref{fig-HE-acd} shows the energy deposited per single tile, together with the HE selection on this quantity.

The \ac{tkr} capability  in identifying the charge of the events is poor; however at sufficiently high energy it can be used to discriminate between electro-magnetic and hadronic showers using the larger spread of hits in the hadronic showers. Therefore in the HE selection a cut is applied on the number of hits comprised in a cylinder around the best track, normalized to the number of hits assigned to the best track.

\begin{figure}[htp!]
\begin{center}

\includegraphics[width=.75\textwidth]{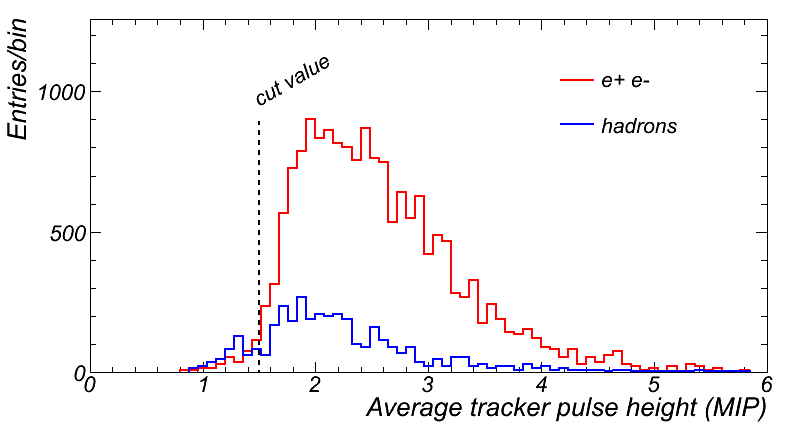}
\caption{Average pulse height in \ac{tkr}, in \ac{mip} units (\ac{tot} on the best track, times 4 and  divided by logarithm of the energy), for electrons and background. The cut value used in the HE selection is shown.}
\label{taglio-tot}
 
\end{center}
\end{figure}

The \ac{tot}, defined in \S~\ref{totref}, is another \ac{tkr} quantity used in the HE selection.  The \ac{tot} gives a rough estimation of the energy deposited in the \ac{tkr} and, at sufficiently high energies, it can be used to separate \acp{mip} from electrons, whose energy losses  are larger (see~\citet[section 30]{pdg}). The \ac{tot} cut  in the HE selection is shown in fig~\ref{taglio-tot}.

Finally a \ac{ct} (see \S~\ref{CT}) was trained for the HE selection using relevant \ac{tkr} quantities. The product of this  \ac{ct} with  similar \ac{ct} trained on \ac{cal} quantities is used  in both  HE and LE selection. In the LE selection  a single \ac{ct} was trained using both \ac{tkr} and \ac{cal} quantities.

The \acf{cal} of the LAT, described in \S~\ref{cal},  is fundamental in the \ac{cre} analysis: its capability to reconstruct the development of the shower is used to discriminate between electromagnetic and hadronic showers.

\begin{figure}[htp!]
\begin{center}

\includegraphics[width=.85\textwidth]{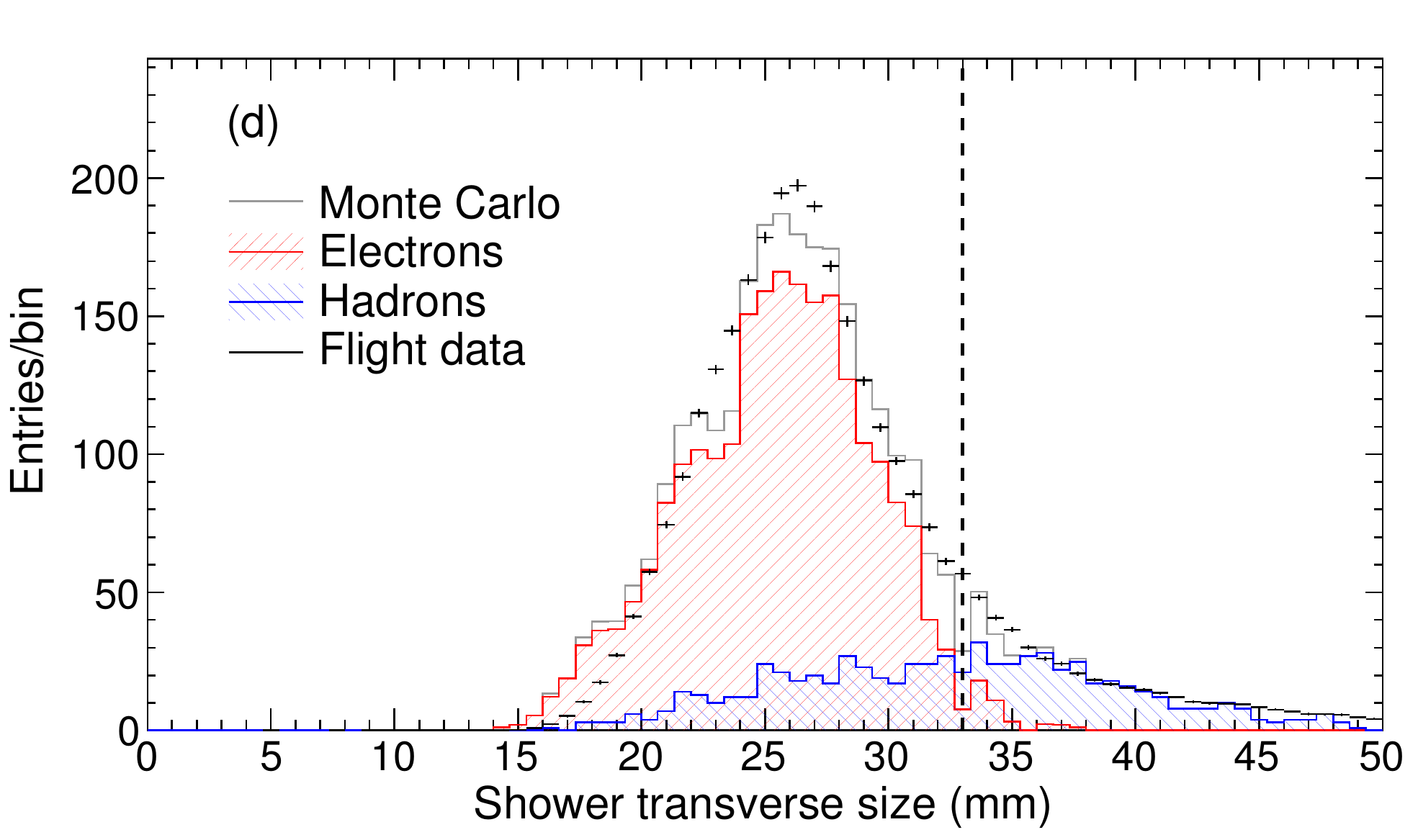}
\caption{Transverse dimension of the energy deposit in \ac{cal}, for electrons and background, in the energy range 133--210~GeV. All HE cuts applied with the exception of the cut on this quantity. Monte Carlo electrons are in red, background in blue, sum of the two Monte Carlo is in grey and flight data are black point. Black dashed line is the cut value.~\cite{electronpaper}}
\label{taglio-transrms}
 
\end{center}
\end{figure}

The quantity with the largest rejection power is the transverse dimension of the energy deposit in the \ac{cal}: because hadronic showers are more dispersed than electromagnetic ones, a simple cut that requires a maximum transverse dimension is present in both HE and LE selection; figure~\ref{taglio-transrms} shows the HE selection and the distribution of the transverse dimension.

Another \ac{cal} quantity used to select electrons is the shower asymmetry. An hadronic shower can have a significantly asymmetric development, therefore the HE selection has a cut on the asymmetry, determined using the two longitudinal moments of the energy deposition. 

Finally, in the HE selection a \ac{ct} was trained using relevant \ac{cal} quantities.

\subsection{Calculation of the Geometric factor}

The Geometric factor G(E), defined at page~\pageref{geometric-factor} of \S~\ref{geometric-factor}, quantifies the capability of the LAT to detect and correctly identify an incoming particle flux. The Geometric factor for electrons therefore includes both the effect of the LAT specifics and of the selections described in  \S~\ref{cut-CRE}, and is binned  as a function of incoming particle energy. To calculate the Geometric factor, the \ac{cre} selections is applied to a simulation of only electrons, including the effect of the on-board filter, and then the fraction of events passing the selections is analyzed.

Because of the power-law spectrum of~\acp{cre}, the geometric factor is binned in an energy logarithmic scale. The Monte Carlo simulation used to calculate the geometric factor generates an isotropic distribution of  electrons with an energy spectrum of $E^{-1}$, which results in equally populated  logarithmic energy bins. 

The Geometric factor is defined as

\begin{equation}
G(E)=A\cdot \left(\frac{N_{pass}(E)}{N_{generated}(E)}\right)
\end{equation}
where $A$ is the area and solid angle over which the Monte Carlo electrons are generated, that in the LAT simulations is $2\pi\cdot 6~\squaren\meter$ or $4\pi\cdot 6~\squaren\meter$. If the energy distribution of generated electrons is $E^{-1}$, the number of generated events in the $i^{th}$ bin is:

 \begin{equation}
 N_{generated}^i=N\frac{\log(E_{i+1})-\log(E_i)}{\log(E_{max})-\log(E_{min})}
 \end{equation} 
where $N$ is the total number of electrons of the simulation, and $E_{max},E_{min}$ the energy bounds of the distribution. 

\begin{figure}[htb!]
\begin{center}
\includegraphics[width=\textwidth]{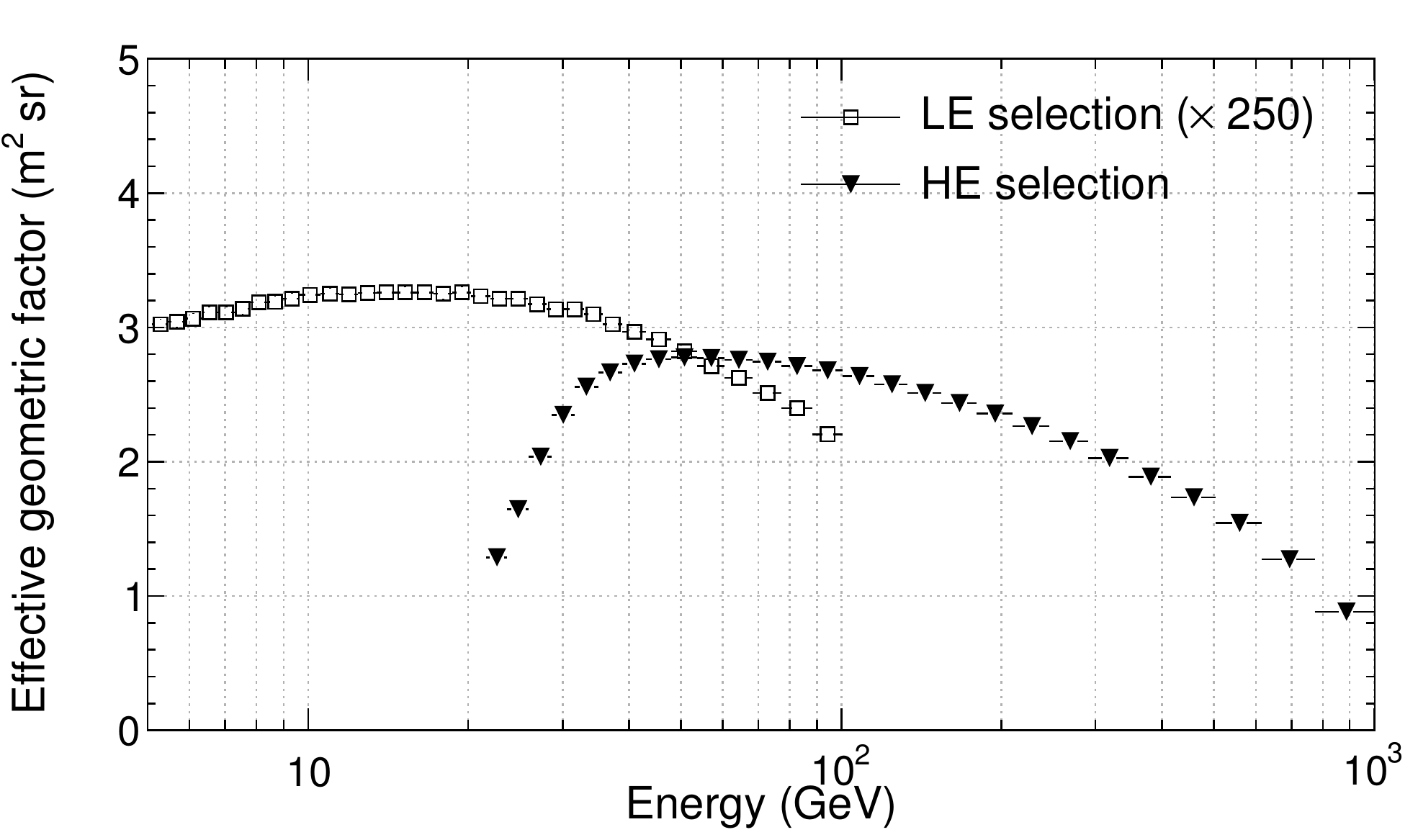}
\caption{Geometric factor for LE and HE electron selections. The LE geometric factor is multiplied by the diagnostic  250 scaling  pre-factor. \cite{electronpaper,melissa}}
	\label{geometric_factor_articolo}

\end{center}
\end{figure}

The geometric factor of the selections is shown in figure~\ref{geometric_factor_articolo}, as a function of reconstructed particle energy. The geometric factor of the LE selection is multiplied by the 250 factor introduced by the diagnostic filter, to make it visually comparable to the HE geometric factor. The HE geometric factor decreases at high energy, for the combined effect of the worsening of the event reconstruction at very high energy and of the increase of the proton background (see fig~\ref{elet-segn-rumore}), that requires a stronger selection on the events. Below 40~\GeV\ the HE geometric factor starts to decrease because of the increasing effect of the on-board gamma filter, that prevents electrons from being sent to ground.

\subsection{Subtraction of the residual background}\label{sottrazione-fondo}
\begin{figure}[htp!]
\begin{center}

\subfigure[Count rate of candidate electrons (black inverted triangles), Monte Carlo estimated residual background rate (open triangles) and electrons rate after the background subtraction (open squares).]{\includegraphics[width=.7\textwidth]{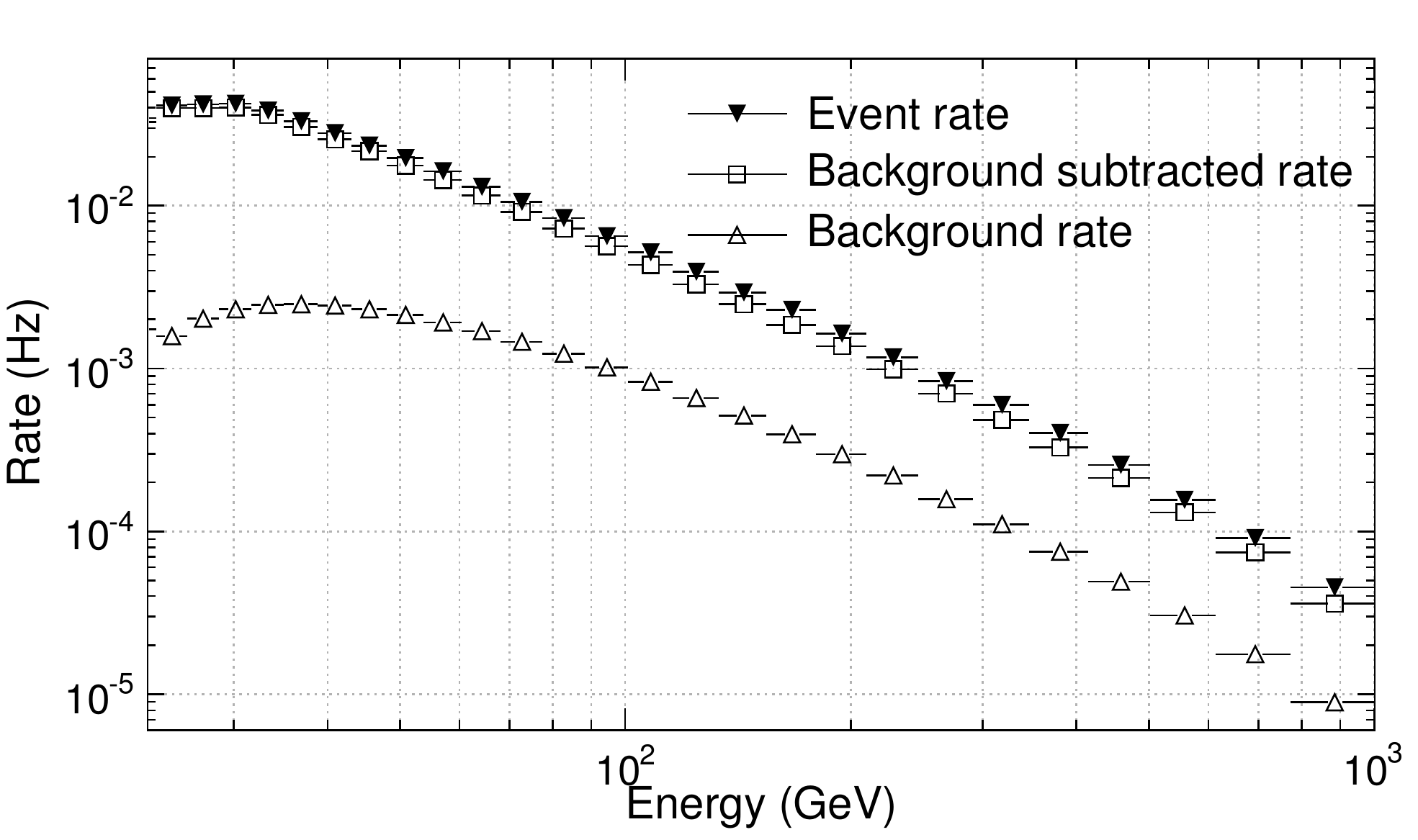}}
\subfigure[Fractional Hadron contamination of the electron selection. ]{\includegraphics[width=.7\textwidth]{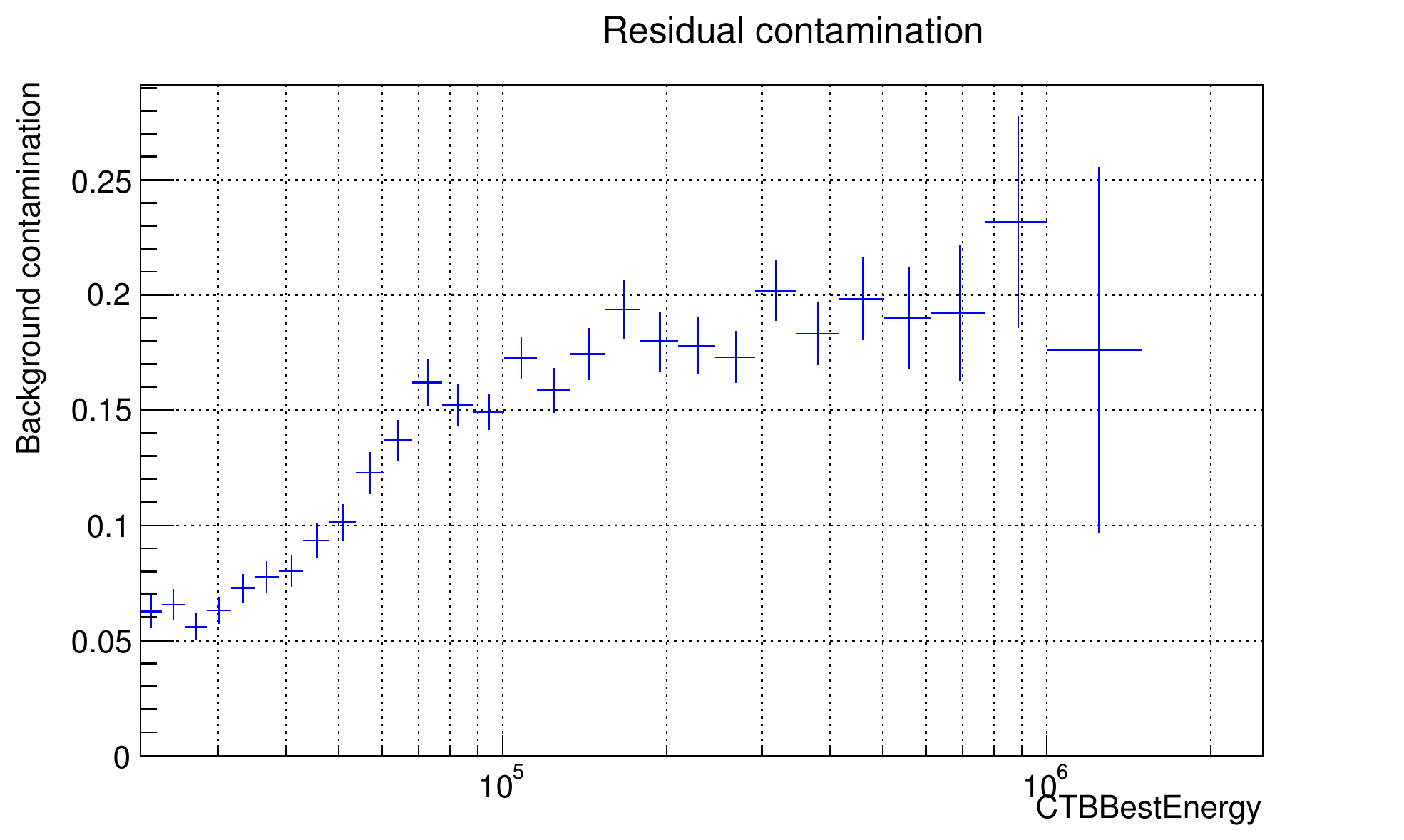}}
\caption{Residual background rate and fractional contamination in the HE \ac{cre} selection.~\cite{electronpaper}}
 \label{residuo-articolo}
\end{center}
\end{figure}

The electron selections have a residual fraction of background that survives all the cuts: this residual background has to be estimated and subtracted from the data before the \ac{cre}  flux calculation.  In the analysis described in~\citet{electronpaper}, the residual background is calculated applying the HE and LE selections to the standard background simulation. The residual background is then re-scaled for the livetime of the analyzed data, therefore obtaining an estimation of the expected background  rate present in the \ac{cre} selection. This  rate is  subtracted from the total  rate of the \ac{cre} selection. Residual background rate and contamination are shown in figure~\ref{residuo-articolo} for the HE data-set, while the LE residual contamination is not fixed but depends on the geomagnetic latitude. We can see that at high energy the combined effect of  background rate increase  and of  worsening of the event reconstruction leads to an increase of the  contamination up to 20\% at 1~TeV.

An alternative way of calculating the residual background at high energy was tested In the analysis presented in~\cite{electronpaper}. Instead of the standard background simulation, which describes the whole particle population on \fermi\ orbit, an isotropic  simulation of  protons with energy spectrum $E^{-1}$ was used, and then re-scaled in energy to match the effective proton flux. A  correction  by a 5\% factor was introduced to keep into account  He and heavier nuclei  flux. This approach grants a large improvement in the statistics of the simulation, particularly at high energy, because of the harder energy spectrum, and a significant reduction of the computational  time requested by the simulation, because the simulation is simpler. This approach led to results in complete agreement with those obtained with the standard background simulation. This method was also used in the positron analysis described  in~\citet{fermi-positroni}.
 
\subsection{Calculation of the CRE spectrum}
After  subtraction of the residual background,  the particle flux is obtained by dividing the electron event rate by the geometric factor and by the width of the energy bins; however,  to obtain a better estimation of the \ac{cre} flux some corrections are further applied.

\label{correzioni-cre}The first effect to correct for is relevant only below some tens of GeV in the HE selection. The on-board gamma filter works on quantities that are calculated on-board, as the deposited energy, whose correct reconstruction in simulations can be difficult. Therefore, the effect of the gamma filter on the selection can be imperfectly estimated by the Monte Carlo simulation used to calculate the geometric factor. To correct this feature,  events selected by the diagnostic filter, and therefore not processed by the gamma filter, are used. The HE selection is applied to a sample of events from the gamma and from the diagnostic filter, and the resultant ratio between the event rates is compared to a similar ratio obtained applying the HE selection to two electrons Monte Carlo samples, before and after the applying of the simulated gamma filter. The difference between the two ratios gives straightforward the correction factor to be applied  to the HE data-set.

The second relevant effect to be corrected is the impact of the energy dispersion, defined in \S~\ref{sec-energy-dispersion}, that can cause events migration  between different energy bins, with a non-negligible  impact on a steep spectrum as that of \acp{cre}. This effect is corrected  by using the iterative unfolding method described in~\citet{unfolding}. This method uses the Bayes theorem to recursively correct the estimation of the \textit{true} number of events in each bin and the related probabilities of event migration between bins. After each iteration, the estimated true number of events is compared with that obtained in the previous step: if the difference is sufficiently small, the iteration is terminated, otherwise the newly estimated number of events and probabilities are used as a prior in the next iteration. \\
The unfolding method leads to a correction on the events per bin  that is relevant only at high energy, and is always less than 5\%.

\begin{figure}[htb!]
\begin{center}
\includegraphics[width=\textwidth]{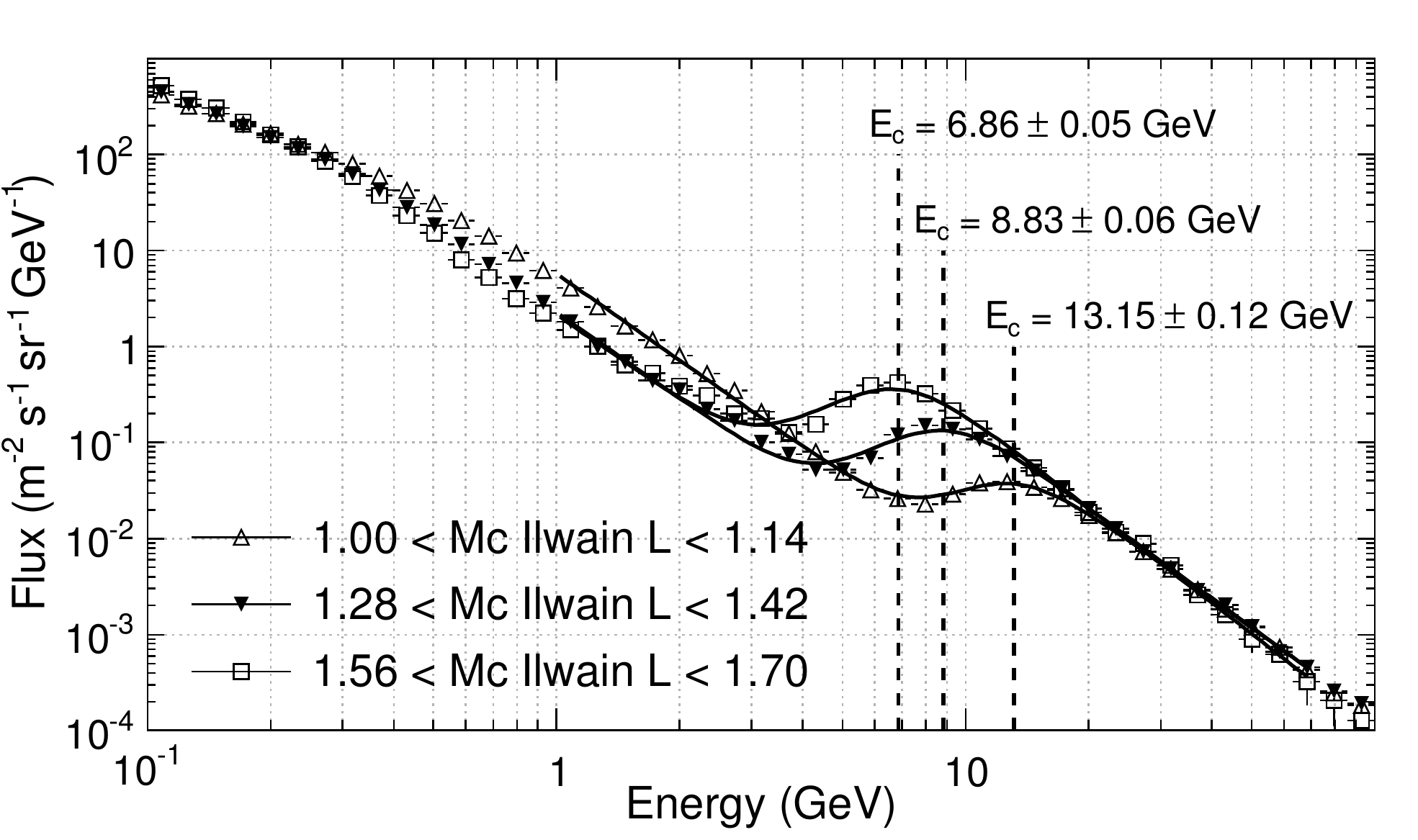}
\caption{Measured LE electron spectrum in three different McIlwain L intervals. The spectrum is fitted with eq~\ref{eq-cutoff}, and the resultant energy cutoff is shown.~\cite{electronpaper}.}
	\label{cutoff-elettroni}

\end{center}
\end{figure}

\label{geomagnetic-cutoff}
Finally, the low part of the measured spectrum has to be calculated keeping into account the effect of the geomagnetic field, that stops events of energy below the local geomagnetic cutoff, a value that in the \fermi\ orbit is in the 6~--~15~GV range. The LE data-set is divided in 10 intervals of McIlwain latitude L from 1.72 to 1.00\footnote{ The McIlwain L parameter, introduced in~\citet{mcilwain} together with its sister quantity B, describes   the strength of the  geomagnetic field: in the approximation described in \citet{cutoff-geomag}, the geomagnetic cutoff rigidity can be easily written as $R_C\propto 1/L^2$.}. For each of these intervals, the energy spectrum was fitted to the function

\begin{equation}\label{eq-cutoff}
\frac{dN}{dE}=c_s E^{-\Gamma_s}\,+\,\frac{c_p E^{-\Gamma_p}}{1+(E/E_c)^{-6}}
\end{equation}
This function  describes the effect of the cutoff on the primary electron flux $c_p E^{-\Gamma_p}$, with the addition of the secondary flux of electrons produced by the Earth atmosphere, $c_s E^{-\Gamma_s}$, and was used in the background simulation code of the LAT. From the fit we obtain the best estimation of the energy cutoff $E_c$. After that, the stability of the flux is studied when varying the value of $E_c$ around that obtained by the fit, to search for  residual effects of the geo-magnetic cut-off. The value $1.15\cdot E_c$, above which the measured flux is stable, is assumed as the lowest energy at which the \ac{cre} spectrum can be measured using that geomagnetic bin. See~\cite[page 129]{melissa} for further details on this correction. Figure~\ref{cutoff-elettroni} shows the spectrum and the measured cut-off for three different McIlwain L intervals, including those with maximum and minimum cut-off.

\begin{figure}[htb!]
\begin{center}
\includegraphics[width=\textwidth]{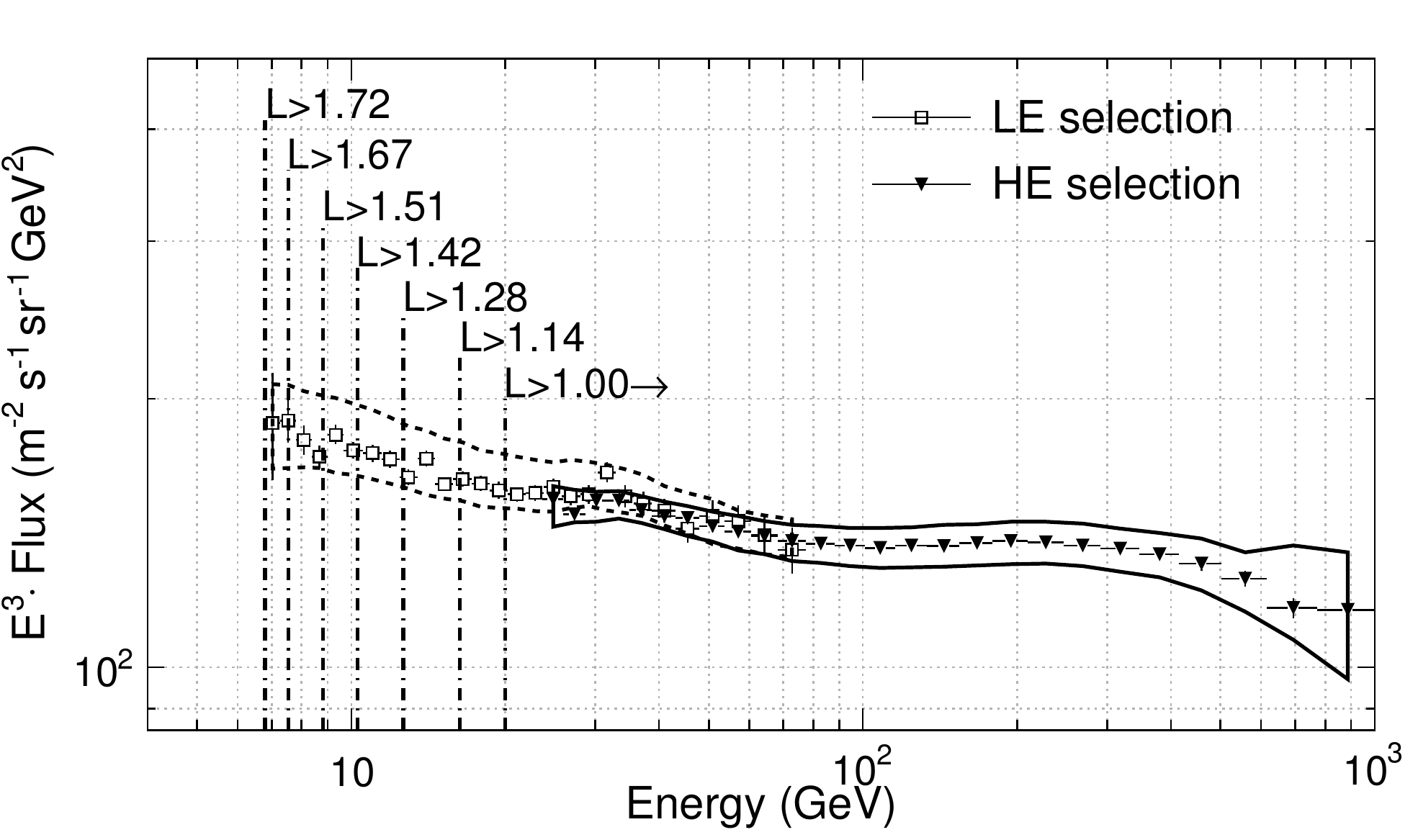}
\caption{CRE spectrum multiplied by $E^3$ from 7~GeV to 1~TeV, as measured by the LAT. Squares are the results of the LE analysis, triangles are the result of the HE analysis. Systematic uncertainties are the  solid and the dashed lines; the two spectra agree in the overlap region 20~--~80~\GeV. Dot-dashed vertical lines show the minimum value of McIlwain L at which data were taken.~\cite{electronpaper}.}
	\label{spettro-con-geomag-bin}

\end{center}
\end{figure}

The measured flux, after the described corrections, is shown in figure~\ref{spettro-con-geomag-bin}, together with the different McIlwain L regions from whose the flux is measured and with the systematic uncertainties, that will be described in the next section.

The spectrum was fitted with a single power-law from 7~\GeV\ to 1~TeV, obtaining a spectral index of $\gamma=-3.08$; to estimate the uncertainty on this value,   the fit was repeated using  several power-laws of fixed index $\gamma$, with $\gamma$ varying around the best-fit result. Assuming that the residual function $\chi^2(\gamma)$ follows a $\chi^2$ distribution, we choose to define the error interval as the   range in $\gamma$ for which $\chi^2(\gamma)$ differs by less than $1\sigma$ from its expectation value, obtaining for the spectral index $\gamma=-3.08\pm0.05$. The fit performed only on HE data~\cite{elettroni} ($E>20~\GeV$) returned a slightly harder spectrum, with $\gamma=-3.04$. \label{misura-spettro}

Some features in the spectral behavior suggested a broken power law with a harder spectrum from 100 to 400~GeV and a steeper spectrum above 400~GeV, but the systematic uncertainties prevented any claim of deviation from a simple power law.~\cite{elettroni-interpr}.

\subsection{Estimation of the systematic uncertainties}\label{p6-systematics}
The error on the measured flux is dominated by the systematic uncertainties because, with the exception of the first two bins and  the last one, the number of data counts per bin ranges from 1000 up to more than 9000 (and can be furthermore increased by using more than 1~\yr\ of data), and the statistical errors are therefore small.

\begin{figure}[htb!]
\begin{center}
\subfigure[Average deposited energy in \ac{acd} tiles, after all cuts but that on the shown  quantity are applied. Vertical line is the used cut value in the HE selection. Geometric factor (dots) and electron flux (triangles) are measured for different cut values. Above 5~MeV the cut is almost ineffective, and the geometric factor reaches its maximum value.]{\includegraphics[width=.76\textwidth]{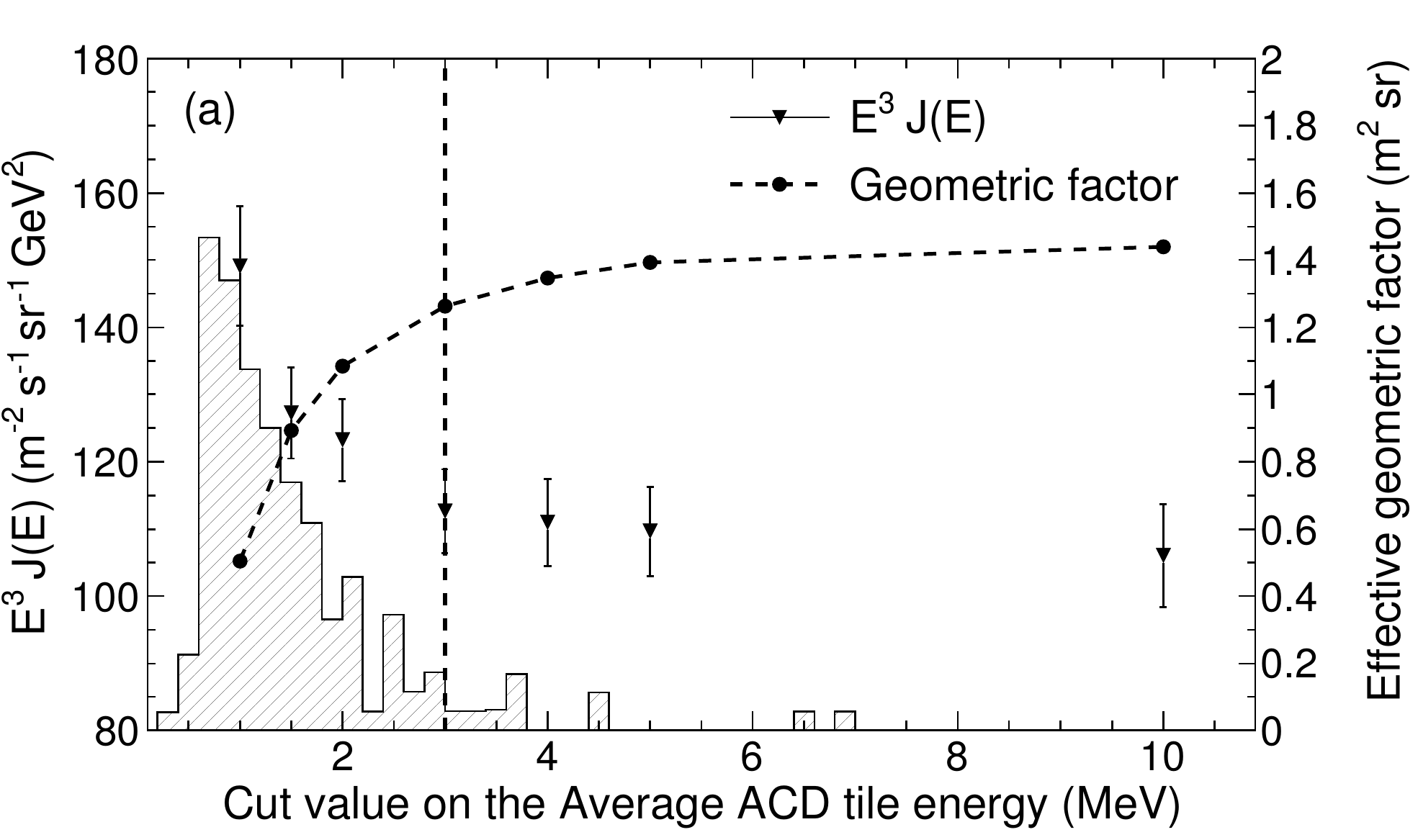}}
\subfigure[Geometric factor vs $E^3\cdot flux$ for measurements shown in previous plot. The linear approximation of the points distribution is completely reasonable. The calculated systematic on the flux $\Delta_{sys}$ is the difference between the flux measured at the standard value (3~MeV) and the value of the flux at 10~MeV geometric factor obtained from the fit.]{\includegraphics[width=.76\textwidth]{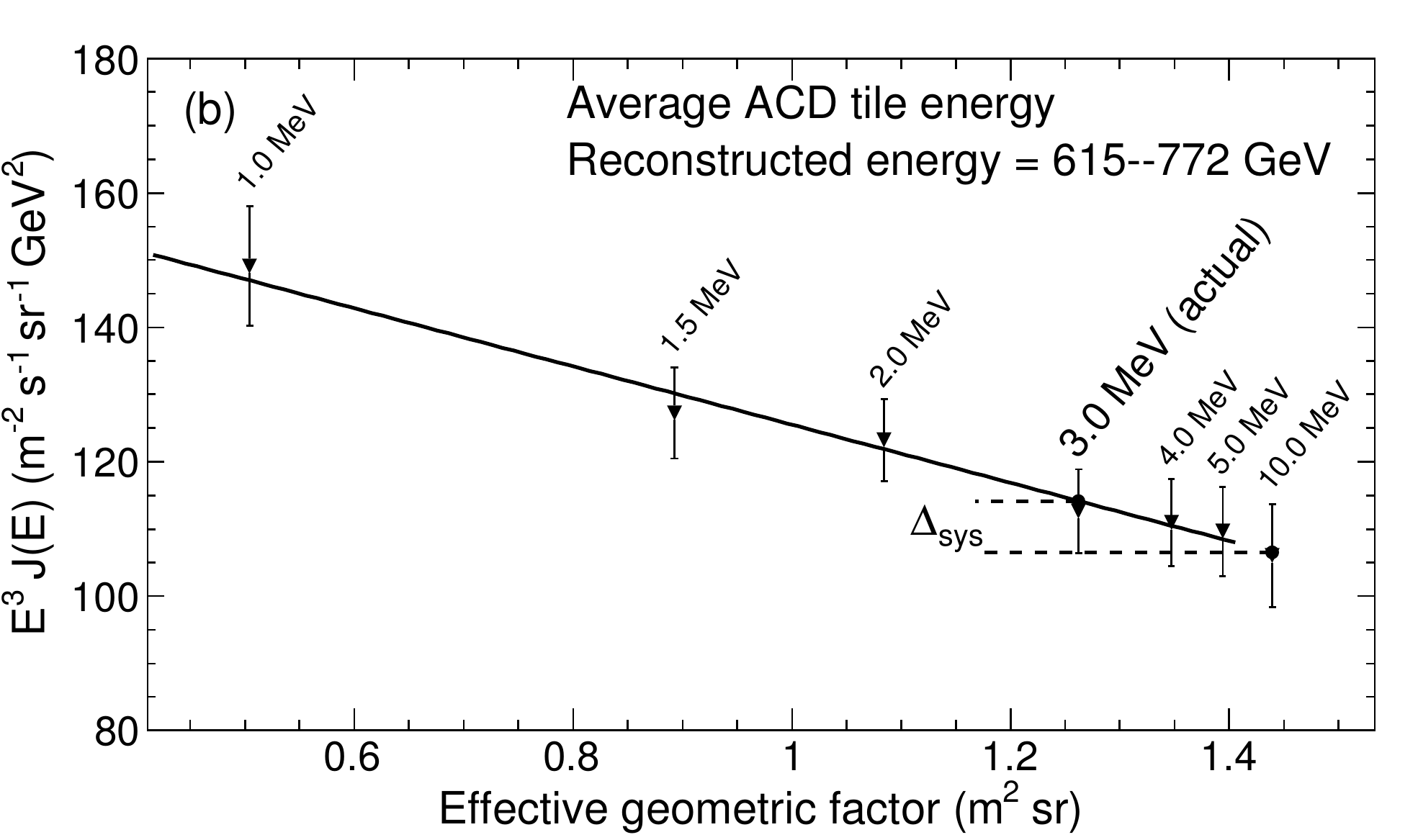}}
\caption{Contribution  to the systematic uncertainty coming from  the cut on the average \ac{acd} tile energy, in the energy bin 615 -- 772~GeV. This quantity gives the largest contribution to the total systematic uncertainty at high energy.~\cite{electronpaper}.}
	\label{sistematiche-articolo}

\end{center}
\end{figure}

The main source of systematic uncertainty is the not perfect reproduction by the Monte Carlo electron simulation of the quantities used in the electron selections, leading to an uncertainty on the geometric factor, with the measured flux that changes when varying the values of the cuts of the electron selections. This uncertainty was then calculated,  for each energy bin,  as the sum of  the single contributions of   the cut in the electron selection. These contributions are defined as the difference between the measured flux when all cuts are applied at their optimal values and the value of flux obtained removing from the selection the cut under study. The value of the flux without applying the studied cut is not  directly measured repeating the analysis without applying the cut, because the residual background fraction would be too large and the measurement untrustworthy: it is indirectly estimated. To do so, the flux is measured for different cut values, while applying all other cuts at their optimal values. The results are then plotted on the plane flux/geometric factor, and fitted with a straight line, that has proved to be a sufficient good fit for all cuts in all energy bins. The fitted line is used to extrapolate the flux at the value assumed by the geometric factor when the cut is not applied, and the difference between the obtained value and the flux measured at the standard cut value is assumed as the systematic associated with the studied cut. This procedure, that is repeated for all the cuts in the selections and for all the energy bins, is exemplified in figure~\ref{sistematiche-articolo}, where  the following factors are shown: the evolution of the geometric factor when relaxing the studied cut until it reaches its asymptotic value, the plot flux/geometric factor, the linear fit, the extrapolated flux and the calculated systematic error.

The  squares of uncertainties  obtained in this way are summed  separately for negative and positive values, therefore obtaining an asymmetric bracketing to the measured flux. 

Another source of systematic uncertainty is the error on the absolute normalization of the background flux, that  leads to an error dependent on the residual background contamination. This uncertainty is conservatively assumed to be $\pm 20\%$ on the background flux normalization, that leads to an error on the electron flux $<4\%$, because the residual contamination is $<20\%$. For each energy bin, the systematic is calculated and summed in square to the previously calculated systematic.

A further source of uncertainty on the flux is the uncertainty on the absolute energy scale of the LAT (see \S~\ref{absolute-energy}), that is the not perfect knowledge of the response of \ac{cal} crystal detectors  to an energy deposition, that causes an uncertainty about the presence of a bias on the reconstructed energy distribution. Because it is  measured  that the error on the absolute energy scale has no significant energy dependence, it  can result only in a rigid shift of the whole spectrum of $(\gamma -1)\Delta E/E$, for a simple power-law  spectrum $E^{-\gamma}$. Inserting the estimated uncertainty of $+5\%/-10\%$ and $\gamma=3$   results in a  $+10\%/-20\%$ uncertainty. Because the shift can only be rigid in energy,  the estimated uncertainty  is not summed to the other systematic errors, but indicated separately.

Finally, the uncertainty caused by the statistical fluctuations in the Monte Carlo simulations used for calculating the geometric factor and the background contamination is negligible, because the used simulations have sufficiently large statistics.
\clearpage
\section{Measurement of the CRE positron fraction using the Large Area Telescope}\label{lat-positroni}
The \ac{lat} does not have a magnet on-board, therefore it is unable to directly discriminate between positive and negative charges, \Pelectron and \APelectron in the case of \ac{cre}. However, it is possible to exploit the effect of the Earth magnetic field on charges of different signs: above the geomagnetic cut-off energy, at which charged particles are completely deflected, the deflection caused by the magnetic field leads to directions which are ``forbidden" to particles of given rigidity, because the Earth blocks that trajectories. This effect, called the ``East-West effect" because it causes differences between  the rates of particles coming from these directions, is the base of the measurement  of the separate \Pelectron and \APelectron fluxes (and therefore of the  \APelectron/\Pelectron fraction) made by the \ac{lat} in the range 20 -- 200~GeV and published in~\citet{fermi-positroni}.

\begin{figure}[htb!]
\begin{center}
\includegraphics[width=\textwidth]{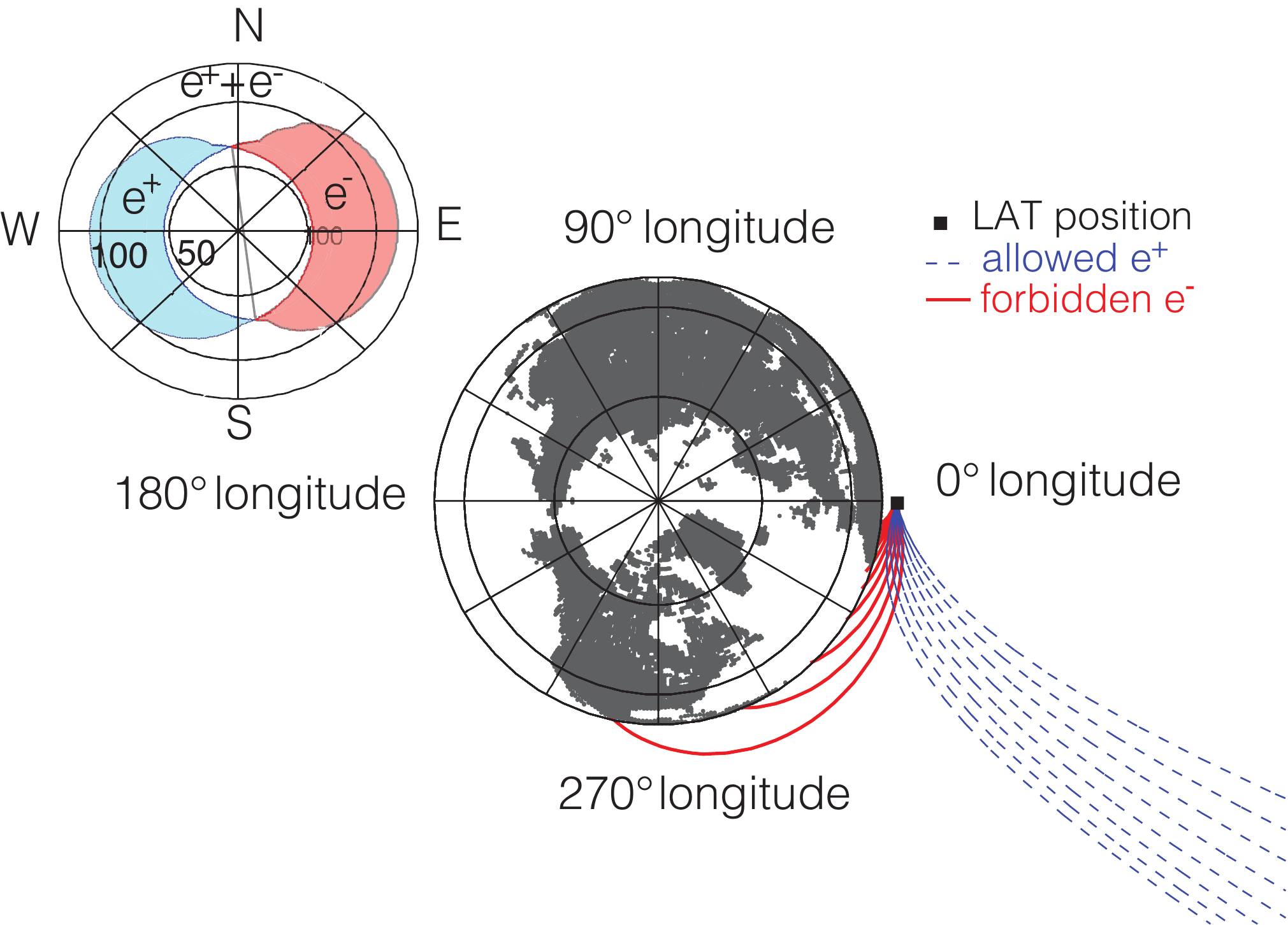}
\caption{Examples of calculated electron (red) and positron (blue) trajectories arriving at the detector, for 28~GeV particles arriving within the Equatorial plane (viewed from the North pole). Forbidden trajectories (reaching the atmosphere) are solid and allowed trajectories are dashed. Inset: the  regions for which only electron or only  positron are allowed, for the same
particle energy and spacecraft position as the trajectory traces (viewed from the instrument position in the Equatorial plane).~\cite{fermi-positroni}.}
	\label{traiettorie-geomag}

\end{center}
\end{figure}

The analysis uses a high-precision geomagnetic field model (the 2010 epoch of the \nth{11} version of the International Geomagnetic Reference Field presented in~\citet{modello-geomag})  to calculate if an event, given its measured energy and the position of \fermi\ at the time of collection, has a direction that is ``forbidden" for \Pelectron of \APelectron, that is if, tracking back the particle, its trajectory crosses the Earth atmosphere (see \cref{traiettorie-geomag}). 

Because the forbidden directions are close to the Earth atmosphere, that  is excluded by the \ac{lat} \ac{fov} during the usual survey mode of operations, the only data available for this analysis were that from the Limb-pointed observations made during the \ac{lat} on-orbit calibration in 2008 and some pointed observations to sources which were sufficiently close to the Earth to include it in the \ac{fov}, for a total of 39.0~d of data  from 2008 to 2011.

The candidate \Pepm were selected using the high-energy selection described in \S~\ref{cut-CRE}; the energy range was chosen because below 20~GeV the on-board filter described in \S~\ref{filtri} rejects a large fraction of charged particles, while introducing significant bias in the data sample, while above 200~GeV the forbidden regions becomes very small (particles deflection is small) and this, together with the steep \ac{cre} spectrum, strongly reduces the statistics.

\begin{figure}[htb!]
\begin{center}
\includegraphics[width=\textwidth]{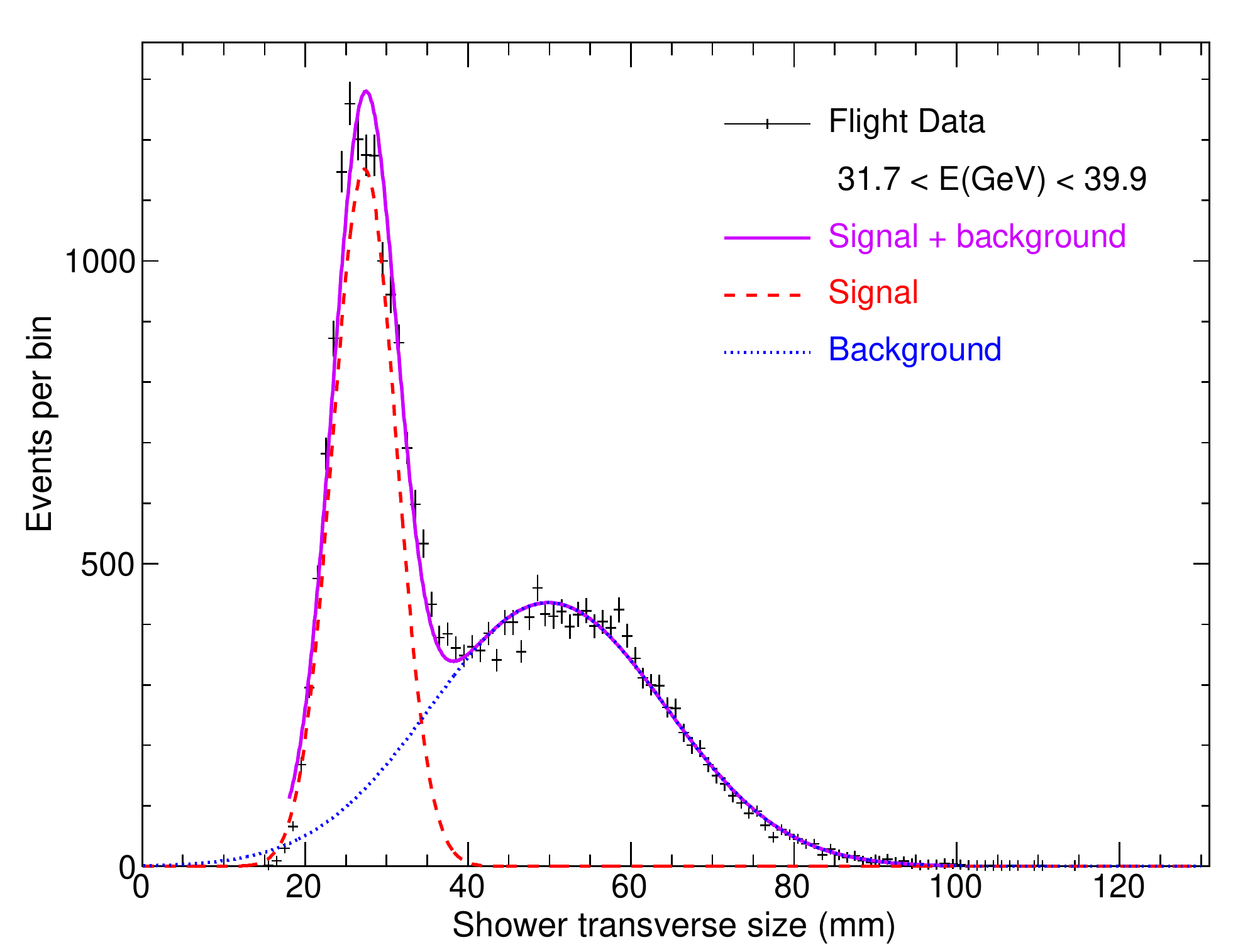}
\caption{Transverse shower size distribution in the electron-only region for flight events (black  points) of energy 31.7 -- 39.9~GeV. The first peak is made mainly by \Pelectron (would be the same for \APelectron), the second by hadrons. Both populations are fit with a gaussian (blue and red lines).~\cite{fermi-positroni}.}
	\label{positroni-calTransRms}

\end{center}
\end{figure}

The residual background was estimated independently using two different approaches: in the first, that uses only flight data without any simulation of the instrument,  the transverse dimension of the shower in the \ac{cal} is used to estimate the two populations, because it can be very well fit with two gaussians centered on the lepton and hadron peaks (see \cref{positroni-calTransRms}). The second approach is described at the end of \S~\ref{sottrazione-fondo}, and  uses a large proton simulation, re-weighted in energy to match the measured spectrum of protons, to estimate the residual hadron contamination. 

The result of this analysis is shown in \ref{frac-positr}, and confirmed the initial claim made by PAMELA~\cite{pamela-positroni} of an increase in the positron fraction. 

\chapter{Recent results in the study  of the Cosmic-ray electrons}\footnote{This chapter is updated to the end of 2014; however, until the time of publishing of this review no significant new measurement has been made neither any new significant theory has been proposed.}\label{interpretazioni-spettro}

Recently, several new experiments  significantly increased our knowledge of \ac{cre}, with new and more precise measurements of their spectrum and composition. The impact of these studies was impressive, first of all because the  \ac{cre}-production paradigm  (primary \Pelectron produced in \acp{snr}, \APelectron produced in interaction between primary \ac{cr} and \ac{ism}) prediction seems to be not in agreement with the new data.  The purpose of this chapter is to summarize  these recent measurements and the new theories they lead to.

Because our knowledge  of the \Pelectron$+$\APelectron spectrum comes from measurements  made using different techniques and covering different energy ranges, I will describe the most recent instruments whose measurements significantly superimpose with  \ac{lat} result: the space-based \ac{cr} experiments PAMELA  and AMS-02 (\S~\ref{ams-sec}) and the balloon-based ATIC (\S~\ref{atic-sec}). Ground \v Cerenkov telescopes like HESS can be used to measure the highest-energy  part of the spectrum, overlapping only with the highest energy bins of the \ac{lat} measurement: a brief description of these instruments can be found in \S~\ref{ground-telescopes}. In \S~\ref{elettroni-tuttemisure} I will then summarize all recent  measurements of the \Pelectron$+$\APelectron spectrum above 20~GeV, discussing the conclusions that can be drawn  and the unresolved discrepancies between different results.

In \S~\ref{status-cre-sec} I will  make a brief summary of the theoretical and experimental status of the \ac{cre} study:  I will summarize recent results in the measurement of the \ac{cre} composition, discussing how they were in disagreement with the \ac{cre}-production paradigm; I will then review the new theoretical solutions  proposed to explain the new data; finally, I will  discuss why it is currently impossible to select one of the proposed solutions while rejecting the others,  and which future results from \ac{cr} and \ac{cre} study could lead to reject or to significantly confirm one of them.

\section{Recent measurements of the  electron plus positron spectrum from space-based experiments}\label{misure-varie-elettroni}

\label{ams-sec}
\begin{figure}[htb!]
\begin{center}
\includegraphics[width=\textwidth]{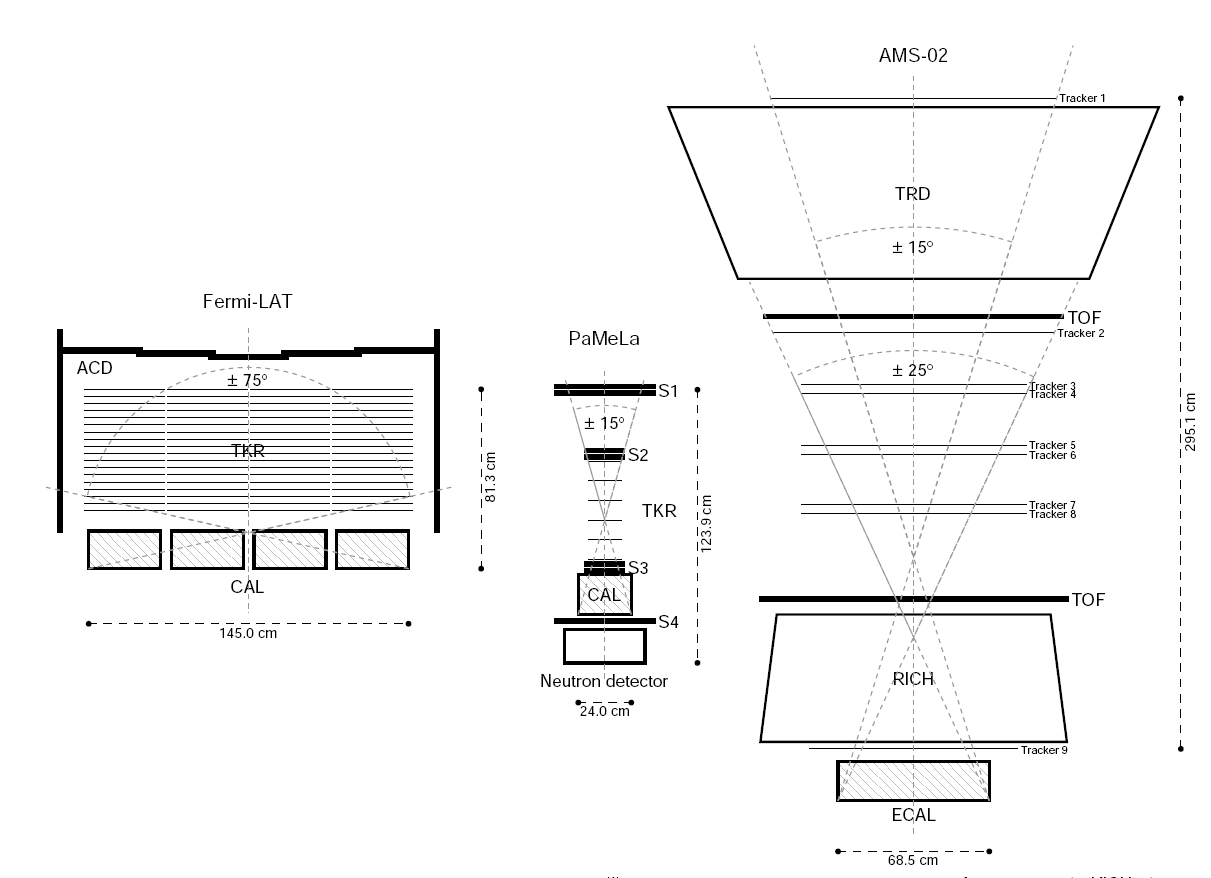}
\caption{Schematic view of some of the most recent space-borne cosmic-ray and gamma-ray detectors (all the dimensions are meant to be in scale). Only the detector subsystems are sketched (note that the magnets for PAMELA and AMS-02 are not represented).~\cite{luca-rivelatori}}
	\label{detectors}

\end{center}
\end{figure}

Both PAMELA and AMS-02 are space-based detectors  mainly designed to study the charged component of the \acp{cr}: their design concept is therefore significantly different from \ac{lat}, because their main task is to identify the charge of incoming particles. The core of both instruments is a magnetic spectrometer, whose magnetic field is fundamental to identify the charge sign. This also leads to the main difference in the instruments design with respect to the \ac{lat}: since the sensitivity of the magnetic spectrometer increases with its height, the base-to-height ratio of these instruments is significantly lower than in the \ac{lat}. This feature, together with the necessity for an event to cross almost the full instrument to be correctly reconstructed, leads to a \ac{fov} significantly smaller that that of the \ac{lat}. Furthermore, charged particles do not need a converting material, and there is no need of an anti-coincidence shield that surrounds the whole instrument. \ref{detectors} compares  PAMELA and AMS with  \ac{lat} design, showing also the maximum angle  a particle must have to be efficiently detected. 

\paragraph{PAMELA}
The  \acf{pamela},  described in \citet{pamela-paper}, is an instrument designed to study \acp{cr}, in particular the anti-particles component (\APproton, \APelectron). Launched on June the \nth{15} 2006 and currently operating, it orbits on a semi-polar elliptic orbit at an angle of 70\degree with respect to the equatorial plane, with an altitude varying between 350 and 600~km. The instrument is 1.3~m high, with a total weight of 470~kg and a consumed power of 355~W.

\begin{figure}[htb!]
\begin{center}
\includegraphics[width=\textwidth]{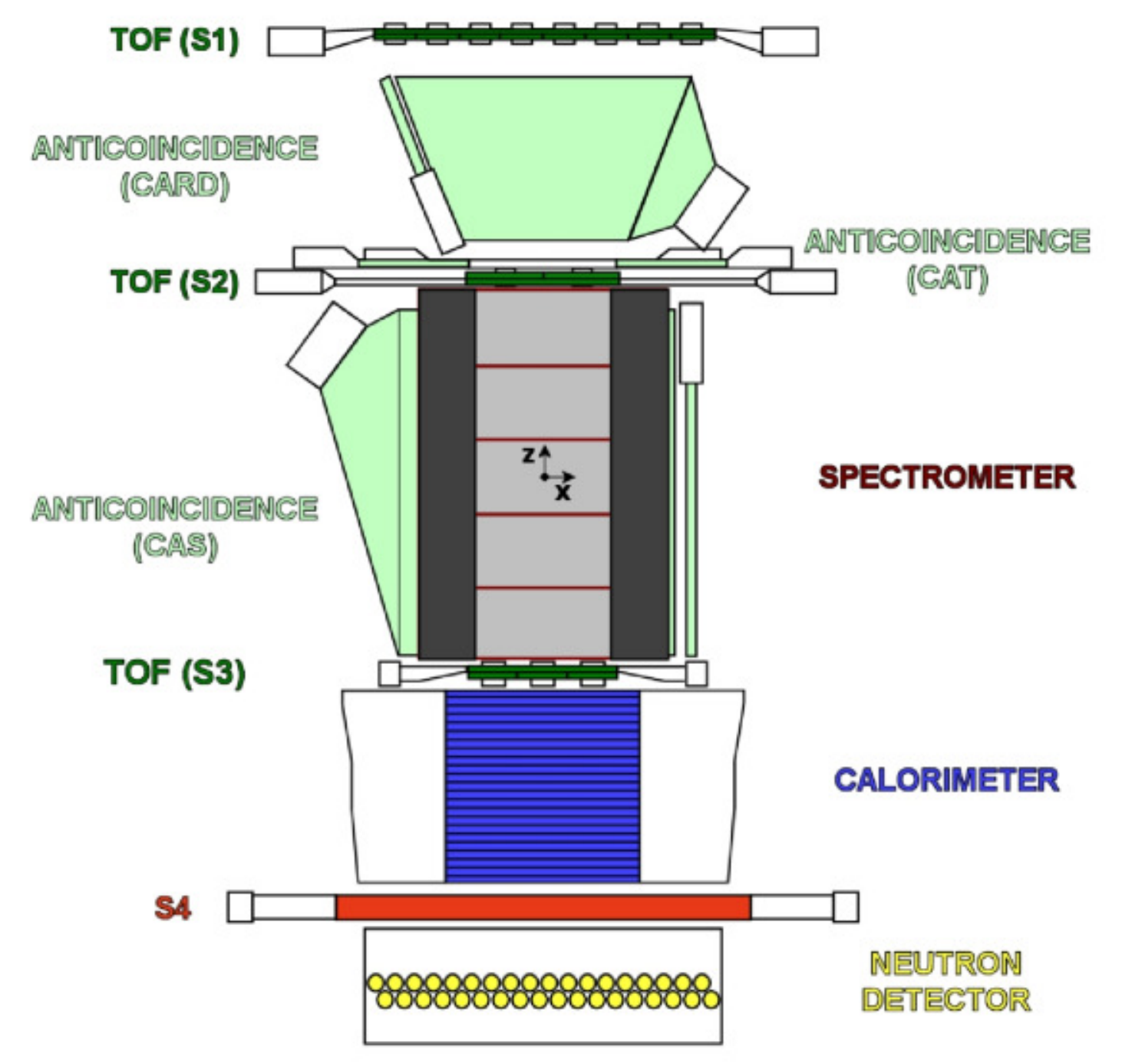}
\caption{Schematic view the \acf{pamela} detector.~\cite{pamela-paper}}
	\label{pamela}

\end{center}
\end{figure}

The instrument, shown in \cref{pamela}, has at his core a magnetic spectrometer, based on a permanent magnet of 0.43~T.  The passage of particles in the magnetic field is detected by six planes of silicon detectors that also measure the ionization energy losses in the silicon so that the spectrometer is able to measure not only the charge sign and the particle rigidity $\mathcal{R}=pc/Ze$, but also its charge ($\frac{dE}{dx}\propto Z^2$  up to $Z=6$. The spectrometer determines the acceptance (defined in \S~\ref{geometric-factor}) of the whole instrument, that is 21.5~\squaren\cm\sr\ (as a comparison, \ac{lat} acceptance   for electrons at 50~GeV is 28000~\squaren\cm\sr).

The main trigger of the instrument is provided by a \acf{tof} device, composed by three planes of plastic scintillator, placed at the top of the instrument, just above the magnetic spectrometer and just below it: the default trigger configuration requires that all three planes have a significant energy deposition.  The \ac{tof} measures the particles velocity and, through the measurement of the deposed energy, also the charge up to $Z=8$; combining these information with the rigidity measured by the spectrometer it is also possible to calculate the particle mass $m\propto \mathcal{R}\sqrt{1-\frac{v^2}{c^2}}Z/v$ ($c$ is the speed of light) to separate \Pepm from \Pproton - \APproton for energies below 1~GeV.

An anti-coincidence system is present, mainly to detect particles entering from the sides of the instrument; the anti-coincidence is composed by different sections, one (CAS) covers the sides of the spectrometer, another (CAT) covers its top and the third (CARD) encloses the space between the first and the second plane of the \ac{tof}: all sub-systems are composed by several plastic scintillators.

Below the spectrometer  an electro-magnetic calorimeter is placed, whose  purpose is not only  energy measurement but also the reconstruction of event shower development, providing the largest contribution in the discrimination between electro-magnetic and hadronic events. Differently from the \ac{lat} calorimeter, this device is a sample calorimeter, with 44 planes of silicon detectors interleaved by 22 planes of Tungsten, for a total thickness of 16.3 radiation lengths ($\sim 0.6$ nuclear interaction lengths). The calorimeter is able to self-trigger, acting as a self-standing detector with an acceptance of $\sim 600~\squaren\cm\sr$.

To increase  \Pepm identification power at high energy, two additional devices are placed below the calorimeter: the shower tail catch scintillator is a plastic scintillator that detects the shower leakage from the bottom of the calorimeter, while the neutron detector, composed by two planes of 18 of $^3$He proportional counters, detects evaporation neutrons (emitted by nuclei during de-excitation) thermalized in calorimeter, that are 10 -- 20 times more numerous for hadronic events with respect to electro-magnetic ones.

During its mission, \ac{pamela} has measured the spectrum of \Pelectron and \APelectron (\citet{pamela-positroni,elettroni-pamela}), of primary \acp{cr} (\citet{pamela-p-and-he}), of \APproton (\citet{pamela-antiP}) and  the fraction of light elements or rare isotopes in \ac{cr} (\citet{pamela-b/c,pamela-b-c,pamela-he3}). Of  particular importance for the study of \ac{cre} was the first claim of an excess in the \APelectron flux (\citet{pamela-eccesso}) in contrast with the assumption of an all-secondary origin of these particles.

\paragraph{AMS-02}   

The  \acf{ams}   is a large acceptance magnetic spectrometer installed on the \acf{iss} on May the \nth{19} 2011; AMS-01 was a prototype tested on a 10 days shuttle flight in June 1998; the orbit of \ac{ams} is therefore that of the \ac{iss}, an almost circular orbit inclined by 51.6\degree with respect to the equatorial plane, with an orbital period of $\sim 93$~minutes. The altitude of the \ac{iss} is comprised between 330 and 430~km. Because of its integration with the \ac{iss}, \ac{ams} is not affected by the typical constraint of rocket-based instruments, having a total mass of 8500~kg (three times the mass of the \ac{lat}), a total power consumption of 2500~W and a total download band of 9 -- 10~Mbps (almost ten times the bandwidth of the \ac{lat}). Its main purpose is the search of \HepAntiParticle{He}{}{}, with an expected sensibility after 20 years of data-taking of $\HepAntiParticle{He}{}{}/\HepParticle{He}{}{}\sim 10^{-10}$, and the measurement of charged \acp{cr} spectrum and composition. 

\begin{figure}[htb!]
\begin{center}
\includegraphics[width=\textwidth]{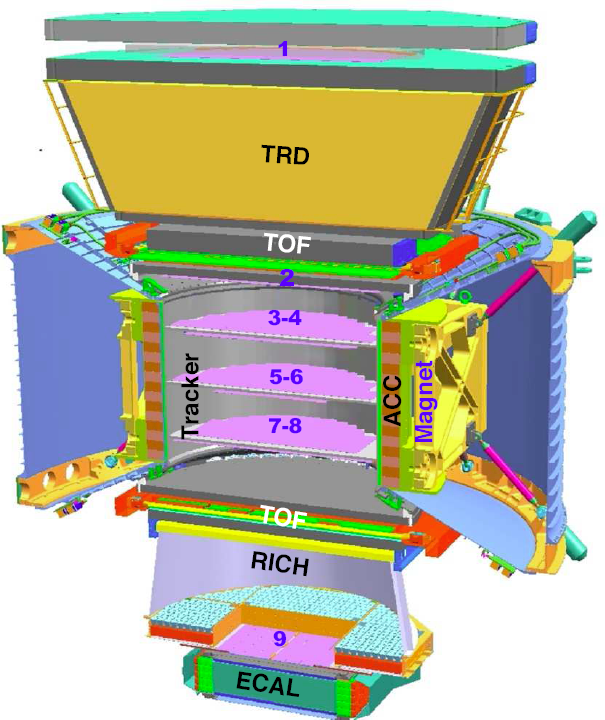}
\caption{Schematic view the \ac{ams} detector, with the various sub-systems. Blue numbers from 1 to 9 mark the planes of the tracker.~\cite{ams-paper2}}
	\label{ams-figura}

\end{center}
\end{figure}

\ac{ams}, described in \citet{ams-paper,ams-paper2}, has at its core a cylindrical permanent magnet of 80~cm of height and 111~cm of inner diameter, with a magnetic field of 0.15~T. Above and below it, different sub-systems make independent measures of quantities like the energy or charge of events, exploiting different techniques (see \cref{ams-figura}). The instrument height is almost 3~m, while the base of the electromagnetic calorimeter at the bottom is 68~cm:  the acceptance, when requesting that an event crosses all the detectors, is  4500~\squaren\cm\sr.

The tracker, which measures the particle trajectory and the deflection caused by the magnetic field, is composed by nine planes of double sided silicon micro-strips detectors, each one able to measure both x and y position of crossing particles.  Six planes are posed inside the magnet, one on its top, while the remaining planes are placed at the top and bottom of the instrument (the last one above the calorimeter), to maximize the lever-arm in the trajectory determination. As in \ac{pamela}, the tracker measures the energy loss for each plane, therefore providing an estimation of the particle charge and, combining this result with the measured rigidity, of its momentum. An anti-coincidence composed of  plastic scintillators is placed around the magnet, to detect particles incoming from the sides.

The main trigger of \ac{ams} is a \ac{tof} system composed by four planes of paddles of plastic scintillators, two planes just before the magnet and two planes just below, with the paddles of a couple of planes disposed orthogonally, so that they can measure both x and y position of the track. The \ac{tof} measures the velocity of incoming particles, identify downward-going nuclei and, by measuring the deposed energy, identifies particle charge up to $Z=6$.

On the top of the instrument, the Transition Radiation Detector (TRD) discriminates leptons from hadrons by measuring the radiation emitted when particles cross the interface between two media with different indexes of refraction. The devices is composed by 20 planes of plastic  material interleaved by a total of 328 gas proportional wire straw tubes. By measuring the deposed energy, also TRD can make an independent estimation of the particle charge up to $Z=6$.

Below the lower plane of \ac{tof}, the Ring Imaging \v Cerenkov detector (RICH) measures the angle of emission of the \v Cerenkov radiation ($\cos{\theta}\propto 1/v$)  and its intensity ($\frac{dN^2}{dxd\lambda}\propto z^2 f(v)$, where $\lambda$ is the light wavelength, $N$ is the number of emitted photons and $f(v)$ is a known function of particle velocity), providing a measurement of the particle velocity and of the charge up to  the iron region. The RICH is composed by a radiator plane, an expansion volume of $\sim 47~\centi\meter$ surrounded by a conical mirror surface and a plane of 680 multi-pixellized photo-tubes.

The electro-magnetic calorimeter (ECAL) is placed at the bottom of the instrument, so that the magnet screens it from particles originated inside the TRD. It is a sampling calorimeter, composed by 9 planes, each one composed by 11 foils of lead interleaved by scintillating fibers, for a total of 17 radiation lengths. Fibers of successive planes are orthogonal, so that ECAL makes a 3-D reconstruction of the shower, making it possible to discriminate between electro-magnetic and hadronic showers. In addition to its main tasks in the \ac{ams} detector, ECAL is able to self-trigger and can be used as a standalone detector, mainly for photons.

The first result from \ac{ams} measurement of \acp{cr} was the measurement of the \APelectron fraction (\citet{positroni-AMS}) with unprecedented precision; furthermore,  measurements of the spectrum of \Pproton, \HepParticle{He}{}{}, \Pelectron, \APelectron (spectrum and anisotropy) and of the  B/C ratio were presented at the \nth{33} International Cosmic Ray Conference in 2013 (see \citet{ams-tutto}) and later published.

\subsection{The ATIC experiment}\label{atic-sec}
Space-based detectors are not the only way to observe \ac{cr}: above some GeV, balloon-borne experiments represent a valid alternative, because these instruments do not have the constraints and technical requests posed by space operation and because the economical effort needed is significantly smaller. Modern balloon detectors  typically operate at an altitude close to 40~km, during flights that can last more than a months. Usually an instrument is used in more flights, with small servicing modifications to the instrument that can be done  between flights.  

The \acf{atic} (\citet{atic-paper},\cite{atic-sito}) is the most recent balloon-experiment that measured the \ac{cre} spectrum   (\citet{atic-elettroni}): it is a calorimetric experiment designed to study \acp{cr} in the energy range 50~GeV -- 100~TeV. Its weight is $\sim 1500~\kilo\gram$, with a total power consumption of $\sim 250~\watt$. After a first test flight of 17 days in 2000 (ATIC-1), during which scientific data were also collected, \ac{atic} successfully completed two scientific 19 days flight in 2002 (ATIC-2) and 2007 (ATIC-4), while the third flight was aborted shortly after launch because of problem to the balloon.

\begin{figure}[htb!]
\begin{center}
\includegraphics[width=\textwidth]{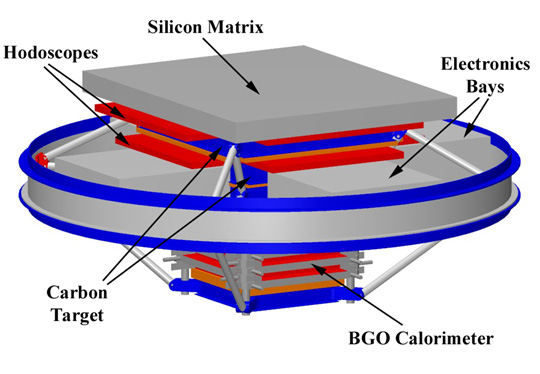}
\caption{Schematic view the \ac{atic} detector.~\cite{atic-sito}}
	\label{atic-fig}

\end{center}
\end{figure}

\ac{atic} (\cref{atic-fig}) main instrument is  a ionization calorimeter that measures particles energy and, reconstructing the axis of the shower, contributes to the measurement of the particle direction: it is composed by several planes (8 in the second flight, 10 in the fourth) of 40 scintillating crystals of Bismuth-Germanate (BGO); the calorimeter, whose thickness  was 17.9 radiation lengths during the second flight and almost 22 radiation lengths and 1.14 nuclear interaction lengths in the fourth flight, is a very sensitive instrument, able to reconstruct the energy of electro-magnetic events with a resolution $\sigma(E)/E$ of the order of 2\% and of hadronic events with a resolution $\sigma(E)/E \sim 35\%$.

Above the calorimeter, a Carbon target is installed to initialize the particle showers that will then develop thought the calorimeter; Carbon was chosen because of its high ratio between nuclear interaction and radiation lengths, to minimize the impact of the target on the electro-magnetic component of showers:  the 30~cm thick target is 1.2 radiation lengths and 0.7 nuclear interaction lengths.

   The trigger to the whole instruments is given by three planes of hodoscopic ladders of plastic scintillators, placed above, inside and below the Carbon target; these devices detect particles in the \ac{fov}, reconstruct the particle charge through its energy deposition and measure the particle direction.  
   
    At the top of the instrument there is a silicon detector matrix, composed by 4480 $2\times 1.5~\squaren\cm$ pixels on two planes, for a total active area of $\sim 1\times1~\squaren\meter$ that precisely measures the particle charge (up to the Iron region), the entry point and discriminates between upward-going and downward-going particles.  
    
    The total acceptance of \ac{atic} ranges from 0.45 to 0.24~\squaren\metre\sr\ depending on the request that particle trajectory crosses the bottom plane of the calorimeter or only the top plane.
    
    Together with the already cited measurement of the \ac{cre} spectrum, \ac{atic} has measured the spectrum of \Pproton and HE (\citet{protoni-atic}) and that of heavier nuclei (\citet{atic-nuclei}).
\subsection{Results and their consequences}
\label{elettroni-tuttemisure}
\begin{figure}[htb!]
\begin{center}
\includegraphics[width=1.03\textwidth]{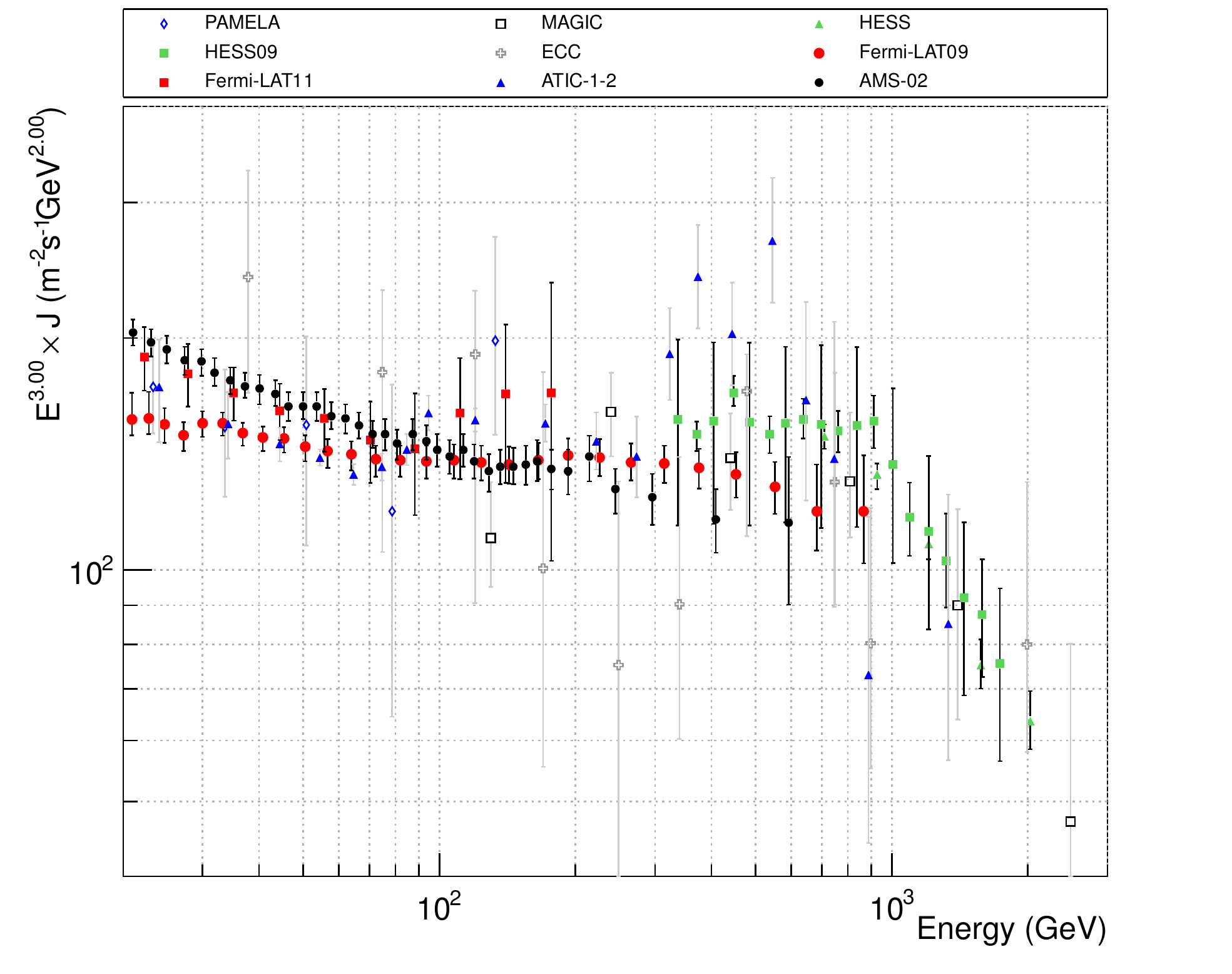}
\caption{CRE spectrum multiplied by $E^3$ measured by recent experiments. \ac{lat}, red filled circles~\cite{electronpaper} and red filled squares~\cite{fermi-positroni}, ATIC 1-2, blue filled triangles~\cite{atic-elettroni},  ECC, grey crosses~\cite{elettroni-ecc},    AMS-02, black filled circles~\cite{elettroni-ams2,ams2-elettroni-positroni}, PAMELA, blue empty diamonds~\cite{elettroni-pamela,pamela-positroni}, MAGIC, black void squares~\cite{elettroni-magic} and HESS, green filled triangles for the HE analysis and green filled squares for the LE analysis~\cite{elettroni-hess, elettroniHess}. Numerical values of ECC and MAGIC measurements were extracted directly from the published graphics, values for AMS-02 were extracted directly from the preliminary graphic and the error  was assumed as the dimension of the points (therefore resulting in a bit larger values than published ones), values for PAMELA are an extrapolation from the published values of the \Pelectron spectrum and of the \APelectron fraction, values for Fermi-11 are the sum of the \Pelectron and \APelectron fluxes measured in~\cite{fermi-positroni}.  All published results were extracted from the data-base presented in~\citet{gcdb}, that was recently replaced by the   database presented in~\citet{new-gcdb}, that allows interactive selection and plot of the collected data. }
	\label{elettroni-tutteMisure}

\end{center}
\end{figure}

During last years, large amount of new measurements of the \Pelectron+~\APelectron spectrum has been published. They are summarized in figure~\ref{elettroni-tutteMisure}, where the \ac{lat} results are shown as red circles. These measurements were made by  different experiments, exploiting different experimental techniques:  balloon-based like ATIC~\cite{atic-elettroni,atic-elettroni-2011}, CAPRICE~\cite{elettroni-caprice}, ECC~\cite{elettroni-ecc}, HEAT~\cite{elettroni-heat}, BETS~\cite{elettroni-bets}  and PPB-BETS~\cite{elettroni-PPB-BETS}, space-based like \fermi-LAT~\cite{electronpaper,fermi-positroni}, AMS-01~\cite{ams1},  AMS-02~\cite{elettroni-ams2,ams2-elettroni-positroni,elettroni-ams-articolo} and PAMELA~\cite{elettroni-pamela,pamela-positroni} and ground gamma telescopes like MAGIC~\cite{elettroni-magic,nuovi-elettroni-magic} and HESS~\cite{elettroni-hess, elettroniHess}.  

 \ac{lat} measurements  cover a large and very important energy range of the \ac{cre} energy spectrum with a precision never reached by previous experiments that,  with the exception of HESS and of  AMS-02, were all dominated by low statistics.

All results  show that the spectrum below 100~\GeV\ is well represented a power-law with a spectral index close to -3; at higher energy  the spectral behavior is not clear: ECC, PPB-BETS and ATIC  measured a bump between 300 and 600~\GeV, that was confirmed by  in-depth analysis of ATIC-4 data (~\citet{atic-elettroni-2011}, the total number of electron events collected by \ac{atic} is 700 above 200~GeV and 70 above 600~GeV), which also raised the possibility of the presence of spectral features, undetected by previous experiments because of lower energy resolution. These measurements are in contrast with the \ac{lat} result, that did not detect any spectral bump. \ac{lat} result,  whose high statistics makes it very significant, was later confirmed by the low energy extension of HESS measurement (\citet{elettroniHess}). Measurement from AMS-02~\cite{elettroni-ams2,ams2-elettroni-positroni,elettroni-ams-articolo} seems also not to have any spectral feature. MAGIC measurement seems to show some spectral feature above 300~\GeV, but the energy resolution is too poor for claiming any significant detection. PAMELA measurement does not reach sufficiently high energy (and with sufficient statistics) to add some information to the issue.

Above 1~TeV, ATIC and HESS show  a sharp cut-off of the spectrum, but for both experiments  errors at high energy become very high, therefore no further study was possible.

\section{Status of Cosmic-ray electrons study: data and current theories}\label{status-cre-sec}

\begin{figure}[htb!]
\begin{center}
\includegraphics[width=\textwidth]{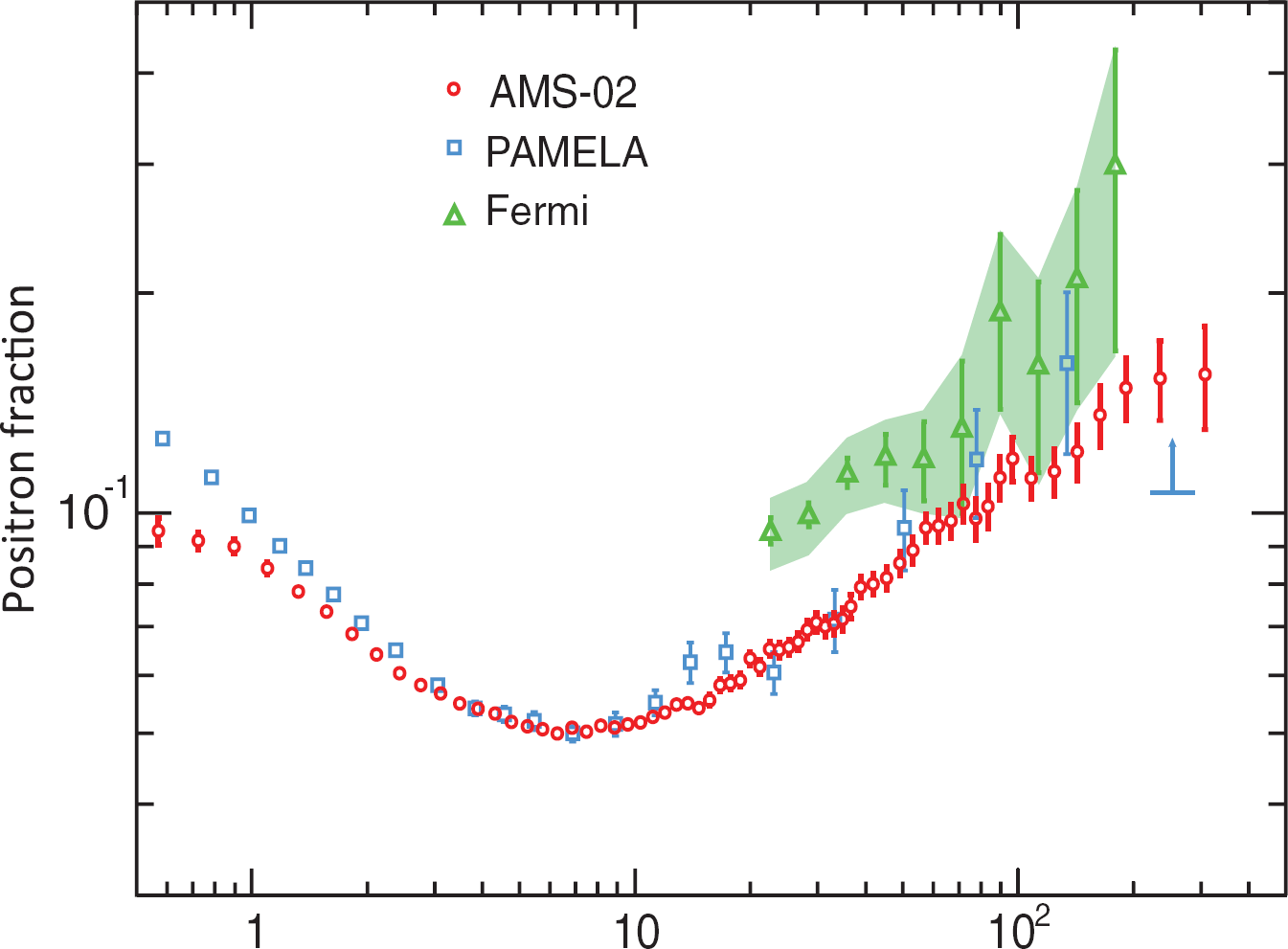}
\caption{The positron fraction ($\APelectron/\Pelectron+\APelectron$) measured by \fermi~\cite{fermi-positroni} (green triangles), PAMELA~\cite{pamela-positroni} (blue squares) and AMS-02~\cite{positroni-AMS}. }
	\label{frac-positr}

\end{center}
\end{figure}

Together with the measurements of the \ac{cre} spectrum  cited in \S~\ref{elettroni-tuttemisure},  the knowledge of the \ac{cre} composition had significant improvements in the past few years. The positron fraction of \ac{cre} defined as $\frac{\APelectron}{\Pelectron+\APelectron}$, was measured by PAMELA~\cite{pamela-positroni,pamela-eccesso}, \fermi-LAT~\cite{fermi-positroni} and recently by AMS-02~\cite{positroni-AMS,positroni-ams-articolo}. These measurements, collected in figure~\ref{frac-positr},  show an increase in the positron fraction above 10~\GeV, and a possible stabilization of the fraction above 200~\GeV. When relating the measurement of the \APelectron fraction  with separate measurements of the \APelectron spectrum (PAMELA~\cite{pamela-positroni} and AMS~\cite{ams2-elettroni-positroni,elettroni-ams-articolo}) and of the \Pelectron spectrum (PAMELA~\cite{elettroni-pamela} and AMS~\cite{ams2-elettroni-positroni,elettroni-ams-articolo}),  an increase at high energy of the \APelectron flux is clear. This is in strong disagreement with standard  theories that assume \APelectron produced only by interaction of primary \acp{cr} with the \ac{ism}. In these models, the secondary production of \APelectron is calculated using the primary flux of \ac{cr} (principally \Pproton and He, measured directly, for example by PAMELA~\cite{pamela-p-and-he},  AMS-02~\cite{ams2-protoni,protoni-ams-articolo,elio-ams-articolo}, CREAM~\cite{protoni-cream,nuclei-cream}, \ac{atic}~\cite{protoni-atic} and indirectly, through  \Pphoton -rays emitted by \acp{cr} interacting with  Earth atmosphere, by \ac{lat}~\cite{protoni-lat}) and  the fractional flux of other secondaries, like B (PAMELA~\cite{pamela-b/c,pamela-b-c}, AMS~\cite{ams-tutto,ams-b/c}) and \APproton (PAMELA~\cite{pamela-antiP}). 

If these fluxes are used as inputs for the solutions of the diffusion equation~\eqref{diffeqCR}, or if it is imposed that they are well reproduced, while assuming a production of \APelectron only in \ac{ism}, the result is a decreasing \APelectron fraction (see for example~\citet{elettroniprefermi}). This is true  either for numerical solutions of the diffusion equation (that use codes like GALPROP~\cite{galprop,galprop2,galprop3} or DRAGON~\cite{dragon,dragon2})  and for semi-analytical solutions (\cite{giuseppe,semi-analitica-soluzione}). Clearly, this is not compatible with the measured increase of the \APelectron fraction. Therefore, even if some publications like~\citet{ams-pro-secondary-positron} claims the compatibility of an increasing \APelectron fraction with the only-secondary scenario,   the existence of sources of primary \APelectron  is widely accepted.  

Studies about the nature of this primary  component of \Pelectron and \APelectron  did not show any evidence of a  spectral break or of a charge asymmetry between positive and negative particles (\citet{model-independent,Pamela-3D-modello,positroni-da-pwn}); however, a  dominance of \APelectron  (\citet{pro-local-snr}) or of \Pelectron  (\citet{Pamela-3D-modello}) can not be excluded. The flux of primary \APelectron should become dominant above 10~\GeV, while at smaller energy the majority of \APelectron should be originated in \ac{ism} (\citet{ams2-elettroni-interpretazione,dragon2,modello-con-radio}).

Three main hypotheses are at now accepted about the origin of the \APelectron primary component. These hypotheses,  all formulated shortly after the identification of the \APelectron excess (\citet{pamela-eccesso}), are: the production and acceleration of secondary particles at \acp{snr} (\citet{Blasi}), the acceleration of \APelectron \Pelectron pairs at \acp{pwn} (\citet{elettroni-pulsar}) and the production of particles in \ac{dm} annihilation or decay (\citet{dm-bergstrom,dm-strumia}). The sources and the physical  mechanisms that  produce  high-energy \APelectron are described in \S~\ref{SNR}, \S~\ref{section-pulsar}, \S~\ref{dm-section}.

A fourth hypothesis, proposed in \citet{elettroni-microquasar}, suggests that the production of \APelectron in micro-quasars (see \S~\ref{microquasar}) could significantly contribute to the primary \APelectron flux. However, this hypothesis is rather new, and the literature about it goes not beyond the demonstration of the plausibility of the explanation.

All the three most accepted models are able to reproduce the spectrum of \acp{cre} coherently with the measured flux  of primary and secondary \acp{cr}, although some evidences have emerged in favor of an astrophysical source rather than of an exotic (\ac{dm}) one. The constraint posed by the \APproton fluxes forces to consider only models for which the annihilation or decay of \ac{dm} particles produces principally leptons and small amounts of hadrons (\citet{serpico}). Furthermore, as already stated in \S~\ref{dm-section}, the \ac{dm} origin of \APelectron should lead to a significant decrease of the spectrum above the mass of the neutralino, feature  that is not detected; this rules out direct production of pairs and forces to consider only \ac{dm} particles with mass $\geq$1~TeV that produce  an intermediate  state that later decays in complex states, like 2\Ppiplus 2\Ppiminus, \APelectron \APmuon 2\Pmuon, 2\APelectron 2\Pelectron, 2\Pmuon 2\APmuon, 2\Ppiplus 2\Ppiminus, as exposed in~\citet{dm-pulsar}. \citet{dm-pulsar} also show that even these \ac{dm} scenarios require a Galactic diffusion halo $>4~\kilo\pc$   to reproduce the \ac{cre} fluxes (at now, the halo estimation is 3 --7~\kilo\pc); this could be checked by more precise measurements of the $^{10}$Be/$^{9}$Be ratio: as already stated at page~\pageref{berillio}, the fraction of this radioactive isotope produced via spallation processes during diffusion can furnish a direct measurement of the diffusion time of \acp{cr} and, combined with measurement of the traversed amount of matter calculated observing secondary abundances, the thickness of the diffusion halo. 

Finally, all \ac{dm} scenarios  require an annihilation cross section or decay time completely incompatible with current estimations, and the constraint  $M_\chi \geq$1~TeV worsen the problem, because the expected spectrum scales as $m_\chi^{-2}$: the introduction of new physic to boost the cross section or the assumption that Earth is near or inside a clump of \ac{dm} is therefore obliged. All the \ac{dm} scenarios require that the \ac{cre} flux has no detectable anisotropy: \citet{dm-anisotropia} demonstrates that such an anisotropy ($\sim 10^{-2}$)  would make gamma production of the \ac{dm} clump at least one order of magnitude higher than \ac{lat} sensitivity.

Under different assumption about the sources distribution and particle diffusion, \ac{snr} and  \ac{pwn} scenarios are both able to explain the existence of a source of primary \APelectron .  The measured fluxes of \Pelectron and \APelectron have been correctly reproduced, consistently with primary and secondary \acp{cr} measured fluxes, when \acp{snr} are described using the known sources from TeVCat~\cite{tevcat} (whose \ac{cre} population  is measured via radio observations of synchrotron emission) plus a far diffuse population (\citet{ams2-elettroni-interpretazione}) or just assumed to be uniformly distributed on galactic arms (\citet{dragon2}); \ac{pwn} were described using position of known pulsar from the ATNF pulsar catalog~\cite{atnf,atnf2} (\citet{ams2-elettroni-interpretazione}), or as the sum of contribution from a few close and powerful sources of known spectrum and emission (the most important being Geminga) and a component distributed on galactic arms (\citet{dragon2,Pamela-3D-modello}).

All models show that, above some tens of \GeV, \APelectron flux is dominated by far sources up to~$\sim 100~\GeV$, where closer sources become dominant. 

\citet{ams2-elettroni-interpretazione}  showed that at high energy the number of dominant \acp{snr} could be very small, down to one in the extreme hypothesis proposed by \citet{pro-local-snr}. Recently, \citet{positroni-B/C}  argued that B/C measurement from AMS-02~\cite{ams-tutto,ams-b/c} and PAMELA~\cite{pamela-b/c} are not compatible with significant  acceleration of secondaries in \acp{snr}, unless the flux on Earth is dominated by a close \ac{snr} with a significantly low ion density; however, this work uses an unrealistic uniform distribution of \acp{snr} in the whole Galactic plane and therefore its conclusion could be a bit less stringent: calculation from~\citet{ams2-snr} using different diffusion models and a self-consistent fitting of the B/C fraction resulted in different conclusions. 

\citet{ams2-elettroni-interpretazione, dragon2, elettroni-interpr} agree that the number of dominating \acp{pwn} should be larger: the Geminga \ac{pwn}, which flux at Earth is probably the largest, should transfer~$\sim 30\%$ of its spin down energy to \acp{cre} to support the hypothesis of just one significant source ~\cite{ams2-elettroni-interpretazione}; \citet{Pamela-3D-modello} calculates an even larger  efficiency necessary to make near pulsars dominating.

Current measurements of \acp{cr} fluxes do not allow to discriminate between these two models. However, it is possible to make  some predictions  that should be verified using future measurement of \ac{cr} and \ac{cre}, either more precise or with a larger energy range.

The major differences between the two models, that could lead to observable differences, arise from the different composition of accelerated particles and from the difference in the acceleration process.

\acp{pwn} accelerate \Pelectron\APelectron pairs, therefore the \APelectron fraction should at high energy reach a maximum, around 50\%, when the \APelectron flux is completely dominated by primary particles (\citet{pro-local-snr,Blasi}). A similar feature could be produced also by \acp{snr}, but the non-observation of such a phenomenon would generally disfavor the \acp{pwn} hypothesis, the exception being an \Pelectron excess generated  by a near source of \Pelectron, as proposed in~\cite{Pamela-3D-modello}. 

\acp{pwn} produce only \Pelectron\APelectron, while in \acp{snr} also other secondary species are in principle accelerated, therefore leading to an increase in the B/C and \APproton/\Pproton ratio; \citet{antip-da-snr2,antip-da-snr,ams-secondari-vicini} predict that the  B/C and \APproton/\Pproton ratios should be almost flat above 100~\GeV/n and up to a few TeV. However, the lack of such an increase at high energy of these ratios would not  lead to  the conclusion  that \acp{snr} are not the source of the \APelectron excess: \citet{positroni-B/C} sustain that a near \ac{snr} with a low ion density would not produce an observable increase in the B/C ratio; furthermore, \citet{serpico,positroni-decad-snr} claim that the acceleration of \APelectron produced in $\beta^+$ decays could justify the \acp{snr} model even in absence of an increase of the \APproton flux.  On the other hand, the detection of an increase in the secondary fluxes at high energy would disfavor the \acp{pwn} model.

It is not easy to relate the spectral index of the \APelectron component to their source, therefore it is difficult to use it as a discriminating quantity. In the \acp{pwn} scenario the correlation between the pulsar observed \Pphoton index and the \APelectron index is uncertain, and no strict prediction can be made.
In the \acp{snr} scenario different calculations lead to significantly different \APelectron: \citet{pro-local-snr}, assuming that particles accelerated in \acp{snr} should have an index similar to primary protons (and harder than particles accelerated in \acp{pwn}), calculate a difference between the \Pelectron and the \APelectron spectra of the order of the diffusion coefficient index (0.4 -- 0.7, that  leads to an \APelectron index 2.3 -- 2.6); \citet{ams2-snr,ams-secondari-vicini} sustain that the source spectral index of \APelectron should be harder than that of primary particles produced in \acp{snr} by one power in momentum,  because in the assumption of Bohm diffusion the diffusion coefficient $D(p) \propto p$, therefore  the probability of diffusing up to the shock and  being subjected to further cycles of acceleration increases with energy.

The maximum energy reached by \APelectron, beyond that a cut-off is expected, could differ in the two scenarios: \APelectron accelerated in \acp{snr} will have the same cut-off of  primary \acp{cre}, that from HESS measurement~\cite{elettroni-hess} seems to be around 2~\TeV\ (a confirm of this value by \fermi\ would be very important), while the cut-off of particles accelerated by \acp{pwn} could extend to the range 1 -- 10~\TeV\ (but a value similar to that of \acp{snr} is expected, see~\cite{ams2-elettroni-interpretazione,elettroni-hess-vela,gamma-crab}). Therefore, a value of the \APelectron cut-off greater than several TeV will strongly favor the \ac{pwn} model, while a lower value would not be decisive. Finally, a value between 1~\TeV\ and 600~\GeV\ (the current limit of the AMS measurement) would be incompatible with both models.

\chapter{Pass 8 new data analysis process}

\label{pass8}
In 2015, the Fermi-LAT collaboration released a new event-level analysis, introducing large improvements in the reconstruction, simulation  and identification of the events: Pass 8.  The main improvements on the performance are an estimated increase of $\sim25\%$ in the effective area, a great improvement in  the spatial resolution radius above a few \GeV\ and an extension of the actual energy range up to 3 \TeV\ and down to less than 30~MeV, with reasonable high-energy resolution (for on-axis events, below 15\% up to 1 \TeV\ and below 30\% in the range 1-3 \TeV).

Pass 8 benefits of the experience accumulated in the first 6 years of the mission, accounting for all the effects that were not initially simulated but that subsequently emerged from on-orbit data. Furthermore, Pass~8 explores new analysis techniques to fully exploit the technical characteristics of the LAT, fully reaching its scientific potential.

In this chapter, I will give a review of the changes introduced with Pass~8 and finally summarize the improvements to the \ac{lat} performances in \S~\ref{p7-p8} .

\section{New tracker pattern recognition}\label{tree-section}
The Kalman track finding-and-fitting, described in \S~\ref{track-rec}, is heavily dependent on the \ac{cal} information used to seed it: a poor seed significantly reduces the power of the method. Furthermore, even if the Kalman filter is designed to keep into account the multiple scattering during the track search, it can be mislead by a single large scatter: in this case, the filter is sometimes unable to identify subsequent hits, and therefore terminates the track, or identifies the hits after the scatter as a new track. Also, the Kalman filter selects the best track(s) from an event, but the  information about the shower development, as for example the deposed energy in the \ac{tkr},  are not fully extracted. 

For these reasons, the track reconstruction algorithm has been completely rewritten using a different approach, briefly described in~\citet{tree-poster}, collecting all the hits belonging to a  shower in a single data structure, named \textit{Tree}, that is created independently from eventual information from \ac{cal}, and inside which tracks and vertex are afterwards  researched. 

\begin{figure}[htb!]
\begin{center}
\subfigure[The construction of links between 3D hits in the \ac{tkr}.]{\includegraphics[width=.6\textwidth]{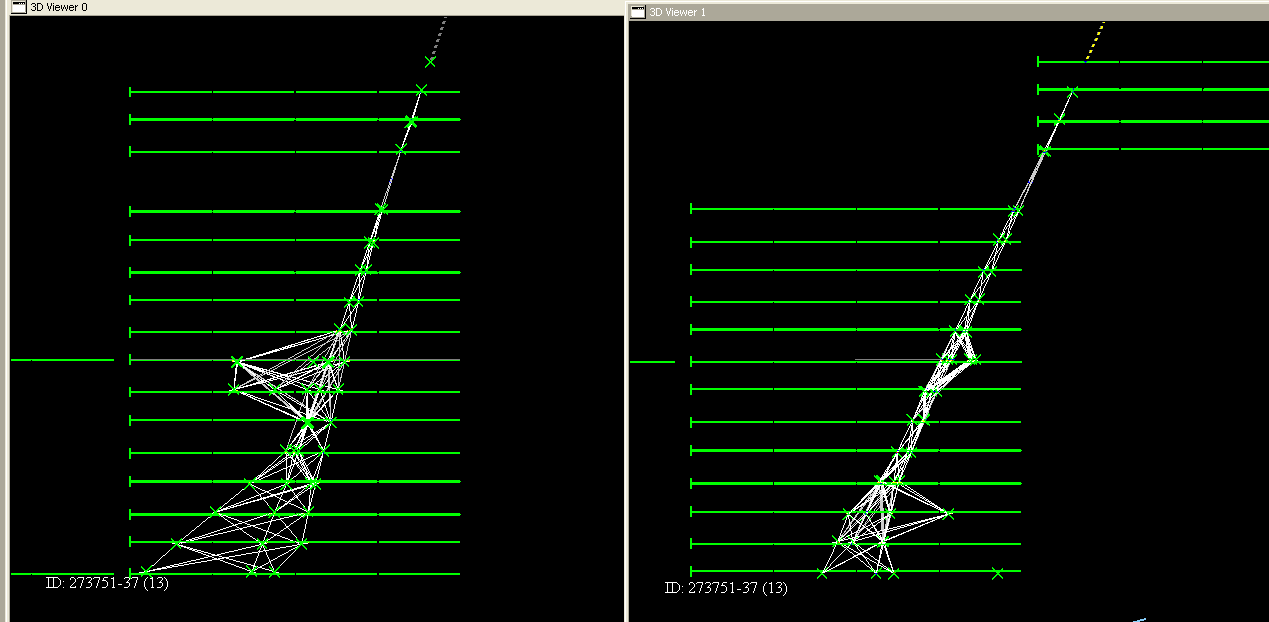}\label{tree1}}
\subfigure[Tree is constructed attaching some of the links that share a 3D hit.]{\includegraphics[width=.6\textwidth]{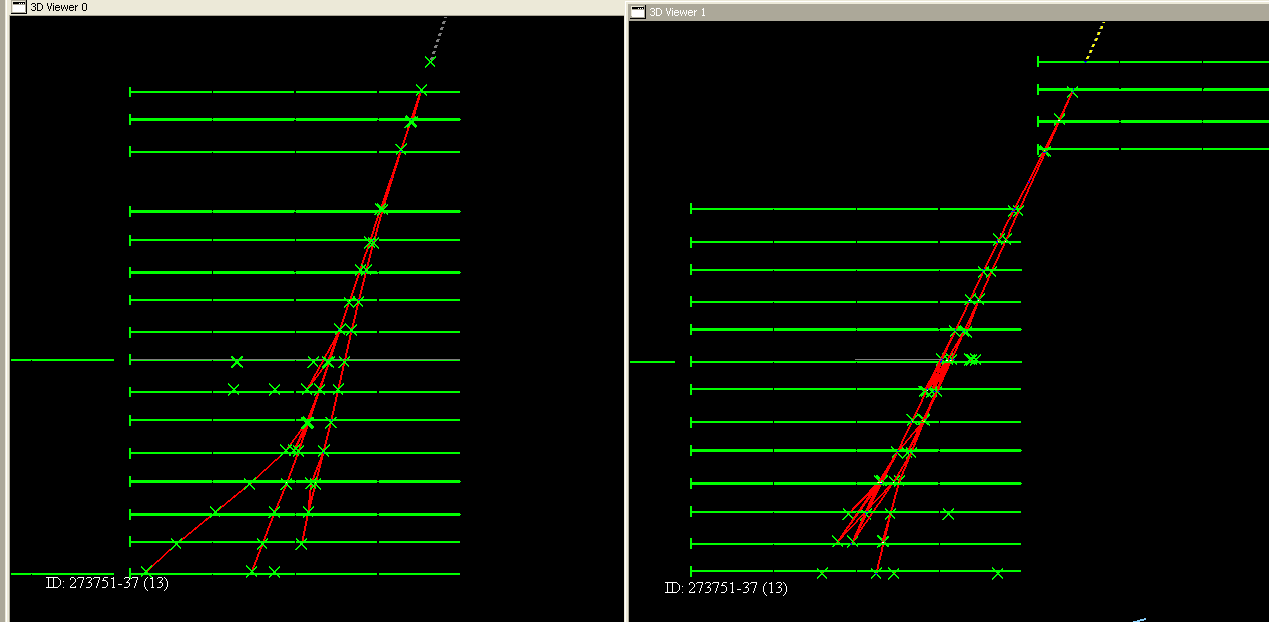}\label{tree2}}
\subfigure[Two tracks are identified inside the Tree.]{\includegraphics[width=.6\textwidth]{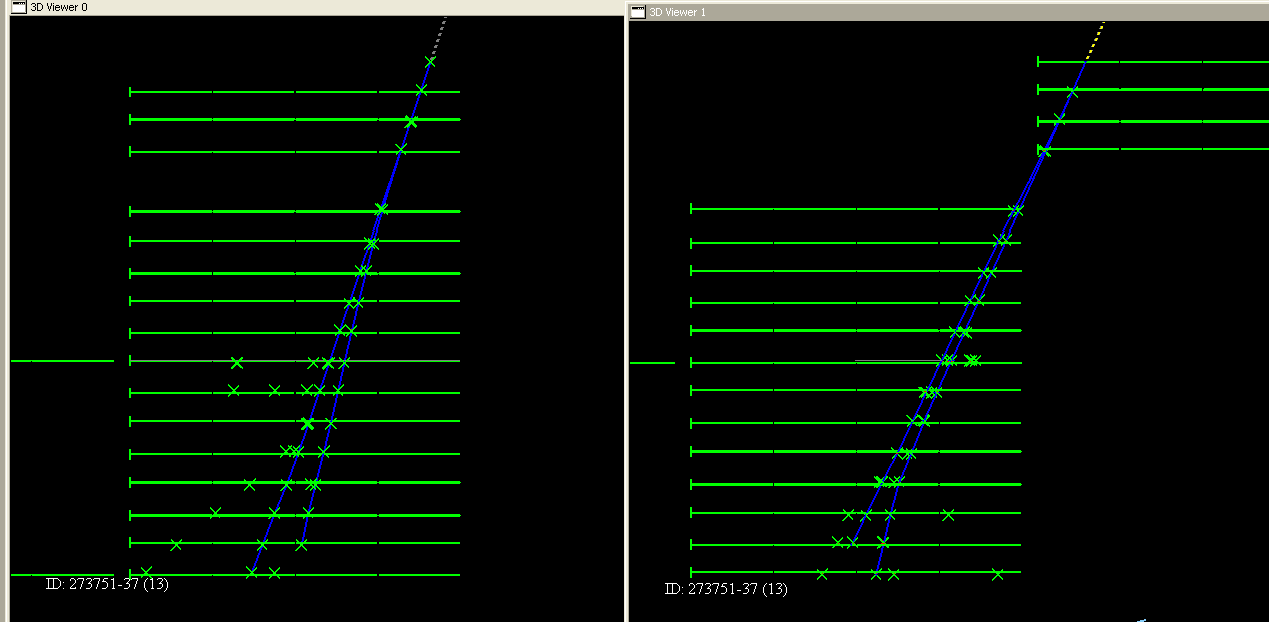}\label{tree3}}

\caption{The tree-based pattern recognition in \ac{tkr}, simulated $\gamma$ event.}

\end{center}
\end{figure}

 The first step of this procedure  is to associate the clusters of strips in adjacent layers to form 3D points. Then (figure~\ref{tree1}), links are created between 3D points separated by less than 3 planes, in two subsequent steps. In the first step, links are formed between 3D points in adjacent layers; to reduce the required computational power, that increases with the number of hits and  is typically larger than in the previous procedure,  if the \ac{cal} moment analysis returns an axis with a sufficient quality, links that are inconsistent with the axis are rejected. At the end of the first step, the Hough filter (\citet{hough-filter1,hough-filter2}) is run on created links, providing a rough but robust estimation of the event axis, that will be used in the next step. In  the second step, links are formed between 3D points separated by up to 3 layers, only if:
\begin{itemize}
\item there is a known reason (inter tower gap, dead strips) to have a gap;
\item links tightly match the axis produced by the Hough filter;
\item there are no close hits in the skipped layer, that could justify the creation of shorter links.
\end{itemize}

Then, the Tree structures are extracted from the collection of links (figure~\ref{tree2}), attaching links that share a 3D point; clearly, not all the links are added to a Tree: a selection is made based on the normalized distance between shared 3D points for given layer (therefore, on the length of links) and on their  consistency, comparing the width of involved clusters and searching for the distance between the links projection.

At the end of these processes, the Trees describe the different showers developed in the \ac{tkr}; Trees are ordered by their quality, with the best Tree being the biggest, longest, straightest one; starting point of Trees are assumed as the (eventual) conversion point, and moments of the Trees are calculated. Tracks are then searched inside the first two Trees (figure~\ref{tree3}), with a maximum of two tracks for each Tree, using two independent methods: selecting the longest and straightest branches of the Tree, and running the Kalman filter seeding it with the tree axis. Both methods return a second track only if it has more than 3 hits and the number of hits common with the first track is small. The Kalman filter is also enabled to search above the Tree for hits not belonging to it, and to attach them to the track. The  produced tracks are then ordered according to their $\chi^2$, and the best two (if present), are associated to the Tree.  If two tracks are found in a  Tree, a vertex between them is also searched.

\begin{figure}[htb!]
\begin{center}
\includegraphics[width=.65\textwidth]{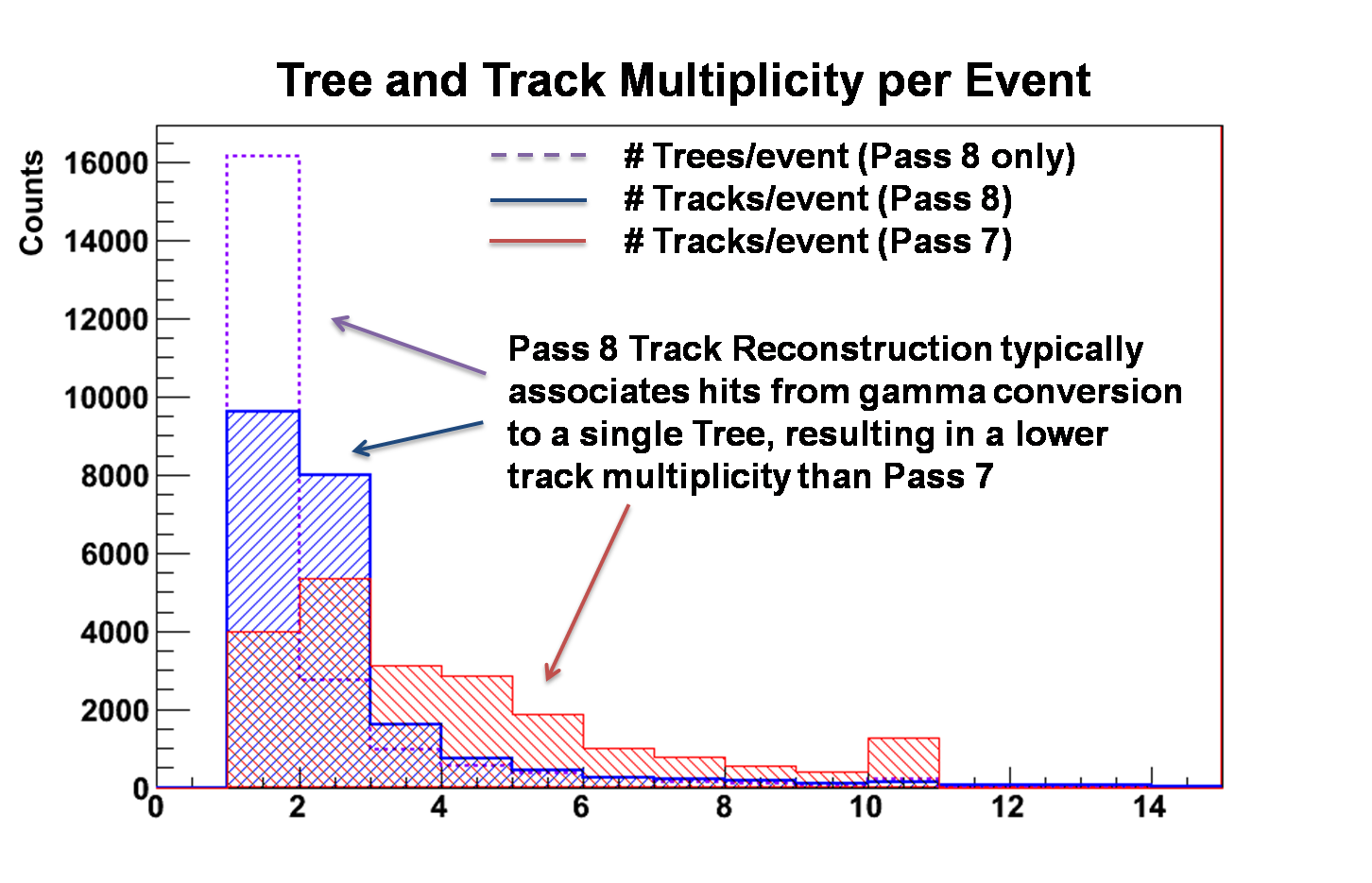}\label{tree-kalman1}
\caption{Number of track and trees reconstructed per event in Pass~7 and in Pass~8. Simulated $\gamma$ with an 1/E energy spectrum from 18~\MeV\ to 562~\GeV, processed either with Pass~7 and Pass~8.~\cite{tree-poster}}
\label{tree-kalman}

\end{center}
\end{figure}

\begin{figure}[htb!]
\begin{center}
\subfigure[Distribution in energy and azimuth angle of events for which a track is reconstructed. Pass~8 is in blue, Pass~7 in red.]{\includegraphics[width=.6\textwidth]{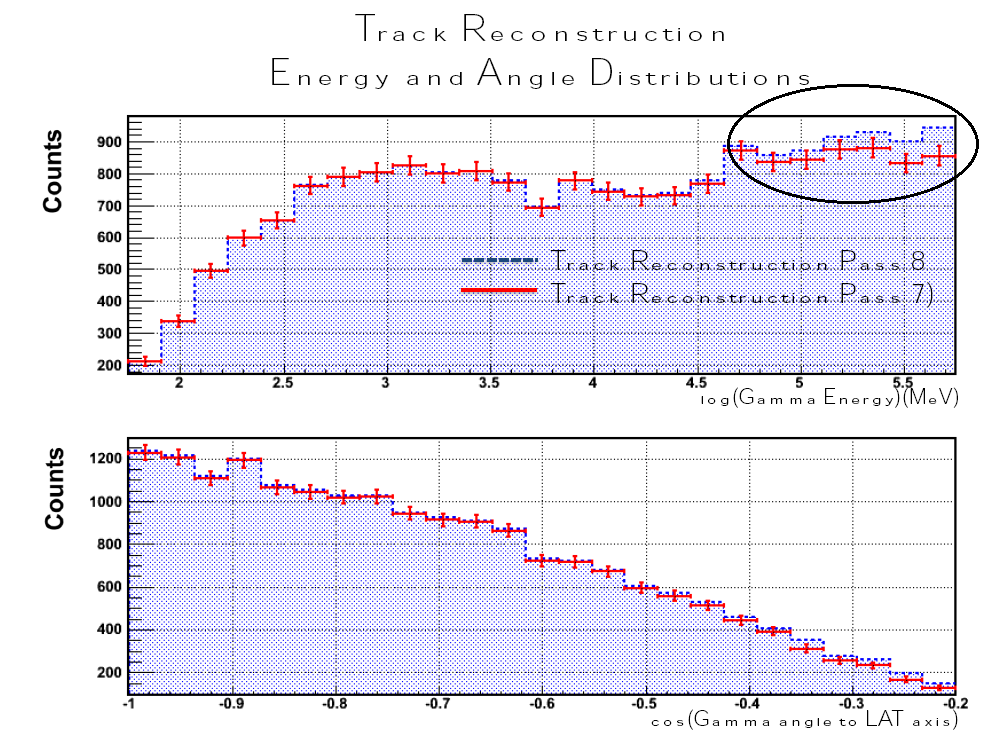}\label{tree-kalman2}}
\subfigure[The 68\% containment radius  for events converting in the two \ac{tkr} sections in Pass~7 and Pass~8.]{\includegraphics[width=.6\textwidth]{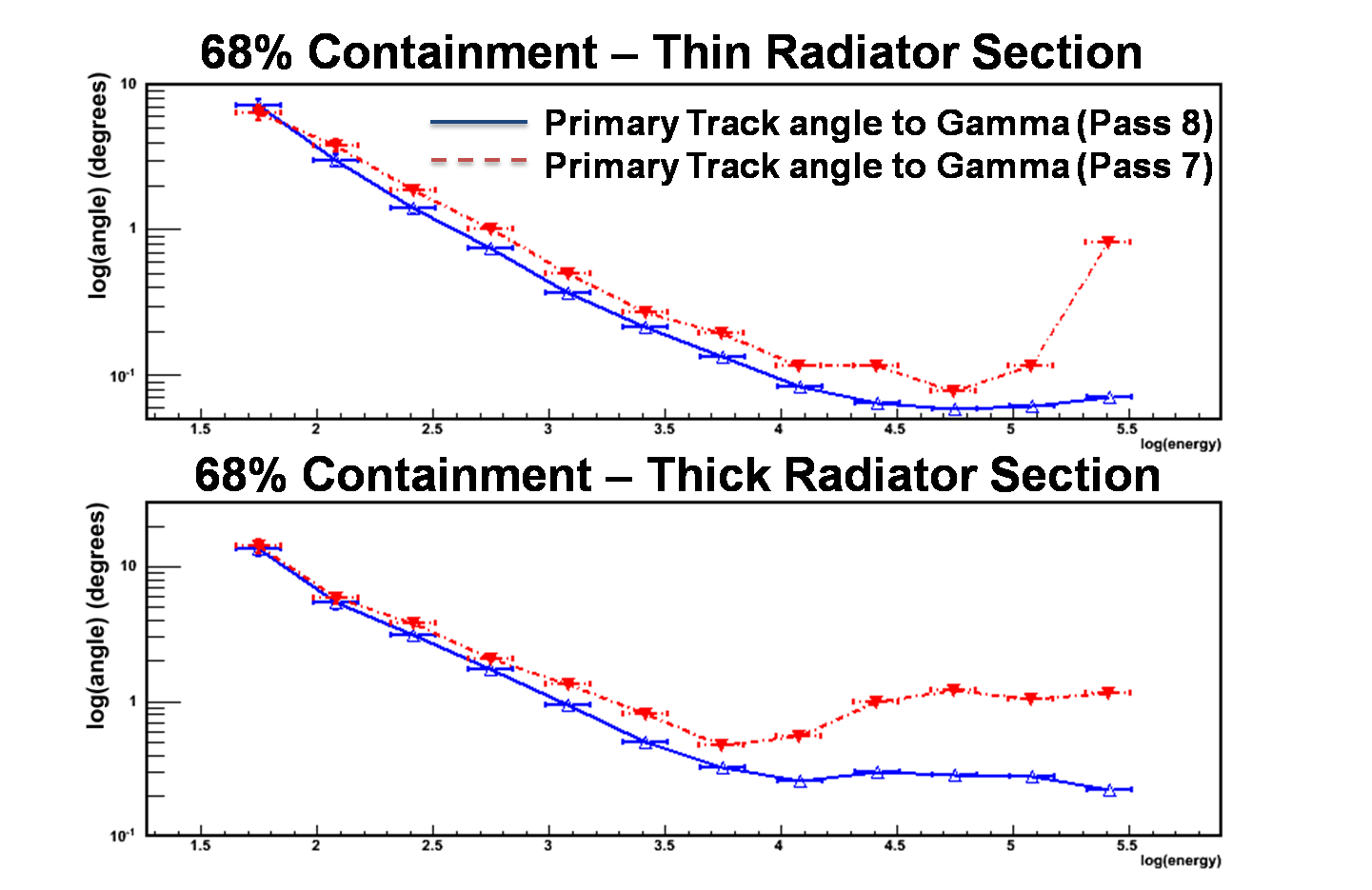}\label{tree-kalman3}}

\caption{Comparison of the tree-based track reconstruction with the Kalman filter used in Pass~7. Simulated $\gamma$ with an 1/E energy spectrum from 18~\MeV\ to 562~\GeV, processed either with Pass~7 and Pass~8.~\cite{tree-poster}}
\label{tree-kalman-bis}

\end{center}
\end{figure}

The comparison of the Tree-based track reconstruction with the previous method is shown in~\cref{tree-kalman,tree-kalman-bis}, where the two methods are compared on a sample of simulated $\gamma$ with an 1/E energy spectrum from 18~\MeV\ to 562~\GeV. It can be seen that the Tree approach, that is designed to associate all the hits from a $\gamma$ conversion into a single data-structure reducing the possibility of splitting it, significantly reduces the number of reconstructed tracks (figure~\ref{tree-kalman1}). Furthermore, the possibility of grouping the hits belonging to the $\gamma$ conversion in a unique structure, using the information contained in it to extract the tracks,  significantly improves the reconstruction capability at high energy (figure~\ref{tree-kalman2} and (figure~\ref{tree-kalman3}). 


\section{Calorimeter clustering}\label{cluster-sec}
As said at page~\pageref{overlay}, the superimposition of out-of-time  background events to gamma-triggered events can significantly affect the event reconstruction, typically resulting in the rejection of the event. In particular this can happen because, even if the \ac{tkr}  correctly reconstructs the \Pphoton event, the \ac{cal} analysis is confused by the presence of two or more deposits of energy: therefore, the energy reconstruction will be in-correct, while the \ac{cal} axis will be largely in disagreement with the reconstructed tracks. 

This problem has been addressed in Pass~8 developing a cluster analysis, which is able to separate different clusters in the \ac{cal}, allowing to associate them to different trees in the \ac{tkr} and therefore to isolate and study the photon event.

\begin{figure}[htb!]
\begin{center}
\subfigure[An \ac{mst} is constructed grouping all the hits in \ac{cal}.]{\includegraphics[width=.6\textwidth]{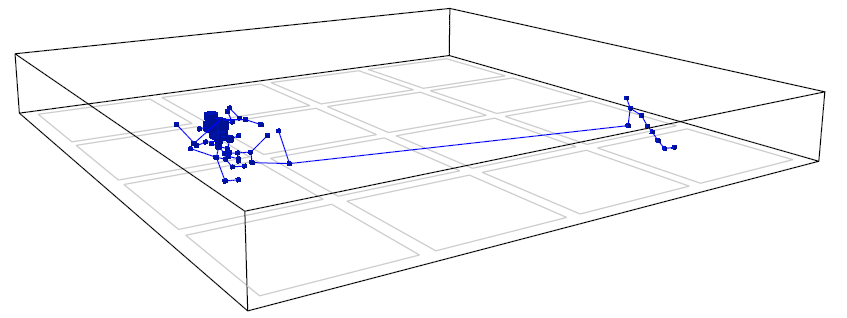}\label{mst1}}
\subfigure[Links exceeding threshold are removed, splitting the tree in clusters.]{\includegraphics[width=.6\textwidth]{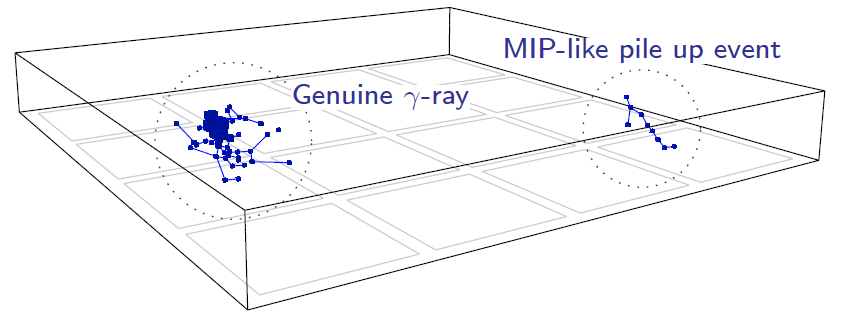}\label{mst2}}

\caption{The \ac{mst} clustering algorithm applied to a simulated \ac{cal} event: crystals on the right are identified as a separate cluster.}

\end{center}
\end{figure}

The method, described in~\citet{cluster-poster}, is based on the construction of a \ac{mst}  on all the \ac{cal} hits. A \ac{mst} is the shortest tree that connects all the given points without loops; after the tree construction (figure~\ref{mst1}), all the links with an euclidean length greater than an energy dependent threshold are removed, possibly splitting the original tree in different sub-trees, each one identified as a cluster (figure~\ref{mst2}).

The identified clusters are then analyzed with a Naive Bayes Classifier~(\citet{nbc}), a simple but really effective classifier based on Bayes theorem, to find the probability that the cluster is gamma-like, \ac{mip}-like, hadron-like or ghost-like (residual of an out-of-time event).

\begin{figure}[htb!]
\begin{center}
\subfigure[Pass~7 reconstruction, the \ac{cal} moment analysis is completely mislead by the overlay.]{\includegraphics[width=.7\textwidth]{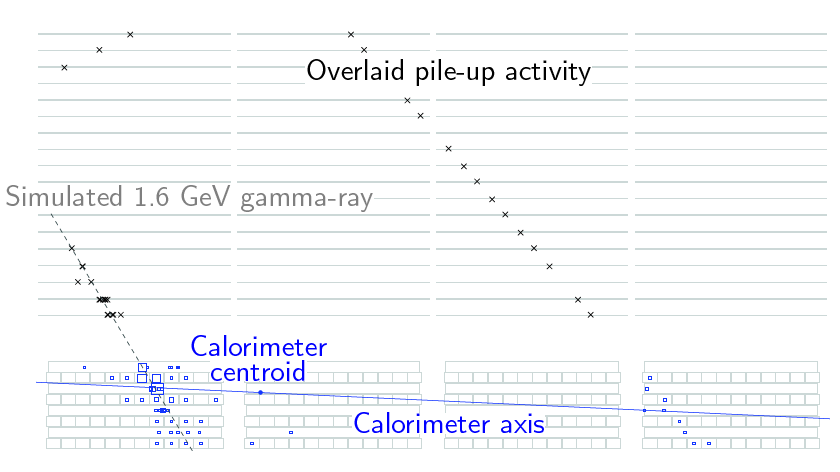}}
\subfigure[Pass~8 reconstruction, the clusters are separated and correctly identified.]{\includegraphics[width=.7\textwidth]{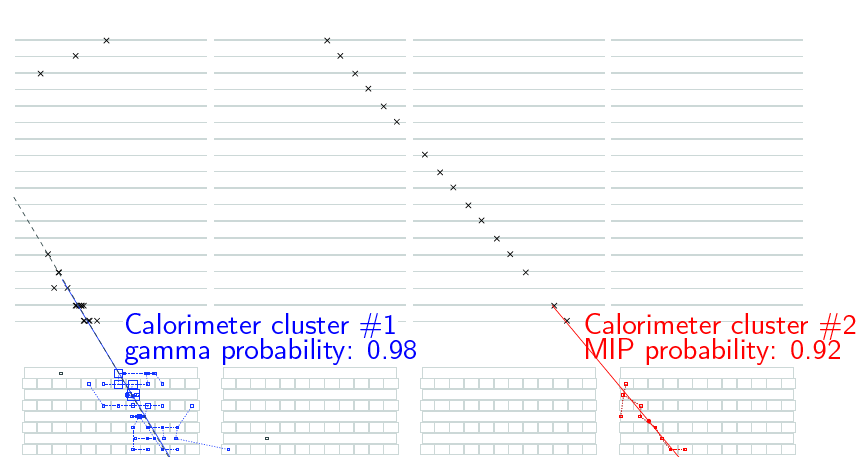}}

\caption{Difference between Pass~7 and Pass~8 reconstruction for a simulated 1.6~GeV photon with significant overlay:  the clustering algorithm correctly resolves the different \ac{cal} clusters, and the photon axis is correctly reconstructed. ~\cite{calorimeter-pass8-poster}}
\label{cluster-ED}

\end{center}
\end{figure}

The effect of this algorithm is shown in figure~\ref{cluster-ED}, taken from~\citet{calorimeter-pass8-poster}, where is it possible to see that events where the \ac{cal} reconstruction was completely mislead by the presence of an overlay are recovered, with the gamma event correctly identified.

In the end, the clustering algorithm makes it possible to associate different clusters with related hit tiles in \ac{acd}.

\section{Identification of overlays}
One of the purposes of Pass~8 was to correctly address  the problem of overlays, that was not forecast before the launch of \fermi\ and typically lead to the rejection of affected gamma events in Pass~6 and Pass~7. 
To do so, methods were developed to identify hits caused by overlays in all the \ac{lat} subsystem. These methods can be used either to remove ghost hits\footnote{Residual of an out-of-time event}  before performing the event reconstruction and to identify ghost events after the reconstruction process.

In the \ac{cal}, the clustering algorithm described in \S~\ref{cluster-sec} is of fundamental importance, because it makes it possible to separate the gamma shower from the hits caused by overlay; the identification of  clusters as ghost is largely based on the comparison of the energy deposed in crystals with the trigger signals: as described in \S~\ref{cal}, the trigger signal is generated by a shaping amplifier with 0.5~\mycro\second\ peaking time, while the read-out of the deposed energy is made by a a shaping amplifier with 3.5~\mycro\second\ peaking time. Therefore, a signal generated by an out-of-time event can be identified if it has deposed energy above the trigger threshold but no trigger signal issued.

The approach in the \ac{acd} (see~\citet{acd-pass8-poster})  is similar to that used to identify ghost hits in the \ac{cal}: the pulse height measure of the tiles output has a shaping time $\sim 4~\mycro\second$, therefore a ghost hit can be identified if the reconstructed deposed energy is above the trigger threshold but a trigger signal is not issued in that tile.

The handling of ghosts in \ac{tkr} is more complex, because trigger information from single strip is not available, and the check for the presence of a trigger signal can be made only at tower level. However, the \ac{tot}, that is calculated for each half-layer (\S~\ref{totref}), can be used to check for ghosts, particularly for that generated by heavy ions, for which the time required to return over threshold can be really long ($\sim 150~\mycro\second$) and the \ac{tot} counter could therefore saturate. 

All the described methods can not identify all the hits that are generated by an overlay, but the hits marked as ghosts in the various sub-systems are then used to identify other ghosts after event reconstruction, by looking at the presence of ghost hits in a reconstructed track or in a \ac{cal} cluster and by looking at the association of \ac{acd} tiles with ghost tracks and clusters.

\clearpage

\section{Improved energy reconstruction}\label{energia-p8}
The energy reconstruction of Pass~8 has been largely rewritten, in order to increase the maximum measurable energy and the energy resolution.

Starting from the observation that in Pass~7 the shower  profile method returns the best estimation above a few GeV (see \S~\ref{energy-reconstruction}), the two energy methods, parametric and shower fit  profile, have been rewritten to improve their low and high energy performance respectively, therefore reducing the overlap in the phase space. For this reason a \ac{ct} is not necessary anymore to chose the best method: below $\sim 630~\MeV$, when the energy deposed in the \ac{tkr} becomes important, and the shower development in the \ac{cal} is short, the energy is calculated using the parametric method, that uses eq~\eqref{energia-tracker} to estimate the deposed energy in \ac{tkr} and corrects the energy deposed in \ac{cal}; above $ 3~\GeV$ the event energy  is calculated using the improved shower profile fit algorithm, described later in this section; in the energy range between 630~MeV and 3~GeV the energy is calculated as the weighted average of the result from the two methods $E=w\times E_{par}+(1-w)\times E_{prof}$, with the weight $w$ increasing linearly with the logarithm of deposed energy from 0 to 1. As in Pass~7, un-biasing is applied on both methods.

The shower profile fit method has been largely rewritten (see~\citet{energia3TeV,energia-p8-poster}) to handle events where:
\begin{itemize} \item the fraction of energy escaping from the \ac{cal} is very large
\item  the maximum of the shower is not contained in the \ac{cal}
\item there are saturated crystals.
\end{itemize}
These are all problematic whose impact increases dramatically with energy, because at 100~GeV the shower maximum is not contained in the \ac{cal} for 25\% of events with incidence angle $<25^\circ$ while at 1~TeV this happens for almost all events with angle less than $45^\circ$, and crystals saturate for an energy deposition larger than 70~GeV, which  happens very often when the event energy is larger than 1~TeV. Typically saturated crystals are in the core of the shower, where the energy deposition is larger, but the partial lack of information  can be recovered using information from near crystals. 

An extensive simulation, using GEANT4~(\citet{geant,geantbis}), has been used to characterize the development of an electro-magnetic shower in a CsI calorimeter for different energies and incidence angles. The longitudinal profile of the shower is described using the formula from~\citet{shower-in-cal}:

\begin{equation}
\frac{dE(t)}{dt}=E\times \frac{(\beta t)^{\alpha-1}\beta e^{-\beta t}}{\Gamma(\alpha)}
\end{equation}
where $t$ is the longitudinal shower depth in units of radiation length. The uncorrelated combinations 

\begin{equation}
\begin{array}{l}
S_0=\ln\alpha \cos{\vartheta_c} + \beta\sin{\vartheta_c}\\
S_1=-\ln\alpha \sin{\vartheta_c} + \beta\cos{\vartheta_c}
\end{array}
\end{equation}
of the shower parameters $\alpha , \beta$, with $\tan{\vartheta_c}=0.5$  are used to characterize the shower longitudinal development, while the traverse development was approximated as constant in $t/T$, with $T=(\alpha-1)\beta$ being the position of the shower maximum.

 This description is used to set-up a step-by-step calculation of the development of the shower in the \ac{lat} \ac{cal}, which  calculates the deposed energy for each layer of crystals, keeping into account possible gaps.  The energy is then calculated by minimizing the quantity

\begin{equation}
\begin{array}{l}
\displaystyle \chi^2(S_0,S_1,E) =\sum^{i=1\rightarrow 8}_{non-sat}\left(\frac{\varepsilon_{m,i}-\varepsilon_{p,i}(E)}{\delta \varepsilon(E)}\right)^2+\sum_{saturated}\left(\frac{\mathrm{max}(0,\varepsilon_{m,i}-\varepsilon_{p,i}(E))}{\delta \varepsilon(E)}\right)^2+ \\ \displaystyle +\sum^2_{i=1}\left(\frac{S_i-\mu_{S_i}(E)}{\sigma_{S_i}(E)}\right)^2
\end{array}
\end{equation}
where the first term sums over the layers the difference between the calculated and measured energies $\varepsilon_p$ and $\varepsilon_m$ excluding  saturated crystals from calculation, the second term runs on the saturated crystals and returns 0 if the predicted deposed energy is larger than 70~GeV (that is, if the saturation is correctly predicted), and the third term forces the calculation to converge on an energy estimation that resembles the measured shower parameters. 

\begin{figure}[htb!]
\begin{center}
\includegraphics[width=\textwidth]{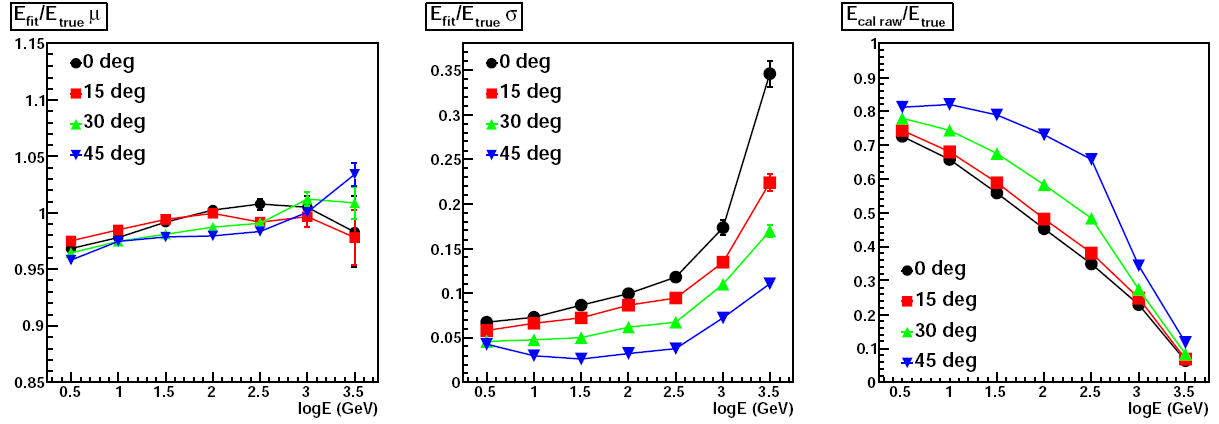}
\caption{Performance of the  Pass~8 shower profile method \cite{energia3TeV}: bias of $\frac{E_{rec}}{E_{true}}$ (left), $\sigma$ of $\frac{E_{rec}}{E_{true}}$ (center) and fraction of energy deposed in the \ac{cal} (right), where it is possible to see that for high energies the largest part of the shower escape from the  detector. }
	\label{energia-p8-philippe}

\end{center}
\end{figure}

The results of the improved profile method are shown in figure~\ref{energia-p8-philippe}, where it is possible to see that the energy can be reconstructed with reasonable precision ($\frac{\Delta E}{E}< 25\%$ for $\theta>15^\circ$) up to 3~TeV; above this energy the fraction of the total energy deposed in the \ac{cal} becomes really low, while the number of saturated crystals becomes too large to reasonably reconstruct the shower. 

\clearpage
\section{New GEANT4 simulation}\label{new-geant-sub}
The GEANT4~\cite{geant,geantbis} software used in \acp{gr} to simulate the interaction of particles with the detector has been updated to version~9.4.1 from previously used 8.01, that was patched on purpose by the \ac{lat} team; this new version was released after the start of LHC, and it uses some of the data from this experiment, introducing significant improvements in both the electro-magnetic sector, where a new multiple scattering model has been introduced, and in hadron interactions, where the simulation of hadron cascades was fully revised. These changes potentially involve almost all the phenomena that take place in the LAT.
\section{Improvement the LAT geometry and mass model}\label{new-mass}
To correctly simulate the interaction of particles with the \ac{lat}, and therefore to calculate its response, it is necessary to have a description of the instrument  as close as possible to the real detector; however, an excessively detailed description (in particular, of the passive materials) would be very difficult to made, and furthermore it would excessively  increase the calculation  power required to run the \ac{lat} simulation, so that a partial simplification in the instrument description is necessary. In Pass~8, the instrument description has been improved to reduce some significant differences that have been observed between the Monte Carlo and the \ac{lat} data/mass measurements.

It was known that some  differences were present between the masses of the model and that effectively measured, that are caused by the neglecting or mis-description of some of the \ac{tkr} components. These differences were reduced by introducing in the \ac{tkr} description components like the data cables, the glue and the tapes present in the trays, by modifying the density of materials like the silicon and by significantly modifying the description of the top and bottom trays, where some structural materials as titanium were neglected. This improvements changes the \ac{tkr} radiation length from 1.36 to 1.38, with an estimated change in the effective area of $\sim 1\%$.

In addition, the \ac{tkr} geometry was revised, with the correction of the towers alignment and of the distance between contiguous \acp{ssd}, that was over-estimated because the thickness of the glue behind the detector was under-estimated. 

\begin{figure}[htb!]
\begin{center}
\includegraphics[width=\textwidth]{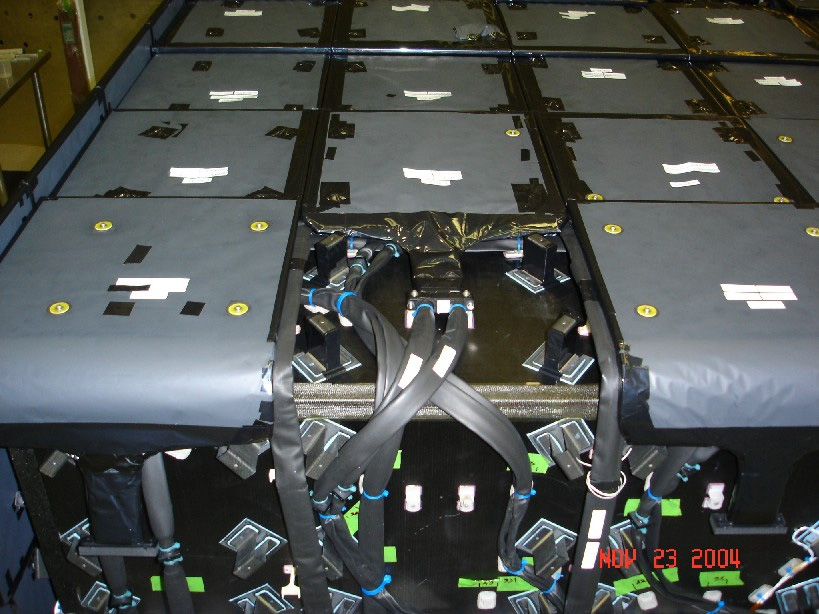}
\caption{Photo of the \ac{acd} showing the elements present below the tiles. }
	\label{cose-in-acd}

\end{center}
\end{figure}

In the \ac{lat} data, an excess with respect to the Monte Carlo was observed in the number of gamma events that started from the top and from the sides of the \ac{tkr}. This excess was caused by the presence of material between the \ac{acd} tiles and the \ac{tkr} that was not included in the simulation, where only the structural material was present (figure~\ref{cose-in-acd} shows that this is not exactly right). The \ac{acd} geometry was therefore updated to account for the presence of cables, fibers, tile wrappings, the \ac{acd} electronic at the base of the \ac{lat} (phototubes, power supply and read-out) were added to the simulation and the materials forming the structure of the \ac{acd} were better characterized. All these changes lead to and increase of the \ac{acd} thickness of 0.005 radiation lengths on the top and 0.003 on the sides.

\section{Improvements in LAT performances}\label{p7-p8}
Pass~8 introduced significant improvements to the \ac{lat} performances, that are clear when comparing the Pass~8 \acp{irf} with those of the previous Pass~7 reconstruction (for a description of the \ac{lat} \acp{irf} see \S~\ref{irf-sec}). In this section \acp{irf}  of P8R2\_V6\_SOURCE class are compared with the similar P7REP\_SOURCE\_V15, that are the last Pass~7 \acp{irf}. All figures in this section are published at~\cite{irf}.

\begin{figure}[htb!]
\begin{center}
\includegraphics[width=\textwidth]{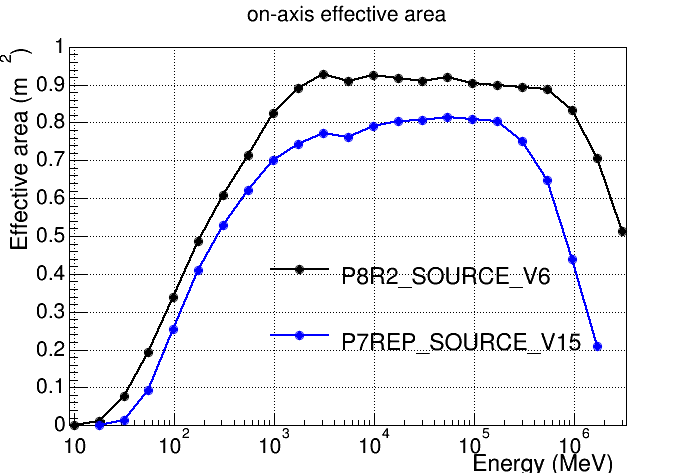}
\caption{Comparison of Pass 7 and Pass 8 source class effective area.}
	\label{aeff-p7-p8}

\end{center}
\end{figure}

Figure~\ref{aeff-p7-p8} shows the comparison on the effective area for on-axis events, where it is clear the significant improvements introduced with Pass~8 both at low and at high energy, and also the extension of the instrument energy range. 

\begin{figure}[htb!]
\begin{center}
\includegraphics[width=\textwidth]{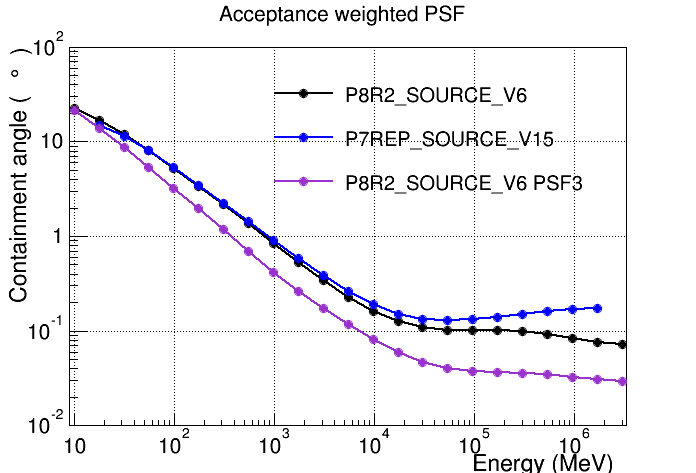}
\caption{Comparison of Pass 7 and Pass 8 68\% containment angle. }
	\label{psf-p7-p8}

\end{center}
\end{figure}

Figure~\ref{psf-p7-p8} shows the comparison of the 68\% containment angle, as a function of event energy and weighted on the \ac{lat} acceptance, where the improvement at high energy is clear. Also shown is the P8\_SOURCE\_PSF3 sub-class, that is the best quality quartile obtained when dividing reconstructed data using a direction-reconstruction quality parameter.

\begin{figure}[htb!]
\begin{center}
\includegraphics[width=\textwidth]{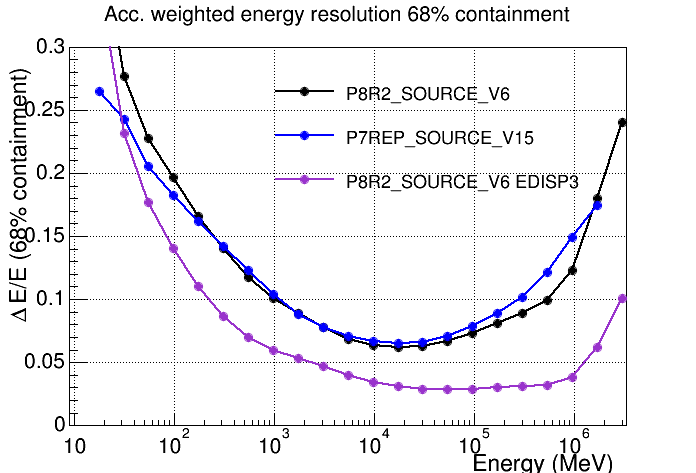}
\caption{Comparison of Pass 7 and Pass 8 energy resolution.}
	\label{edisp-p7-p8}

\end{center}
\end{figure}

Figure~\ref{edisp-p7-p8} shows the comparison of the energy resolution, defined as half width of the 68\% containment divided by the most probable energy value, as a function of event energy and weighted on the \ac{lat} acceptance. The improvements at high energy are clear. Also shown is the P8\_SOURCE\_EDISP3 sub-class, that is the best quality quartile obtained when dividing reconstructed data using an energy-reconstruction quality parameter.

\begin{figure}[htb!]
\begin{center}
\includegraphics[width=\textwidth]{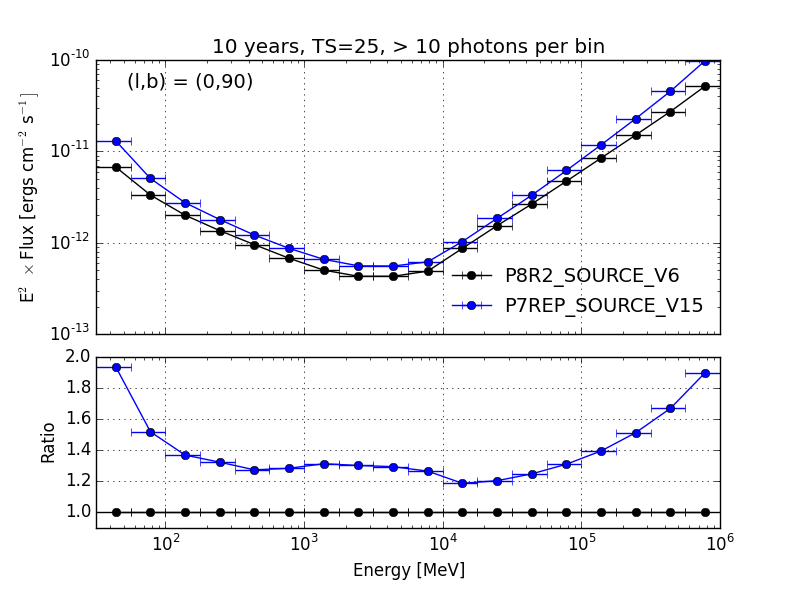}
\caption{Comparison of Pass 7 and Pass 8 differential sensitivity (top) and ratio between differential sensitivities (bottom).}
	\label{sensibility-p7-p8}

\end{center}
\end{figure}

Figure~\ref{sensibility-p7-p8} shows the comparison of the differential sensitivity for a point source at the north galactic pole, after ten years of standard survey observations. In the plot, a source with a power-law spectrum with spectral index 2 is assumed, with only diffuse background around it, without unresolved sources. Counts are grouped in 4 bins per energy decade, and a minimum of 10 photons per energy bin is required. The increase in the \ac{lat} sensitivity can be seen on the whole energy range of the instrument.
\clearpage

\chapter*{Conclusion\markboth{Conclusion}{}}
\addcontentsline{toc}{chapter}{Conclusion}

\acf{cre} knowledge saw important changes in recent years, with models where \APelectron are produced only via interactions of \acl{cr} with \acl{ism} strongly disfavored by recent measurements of the \ac{cre} spectrum and of related quantities.

However, the nature of the sources of primary \APelectron is still under debate, because experimental uncertainties and disagreement between recent experiments do not permit to discriminate between different models, that involve Dark Matter, Super Nova Remnants and Pulsar Wind Nebulae.

This review highlighted the fundamental role played by the \ac{lat} measurement in the \ac{cre} study, underlying how, with the release of Pass~8 reconstruction in 2015 and the ensuing significant improvement in \ac{lat} performance, its future role in the field will be determinant.

\section*{Acknowledgments}

The idea of this review first came during a discussion with Dott. Marco Strambi, who encouraged me during all the writing. Many thanks to my wife Giovanna and my mother Nicoletta for reading and correcting this work, and for the support they gave me. Thanks to J.S. Bach, A. Vivaldi, B. Marcello and all who composed the soundtrack for the writing. 
\markboth{List of acronyms}{List of acronyms}

\chapter*{List of acronyms}
\addcontentsline{toc}{chapter}{List of acronyms}
\begin{acronym}[acd]
\acro{acd}[ACD]{Anti-Coincidence Detector}
\acro{act}[ACT]{Air \v Cerenkov Telescope}
\acro{agile}[AGILE]{Astro-rivelatore Gamma a Immagini LEggero}
\acro{ams}[AMS-02]{Alpha Magnetic Spectrometer}
\acro{atic}[ATIC]{Advanced Thin Ionization Calorimeter}
\acro{batse}[BATSE]{Burst And Transient Source Experiment}
\acro{bspr}[BSPR]{Blind Search Pattern Reconstruction}
\acro{cal}[CAL]{Electromagnetic Calorimeter}
\acro{cart}[CART]{Classification And Regression Tree}
\acro{cmb}[CMB]{Cosmic Microwave Background}
\acro{cr}[CR]{Cosmic Ray}
\acro{cre}[CRE]{Cosmic Ray Electron}
\acro{csm}[CSM]{Circum-Stellar Matter} 
\acro{cspr}[CSPR]{Cal Seeded Pattern Reconstruction}
\acro{ct}[CT]{Classification Tree}
\acro{digi}[\textit{digi}]{digitized electronic outputs  from the \ac{lat} subsystems}
\acro{dm}[DM]{Dark Matter}
\acro{dsa}[DSA]{Diffusive Shock Acceleration} 
\acro{egret}[EGRET]{Energetic Gamma-Ray Experiment Telescope}
\acro{fov}[FoV]{Field of View}
\acro{fssc}[FSSC]{Fermi Science Support Center}
\acro{gbm}[GBM]{Gamma-ray Burst Monitor}
\acro{glast}[\hbox{GLAST}]{Gamma-ray Large Area Space Telescope}
\acro{gr}[GR]{GlastRelease}
\acro{grb}[GRB]{Gamma-Ray Burst}
\acro{ic}[IC]{Inverse Compton scattering}
\acro{irf}[IRF]{Instrument Response Function}
\acro{ism}[ISM]{Interstellar Medium}
\acro{isrf}[ISRF]{Interstellar Radiation Field} 
\acro{iss}[ISS]{International Space Station}
\acro{l1}[L1~Proc]{Level~1 processing pipeline}
\acro{lat}[LAT]{Large Area Telescope}
\acro{mcm}[MCM]{Multi Chip Module}
\acro{mcs}[MCS]{Multiple Coulomb Scattering}
\acro{merit}[\textit{merit}]{``figures of merit'' quantities}
\acro{mhw}[MHD wave]{Magneto-HydroDynamic wave}
\acro{mip}[MIP]{Minimum Ionizing Particle}
\acro{mms}[MMS]{Micro-Meteoroid Shield}
\acro{mst}[MST]{Minimum Spanning Tree}
\acro{pamela}[PAMELA]{Payload for Antimatter Matter Exploration and Light Nuclei Astro\-physics}
\acro{pda}[PDA]{Photo-Diode Assembly}
\acro{psf}[PSF]{Point Spread Function}
\acro{pwn}[PWN]{Pulsar Wind Nebula} 
\acro{recon}[\textit{recon}]{whole output of the \ac{cal}, \ac{tkr} and \ac{acd} reconstruction}
\acro{saa}[SAA]{South Atlantic Anomaly}
\acro{sn}[SN]{SuperNova}
\acro{snr}[SNR]{SuperNova Remnant}
\acro{ssd}[SSD]{Silicon micro-Strip Detector}
\acro{tem}[TEM]{Tower Electronic Module}
\acro{tkr}[TKR]{Tracker/converter}
\acro{tof}[ToF]{Time of Flight}
\acro{tot}[ToT]{Time over Threshold}
\acro{vhe}[VHE]{Very High Energy}
\acro{wls}[WLS fiber]{WaveLength Shift fiber}
\acroplural{cr}[CRs]{Cosmic Rays}
\acroplural{irf}[IRFs]{Instrumental Response Functions}
\acroplural{mhw}[MHD waves]{Magneto-HydroDynamic waves}
\acroplural{pwn}[PWNe]{Pulsar Wind Nebulae} 
\acroplural{sn}[SNe]{SuperNovae}
\acroplural{wls}[WLS fibers]{WaveLength Shift fibers}
\end{acronym}

\cleardoublepage

\phantomsection
\renewcommand{\chapter}[2]{Bibliography}
\markboth{Bibliography}{Bibliography}
\addcontentsline{toc}{chapter}{Bibliography}

\nocite{leo, knoll, pdg, pivato,vietri}
\bibliographystyle{unsrtnatMOD}

\bibliography{main-cre-review}
\end{document}